\documentclass[10pt,a4paper]{article}
\newcommand{\llangle}{\langle\!\langle}
\newcommand{\rrangle}{\rangle\!\rangle}
\newcommand{\LLangle}{\left\langle\!\!\!\left\langle}
\newcommand{\RRangle}{\right\rangle\!\!\!\right\rangle}
\begin{document}
\title{\bf Dynamical Reduction Models}

\author{Angelo Bassi\footnote{e-mail: bassi@ictp.trieste.it}\\
{\small The Abdus Salam International Centre for
Theoretical Physics, Trieste, and}\\{\small Istituto Nazionale di Fisica
Nucleare, Sezione di Trieste, Italy.}\\ and\\
\\ GianCarlo Ghirardi\footnote{e-mail: ghirardi@ts.infn.it}\\
{\small Department of Theoretical Physics of the University of
Trieste, and}\\ {\small the Abdus Salam International Centre for
Theoretical Physics, Trieste, and}\\{\small Istituto Nazionale di
Fisica Nucleare, Sezione di Trieste, Italy.}}

\date{}

\maketitle

\begin{abstract}

The report presents an exhaustive review of the recent attempt to
overcome the difficulties that standard quantum mechanics meets in
accounting for the measurement (or macro--objectification)
problem, an attempt based on the consideration of nonlinear and
stochastic modifications of the Schr\"odinger equation. The
proposed new dynamics is characterized by the feature of not
contradicting any known fact about microsystems and of accounting,
on the basis of a unique, universal dynamical principle, for
wavepacket reduction and for the classical behavior of macroscopic
systems. We recall the motivations for the new approach and we
briefly review the other proposals to circumvent the above
mentioned difficulties which appeared in the literature. In this
way we make clear the conceptual and historical context
characterizing the new approach. After having reviewed the
mathematical techniques (stochastic differential calculus) which
are essential for the rigorous and precise formulation of the new
dynamics, we discuss in great detail its implications and we
stress its relevant conceptual achievements. The new proposal
requires also to work out an appropriate interpretation; a
procedure which leads us to a reconsideration of many important
issues about the conceptual status of  theories based on a
genuinely Hilbert space description of natural processes.
Attention is also paid to many problems which are naturally raised
by the dynamical reduction program. In particular we discuss the
possibility and the problems one meets in trying to develop an
analogous formalism for the relativistic case.  Finally we discuss
the experimental implications of the new dynamics for various
physical processes which should allow, in principle, to test it
against quantum mechanics. The review covers the work which has
been done in the last fifteen years by various scientists and the
lively debate which has accompanied the elaboration of the new
proposal.
\end{abstract}
\makeatletter \@addtoreset{equation}{section} \makeatother
\renewcommand{\theequation}{\thesection.\arabic{equation}}
\tableofcontents
\newpage

\section{Introduction} \label{intro}

The twentieth century has seen the birth of what are unanimously
considered the two basic pillars of modern science: relativity
theory and quantum mechanics. Both these theoretical constructions
have met an unprecedented predictive success in accounting for the
results of  the incredibly refined experiments which have been
made possible by recent technological improvements. Both schemes
imply radical changes concerning the (classical) views about
natural phenomena.

Relativity theory has required a drastic modification of our views
concerning space and time, quantum mechanics has compelled the
scientific community to attribute a prominent role to chance in
physics and to accept the existence of unavoidable limitations to
the attainable knowledge about physical systems. However, the
conceptual status of the two theories is remarkably different. If
one accepts that instantaneous communication is impossible and
that the velocity of light represents an upper limit to the
propagation of any physical action, one must reconsider the
problem of synchronizing clocks and is led to the conclusion that
the space-time continuum is the correct framework for the
description of natural processes. The ensuing theory, special
relativity, is an example of a precisely formulated and internally
consistent theory, one which, to use A. Shimony's words
\cite{shipcqf}, allows to {\it close the circle} and to base on it
a coherent worldview.

The situation is quite different with the other pillar of modern
science, quantum mechanics, as is evident if one takes into
account the lively debate about its interpretation which started
soon after its formulation and which is still going on. This
debate concerns one of the most peculiar aspects of the theory,
generally known as {\it the measurement problem}, even though a
more appropriate term to characterize it would be {\it the
macro--objectification problem}. It stems directly from the linear
nature of the theory and the way in which it connects the
mathematical entities which are claimed to represent the most
accurate specification which is in principle possible of the state
of a system and the outcomes of prospective measurement processes.
>From this point of view, quantum mechanics is an extremely
successful and powerful mathematical device yielding the
probabilities of the  results of any conceivable measurement
procedure. But, in contrast with this unprecedented efficiency in
telling us everything about {\it what we find} the theory is
silent, to use J.S. Bell's words \cite{bellam}, about {\it what it
is}. Actually, just due to its linear nature, the quantum
description of the measurement process and of all those
measurement--like processes \cite{bellam} {\it we are obliged to
admit ... are going on more or less all the time, more or less
everywhere} and lead to the definite perceptions which
characterize our experience, contradicts the idea that  all
natural processes and, in particular, the micro--macro
interactions occurring in the situations we have just mentioned
are governed by quantum mechanics itself. In brief, the theory
contains two incompatible dynamical principles, the linear
Schr\"odinger evolution and the wavepacket reduction  process
which is associated to micro--macro interactions.

This serious limitation would not represent by itself a deadlock
for the theory: one could simply accept that it has only a limited
field of validity. However, if one takes such a position one must
pretend that the theory itself allows the identification of a
phenomenological area in which the transition from micro to macro,
from reversible to irreversible, from deterministic to stochastic,
in short, from quantum to classical, takes place. But this is not
the case. The borderline between these two different regimes is by
no means precisely identifiable. There are for sure many
macro--systems which require a fully quantum treatment.

A significant  indication of the peculiar situation we have just
outlined comes from reconsidering the historical debate about the
meaning and the interpretation of the formalism. Such a debate has
seen the successive identification of different levels which,
according to the various thinkers, should mark the place at which
one has to pose the split, which is then characterized by a
certain shiftiness.  For instance, in the famous Bohr--Einstein
debate, Bohr, who had repeatedly claimed that the split should be
associated to the micro or macroscopic nature of the physical
systems under consideration, was compelled, in order to reject the
pressing criticisms by Einstein,  to accept that macroscopic parts
of the apparatus (such as macro shutters and macro pointers)
require a fully quantum treatment. An analogous situation occurred
subsequently: London, Bauer and Wigner were led to identify the
split with the borderline separating physical from conscious
processes. But, once more, our present knowledge does not give a
clear indication about what is conscious, so that a remarkable
vagueness characterizes also such a proposal.

The recent years have seen a noticeable and renewed interest about
the macro--objectification problem. We can quite confidently state
that, nowadays, there is a large consensus among scientists
interested in the foundational aspects of the theory that the so
called orthodox interpretation (a rather ambiguous expression
which however encompasses similar positions concerning the
measurement problem) has completely failed in yielding a
consistent and coherent account of natural phenomena. In
connection with this renewed interest in the field, new and
original attempts to overcome the difficulties have appeared. What
one wants is a quantum mechanical model of measurements as
dynamical processes governed by precise rules agreeing, on the one
side, with the quantum description of microscopic systems, and, on
the other, with the classical aspects characterizing the
macroscopic world.

The present report is devoted to a detailed presentation of recent
proposals (the dynamical reduction theories) aimed at overcoming
the difficulties of the macro--objectification process, proposals
which stick to the idea that the knowledge of the statevector
represents the maximal information one can have about the state of
a physical system. Accordingly, these attempts differ radically
from all those, like hidden variable theories, which invoke the
incompleteness of the Hilbert space description to solve the
measurement problem. Within the dynamical reduction framework, the
basic idea to get the desired result consists in accepting that
the linear and deterministic Schr\"odinger equation has to be
modified by the addition of non linear and stochastic terms. As we
will see, this program, which Einstein himself \footnote{We recall that
Einstein, in his {\it Reply to critics} \cite{eini}, has explicitly
considered the possibility of giving up, for micro systems, the request
that they possess objectively all properties, i.e., he was prepared to
accept the linear nature of their state space. But he has stressed that he
could not renounce to his realistic requirements at the macro
level, so that macro objects cannot be in superpositions of
macroscopically different states. In accordance with this position
he has contemplated the idea of abandoning the superposition
principle at the macro level. His concluding remarks are of great
relevance for the dynamical reduction program: {\it the
macroscopic and the microscopic are so inter--related that it
appears impracticable to give up this program} [scientific realism
in the classical sense, as requiring that all systems possess
objective properties] {\it in the microscopic alone}. The
dynamical reduction program proves that the line that Einstein
considered impracticable is actually consistently viable. More
recently analogous remarks --- which occurred repeatedly during
all the history of quantum mechanics --- have been put forward
once more by Shimony. In contemplating the possibility of
reduction at an appropriate level, he has stated that \cite{shimi}
{\it reasonable desiderata} [for the dynamical reduction program]
{\it pull in opposite directions}.}  considered unviable, can be
consistently followed leading to a fully satisfactory model at the non
relativistic level.

Obviously, we do not think that the  dynamical reduction theories
we are going to discuss in this report might represent instances
of the final theory of natural processes. However, the very fact
that they can be consistently developed throws a new light on the
subject, allows us to identify some basic features of any dynamics
inducing reductions, makes precise and allows us to better
understand the action at a distance of the standard theory and
might also suggest  interesting experiments aimed to identify
possible violations of the superposition principle.

The paper is organized as follows. First of all we discuss in all
details the macro--objectification problem to make clear how it
gives rise to a physically unacceptable situation which must be
faced (part I). We also briefly review many other proposals to
overcome the difficulties of the formalism, to make clear  the
differences of the dynamical reduction point of view with respect
to  other attempts to avoid the inconsistencies of the orthodox
position. We then come to the core of the report by reviewing the
development of the dynamical reduction program and by discussing
all the most relevant mathematical features and physical
implications of such an approach (part II). We pass to analyze the debate
about the interpretation of  dynamical reduction models (part III).
This analysis will allow the reader to understand and to fully
appreciate the crucial innovative implications of the approach. Part IV
of the report is devoted to an important and to a large extent still open
problem, i.e. the one of finding a fully satisfactory relativistic
model inducing reductions.  We
will conclude our analysis with a short discussion of some
experimental situations which will make clear, on the one side,
how it is difficult to devise {\it experimenta crucis} allowing to
discriminate such models from quantum mechanics, and, on the
other, will give some indications about the experiments which seem
most appropriate to identify violations (if they are there) of the
linear nature of quantum theory (part V).

\part{The Measurement Problem of Quantum Mechanics}

\section{Basic principles of Quantum Mechanics} \label{sec1}

We review here the general mathematical structure of Quantum
Mechanics, with special attention to the postulate of wavepacket
reduction, the one which gives rise to the measurement problem. We
also introduce formal tools like the statistical operator
formalism, calling attention to some subtle features which can
give rise to misunderstandings concerning some fundamental
questions.

\subsection{The axioms of Quantum Mechanics} \label{sec11}

Standard Quantum Mechanics can be synthetically summarized by the
following set of rules:
\begin{enumerate}
\item Every physical system $S$ is associated to an Hilbert space
${\mathcal H}$; the physical states of $S$ are represented by
normalized vectors (called ``statevectors'') $|\psi\rangle$ of
${\mathcal H}$. Physical observables\footnote{In the following, we
will denote with $O$ both the observable and the corresponding
operator in ${\mathcal H}$; when confusion arises, we will specify
whether $O$ refers to the observable or to the operator.} $O$ of
the system are represented by self--adjoint operators in
${\mathcal H}$: the possible outcomes of a measurement of $O$ are
given by the eigenvalues $o_{n}$ of the corresponding operator,
which we assume here, for simplicity, to have a discrete and non
degenerate spectrum:
\begin{equation}
O\,|o_{n}\rangle\quad =\quad o_{n}\,|o_{n}\rangle.
\end{equation}
Since $O$ is a self--adjoint operator, its eigenvalues $o_{n}$ are
real and the eigenvectors $|o_{n}\rangle$ form a complete
orthonormal set in the Hilbert space ${\mathcal H}$.
\item To determine the state $|\psi(t_{0})\rangle$ of
the system $S$ at a given initial time $t_{0}$, a complete set of
commuting observables for $S$ is measured: the initial statevector
is then the unique common eigenstate of such observables. Its
subsequent time evolution is governed by the Schr\"odinger
equation:
\begin{equation} \label{sce}
i\hbar\,\frac{d}{dt}|\psi(t)\rangle \quad = \quad H\,
|\psi(t)\rangle,
\end{equation}
which uniquely determines the state at any time once one knows it
at the initial one. The operator $H$ is the Hamiltonian of the
system $S$.
\item The probability of getting, in a measurement at time $t$,
the eigenvalue $o_{n}$ in a measurement of the observable $O$ is
given by:
\begin{equation}
P[o_{n}] \quad = \quad |\langle o_{n} |\psi(t)\rangle |^{2},
\end{equation}
$|\psi(t)\rangle$ being the state of the system at the time in
which the measurement is performed.
\item The effect of a measurement on the system $S$ is to
drastically change its statevector from $|\psi(t)\rangle$ to
$|o_{n}\rangle$, $o_{n}$ being the eigenvalue obtained in the
measurement:
\[
|\psi(t)\rangle \;\;\makebox{before measurement} \qquad
\longrightarrow \qquad |o_{n}\rangle \;\;\makebox{after
measurement}.
\]
This is the famous {\bf postulate of wavepacket reduction (WPR)}.
\end{enumerate}
These, in short, are the postulates of Quantum Mechanics. Of
course, we have somewhat simplified the exposition; for example we
have ignored the possibility of the operator $O$ having a
continuous spectrum besides or in place of the discrete one; we
have not discussed the case of degenerate eigenvectors, and so on.
All such features, even though important, are not crucial for
understanding the measurement problem.

\subsection{Schr\"odinger evolution and wavepacket reduction}
\label{sec12}

The Schr\"odinger equation (\ref{sce}) has two basic properties.
First, it is a first order differential equation in the time
variable; this means that once the initial state of the system
$|\psi(t_{0})\rangle$ is known, its future evolution is completely
determined. The evolution of the statevector of any physical
system is thus perfectly {\bf deterministic}, like in classical
mechanics.

The solution of equation (\ref{sce}) can be written as follows:
\begin{equation}
|\psi(t)\rangle \quad = \quad U(t, t_{0})\, |\psi(t_{0})\rangle,
\end{equation}
where $U(t, t_{0})$ is the evolution operator. This is a unitary
operator, a necessary requirement in order that it preserves the
norm of the statevector and, accordingly, to make tenable the
probabilistic interpretation of the theory. Its formal expression
is:
\begin{equation}
U(t, t_{0}) \quad = \quad e^{\displaystyle -i\, H\,
(t-t_{0})/\hbar}.
\end{equation}

The second important feature of the Schr\"odinger equation is that
it is {\bf linear}: if $|\psi_{1}(t)\rangle$ and
$|\psi_{2}(t)\rangle$ are two possible solution of (\ref{sce}),
then also $\alpha\, |\psi_{1}(t)\rangle \, + \, \beta\,
|\psi_{2}(t)\rangle$ is a possible solution, where $\alpha$ and
$\beta$ are two arbitrary complex numbers\footnote{If the original
vectors are normalized and the two states $|\psi_{1}(t)\rangle$
and $|\psi_{2}(t)\rangle$ are orthogonal, their linear combination
is normalized when $|\alpha|^{2} + |\beta|^{2} = 1$.}. This is the
mathematical formulation of the celebrated\footnote{To be precise,
the superposition principle includes also the assumption that all
states of the Hilbert space can actually occur and thus in
particular that if $|\psi_{1}\rangle$ and $|\psi_{2}\rangle$ are
possible states for $S$, then also $\alpha\, |\psi_{1}\rangle \, +
\, \beta\, |\psi_{2}\rangle$ is a possible state.} {\bf
superposition principle}.

The postulate of  wavepacket reduction exhibits features which are at
odds with Schr\"odinger's evolution. First of all it describes a
{\bf nonlinear} evolution of the statevector, since it transforms
the state $\alpha\, |\psi_{1}(t)\rangle \, + \, \beta\,
|\psi_{2}(t)\rangle$ into either $|\psi_{1}(t)\rangle$ or
$|\psi_{2}(t)\rangle$ with probabilities $|\alpha|^{2}$ and
$|\beta|^{2}$ respectively. The second important property of
wavepacket reduction is its genuinely {\bf probabilistic} nature:
in general we cannot know to which one of the eigenstates of $O$
the statevector of $S$ will be reduced as a consequence of a
measurement process; the theory determines only the {\it
probability} of the reduction to any particular eigenstate. This
is where probability and indeterminacy enter into play, making
Quantum Mechanics so different from classical theories.

Summing up, we have: \vspace{0.5cm}

\begin{center}
\begin{tabular}{|c|c|} \hline
& \\
Schr\"odinger evolution & Wavepacket reduction \\
& \\ \hline
& \\
linear & nonlinear \\
& \\
deterministic & stochastic \\
& \\ \hline
\end{tabular}
\end{center}
\vspace{0.5cm}

\noindent The above table shows clearly that standard Quantum
Mechanics has some peculiar features: it contains {\bf two
dynamical evolution principles}, one governed by the Schr\"odinger
equation and the other taking place when wavepacket reduction
occurs. They are radically different, and they contradict each
other. This fact gives rise to the measurement problem of the
theory, which represents the starting point which has led to the
elaboration of the theories which are the subject of the present
report.

\subsection{The statistical operator} \label{sec13}

The quantum mechanical rules sketched in section \ref{sec11} refer
only to {\bf pure states}, i.e. to physical systems whose
statevector is perfectly known. As already remarked, the state of
a system can be determined exactly only by measuring a complete
set of commuting observables.

In practice  it is not always possible to perform such kinds of
measurements; the real experimental situation could not even
require it. It may very well happen that we have (or need) only a
partial knowledge of the state of the system, pretty much like in
classical statistical mechanics: the statistical operator
formalism has been designed to deal with these situations.

Suppose we know that the statevector describing an individual
system is one among a set $\{|\psi_{i}\rangle\}$ of vectors, but
we do not know which one it actually is: we only know the
probability $p_{i}$ that $|\psi_{i}\rangle$ is the correct
statevector. Equivalently, we can suppose that we have an ensemble
of $N$ systems, a fraction $N_{1}$ of which is described by the
vector $|\psi_{1}\rangle$, a fraction $N_{2}$ by the vector
$|\psi_{2}\rangle$ and so on. Taking a system out of the ensemble,
the probability that $|\psi_{i}\rangle$ is its associated
statevector is $p_{i} = N_{i}/N$. We call such ensembles {\bf
statistical mixtures}; they are characterized by the vectors
$\{|\psi_{i}\rangle\}$ together with their probability
distribution $p_{i}$.

The {\bf statistical operator} is then defined as follows:
\begin{equation}
\rho \quad = \quad \sum_{i} p_{i}\, |\psi_{i}\rangle
\langle\psi_{i}|.
\end{equation}
The operator $\rho$ is a trace class, trace one semi--positive
definite operator, and replaces completely the statevector when
only a partial knowledge of the state of the system is available.
Of course, for a pure state $|\psi\rangle$, i.e. for one
representing the most accurate knowledge that the theory considers
as possible about a system, the statistical operator reduces to
the projection operator $|\psi\rangle\langle\psi|$.

A nice feature of statistical operators is that they allow us to
use a compact formalism to deal both with pure states and with
genuine statistical mixtures and, at the same time, they allow us
to distinguish between them. In fact the following property holds:
\[
\begin{array}{ll}
\rho^{2} \; = \; \rho \qquad & \qquad \makebox{for pure states} \\
\rho^{2} \; \neq \; \rho \qquad & \qquad \makebox{for statistical
mixtures.}
\end{array}
\]

The quantum axioms 2--4 are expressed  as follows in this new
language. The evolution equation for $\rho$ is:
\begin{equation}
i\hbar\,\frac{d}{dt} \rho(t) \quad = \quad \left[ H, \rho(t)
\right],
\end{equation}
which is again a linear first order differential equation to be
solved taking into account the initial condition $\rho(t_{0})$.
The probability that the outcome of a measurement of an observable
$O$ is one, $o_{n}$, of its eigenvalues is given
by\footnote{``$\makebox{Tr}$'' denotes the trace of the
operator.}:
\begin{equation}
P[o_{n}] \quad = \quad \makebox{Tr}[P_{n}\,\rho(t)],
\end{equation}
$P_{n}$ being the operator which projects onto the linear manifold
associated to the eigenvalue $o_{n}$. Finally, at the end of a
measurement process giving $o_{n}$ as its outcome, the statistical
operator changes in the following way:
\[
\rho \;\;\makebox{before measurement} \qquad \longrightarrow
\qquad \frac{P_{n}\,\rho\,P_{n}}{\makebox{Tr}[P_{n}\,\rho\,P_{n}]}
\;\;\makebox{after measurement}.
\]
This is of course the appropriate expression for the wavepacket
reduction postulate.

A final remark. The statistical operator formalism allows us to
handle also {\it non selective measurements}, which we are now
going to define, in a quite natural way. Suppose we perform a
measurement on an ensemble of systems in the same state (pure or
mixed); in general each of them will give different outcomes and,
after the measurement, due to wavepacket reduction, they will be
described by different statevectors. If we decide to keep all the
systems, independently of the outcomes we have obtained, we
perform a {\it non selective} measurement: we do not have a pure
state, even when the state is pure before the measurement is
performed. The measurement process turns it into the following
statistical mixture:
\[
\rho \;\;\makebox{before measurement} \qquad \longrightarrow
\qquad \sum_{n} P_{n}\,\rho\,P_{n} \;\;\makebox{after
measurement},
\]
$P_{n}$ being the projection operators associated to the
eigenmanifolds of the observable which has been measured. Since
the measurement is non selective, the operators $P_{n}$ sum up to
the identity: it is easy to check that all the mathematical
properties of $\rho$ are preserved.

\subsection{Property attribution in standard quantum mechanics}
\label{sec14}

As already remarked, the characteristic trait of standard Quantum
Mechanics is that, in general, it allows only probabilistic
predictions about the possible outcomes of measurements. Such
probabilities have a truly {\bf nonepistemic} character, i.e. they
are not due to our ignorance about the precise state of the
system, like in classical statistical mechanics: rather, quantum
theory is such that physical systems by themselves do not possess
all properties one can think of. Another way to put it ---
following the Copenhagen doctrine --- is to say that, in general,
we are not allowed to speak of the properties of a system
concerning most of its observables: the best we can do is to speak
of what we can {\it observe} about the system.

However, it is possible to rescue, still remaining within the
standard formalism, a sort of ``minimal ontology'' for physical
systems. This was first formulated  by Einstein, Podolski and
Rosen \cite{epr}:
\begin{quotation}
If, without in any way disturbing the system, we can predict with
certainty (i.e. with probability equal to unity) the value of a
physical quantity, then there exists an element of physical
reality corresponding to this physical quantity.
\end{quotation}
Following axiom 3, an eigenvalue $o_{n}$ of an observable $O$ has
probability $1$ of being found in a measurement of $O$ if and only
if the statevector $|\psi(t)\rangle$ of the system is an
eigenstate of $O$ pertaining to the considered eigenvalue. We can
then say that {\bf a system possesses a property if and only if
its statevector belongs to the eigenmanifold associated to the
eigenvalue corresponding to that property}. Any measurement aimed
to test this statement will give a positive result. Of course, we
can develop a similar argument when the probability of an outcome
is equal to zero and we can claim that the corresponding property is
certainly not possessed by the system.

It is not difficult to understand that within this ``minimal
ontology'' physical systems do not possess all the properties one
would be inclined to attach to them (like position and momentum at
any given time), because a vector cannot be a simultaneous
eigenvector of too many operators, due to the non--abelian
character of the algebra of the operators. Worse than that,
individual systems in entangled states possess in general no
properties at all \cite{gm}.

In particular, one can almost never attach definite
macro--properties, specifically precise locations in space, even
to macroscopic systems, so that they cannot be considered as being
in a definite region of space. This fact gives rise to the so
called {\bf macro--objectification problem}, i.e. to the necessity
of accounting for the emergence of the properties corresponding to
our definite perceptions for such systems. This crucial point,
another aspect of the quantum measurement problem, will be
extensively analyzed in the following section.

When dealing with statistical mixtures, property attribution is
rather delicate since there is an interplay between {\it
epistemic} (classical--like) and {\it nonepistemic}
(quantum--like) probabilities. To be precise, let us consider an
arbitrary mixture of states $\{|\psi_{i}\rangle\}$, with
probabilities $p_{i}$; in analogy with classical statistical
mechanics, the state of any system of the ensemble is described by
a precise vector of the set $\{|\psi_{i}\rangle\}$, but we are
ignorant about which is the correct one: we only know the
probability characterizing each of them. Accordingly, the
probability distribution $p_{i}$ has an epistemic character. On
the other hand, each such vector has a probabilistic physical
content which, as before, is genuinely nonepistemic.

\section{The quantum measurement, or
macro--objecti\-fi\-ca\-ti\-on, problem} \label{sec2}

This section is devoted to a general and detailed discussion of
the measurement, or macro--objectification, problem of Quantum
Mechanics. The nature of the problem has already been anticipated:
the linear nature of Quantum Mechanics allows the occurrence of
superpositions of macroscopically different states of a
macro--object, e.g. concerning their location, in spite of the
fact that macroscopic systems are always located, or at least we
perceive them as being located, in a well defined region of space.

We will first analyze the measurement problem within the framework
of the von Neumann scheme for an ideal measurement process
(subsection \ref{sec21}): this is a very simple and elegant
measurement model which goes directly to the root of the problem.
Nontheless, von Neumann's argument has been repeatedly criticized
for the over--simplified assumptions on which it is based. In
subsection \ref{sec22}, we will show that even by adopting a very
general and realistic measurement model, one can derive the same
conclusions reached by von Neumann: superpositions of different
macroscopic states cannot be avoided within the quantum framework.

\subsection{The von Neumann measurement scheme} \label{sec21}

The first explicit example of the quantum description of a
measurement process was presented by John von Neumann  \cite{vn}
and is usually referred to as the ``ideal'' measurement scheme. It
gained great popularity since, due to its simplicity, it allows us
to grasp immediately the key points of the problem; nowadays
almost all textbooks on the foundations of Quantum Mechanics make
reference to it. The von Neumann argument goes as follows.

Let us consider a microscopic system $S$ and one of its
observables $O$. Let $o_{n}$ be the eigenvalues of $O$ (we assume,
for simplicity, that its spectrum is purely discrete and
non--degenerate) and $|o_{n}\rangle$ the corresponding
eigenvectors. Let us call $M$ the apparatus devised to measure the
observable $O$ of the microsystem $S$.  $M$ has a ready--state
$|A_{0}\rangle$, i.e. a state in which the apparatus is ready to
measure the considered property, plus a set of {\it mutually
orthogonal} states  $|A_{n}\rangle$ corresponding to {\it
different macroscopic configurations} of the instrument, like,
e.g., different positions of a pointer along a scale.

Finally, we assume that the interaction between the microsystem
$S$ and the apparatus $M$ is {\it linear} (since the Schr\"odinger
equation is supposed to govern all natural processes) and that it
yields a {\it perfect  correlation} between the initial state of
$S$ and the final state of  the apparatus, i.e.
\begin{equation} \label{eq0}
\makebox{Initial state:
$\;\;|o_{n}\rangle\otimes|A_{0}\rangle$}\qquad
\longrightarrow\qquad \makebox{Final state:
$\;\;|o_{n}\rangle\otimes|A_{n}\rangle$};
\end{equation}
in this way we are sure that if the final state of the apparatus
is $|A_{n}\rangle$ (i.e. the pointer for example is in the $n$--th
position along the scale), the final state of the particle is
$|o_{n}\rangle$ and the observable $O$ has the value $o_{n}$, in
accordance with the property attribution discussed in section
\ref{sec14}.

The measurement problem arises when the initial state of the
particle, previous to the measurement, is not just one of the
eigenvectors $|o_{n}\rangle$ like in equation (\ref{eq0}), but a
superposition of them, for example:
\[
|m + l\rangle\quad =\quad \frac{1}{\sqrt{2}}[ |o_{m}\rangle +
|o_{l}\rangle],
\]
which can be very easily prepared in our laboratories. In this
case, if the {\it linear} evolution equation of the theory is
assumed to govern all physical processes, the final state of the
microsystem+apparatus will be:
\begin{eqnarray} \label{eq01}
|m + l\rangle\otimes|A_{0}\rangle & = & \frac{1}{\sqrt{2}}\;[
|o_{m}\rangle + |o_{l}\rangle]\otimes|A_{0}\rangle \longrightarrow
\nonumber \\
& \longrightarrow & \frac{1}{\sqrt{2}}\,[|o_{m}\rangle\otimes
|A_{m}\rangle + |o_{l}\rangle\otimes|A_{l}\rangle ].
\end{eqnarray}
Such a state is an entangled state of the microscopic system and
of the apparatus, which is not an eigenstate of the relevant
observable  $M$ of the apparatus, i.e. the position of the pointer.
In situations like this, as already discussed, it is not
legitimate, {\it even in principle}, to state that the properties
associated to the states $|A_{m}\rangle$  or $|A_{l}\rangle$ are
possessed by the apparatus: as a consequence the apparatus is not
in any of the macroscopic definite configurations we perceive it
to be. This is the first part of the quantum measurement problem.

The {\it standard} way out from this difficulty is given by the
{\it wavepacket reduction postulate} (axiom 4 listed in section
\ref{sec11}), which states that ``at the end of the {\it
measurement} process'' the final vector in equation (\ref{eq01})
reduces to one of its terms:
\[
|o_{m}\rangle\otimes|A_{m}\rangle \qquad \makebox{or} \qquad
|o_{l}\rangle\otimes|A_{l}\rangle,
\]
with a probability given by the square modulus of the coefficient
associated to that term ($1/2$ for both cases, in our example).

We have already mentioned, and we have proved now, that the
postulate of wavepacket reduction contradicts the assumption of
the general validity of the Schr\"odinger equation; this of course
is a very unsatisfactory feature of standard Quantum Mechanics: it
incorporates two contradictory dynamical evolutions, something we
cannot accept for a physical theory. Moreover, the real physical
difficulty is not only the one of the consistency of the
Schr\"odinger evolution and wavepacket reduction\footnote{Such a
difficulty could be circumvented by assuming that the theory has
only a limited field of validity and, in particular, it does not
apply to macro--systems.}; the even more serious problem is that
the theory does not tell us in which precise cases the linear
hamiltonian evolution has to be suspended and wavepacket reduction
takes place. As we will see, dynamical reduction models offer a
natural, precise and unambiguous solution to both problematic
aspects of the measurement problem.

Coming back to the standard theory, we mention that, in spite of
the above difficulties, various authors \cite{pri,zur1,zu,lib}
have maintained that the measurement problem does not derive from
the structure of quantum mechanics (in  particular from the linear
character of the quantum evolution), or from the postulate of
wavepacket reduction, but from adopting the over--simplified model
of measurement processes put forward by von Neumann. If one takes
into account more realistic models, they argue, the measurement
problem turns into a false one: the postulate of wavepacket
reduction is not anymore necessary and, consequently, there is no
need to modify the interpretation of the theory, or even to put
forward a new theory.

In particular, the following assumptions have been criticized:
\begin{itemize}
\item That the measuring apparatus can be prepared in a
precise state $|M_{0}\rangle$: since the instrument is a
macroscopic object with many degrees of freedom, it is impossible
to know its precise state at any given time.

\item That one can safely neglect the interactions between the
apparatus and the surrounding environment. Such interactions with
the environment (which are referred to as decoherence) produce
essentially a randomization of the phases associated to the
different components of the wavefunction, a process which can be
seen as an {\it apparent} collapse of the  wavefunction into one
of its components.

\item  That the final states of the apparatus, corresponding to
perceptively different macroscopic configurations of the apparatus
itself, are orthogonal: actually, different states usually
correspond to different {\it positions} of some part of the
instrument, and since no wavefunction can have compact support in
configuration space (because of the quantum evolution),
wavefunctions corresponding to different states cannot, in
general, be orthogonal.

\item That the final state of the apparatus gets perfectly
correlated to the initial state of the microscopic system: this is
an highly idealized characteristic which is not shared by any
realistic physical instrument.

\end{itemize}

In the next subsection we will consider a very general measurement
scheme \cite{bg1} which takes into account all the above
criticisms. We will show that superpositions of states
corresponding to different macroscopic configurations of
macro--objects cannot be avoided within a strict quantum
mechanical context. Correspondingly, the appearance of macroscopic
situations which are incompatible with our definite perceptions
about such objects is inescapable. This ``empasse" can only be
circumvented either by adopting a precise and unambiguous
interpretation which differs from the  orthodox one, or by
modifying the theory itself\footnote{An explicit  proof that
releasing the request of an ideal measurement does not allow us to
circumvent the measurement problem  can be found in the well known
book by d'Espagnat \cite{de}; however, his proof is much more
complex and much  less general than the one we are going to
present here.}.

\subsection{A completely general measurement scheme} \label{sec22}

In this subsection we re--derive von Neumann's conclusions on the
basis of what, in our opinion, is the most general possible
description of a measurement instrument and of a measurement
process. We begin first by defining the microscopic system whose
properties we want to measure.

\subsubsection{The Microscopic System} \label{sec221}

For simplicity we consider the simplest system upon which
non--trivial measurements can be performed, i.e., a system $S$
whose associated Hilbert space ${\mathcal H}_{S}$ is
two--dimensional  --- like the one describing the spin of an
electron, or the polarization states of a photon --- and we
consider  an observable $O$  having two different eigenvalues; let
us call $|\makebox{u}\rangle$ and $|\makebox{d}\rangle$ the
eigenstates associated to these eigenvalues. For definiteness, we
will consider an individual such system and we will call ``spin''
its degree of freedom;  we will say that the particle has ``spin
Up'' when it is in state $|\makebox{u}\rangle$, and that it has
``spin Down'' when it is in state $|\makebox{d}\rangle$. Besides
these two states, also their superpositions can be taken into
account, in particular the following one:
\[
| \makebox{u + d}\rangle \quad =\quad \frac{1}{\sqrt{2}}\,\,[\,
|\makebox{u}\rangle\, +\, |\makebox{d}\rangle\,],
\]
a vector describing a new state, ``spin Up + spin Down'', of the
particle. Without any loss of generality, we will assume that, by
resorting to appropriate procedures, one can ``prepare'' the
system $S$ in any one of the three above considered states
$|\makebox{u}\rangle$, $|\makebox{d}\rangle$ and $|\makebox{u +
d}\rangle$.

We remark that we could have considered more general physical
systems, like compound ones, and observables having a more
complicated spectrum. However, in accordance with the generally
accepted position  that microsystems can be prepared in  a precise
quantum state and with the nowadays common experimental practice
to  handle single particles and to measure their discrete
properties, we have chosen to work with a very simple microsystem
like the one we are considering here.

Accordingly, after the preparation, the system is in a precise and
known state, and it can be treated as isolated from the rest of
the universe, at least until  the measurement process
begins\footnote{In mathematical terms,  we assume that, {\it prior
to the measurement process},  the wavefunction of the universe
factorizes into the wavefunction of the particle times the
wavefunction of the rest of the world.}. We stress that if one
denies these assumptions it is not even clear what he takes
quantum theory to be about.

\subsubsection{The Measuring Apparatus} \label{sec222}

A measuring apparatus is a {\it macroscopic} system which,
interacting with the microsystem whose properties one is
interested in ascertaining, ends up into a state more or less
strictly correlated with the eigenstates of the observable it is
devised to measure. The different possible outcomes of the
measurement are supposed to be associated to {\it perceptively
different macroscopic configurations} of a part of the apparatus,
e.g. different positions of the pointer (for analogic
instruments), different numbers on a display (for digital ones),
different spots on a photographic plate, different plots on a
screen, and so on. For simplicity, in what follows we will assume
that the apparatus has a pointer  movable along a scale, whose
position registers the result of the measurement.

Contrary to microsystems, the measuring apparatus, being a
macroscopic object, has many degrees of freedom, most of which ---
in particular the microscopic ones --- we cannot control at all;
and of the macroscopic ones, like the position of the pointer, we
can have only a very limited control. Moreover, due to its
dimensions, the apparatus is always interacting with the
environment, whose degrees of freedom are also essentially out of
control. Following this line of reasoning, one can remark that the
apparatus --- or at least its constituents --- existed quite a
long time before the measurement is performed, so it had all the
time to interact, even if only weakly, with a large part of the
universe. All these interactions make to a large extent unknown
and uncontrollable the state of the macroscopic system which
enters into play. In spite of this difficulty, in order to keep
our analysis as general as possible, we will take into account all
the above mentioned facts which make the measurement non--ideal.

With reference to the above discussion, we should in general speak
of different situations of the ``whole universe'', even though our
``reading'' refers only to the degrees of freedom of the  pointer;
accordingly, we will indicate the statevectors we will deal with
in the following way:
\[
        |A\; \alpha\rangle.
\]
These vectors belong to the Hilbert space  associated to the
apparatus,  the environment, and, in the most general case, to the
whole universe. $A$ is a label indicating that the pointer of the
apparatus is in a specific macroscopic configuration, i.e. one
which we perceive and we identify with a  specific position along
the scale. In first  approximation, we could say that $A$ is
essentially the value $x$  characterizing the ``projection
operator'' $|x\rangle\langle x|$ ($|x\rangle$ being an
``improper'' statevector of the Hilbert space of the pointer)
giving the exact  position of e.g. the centre of mass of the
pointer along the scale. But it is evident that no system  can be
prepared in such a state since it is impossible to measure a
continuous variable with perfect accuracy; and even if it were
possible to do so, the hamiltonian evolution would immediately
change that state.

We could try to improve our description by taking into account, in
place of precise positions along the scale, small intervals
$\Delta(x) = [ x - \delta, x + \delta ]$, and to claim that ``the
pointer is at position $x$'' when the wavefunction is an
eigenstate of the projection operator onto the
interval\footnote{Of course, here we are considering for
simplicity a one--dimensional situation; the argument can be
easily generalized to the three--dimensional case.} $\Delta(x)$ of
the center  of mass position. If one makes such a choice, the
label $A$ characterizing our general state $|A\; \alpha\rangle$
refers to any wavefunction having  such a property, of course with
the interval $\Delta(x)$ replaced by the interval $\Delta(A)$: as
a consequence, for the considered state we could claim that ``the
pointer is at position A''. However, also this approach is not
viable since the hamiltonian evolution transforms any wavefunction
with compact support into a wavefunction with a non--compact one;
this fact gives rise to what has been called the ``tail problem'',
a problem which cannot be avoided, and which  renders rather
delicate the task of making precise the idea of ``an object being
somewhere'' within a quantum mechanical framework. More about this
in what follows.

Following the above analysis, we consider a very general physical
situation: we call $V_{A}$ the set of all (normalized) vectors $
|A\; \alpha\rangle$ for which we are {\it allowed} to say that
``the pointer of the apparatus is at position $A$'' or, stated
differently, that ``the universe is in a configuration which we
perceive as one corresponding to the statement: the pointer is at
$A$''. We do not put any restriction to the vectors belonging to
$V_{A}$: they can represent wavefunctions with or without tails,
more or less localized in space, and so on; we do not even resort
to projection operators to characterize these states. The only
physical requirement we put forward is that, if the pointer admits
two macroscopically and perceptively different ``positions'' along
the scale (let us call them $A$ and $B$), then any two vectors
corresponding to such different configurations must be ``almost
orthogonal''. This requirement can be translated into the
following mathematical relation: denoting by $V_{B}$ the set of
all normalized vectors corresponding to the statement ``the
pointer is at $B$'' while $V_{A}$, as before, contains all the
vectors corresponding to the statement ``the pointer is at $A$'',
we must have:
\begin{equation} \label{req}
\inf_{\begin{array}{l}
\makebox{\footnotesize $|A\;\alpha\rangle \,\in\, V_{A}$} \\
\makebox{\footnotesize $|B\;\beta\rangle \,\in\, V_{B}$}
\end{array}}\, \| |A\;\alpha\rangle -
|B\;\beta\rangle \|\, \geq\, \sqrt{2} - \eta \qquad\qquad \eta \ll
1,
\end{equation}
i.e. the minimum distance between the vectors of the two above
sets cannot differ too much from $\sqrt{2}$, which is the distance
between two orthogonal normalized states. We recall that the
orthogonality request of the standard measurement theory is done
to be sure to be dealing with strictly mutually exclusive
situations. Obviously such a request can be partially released (as
we are doing here) but not given up completely if one wants to be
able to ``read'' the outcome in a fundamentally non ambiguous way.
It is evident that  (\ref{req}) is a necessary requirement if one
pretends that {\it different} macroscopic positions of the pointer
(and of any other system) represent {\it practically mutually exclusive}
configurations of the object\footnote{Obviously, here we are
making reference to a genuinely quantum description (with the
completeness assumption). In alternative interpretations or
formulations of the theory, orthogonality is not necessary  to
guarantee macroscopic differences. Typically, in hidden variables
theories one could have non orthogonal wavefunctions and different
values for the hidden variables such that the associated physical
situations are macroscopically different and mutually exclusive.}
(see also the final remark of this subsection).

Let us now comment on the second parameter $\alpha$ characterizing
our states: this is an index which takes into account all other
degrees of freedom that are out of control\footnote{From the
mathematical point of view, $\alpha$ stands for the eigenvalues of
a  complete set of commuting observables for the whole universe,
exception made for the ``location" of the pointer.}; thus two vectors
labeled by $A$ but with different values for $\alpha$, refer to
the ``same'' macroscopic configuration for the pointer (or, in
general, of the ``part of the universe we perceive"), while they
describe two different states for the rest of the universe (e.g.,
given a certain atom of the pointer, it might be in the ground
state when the state is $|A\; \alpha\rangle$, while it might be in
an excited state when it is $|A\;\beta\rangle$).

Since we are interested in the two spin states of the microscopic
particle, if we want to use the apparatus to distinguish them we
have to assume that the pointer admits at least two
macroscopically different positions ($U$ and $D$) along the
scale\footnote{The idea is that, if we perform the measurement and
we find the pointer in the position labeled by $U$, then we can
claim the ``the result of the measurement is that the spin of the
particle is Up''; similarly, if we find the pointer in $D$, then
we can say that ``the spin of the particle is Down''.}. The
previous argument requires then that there exist two sets $V_{U}$
and $V_{D}$, the first one containing all the vectors
corresponding to the situation in which the pointer can be said to
point at ``U'', the second all those vectors associated to the
statement ``the pointer is at D''. Moreover, these two sets must
be almost orthogonal in the sense of (\ref{req}):
\begin{equation} \label{req2}
\inf_{\begin{array}{l}
\makebox{\footnotesize $|U\;\alpha\rangle \,\in\, V_{U}$} \\
\makebox{\footnotesize $|D\;\beta\rangle \,\in\, V_{D}$}
\end{array}}\, \| |U\;\alpha\rangle -
|D\;\beta\rangle \|\, \geq\, \sqrt{2} - \eta \qquad\qquad \eta \ll
1,
\end{equation}
One interesting property of $V_{U}$ and $V_{D}$ (which is shared
by any pair  of sets for which (\ref{req2}) is satisfied) is that
they have no vectors in common: in fact, it is easy to see that if
$V_{U}$ and $V_{D}$ had such a common vector, then the left hand
side of (\ref{req2}) would take the value zero, a fact which would
contradict (\ref{req2}). From the physical point of view, this
property is obvious since a vector belonging both to $V_{U}$ and
to $V_{D}$ would be a vector for which we could claim both that
``the pointer points at U'' and that ``the pointer points at D'',
a contradictory situation since ``U'' and ``D'' correspond to
macroscopically and perceptively different situations.

\subsubsection{The Preparation of the Apparatus} \label{sec223}

A measuring instrument must be prepared before one performs a
measurement, i.e. one has to arrange the apparatus in such a way
that it is ready to interact with the microscopic system and give
a result; following the discussion of the previous subsection, it
is evident that the initial statevector must carry an index
$\alpha$ which takes into account the state of the rest of the
universe. Accordingly, we will denote the initial statevector as
$|A_{0}\;\alpha\rangle$, where $A_{0}$ indicates that the pointer
``is'' in the ready ($A_{0}$) state.

However, we note that, besides the measuring instrument, we have
also to prepare the microsystem in a precise state, and moreover
we have assumed that after the preparation and immediately before
the measurement process, the microsystem itself is isolated from
the rest of the universe; the initial statevector for the whole
universe can then be written as:
\[
|A_{0}\;\alpha\rangle \quad =\quad
|\makebox{spin}\rangle\otimes|A_{0}\;\overline{\alpha}\rangle,
\]
where $\overline{\alpha}$ specifies the state of the whole
universe, with the exception of the initial state of the
micro--particle and the initial ``position" of the pointer;
$|\makebox{spin}\rangle$ is the initial statevector of the
particle.

Obviously, also in the process of preparing the apparatus we
cannot control the state of the universe, so that we do not know
the precise initial state $|A_{0}\;\overline{\alpha}\rangle$: in
practice, in any specific situation any value for the index
$\overline{\alpha}$ will occur with a given probability
$p(\overline{\alpha})$, which in general is unknown to us --- but
of course it has to satisfy appropriate requirements  we will
discuss in what follows. Accordingly, the initial setup for the
apparatus and the microscopic particle will be described as
follows:
\[ \makebox{Initial Setup} \quad =\quad \left\{\,
|\makebox{spin}\rangle\otimes|A_{0}\;\overline{\alpha}\rangle,
\;\;\; p(\overline{\alpha}) \right\},
\]
where $p(\overline{\alpha})$ gives the probability distribution of
the remaining, uncontrollable, degrees of freedom.

\subsubsection{The Measurement Process} \label{sec224}

If one assumes that Quantum Mechanics governs all physical
systems, the measurement process, being an interaction between two
quantum systems, is governed by a unitary operator $U(t_{I},
t_{F})$. Suppose the initial state of the microsystem is
$|\makebox{u}\rangle$ and the one of the apparatus (plus the rest
of the universe) is $|A_{0}\; \overline{\alpha}\rangle$; then,
during the measurement, the whole universe evolves in the
following way:
\begin{equation} \label{evu}
|\makebox{u}\rangle\otimes|\makebox{$A_{0}$}\; \overline{\alpha}
\rangle \quad \longrightarrow \quad U(t_{F}, t_{I})\left[\,
|\makebox{u}\rangle\otimes| \makebox{$A_{0}$}\; \overline{\alpha}
\rangle \right]\quad =\quad |\makebox{F u}\; \overline{\alpha}
\rangle,
\end{equation}
while, if the initial state of the microsystem is
$|\makebox{d}\rangle$,  one has:
\begin{equation} \label{evd}
|\makebox{d}\rangle\otimes|\makebox{$A_{0}$}\; \overline{\alpha}
\rangle \quad \longrightarrow \quad U(t_{F}, t_{I})\left[\,
|\makebox{d}\rangle\otimes| \makebox{$A_{0}$}\; \overline{\alpha}
\rangle \right]\quad =\quad |\makebox{F d}\; \overline{\alpha}
\rangle.
\end{equation}
Some comments are needed.
\begin{itemize}
\item Note that in the above equations (\ref{evu}) and (\ref{evd})
the index $\overline{\alpha}$ distinguishes various possible and
uncontrollable situations of the measuring apparatus in its
``ready'' state. Once the initial state is fully specified also
the final one, since the evolution is unitary, is perfectly and
unambiguously determined. Accordingly, such a state is
appropriately characterized by the same index $\overline{\alpha}$.
Note also that, while the state $|\makebox{$A_{0}$}\;
\overline{\alpha} \rangle$ belongs to the Hilbert space of the
whole universe exception made for the micro--particle, the state
$|\makebox{F d}\; \overline{\alpha}\rangle$ now includes also the
particle.

\item Contrary to what one does in the ideal measurement
scheme of von Neumann, we do not assume that  the final state is
factorized; thus, in general
\[ |\makebox{F u}\; \overline{\alpha} \rangle \quad \neq \quad
|\makebox{u}\rangle \otimes |A_{U}\; \overline{\alpha}\rangle.
\]
\item In particular, we do not suppose that the final state of the
microsystem be the same as the initial one: we allow the
measurement process to modify in a significant way the state of
the particle; it could even destroy the particle.
\end{itemize}

The only thing we require is that {\bf the measuring apparatus is
reliable to a high degree}, i.e. that it can safely be used to
measure the state of the microsystem since in most cases it gives
the correct answer. This means that if the initial state of the
microsystem (prior to the measurement) is $|\makebox{u}\rangle$,
then the final state $|\makebox{F u}\; \overline{\alpha}\rangle$
must belong {\it in most of the cases} to $V_{U}$, while, if the
initial state of the particle is $|\makebox{d}\rangle$, the final
state $|\makebox{F d}\; \overline{\alpha}\rangle$ must {\it almost
always} belong to $V_{D}$. Note that by not requiring full
reliability, we take into account also the possibility that the
measuring  instrument  gives the wrong results, though pretending
that such mistakes occur quite seldom.

It is possible to formalize the above reliability requests in the
following way. Let us consider the set $K$ of all subsets $J$ of
the possible values that the index $\overline{\alpha}$ can assume
and let us equip it with the following (natural) measure:
\[
\mu(J)\quad =\quad \sum_{\overline{\alpha}\in J}
p(\overline{\alpha}).
\]
Let us also define the two following sets:
\[
\begin{array}{lcl}
J^{-}_{U} & = & \left\{\overline{\alpha}\;\; \makebox{such that:}
\;\; |\makebox{F u}\; \overline{\alpha}\rangle \not\in V_{U}
\right\},
\end{array}
\]
\[
\begin{array}{lcl}
J^{-}_{D} & = & \left\{\overline{\alpha}\;\; \makebox{such that:}
\;\; |\makebox{F d}\;\overline{\alpha}\rangle \not\in V_{D}
\right\}.
\end{array}
\]
$J^{-}_{U}$ is the sets of all the indices $\overline{\alpha}$
such that the states $|\makebox{F u}\;\overline{\alpha}\rangle$ do
not correspond to the outcome {\it ``the pointer is at position
U''}, despite the fact that prior to the measurement the state of
the particle was $|\makebox{u} \rangle$. Similarly, $J^{-}_{D}$
corresponds to the states $|\makebox{F d}\; \overline{\alpha}
\rangle$ for which we cannot claim that {\it ``the pointer points
at D''}, even if the initial state of the system was
$|\makebox{d}\rangle$. Let also $J^{+}_{U} = {\mathcal
C}J^{-}_{U}$ be the complement of $J^{-}_{U}$, and $J^{+}_{D} =
{\mathcal C} J^{-}_{D}$ the complement of $J^{-}_{D}$.

Given this, the requirement that the instrument is reliable can be
mathematically expressed in the following way:
\begin{equation} \label{req3}
\mu(J^{-}_{U})\;\; \leq\;\; \epsilon \qquad\quad
\mu(J^{-}_{D})\;\; \leq\;\; \epsilon, \qquad\quad \epsilon \ll 1,
\end{equation}
i.e. the probability that the final position of the pointer does
not match the initial spin--value of the particle is very small,
this smallness being controlled by an appropriate parameter
$\epsilon$ expressing the efficiency of the measuring device and
which, as such, can change (always remaining very small) with the
different actual measurement procedures one can devise.

Of course, it is easy to derive also limits on the measure for the
complements of the above sets:
\[
\mu(J^{+}_{U}) \;\; \geq \;\; 1 - \epsilon \qquad\qquad
\mu(J^{+}_{D}) \;\; \geq \;\; 1 - \epsilon.
\]
We need to take into account also the two sets: $J^{-} = J^{-}_{U}
\cup J^{-}_{D}$ and $J^{+} = {\mathcal C}J^{-} = J^{+}_{U} \cap
J^{+}_{D}$; they satisfy the following relations:
\[
\mu(J^{-}) \;\; \leq \;\; 2\epsilon \qquad\qquad \mu(J^{+}) \;\;
\geq \;\; 1 - 2\epsilon.
\]
Again, all these limits simply state that, since the apparatus is
reliable, the probability that --- at the end of the measurement
process --- the pointer is not in the correct position is very
small, {\it if} the initial state of the particle is either
$|\makebox{u}\rangle$ or $|\makebox{d}\rangle$.

It is useful to remark that, having taken into account the
possibility that the measuring instrument can make mistakes, we
can easily include also the possibility that it fails to interact
at all with the microsystem, thus giving no result: in such a
case, the pointer remains in the ``ready--state'', and the
corresponding vector belongs to the set $J^{-}$. In fact, let us
consider the set $V_{0}$ associated to the ``ready--state'', as we
did for the two  sets $V_{U}$ and $V_{D}$ referring to the ``U''
and ``D''  positions of the pointer. By the same argument as
before,  $V_{0}$ is disjoint from the two sets $V_{U}$ and
$V_{D}$, since the ``ready--state'' is assumed to be
macroscopically different from the ``U'' and ``D'' states;
consequently if the vector at the end of the measuring process
belongs to $V_{0}$, it cannot belong either to $V_{U}$ or to
$V_{D}$.

We have mentioned the possibility that the apparatus misses to
detect the particle because such an occurrence affects, in some
case in an appreciable way, many experimental situations; for
example the efficiency of photodetectors is usually quite low.
This does not pose any problem to our treatment: we can easily
circumvent this difficulty by simply disregarding, just as it is
common practice in actual experiments, all cases in which a
detector should register something but it doesn't. The previous
analysis and the sets we have identified by precise mathematical
criteria must then be read as referring exclusively to the cases
in which the apparatus registers an outcome.

\subsubsection{The Measurement Problem} \label{sec225}

We recall the two basic assumptions we discussed in the previous
subsections:
\begin{enumerate}
\item The quantum evolution of any physical system is {\it linear}
and unitary, since it is governed by the Schr\"odinger equation;
\item Any two sets, like $V_{U}$ and $V_{D}$, containing vectors
corresponding to {\it macroscopically different configuration} of
a macro--object are almost orthogonal:
\begin{equation} \label{req4}
\inf_{\begin{array}{l}
\makebox{\footnotesize $|U\;\alpha\rangle \,\in\, V_{U}$} \\
\makebox{\footnotesize $|D\;\beta\rangle \,\in\, V_{D}$}
\end{array}}\, \| |U\;\alpha\rangle -
|D\;\beta\rangle \|\, \geq\, \sqrt{2} - \eta \qquad\qquad \eta \ll
1.
\end{equation}
\end{enumerate}
We think that everybody would agree that {\it any real}
measurement situation, if it has to be described in  quantum
mechanical terms, shares the above two properties. Starting with
these very simple premises we can now easily  show that quantum
mechanics must face the problem of the occurrence of
superpositions of macroscopically different states of the
apparatus, and in general of a macro--system\footnote{As already
remarked, request 2) can be violated in hidden variables theories.
On the other hand, request 1) is purposedly violated in dynamical
reduction theories. Since both theories account for the
objectification of macroscopic properties, they must necessarily
violate one of the two requests.}.

In our terms, the ``measurement problem'' arises (as usual) when
the initial spin--state of the particle is not
$|\makebox{u}\rangle$ or $|\makebox{d}\rangle$, as we have
considered in the previous subsections, but a superposition of
them, like the state $|\makebox{u + d}\rangle$ of section
\ref{sec221}, which can be easily prepared in the laboratory. In
such a case, due to the linearity of the evolution, the final
state of the particle+apparatus system will be:
\[
\begin{array}{lcl}
|\makebox{u +
d}\rangle\otimes|\makebox{$A_{0}$}\;\overline{\alpha} \rangle
\quad & \longrightarrow & \quad U(t_{F}, t_{I})\left[\,
|\makebox{u + d}\rangle\otimes|
\makebox{$A_{0}$}\;\overline{\alpha} \rangle \right]\quad = \quad
|\makebox{F u+d}\;\overline{\alpha}
\rangle\\
& & \quad= {\displaystyle \frac{1}{\sqrt{2}}\left[ |\makebox{F
u}\;\overline{\alpha}\rangle + |\makebox{F
d}\;\overline{\alpha}\rangle \right]}.
\end{array}
\]
It is now very simple to prove that for each $\overline{\alpha}$
belonging to $J^{+}$, $|\makebox{F u+d}\;\overline{\alpha}\rangle$
{\it cannot} belong either to $V_{U}$ or to $V_{D}$. In fact, let
us suppose that it belongs to $V_{U}$; the proof in the case in
which it is assumed to belong to $V_{D}$ is analogous. Since the
distance between $|\makebox{F u+d}\;\overline{\alpha}\rangle$  and
$|\makebox{F d}\;\overline{\alpha}\rangle$ is:
\begin{eqnarray*}
\| |\makebox{F u+d}\;\overline{\alpha}\rangle - |\makebox{F
d}\;\overline{\alpha}\rangle \| & = & \| 1/\sqrt{2}\, |\makebox{F
u}\;\overline{\alpha}\rangle + \left( 1/\sqrt{2} - 1\right) \,
|\makebox{F d}\;\overline{\alpha}\rangle \| \quad \leq \quad \\
& \leq & \frac{1}{\sqrt{2}} + 1 - \frac{1}{\sqrt{2}} \quad = \quad
1,
\end{eqnarray*}
we get a contradiction, because $|\makebox{F u+d}\;
\overline{\alpha}\rangle$ is supposed to belong to $V_{U}$ while
$|\makebox{F d}\;\overline{\alpha}\rangle$ belongs to $V_{D}$, and
relation (\ref{req4}) must hold between any two vectors of these
two sets. This completes our proof. Of course, by the same
argument we can also prove that, for all
$\overline{\alpha}\,\in\,J^{+}$, the index of the apparatus cannot
be in any other macroscopic position different from ``U'' and
``D''.

The conclusion is: for all $\overline{\alpha}\,\in\,J^{+}$ and for
all measurements processes in which the apparatus registers an
outcome, the vector $|\makebox{F u+d}\;\overline{\alpha}\rangle$
{\bf does not allow us to assign  any macroscopic definite
position to the index  of the apparatus}, not even  one different
from ``U'' or ``D''. Stated differently, the large majority of the
initial apparatus states, when they are triggered by the
superposition $|\makebox{u + d}\rangle$, end up in a state which
does not correspond to any definite position or, in our general
language, to any definite situation of the part of the universe we
perceive, i.e. one paralleling our definite perceptions.

We believe that the above argument represents the most general
proof of the unavoidability of  the macro--objectification problem
for the absolutely minimal and physically necessary requests on
which it is based: that  one can prepare microscopic systems in
well defined states which are eigenstates of a quantum observable
and that when this is done and the considered observable is
measured, one can  get reliable information about the eigenvalue
of the observable itself, by appropriate amplification procedures
leading to perceivably different macroscopic situations of the
universe.

In the next section we will analyze the various proposals which
have been put forward to overcome the macro--objectification
problem; we will briefly describe them and discuss their pros and
cons.

\section{Possible Ways Out of the Macro--Objecti\-fi\-ca\-ti\-on
Problem} \label{sec3}

 Various ways to overcome the measurement problem have
been considered in the literature: in this section we briefly
describe and discuss them. It is useful to arrange the various
proposals in a hierarchical tree--like structure \cite{ghirc},
taking into account the fundamental points on which they differ:
in the figure below we present a diagram which may help in
following the argument. Subsequently we will comment on the
various options.
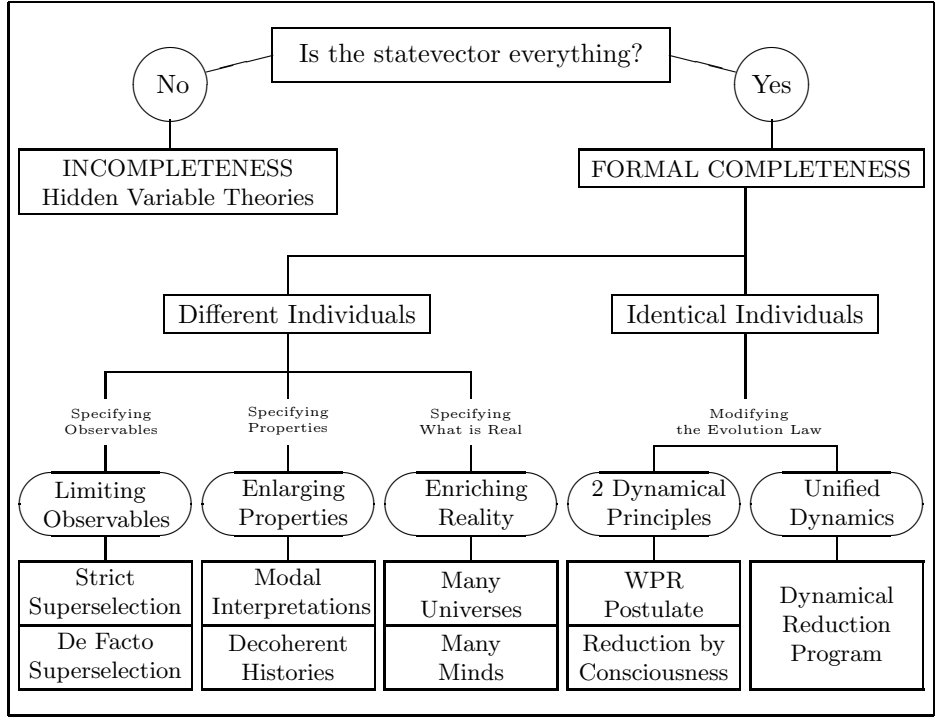
\begin{center}
\begin{picture}(350,270)(0,0)
\put(0,0){\line(1,0){350}} \put(0,270){\line(1,0){350}}
\put(0,0){\line(0,1){270}} \put(350,0){\line(0,1){270}}
\put(100,240){\framebox(150,20){Is the statevector everything?}}
\put(100,250){\line(-4,-1){25}} \put(250,250){\line(4,-1){25}}
\put(61.24,239){\circle{27}} \put(47.24,229){\makebox(30,20){No}}
\put(288.72,239){\circle{27}}
\put(273.72,229){\makebox(30,20){Yes}}
\put(61.24,225.5){\line(0,-1){11}}
\put(288.72,225.5){\line(0,-1){11}}
\put(4.16,190){\framebox(120,24){\parbox{4in}{\small \centering
INCOMPLETENESS \\ Hidden Variable Theories}}}
\put(216,200){\framebox(130,14){\small \centering FORMAL
COMPLETENESS}}
\put(105.82,159){\line(0,1){15}}
\put(105.82,174){\line(1,0){172.9}}
\put(278.72,159){\line(0,1){41}}
\put(59.16,145){\framebox(100,14){\centering Different
Individuals}} \put(228.64,145){\framebox(100,14){\centering
Identical Individuals}}
\put(10,109){\parbox{0.8in}{\tiny\centering Specifying \\
Observables}}
\put(77,109){\parbox{0.8in}{\tiny\centering Specifying \\
Properties}}
\put(146,109){\parbox{0.8in}{\tiny\centering Specifying \\
What is Real}}
\put(244,109){\parbox{1in}{\tiny\centering Modifying \\
the Evolution Law}}
\put(36.66,130){\line(1,0){138.32}} \put(36.66,92){\line(0,1){10}}
\put(36.66,120){\line(0,1){10}} \put(105.82,92){\line(0,1){10}}
\put(105.82,120){\line(0,1){25}} \put(174.98,92){\line(0,1){10}}
\put(174.98,120){\line(0,1){10}} \put(244.14,92){\line(0,1){10}}
\put(313.3,92){\line(0,1){10}} \put(278.72,120){\line(0,1){25}}
\put(244.14,102){\line(1,0){69.16}}
\put(36.66,58){\line(0,1){10}} \put(105.82,58){\line(0,1){10}}
\put(174.98,58){\line(0,1){10}} \put(244.14,58){\line(0,1){10}}
\put(313.3,58){\line(0,1){10}}
\put(36.66,80){\oval(65,24)}
\put(8.66,68){\makebox(52,24){\parbox{0.6in}{\small \centering
Limiting\\ Observables}}} \put(105.82,80){\oval(65,24)}
\put(81.82,68){\makebox(52,24){\parbox{0.6in}{\small \centering
Enlarging\\ Properties}}} \put(174.98,80){\oval(65,24)}
\put(150.98,68){\makebox(52,24){\parbox{0.6in}{\small \centering
Enriching \\ Reality}}} \put(244.14,80){\oval(65,24)}
\put(220.14,68){\makebox(52,24){\parbox{0.9in}{\small \centering 2
Dynamical \\ Principles}}} \put(313.3,80){\oval(65,24)}
\put(289.3,68){\makebox(52,24){\parbox{0.6in}{\small \centering
Unified \\ Dynamics}}}
\put(4.16,34){\framebox(65,24){\parbox{0.8in}{\small \centering
Strict \\ Superselection}}}
\put(4.16,10){\framebox(65,24){\parbox{0.8in}{\small \centering De
Facto \\ Superselection}}}
\put(73.32,34){\framebox(65,24){\parbox{0.8in}{\small \centering
Modal \\ Interpretations}}}
\put(73.32,10){\framebox(65,24){\parbox{0.8in}{\small \centering
Decoherent \\ Histories}}}
\put(142.48,34){\framebox(65,24){\parbox{0.8in}{\small \centering
Many \\ Universes}}}
\put(142.48,10){\framebox(65,24){\parbox{0.8in}{\small \centering
Many \\ Minds}}}
\put(211.64,34){\framebox(65,24){\parbox{0.8in}{\small \centering
WPR \\ Postulate}}}
\put(211.64,10){\framebox(65,24){\parbox{0.8in}{\small \centering
Reduction by \\ Consciousness}}}
\put(280.8,10){\framebox(65,48){\parbox{0.8in}{\small \centering
Dynamical \\ Reduction \\ Program}}}
\end{picture}
\end{center}

\subsection{Listing the possible ways out} \label{sec31}

A first distinction among the alternatives which have been
considered in the literature derives from taking into account the
role which they assign to the statevector $|\psi\rangle$ of a
system. This leads to the Incompleteness versus Formal
Completeness option:
\begin{description}
\item[Incompleteness:] this approach rests on the assertion  that
the specification of the state $|\psi\rangle$ of the system is
insufficient: further parameters, besides the wavefunction, must
be considered, allowing us to assign definite properties to
physical systems
\item[Formal Completeness:] it is assumed that the assignment of
the statevector represents the most accurate possible
specification of the state of a physical system.
\end{description}
When the assumption of Formal Completeness is made, two
fundamentally different positions can be taken about the status of
an ensemble --- a pure case in the standard scheme --- all
individuals of which are described by the same wavefunction:
\begin{description}
\item[Formal Completeness with Different Individuals:] the same
wavefunction de\-scri\-bes individuals which can have different
properties, even though there is no further element in the formal
theory that specifies such properties.
\item[Formal Completeness with Identical Individuals:] all
individuals associated to the same statevector have the same
properties. Pure cases correspond to genuinely homogeneous
ensembles.
\end{description}
The two options we have just mentioned require different
strategies to circumvent the difficulties related to the
objectification problem. The first case can be analyzed in greater
detail by considering the three following alternatives:
\begin{description}
\item[Limiting the Observables:] The specification of what is
actually observable in the case of a macro--system has to be
reconsidered: by taking into account appropriate and unavoidable
limitations of the class of observables one can legitimately
consider the macro--properties to be actual.
\item[Enlarging the Criteria for the Attribution of Properties:] The
possibility of considering a property actual is related in a more
subtle way than in standard Quantum Mechanics to the statevector;
in particular, an individual system can possess a property even
though it is not in an eigenstate of the corresponding observable.
\item[Enriching Reality:] Many real happenings can occur together;
all potentialities of the statevector become actual.
\end{description}
When the option of Formal Completeness with Identical Individuals
is chosen the strategy to circumvent the difficulties consists in
reconsidering the dynamics of the theory. One contemplates the
possibility of a Modified Dynamics: the unitary evolution law of
the theory is not always or not exactly right; the modifications
which have to be taken into account make the potentialities
actual. This case too leads to further alternatives:
\begin{description}
\item[Two Dynamical Principles:] different physical situations
require different evolution laws.
\item[Unified Dynamics:] the evolution equation of Quantum
Mechanics has to be modified. The new dynamical principle does not
lead to a violation of tested quantum predictions for microsystems
but it is able to induce the dynamical objectification of
macro--properties.
\end{description}

\subsection{Incompleteness: the specification of the state is
insufficient} \label{sec32}

This option corresponds to challenging the completeness of the
quantum description of physical systems: the statevector is not
all. To complete the theory, new ``hidden variables" besides the
statevector $|\psi\rangle$ are introduced: these are putative
parameters related to properties of a physical system which are
not specified by the statevector. The intended aim is that of
making legitimate an epistemic interpretation of quantum
probabilities.

The best known example of a hidden variable theory is Bohmian
Mechanics \cite{b1,bh,hol,dgz1,dgz2,dgz3}, where the new variables
are the positions ${\bf x}_{i}$ of the particles. The basic rules
are:
\begin{enumerate}
\item The state of a physical system $S$ at an initial time $t_{0}$
is given by the wavefunction $\psi({\bf q}_{1}, {\bf q}_{2},
\ldots {\bf q}_{n}; t_{0})$ together with the positions ${\bf
x}_{1}(t_{0})$, ${\bf x}_{2}(t_{0})$, $\ldots {\bf x}_{n}(t_{0})$
of all the particles of $S$.
\item The wavefunction evolves according to the Schr\"odinger
equation:
\[
i\hbar\, \frac{\partial\,\psi({\bf q}_{1}, {\bf q}_{2}, \ldots
{\bf q}_{n}; t)}{\partial t} \quad = \quad H\, \psi({\bf q}_{1},
{\bf q}_{2}, \ldots {\bf q}_{n}; t),
\]
while the equations of motion for the positions ${\bf x}_{i}(t)$
of the particles are:
\[
\frac{d {\bf x}_{i}(t)}{dt} \quad = \quad \frac{\hbar}{m_{i}}\,
\makebox{Im}\, \left.\frac{\psi^{*}({\bf q}_{1}, {\bf q}_{2},
\ldots {\bf q}_{n}; t)\, {\bf \nabla}_{i}\, \psi({\bf q}_{1}, {\bf
q}_{2}, \ldots {\bf q}_{n}; t)}{|\psi({\bf q}_{1}, {\bf q}_{2},
\ldots {\bf q}_{n}; t)|^{2}}\right|_{{\bf q}_{i} = {\bf x}_{i}}
\]
\item The Schr\"odinger equation can be solved with the given
initial condition. Once the solution has been found, it is used to
solve the equations of motion for the ``hidden'' variables ${\bf
x}_{i}(t)$.
\end{enumerate}

Bohmian Mechanics has two basic features. First of all, the theory
assigns always a definite position in space to all particles; in
particular, macroscopic objects have definite properties, and they
are where we see them to be: this is how Bohmain mechanics solves
the measurement problem of quantum mechanics. The second basic
feature is the following: let us consider an ensemble of physical
systems described by the same wavefunction $\psi({\bf q}_{1}, {\bf
q}_{2}, \ldots {\bf q}_{n}; t)$, each containing $n$ particles
whose positions are ${\bf x}_{1}, {\bf x}_{2}, \ldots {\bf
x}_{n}$. Let us also suppose that the probability distribution
$\rho({\bf x}_{1}, {\bf x}_{2}, \ldots {\bf x}_{n}; t_{0})$ of the
positions of the particle in the ensemble, at a given initial time
$t_{0}$ is:
\[
\rho({\bf x}_{1}, {\bf x}_{2}, \ldots {\bf x}_{n}; t_{0}) \quad =
\quad |\psi({\bf x}_{1}, {\bf x}_{2}, \ldots {\bf x}_{n};
t_{0})|^{2}.
\]
It follows that the trajectories followed by the particles of the
systems in the ensemble are such that, at any later time $t$:
\[
\rho({\bf x}_{1}, {\bf x}_{2}, \ldots {\bf x}_{n}; t) \quad =
\quad |\psi({\bf x}_{1}, {\bf x}_{2}, \ldots {\bf x}_{n}; t)|^{2}.
\]
This means that the theory is predictively equivalent to standard
Quantum Mechanics concerning the positions of all the particles of
the universe.

However, one has to call attention to a peculiar aspect (shared by
all hidden variable theories) of Bohmian Mechanics, i.e. to its
{\bf contextual} nature.  Various authors \cite{gl,ks} have
exhibited general proofs showing that the very algebraic structure
of quantum formalism implies\footnote{For a Hilbert space of more
than two dimensions.} that any complete specification of the state
of a system can assign, in general, a definite truth value to most
of the propositions concerning its properties only with reference
to a specified context. This means that within such a framework,
the most complete specification of the state of an individual
physical system is not sufficient, by itself, to determine the
outcome of a measurement process for most of the observable
quantities one can consider, but that such an outcome depends from
the overall factual situation. For instance within Bohmian
Mechanics, a system with a precise wavefunction and a precise
position, when subjected, e.g., to a measurement  of its momentum,
may give one or the other of the outcomes compatible with its
wavefunction, depending on the specific apparatus one chooses to
perform the measurement.

This situation, which at first sight might be considered as
puzzling, in reality  gives simply important indications about the
ontology which is appropriate for the theory. The way out derives
from taking the attitude that the only physical entities the
theory is about are the noncontextual ones. In Bohmian Mechanics
the positions of the particles play such a privileged role: they
are the only non contextual, objective, {\it real} variables (the
``local beables'' \cite{bell1,bell1b}) of the theory\footnote{Of
course at the formal level one can easily work out models making
e.g. the momentum variables non contextual. However, the ensuing
contextual nature of positions makes it quite difficult to build
up a coherent description of natural phenomena based on such a
scheme.}. What about the other observables? \cite{dgz3}
``Properties that are merely contextual are not properties at all;
they do not exist, and their failure to do so is in the strongest
sense possible''.

A weakness, in our opinion, of the theory is that one can exhibit
\cite{dg} infinitely many inequivalent hidden variable theories
--- whose hidden variables are the position of the particles of
the universe --- different from Bohmian Mechanics. They are all
perfectly consistent, differing among themselves only for the
trajectories they assign to the particles.

Of course, this is not a mathematical fault of Bohmian Mechanics;
however, it casts some shadow over the ``ontological'' basic
position of the theory: that particles have always definite
positions and follow precise trajectories. If many inequivalent
Bohmian--like theories assigning different trajectories to
particles are possible, which trajectories are the correct ones?
Is there a criterion to choose only one among them? Some authors
\cite{rim1} have tried to identify such criterion with the so
called ``request of compoundational invariance'' of the theory.
However, such a request does not seem logically
necessary\footnote{S. Goldstein [private communication], has
raised the objection that also within dynamical reduction models
``there is a good deal of arbitrariness, in the choice of parameters,
of smoothing functions, of basic observables, and the like'',
suggesting that if this arbitrariness is not a problem for
collapse models, it should not be a problem also for Bohmian
mechanics. We do not agree on this point, since
the situation is radically different in the two theories. Within dynamical
reduction models, changing the values of the parameters or of
the smoothing functions one modifies the theory in a --- at least
in principle --- {\it testable} way (this will become clear after
the analysis of the following sections). In Bohmian mechanics, on
the other hand, the different formulations are equivalent and lead
exactly to the same physical predictions.}.

In spite of this difficulty, Bohmian Mechanics is undoubtedly one
of the (few) promising and consistent theories solving the
measurement problem of Quantum Mechanics; of course, the great
challenge is to formulate a relativistic invariant version of it.

\subsection{Limiting the class of observables} \label{sec33}

The attempts to get objectification through a limitation of the
class of observables have received great attention
\cite{ja,dlp,got,zur1,zu,jz}. We consider it appropriate to
distinguish two different positions which have been taken when
trying to implement such an approach:
\begin{description}
\item[Strict Superselection Rules:] the set of the observables
which can actually be measured for any macro--system does not
coincide with the set of self--adjoint operators of the associated
Hilbert space; it admits superselection rules. In particular, the
eigenmanifolds corresponding to different macroscopic properties
are superselected.
\item[De Facto Superselection Rules:] The impossibility of putting
into evidence macroscopic coherence is not a matter of principle
but derives from practical, but practically insurmountable,
limitations.
\end{description}
Notice that both the above programs require the consideration of
the dynamics of the theory; the possibility and the consistency of
assuming limitations of measurability cannot be analyzed at the
kinematical level only. We turn now to discuss the two cases.

\subsubsection{Strict Superselection Rules} \label{sec331}

Suppose that, at a certain level --- in the present case, the
macroscopic one --- the set of observables of the system admits
strict superselection rules, i.e. that the set of operators associated to
all physical quantities which are {\it actually} measurable is an abelian
set. In such a case, as well known, the phase relations between
components of the statevector belonging to different superselected
manifolds become physically irrelevant. This amounts to saying that what
actually characterizes the states of physical systems (ensembles) are not
the statevectors or the statistical operators, but the {\it equivalence
classes} $[\rho]$ of statistical operators with respect to the allowed
observables $\Omega$, i.e.:
\begin{equation}
[\rho] \quad = \quad \left\{\rho^{*}\in\, T_{1} \, :\,
\makebox{Tr}(\rho\,\Omega) = \makebox{Tr}(\rho^{*}\Omega),
\;\forall\;\; \Omega\right\}
\end{equation}
where we have denoted by $T_{1}$ the set of positive trace--class
operators of trace $1$. The basic idea for circumventing the
difficulties of Quantum Mechanics goes as follows. Consider, e.g.,
the evolution characterizing an ideal measurement process:
\begin{equation} \label{se}
\frac{1}{\sqrt{2}}\,\left[ |\makebox{u}\rangle +
|\makebox{d}\rangle \right]\otimes |A_{0}\rangle
\quad\longrightarrow\quad \frac{1}{\sqrt{2}}\,\left[
|\makebox{u}\rangle \otimes |A_{u}\rangle + |\makebox{d}\rangle
\otimes |A_{d}\rangle \right],
\end{equation}
where $|A_{0}\rangle$, $|A_{u}\rangle$ and $|A_{d}\rangle$ denote
the ``ready", ``points at u" and ``points at d" states of the
macroscopic pointer. The states $|A_{u}\rangle$ and
$|A_{d}\rangle$ are macroscopically different and therefore,
according to our assumptions, belong to superselected
eigenmanifolds; one can then legitimately assert that the final
situation consists of the equal weights statistical mixture
$E=(E_{u})\cup (E_{d})$ of the pure cases
$|\makebox{u}\rangle\otimes |A_{u}\rangle$ and
$|\makebox{d}\rangle\otimes |A_{d}\rangle$.

The program is appealing, even though to take it seriously as a
candidate for a coherent worldview one should make it more
precise. In particular, one should exhibit the formal elements
accounting for the way in which the superselection rules emerge
(the location of the split between two types of physical systems,
i.e., those for which no limitation of observability occurs and
those for which it does), allowing the precise identification of
the superselected manifolds (the ``preferred basis problem").
There is, however, a more fundamental reason which forbids us to
take it seriously: it meets insurmountable difficulties when one
takes into account the dynamics \cite{blm}. This is easily proved
by taking into account that the initial and final system+apparatus states
in a measurement process, being necessarily macroscopically
distinguishable, must belong to different equivalence classes. As
a consequence, the hamiltonian itself, since it connects different
superselected manifolds, is not an allowed observable: this is
quite peculiar.

Another related problem derives from considering the
``reversibility" of the process. To discuss this let us consider,
e.g., the final state (\ref{se}) and the equivalent statistical
mixture $E=(E_{u})\cup (E_{d})$. Suppose then one can ``evolve
back" or ``reverse" the measurement process. According to whether
one starts from the state (\ref{se}) or from the statistical
mixture, one goes back either to the state
$1/\sqrt{2}\,[|\makebox{u}\rangle +
|\makebox{d}\rangle]\otimes|A_{0}\rangle$ or to the equal weights
mixture of the states $|\makebox{u}\rangle\otimes |A_{0}\rangle$
and $|\makebox{d}\rangle\otimes |A_{0}\rangle$. Should one then
perform, by means of another apparatus, a measurement process to
ascertain the value of the observable $\sigma_{x}$, he would, in
the first case, get the result $+1$ with certainty, while in the
second case there is a probability $1/2$ of getting the result
$-1$. The combined ``reversal of the process" and ``measurement of
$\sigma_{x}$" would then constitute a measurement process which
allows us to distinguish the final pure state from the equivalent
statistical mixture, contradicting the assumption that only
superselected observables are allowed\footnote {Note that the only way
out from this problem would be to deny the reversible nature of quantum
evolution at least for processes involving macroscopic systems. But this
would amount to accept that \cite{bells} ``Scrh\"{o}dinger's equation
is not always right".}.

To conclude, the previous analysis should have made clear why the
strict superselection  program cannot be fulfilled.

\subsubsection{The De Facto Superselection Rules option}
\label{sec332}

Recall that what makes the strict superselection program not
viable is the fact that the Hamiltonian connects eigenmanifolds
corresponding to different macroscopic properties, and the related
fact that the possibility of reversing the evolution leading to
superpositions of macroscopically different states contradicts the
assumption that the superposition is in the same equivalence class
as the corresponding statistical mixture.

This provides the basic idea of the de facto superselection
option. Since, when macroscopic objects are involved it is
practically impossible to distinguish pure states from statistical
mixtures or to ``undo" a measurement process, one could be tempted
to assert that for such systems a {\it de facto} limitation of
observability must be recognized. Such a position has actually
been taken in many interesting papers \cite{dlp,got,zur1,zu,jz}.
In these papers attention has been called to various features and
mechanisms inducing the de facto impossibility we have just
mentioned: the extreme complexity of a macroscopic object, its
unavoidable and uncontrollable interactions with the environment,
and so on.

In the previous section we have shown that such proposals cannot
overcome, in principle, the measurement problem of Quantum
Mechanics; however  a deeper analysis may be helpful to clarify
the matter. We begin with a digression. We have used the
expression ``de facto", in place of the fashionable acronym
(introduced by J. S. Bell \cite{bellam}) FAPP (for all practical
purposes), for a precise reason. It seems to us that describing
this position as FAPP suggests accusing people following this line
of taking an instrumentalist position about science. We do not
think that most of the proposals for a de facto superselection
solution to the objectification problem require such
instrumentalism. Most people taking the de facto attitude would
claim that this is as legitimate as accepting the de facto
validity of the second law of thermodynamics, in spite of the
reversibility of the basic mechanical laws. Obviously it would be
inappropriate to maintain that accepting thermodynamics involves
taking an instrumentalist position.

Having stated this we would like, however, to call attention to
the fundamental conceptual differences between the case of
thermodynamics in relation to classical mechanics and the case of
the de facto superselection assumption in relation to the unitary
evolution. To do this we start by considering the two premises:
\begin{description}
\item[1C] The reversible classical laws are the ``correct" laws
of nature;
\item[1Q] The superposition principle has unlimited validity;
\end{description}
and the legitimate classical statement:
\begin{description}
\item[2C] Under appropriate circumstances the irreversible
thermodynamical laws are ``de facto" correct.
\end{description}
Taking the de facto superselection position amounts to claiming
that the corresponding quantum statement:
\begin{description}
\item[2Q] Under appropriate circumstances the irreversible
process of wavepacket reduction and the replacement of a pure
state with a statistical mixture are ``de facto" correct
\end{description}
is equally legitimate.

Can we take such a position consistently? To answer this question
let us consider the classical case. It is obviously true that the
irreversible thermodynamical laws cannot describe correctly the
behaviour of e.g. a gas for arbitrarily large times since it is a
consequence of assumption 1C that the point representing the
system in phase space will, after Poincar\`e recurrence times,
return as close as one wants to its present value. One could then
raise the question: does this fact imply that the assertion that
``de facto, in an ensemble of gases almost all of them are {\bf
now} evolving irreversibly towards equilibrium'' will be falsified
by the {\bf future} behaviour? Surely not.

Zurek \cite{zur1,zu}, in his detailed analysis seems to suggest
that, since the situation in the quantum case is analogous to the
classical one, statement 2Q has the same conceptual status as 2C.
To this purpose he proves that, due to the unavoidable
interactions with the environment, in the case of a macroscopic
system in a superposition of macroscopically distinguishable
states the off--diagonal elements of the reduced statistical
operator (i.e. the one obtained by tracing out the environment
variables) become rapidly negligibly small and remain so for times
comparable to the Poincar\`e recurrence times for a gas. This is
certainly true; but does it prove that the situation is
conceptually analogous to that of thermodynamics? We think not. In
fact in the quantum case the assumption 1Q that the linear laws of
Quantum Mechanics are correct and have universal validity implies
that the result of a prospective measurement on an ensemble {\it
in the very far future} would falsify not only our statements
about future events but also the assertion that {\bf now} the
ensemble is the union of the pure subensembles corresponding to
definite macroscopic positions. Such an assertion would turn out
to be in no sense, even approximately, correct. The argument we
have presented briefly has been expounded with great clarity and
precision by d'Espagnat in \cite{be1}, to which we refer the
reader for a deeper analysis.

We can further clarify the matter by repeating the previous
analysis within the context of the pilot wave theory; which, we
recall, is fully equivalent to Quantum Mechanics in its physical
predictions and which assigns definite positions to all particles
of a system at all times. The approximation which corresponds to
assuming that wavepacket reduction occurs, consists in
disregarding, in the description of the evolution of an ``up"
(``down") pointer position (after the measurement is over), the
contribution to the wave function coming from the term
corresponding to the ``pointer down" (``pointer up") in the
statevector. Again, such an assumption will surely be proved false
by events in the very far future. However, both the approximate
and the ``true" versions of the theory assert that presently the
pointer is either up or down, or equivalently that all pointers of
the ensembles are in one of the two positions, and the future
happenings neither falsify this statement nor deny that the
approximate description is extremely accurate for extremely long
times. Therefore, within the pilot wave framework, the analog of
assertion 2Q has the same conceptual status as 2C. As discussed
above, this is not the case for the de facto superselection
program.

Other significant differences between the thermodynamical and the
de facto superselection situations deserve to be mentioned. For
instance, in the classical case the following three statements are
correct:
\begin{description}
\item[3C] The assignment of the phase--space distribution identifies
with sufficient precision the corresponding physical ensemble.
\item[4C] The approximation made in using thermodynamical equations
to describe the behaviour of a system is under control. The split
between mechanical and thermodynamical systems is not shifty. Just
to give an example, while few molecules are not a gas, an
Avogadro's number of them is a gas.
\item[5C] The exact (mechanical) and the approximate
(thermodynamical) laws both make sense and both allow simple and
sensible assumptions about the psycho--physical correspondence
allowing us ``to close the circle" for the appropriate classes of
phenomena.
\end{description}
But the corresponding statements in the quantum case are fully
inappropriate. In fact:
\begin{description}
\item[3Q] The same statistical operator corresponds to completely
different physical ensembles.
\item[4Q] The approximation made in breaking linearity is shifty:
macroscopic systems exist which require a genuine quantum
treatment (more on this in section \ref{sec41}).
\item[5Q] The correct (linear evolution) law leads to a situation
which does not make sense from the point of view of our (definite)
perceptions, only the approximation allows a sensible
psycho--physical correspondence.
\end{description}
Statement 3Q further emphasizes the difficulties in relating the
states of the system to our perceptions. Even ensembles
corresponding to the same statistical operator can be very
different in their compositions in pure subensembles \cite{bcas}.
This proves once more that the simple recognition that two
ensembles can be de facto in the same equivalence class is not
sufficient to explain why our perceptions unavoidably correspond
to a specific composition, i.e. the one whose subensembles have
definite macroscopic properties. As recognized by Joos and Zeh
\cite{jz} who have presented one of the most interesting proposals
along these line: {\it ``perhaps (this fact) can be justified by a
fundamental (underivable) assumption about the local nature of the
observer"}.

Our conclusion is that one cannot consider the de facto
superselection proposals as yielding a consistent way of ``closing
the circle". We will come back again to  this point in what follows since,
from the mathematical point of view, it has strict connections with
Dynamical Reduction Models.

\subsection{The Modal Interpretations of Quantum Mechanics}
\label{sec34}

These approaches \cite{mod0,mod1,mod2,mod3,mod4,mod5,vd,bc} rest
on the introduction of appropriate rules which allow us to
attribute some properties to the subsystems of a composite system
even when there is no observable whose outcome can be predicted
with certainty. To illustrate the general lines of the program we
make reference to the proposal by Dieks \cite{mod2}.

Consider a composite system containing several (let us say $N$)
constituents and suppose that we are dealing with a pure case
associated to an entangled statevector. The proposal goes as
follows. Any subsystem of the whole system has at any time
definite properties, identified by the following procedure.
Suppose we are interested in the subsystem $S^{M}$ constituted by
a group of particles (let us say the first $M < N$); the case of
only one particle or even of a specific degree of freedom of a
particle is not excluded. We also denote by $S^{N-M}$ the system
of the remaining particles. One then considers the whole Hilbert
space as the direct product of the Hilbert space referring to the
considered group and to the rest:
\begin{equation}
{\mathcal H}\; \equiv \; {\mathcal H}(1)\otimes{\mathcal H}(2)
\otimes ...\, {\mathcal H}(N) \; = \; {\mathcal H}(1, ... , M)
\otimes {\mathcal H}(M+1, ... , N)
\end{equation}
Accordingly, one takes into account the biorthonormal
decomposition of the statevector:
\begin{equation} \label{bd}
|\psi(1, ... , N)\rangle \quad = \quad \sum_{i} \sqrt{p_{i}}\;
|\chi_{i}(1, ... , M)\rangle\otimes |\Omega_{i}(M+1, ... ,
N)\rangle;
\end{equation}
in the above equation, the parameters $p_{i}$ are positive
constants summing up to $1$: they are the eigenvalues of the
reduced statistical operators obtained by taking the partial trace
of $|\psi\rangle\langle\psi|$ either on ${\mathcal H}(1, ... , M)$
or on ${\mathcal H}(M+1, ... , N)$. The states $|\chi_{i}\rangle$
and $|\Omega_{i}\rangle$ satisfy:
\begin{equation}
\langle\chi_{i}|\chi_{j}\rangle \quad = \quad
\langle\Omega_{i}|\Omega_{j}\rangle \quad = \quad \delta_{ij}.
\end{equation}
As proved by von Neumann \cite{vn}, such a decomposition is
uniquely determined by $|\psi(1, ... , N)\rangle$ except in the
case of degeneracy of the above eigenvalues. Ignoring the
complications arising from accidental degeneracy, we can now state
the rule for assigning properties to the subsystems $S^{M}$ and
$S^{N-M}$: {\it when dealing with a pure case associated to the
state (\ref{bd}) the subsystems $S^{M}$ and $S^{N-M}$ have
definite properties. They are those associated to the observables
having the states $|\chi_{i}\rangle$ and $|\Omega_{i}\rangle$ as
eigenvectors. The probability of the i-th property to be actually
possessed (or better: the fraction of systems in the ensemble
which have such a property) is given by $p_{i}$} (in other words,
the model is basically a hidden variable model \cite{bc}, whose
hidden variables are identified via the biorthonormal
decomposition (\ref{bd})).

As usual we indicate the way in which the proposal circumvents the
problems of the theory of measurement. According to the ideal von
Neumann measurement scheme the final state in a measurement
process consists, typically, of a superposition of states each
term of which involves an eigenvector referring to a different
reading of the apparatus (compare equation (\ref{eq01})): this
final state already gives the von Neumann biorthonormal
decomposition for the system+apparatus so that, according to the
previous criterion one can assert that the appropriate fractions
of the apparata have their pointers in precise and different
positions.

Such proposals are surely interesting but they meet various
difficulties which have been discussed e.g. in
\cite{blm,al2,sq1,alo1}. We refer the reader to the above papers
for details. Apart from this, we would like to call attention on
the fact that the proposal raises other problems, in particular,
it lacks what we might call {\it ``structural completeness"}. The
situation can be summarized in the following simple terms. Quantum
mechanics, in its general formulation, allows the treatment of
statistical ensembles. We are considering now theoretical models
which accept that systems associated to the same statevector have
different properties.

Suppose now we are dealing with an ensemble which is a pure case
associated to the state (\ref{bd}). What meaning can be attached
to the statement that the considered subsystems have properties?
Even within very weak varieties of realism, this amounts to
asserting that the ensemble is inhomogeneous, that it actually is
the union $E = \cup_{i} E_{i}$ of different subensembles. One can
then raise the question: can we prepare such an ensemble in the
standard way, i.e. by taking (for all $i$) a fraction $p_{i}$ of
subensembles in the state $|\chi_{i}\rangle |\Omega_{i}\rangle$?
Obviously not. In fact, if we were to prepare the subsystems in
the states $|\chi_{j}\rangle |\Omega_{j}\rangle$ it would be false
to assert, contrary to what this interpretation holds true, that
the composite system has with certainty the properties associated
according to the above procedure to an ensemble described by
$|\psi\rangle$. This means that the model deals with two kinds of
statistical ensembles: the one associated to the state (\ref{bd})
and the one prepared by taking a fraction $p_{i}$ of subsystems in
the state $|\chi_{j}\rangle |\Omega_{j}\rangle$. They both are the
union $E = \cup_{i} E_{i}$ of the same kind of subensembles (each
taken with the same probability $p_{i}$), but they are {\it
structurally different} since one of them has specific hidden
features to which one does not have access.

\subsection{The Decoherent Histories approach} \label{sec35}

The main purpose of this proposal \cite{gri1,gri2,omn1,omn2,gel1}
is to assign definite probabilities to alternative histories of a
physical system, which may also be the whole universe. The idea
goes as follows. One defines histories as sequences of events
yielding a sort of motion picture of the evolution of a system,
and attaches appropriate probabilities to them. Let us first
define the notation we will use.

We consider, for a given $k$, a set of orthogonal projection
operators $P^{k}_{\alpha_{k}}$ yielding a resolution of the
identity:
\begin{equation} \label{ri}
\sum_{\alpha_{k}} P^{k}_{\alpha_{k}} \; = \; 1; \qquad
P^{k}_{\alpha_{k}} P^{k}_{\beta_{k}} \; = \;
\delta_{\alpha_{k}\beta_{k}}
\end{equation}
The index $k$ labels the ``question" or ``property" we are
interested in, while the parameter $\alpha_{k}$, which runs over
an appropriate range, labels a set of ``alternative values for the
considered property" which are, according to (\ref{ri}),
exhaustive and mutually exclusive. It is important to have clear
the meaning of the considered projection operators. For this
purpose we refer, for simplicity, to the spin space of a spin
$3/2$ particle. In such a case, specifying $k$ could mean, e.g.,
to specify a spin component, so that $k=1$ could be related to
$S_{z}$ and $k=2$ to $S_{x}$. For fixed $k$ the index $\alpha_{k}$
identifies  a set of mutually orthogonal manifolds (each of them
being either one or the direct sum of various eigenmanifolds of
the $k$--th operator) whose direct sum is the whole space. So one
could have, for $k=1$, $\alpha_{1}$ taking e.g. $3$ values:
\begin{itemize}
\item $P^{1}_{1_{1}}$ projects on the eigenmanifold spanned by the
eigenstates belonging to $S_{z} = 3/2$ and $S_{z} = 1/2$.
\item $P^{1}_{2_{1}}$ projects on the eigenmanifold spanned by the
eigenstate belonging to $S_{z} = -1/2$.
\item $P^{1}_{3_{1}}$ projects on the eigenmanifold spanned by the
eigenstate belonging to $S_{z} = -3/2$.
\end{itemize}
Analogously $P^{2}_{\alpha_{2}}$ could represent, for $\alpha_{2}$
taking only two values, the two projection operators on the
positive and negative parts of the spectrum of $S_{x}$. We will
denote by $P^{k}_{\{\alpha_{k}\}}$ the set of all the projection
operators associated to the ``property $k$"  when
$\alpha_{k}$ runs through its range. One then considers the corresponding
projectors in the Heisenberg picture at time t:
\begin{equation}
P^{k}_{\alpha_{k}}(t) \quad = \quad e^{iHt}\, P^{k}_{\alpha_{k}}\,
e^{-iHt}
\end{equation}

A {\bf history} is defined by a succession of times $t_{1} < t_{2}
... < t_{n}$ and a sequence of projection operators. It will be
denoted by $(\alpha_{n},t_{n}) ... (\alpha_{2},t_{2})\,
(\alpha_{1},t_{1})$. If the initial conditions are fixed by
specifying the initial statevector $|\psi(0)\rangle$, one
attributes to the above history the probability:
\begin{eqnarray}
\lefteqn{P[\alpha_{n},t_{n}; ... \,\alpha_{2},t_{2};\,
\alpha_{1},t_{1}] \; = \;} \nonumber \\ && \makebox{Tr} \left[
P^{n}_{\alpha_{n}}(t_{n}) ...
P^{1}_{\alpha_{1}}(t_{1})\,|\psi(0)\rangle\langle\psi(0)|
P^{1}_{\alpha_{1}}(t_{1})... P^{n}_{\alpha_{n}}(t_{n})\right]
\end{eqnarray}
One then considers the set of all alternative histories, which we
will denote with self--explanatory notation as
$(\{\alpha_{n}\},t_{n}) ... (\{\alpha_{2}\},t_{2})\,
(\{\alpha_{1}\},t_{1})$, i.e., the set of all quantum histories
obtained by letting each $\alpha_{j}$ take all values in its
range. Due to quantum interference, the probabilities of the
histories of this set turn out not to satisfy, in general, the
additivity conditions which are necessary in order that they could
be interpreted as true probabilities. For instance one usually
has:
\begin{equation}
P(\alpha_{2}, t_{2}) \quad \neq \quad \sum_{\alpha_{1}}
P(\alpha_{2}, t_{2}; \alpha_{1}, t_{1})
\end{equation}

To circumvent this difficulty one introduces the idea of a
decoherent set of alternative histories. This can be implemented
mathematically by defining the {\bf decoherence functional}:
\begin{eqnarray}
\lefteqn{D[\alpha_{n},t_{n}; ...\, \alpha_{1},t_{1}\, |\,
\beta_{n},t_{n}; ...\, \beta_{1},t_{1}]
\; = \;} \nonumber \\
& & \makebox{Tr} \left[ P^{n}_{\alpha_{n}}(t_{n}) ...\,
P^{1}_{\alpha_{1}}(t_{1})\,|\psi(0)\rangle\langle\psi(0)|
P^{1}_{\beta_{1}}(t_{1})...\, P^{n}_{\beta_{n}}(t_{n})\right]
\end{eqnarray}
If such a functional vanishes whenever at least one of the
$\beta_{k}$ differs from $\alpha_{k}$, one says that the
considered set of alternative histories is consistent since the
associated probabilities satisfy all necessary requirements. For a
given set of histories one may construct coarser--grained
histories by summing over the finer--grained projections.

We do not want to be more specific about this program. We refer
the reader to the references quoted above, particularly to the
book \cite{omn2} by Omn\`es for a thorough analysis. We prefer to
comment about the relations between the Decoherent Histories
approach and some of the approaches we have already discussed.

Let us look at the Decoherent Histories approach from the point of
view of the strict superselection  option. If the conditions
presupposed by the strict superselection rules were satisfied,
i.e. if there were a level at which macroscopically different
eigenmanifolds are strictly superselected and are not connected by
the Hamiltonian, all histories attributing macroscopic properties
to the physical system at the considered level would decohere: one
could then truly describe the unfolding of the evolution by a
consistent snapshot--like motion picture. This comparison with the
strict superselection case immediately reveals an interesting
advantage of the Decoherent Histories approach. Namely, within the
strict superselection scheme the assignment of the statevector at
various times tells us which fractions of systems have various
macro--properties but it does not attach probabilities to time
sequences of events; the snapshots at different times cannot be
organized in motion pictures as in the case of Decoherent
Histories.

Since, as already remarked, the assumptions of strict
superselection rules cannot hold consistently, it becomes quite
natural to look at Decoherent Histories from the point of view of
the de facto superselection rules. In particular, by taking
advantage of the many proofs that the environment induces ``de
facto" superselection rules associated to macroscopic position
variables, one could limit one's considerations, within the
Decoherent Histories approach, to alternative histories specifying
e.g. the intervals in which the macroscopic pointer lies at
various times. Again this point of view represents an interesting
improvement with respect to the simple de facto superselection
program since it allows the consideration of a time--chain of
events. Moreover, it gives precise criteria to select decoherent
histories from non decoherent ones.

Decoherent Histories supporters  maintain that the decoherent sets
of histories can be considered completely in general, i.e. with
reference both to macroscopic and microscopic systems, that one
can assign probabilities to them if the consistency conditions are
met, and  that within the scheme decoherence replaces the notion
of measurement. This seems to suggest that some objective meaning
is given to consistent alternative histories. Here a serious
problem arises. It can easily be seen that alternative histories
involving only one time always decohere. One can then consider, at
a given time, different sets of incompatible alternative
histories. If the probabilities are related to possessed
properties, then one should assign objective meaning to the
different possible incompatible sets of decoherent histories. In
references \cite{bg2,bgp98}\footnote{See also \cite{gri5,bg2b} for
further discussion on this issue.} we have proved that this cannot
be done in a consistent way. The problem is the same as the one of
contextuality in hidden variable theories: not too many properties
can be assigned to quantum systems. As a consequence it is not
clear the meaning which should be given to decoherent histories.

A last remark: decoherent histories can also be considered as
strictly referring only to the universe as a whole. When one takes
such an attitude one invokes the natural de facto decoherence of
histories about the universe. As appropriately remarked in
\cite{rim2} one can then raise the question: what is the status of
two histories belonging to incompatible sets of alternative
histories? References \cite{rim2} and \cite{de3} have also called
attention to peculiar difficulties that the Decoherent Histories
approach meets concerning the future--past relation.

We do not pursue the analysis further. Concluding, it seems to us
that even Decoherent Histories  do not allow us  to attribute
consistently an objective meaning to statements about possessed
properties.

\subsection{Enriching reality} \label{sec36}

Such proposals \cite{ev1,dw1,dw2,de1,al7,al5} maintain that the
unitary evolution holds in all circumstances and dispose of the
embarrassment arising from the occurrence of superpositions of
perceptively different states by assuming that in a sense, all
potentialities of the wavefunction become actual. The most widely
known proposal of this type is usually referred to as ``The Many
Universes Interpretation of Quantum Mechanics''
\cite{de1,al7,al5}.

According to this proposal, each time an interaction leading to
superpositions of macroscopically distinguishable situations
occurs, the universe literally splits into (in general infinitely
many) replicas of itself: each replica corresponds to one of the
terms in the superposition and occurs with the appropriate
probability. So, in a situation like the one of equation
(\ref{se}) one would state that, after the measurement is over,
there are actually two types of universes; in those of the first
type there is an apparatus whose pointer ``points at u" and in the
second an apparatus whose pointer ``points at d". Needless to say,
if one wants to describe the situation at later times one has to
go on with the unitary evolution taking into account all
interactions which take place and then, having expressed the final
statevector as a superposition of states in each of which all
macro---objects have definite macroscopic properties, associate
each term of the superposition to universes of a different type.

A serious limitation for the proposal comes from the fact that it
leaves largely undefined how and when the multifurcation of the
universe takes place. This ambiguity reflects the basic difficulty
that Quantum Mechanics meets in locating the ``shifty split"
between micro and macroscopic phenomena.

Detailed analyses of the many universes theories have been
presented \cite{de1,al7,al5}. Here we want to stress that, since
to close the circle one needs also some assumptions about the
process of perception, there are at least two choices for this,
which give rise to two quite different alternatives. If one makes
simple assumptions about the psycho--physical correspondence one
has the ``genuine" many world interpretation: in the process of
replicating the universe also the perceiving subject is
replicated, so that in the above example there will be universes
in which we perceive that the pointer points up and universes with
replicas of ourselves having the other perception. Within each
universe the perception is strictly correlated to statevectors
corresponding to different macroscopic situations.

On the other hand, one can take the attitude that it is the
perception mechanism which is more complex than we usually assume;
this leads to what has been referred to as ``the many minds
interpretation" of the theory. Such a formulation has the
advantage of allowing us to circumvent the ambiguities about the
branching of the universe; there is only one universe and there
are many minds (i.e. each mind exhibits some sort of a full
spectrum of perceptions reflecting the macroscopically different
states in the superposition). Many interesting problems arise when
one takes this attitude, the most relevant ones having to do with
the intersubjective agreement and with the reliability of our
beliefs. We will not discuss, for lack of space, the details of
these approaches.

We conclude this subsection by stating that, even though we
consider these ``enriching reality'' proposals interesting, they
seem to require a too radical change in our views about reality
and the adoption of a rather strange ontology. For, according to
them science does not deal any longer with the one world we live
in or the perceptual processes we experience, but at the same time
with the totality of all possible worlds and all possible
perceptions.

\subsection{Modifying the dynamics} \label{sec37}

When one assumes that the theory is complete and that pure cases
describe genuinely homogeneous ensembles, the only way to dispose
of the embarrassing superpositions is to say that in one way or
another the dynamical equations of the theory are not always or
not exactly right. At this point, two completely different
positions can be and have actually been taken. We will briefly
discuss them in the two following subsections.

\subsubsection{Two dynamical principles} \label{sec371}

This line of thought plainly accepts that there are two dynamical
principles which must be used for describing different physical
situations. The best known example of this attitude consists in
accepting wavepacket Reduction: the evolution of microscopic
quantum systems is governed by the unitary and reversible linear
Schr\"odinger equation; the measurement process is governed by the
nonlinear process of wavepacket reduction transforming, in
general, pure states into statistical mixtures. The reasons for
this different treatment of physical systems are traced back to
the recognition that there are two classes of phenomena in nature,
the quantum and the classical, the reversible and the
irreversible, the microscopic and the macroscopic ones. Further
support to this attitude is given by saying that, in a sense,
classical concepts are a prerequisite for the very formulation of
quantum formalism.

One could find many reasons for considering legitimate such a
position; after all, all physical theories have a limited range of
applicability. In this respect Quantum Mechanics would be claimed
to find its limit in the description of the micro--macro
interactions taking place in the measurement process. However, as
repeatedly stressed by J. Bell and as already discussed in the
previous section, the real difficulty which this line meets does
not stem from its dualistic attitude about our understanding of
natural phenomena, but derives from the fact that there is nothing
in the theory which allows us to locate the ``split" between the
two considered classes of phenomena. When trying to follow this
line, as J. Bell has stated \cite{bellam}, {\it are we not obliged
to admit that measurement like processes are going more or less
all the time, more or less everywhere?}

The remark is so appropriate that E.P. Wigner \cite{wig1}, having
recognized the unavoidability of accepting two dynamical
principles, felt the necessity of following von Neumann's
proposal: to solve the problem one has to go to the extreme end of
the chain of observation and to assume that reduction does not
take place until somebody knows that it must, i.e. up to when
conscious observers are involved. This position leads to a quite
peculiar conclusion, i.e. that the world as we know it, is very
much a product of conscious mind. In spite of this, one could say
that such a position represents a simple and effective solution
(reduction actually takes place) to the problems we are debating
except that it suffers once again from an intrinsic ambiguity. For
the question: ``what is conscious?" does not admit any unambiguous
answer on the basis of our present knowledge about nature and
human beings.

The impossibility of locating the split between the two types of
physical systems (quantum--classical) which should be governed by
different laws as well as the impossibility of clearly identifying
the processes involving consciousness clearly show that also the
program outlined here does not allow one to ``close the circle".
We pass then to consider the other option: the dynamical equation
of the theory is not right.

\subsubsection{Unified dynamics: Dynamical Reduction Models}
\label{sec372}

The program seeks a modification of the evolution law in such a
way that measurement--like processes have definite outcomes as a
consequence of the unified dynamics governing all physical
processes \cite{heid,grw,csl,rel,cc}. In this search some guidance is
of course given by the fact that the modified dynamics should
imply wavepacket reduction as a consequence of the interaction of
the microsystem and the macro--apparatus and, more generally,
forbid the persistence of linear superpositions of macroscopically
different states. With this in mind, one remarks that the
characteristic features distinguishing quantum evolution from
wavepacket reduction are that, while Schr\"odinger equation is
linear and deterministic, wavepacket reduction is nonlinear and
stochastic. It is then natural to entertain the idea of nonlinear
and stochastic modifications of the standard Hamiltonian dynamics.

Obviously such a program must respect strict constraints, in
particular, it must not contradict any known fact about
micro--phenomena. Secondly, to meet the requests we have
repeatedly mentioned in this paper, it must allow a clearcut
identification of the split between phenomena for which standard
Quantum Mechanics holds (obviously this has now to be read: for
which the approximation consisting in disregarding the nonlinear
terms of the ``exact" theory is legitimate and under control) and
those for which the new dynamics leads to relevant differences
with respect to the standard theory, more specifically to a
``classical behaviour". The analysis of proposals of this type, of
what they have accomplished and of the difficulties they meet will
be the subject of the rest of the report.

\part{Non Relativistic Dynamical Reduction Models}

\section{Preliminary considerations} \label{sec4}

The aim of Dynamical Reduction Models is to account for the
process of wavepacket reduction and the Schr\"odinger evolution in
terms of a unique dynamical equation leading to the spontaneous
suppression of the superpositions of different macroscopic states
of a macro--system; at the same time, the new dynamics must not
change in any appreciable way all the known properties of
microscopic quantum systems. As already stated, one tries to
achieve this goal by introducing nonlinear and stochastic
modifications of the standard Hamiltonian dynamics.

In this section we want to prove that both non linearity and
stochasticity are necessary ingredients in order to account for an
acceptable spontaneous reduction mechanism. More specifically, we
will show that neither a nonlinear but deterministic modification
nor a stochastic but linear one, can lead to a consistent theory
of dynamical reductions. A linear and stochastic modification
induces at most an {\it apparent} collapse of the wavefunction; a
nonlinear but deterministic modification, on the other hand,
unavoidably violates basic relativistic constraints. Before
discussing these issues, we will answer the following question:
should the localization mechanism act at the wavefunction level,
or is it sufficient, as suggested by some authors
\cite{zur1,zu,lib}, that it suppresses the the off--diagonal
elements of the statistical operator? The answer will be clear: a
consistent dynamical reduction theory must induce localizations
directly at the wavefunction level (we speak in this case of
individual or {\bf Heisemberg reductions} \cite{std}), and not
only at the statistical operator level (which we refer to as
ensemble or {\bf von Neumann} reductions).

\subsection{Individual and ensemble reductions} \label{sec41}

We have widely discussed in the previous section the fact that the
macro--objectification problem arises when a superposition of
macroscopically different states of a macroscopic object --- for
example the superposition with equal weights of two such states
$|\makebox{Here}\rangle$ and $|\makebox{There}\rangle$ --- occurs,
e.g. as the result of a measurement process:
\begin{equation} \label{smd}
|\psi\rangle \quad = \quad \frac{1}{\sqrt{2}}\,\left[\,
|\makebox{Here}\rangle \; + \; |\makebox{There}\rangle\, \right].
\end{equation}
In the language of the statistical operator, state (\ref{smd}) is
represented by:
\begin{eqnarray}
\rho \; \equiv \; |\psi\rangle\langle\psi| & = &
\frac{1}{2}\,\left[\, |\makebox{Here}\rangle\langle\makebox{Here}|
\; + \; |\makebox{There}\rangle\langle\makebox{There}| \right.\; +
\nonumber \\ & &
\quad\left.|\makebox{Here}\rangle\langle\makebox{There}| \; + \;
|\makebox{There}\rangle\langle\makebox{Here}|\, \right],
\end{eqnarray}
whose matrix representation with respect to the basis\footnote{ Of
course, here we make a gross simplification, treating a
macroscopic object like a simple two--dimensional system; however,
this does not invalidate the basic conclusions of our analysis.}
$|\makebox{Here}\rangle$ and $|\makebox{There}\rangle$ is:
\begin{equation}
\rho \quad = \quad \frac{1}{2} \left(
\begin{array}{cc}
1 & 1 \\ 1 & 1
\end{array}
\right).
\end{equation}

Let us consider now $N$ identical macroscopic systems whose state
is (\ref{smd}). Our declared goal is to find a universal mechanism
which transforms such an ensemble into the statistical mixture in
which half of the systems are in the state
$|\makebox{Here}\rangle$, and the other half in the state
$|\makebox{There}\rangle$, in accordance with the wavepacket
reduction postulate:
\begin{equation} \label{sm1}
\frac{1}{2}\;\makebox{systems in state $|\makebox{Here}\rangle$
and}\;\; \frac{1}{2}\;\makebox{systems in state
$|\makebox{There}\rangle$}.
\end{equation}
The ensemble (\ref{sm1}), in which all systems have definite
macro--properties, can be easily described within the statistical
operator formalism:
\begin{equation} \label{sop}
\rho' \quad = \quad \frac{1}{2}\,\left[\,
|\makebox{Here}\rangle\langle\makebox{Here}| \; + \;
|\makebox{There}\rangle\langle\makebox{There}| \right];
\end{equation}
the corresponding density matrix is:
\begin{equation} \label{ppp}
\rho' \quad = \quad \frac{1}{2} \left(
\begin{array}{cc}
1 & 0 \\ 0 & 1
\end{array}
\right).
\end{equation}
Thus we see that, in order to eliminate the embarrassing
superposition of different macroscopic states, the dynamics we are
looking for must induce the following change of the statistical
operator:
\begin{equation} \label{deso}
\frac{1}{2} \left(
\begin{array}{cc}
1 & 1 \\ 1 & 1
\end{array}
\right) \qquad \stackrel{\makebox{\tiny
evolution}}{\longrightarrow} \qquad \frac{1}{2} \left(
\begin{array}{cc}
1 & 0 \\ 0 & 1
\end{array}
\right),
\end{equation}
i.e. it must suppress the off--diagonal terms of the density
matrix, corresponding to  matrix elements connecting different
macroscopic states.

Here comes the crucial point: does a dynamical evolution like
(\ref{deso}) really guarantee by itself the suppression of
superpositions of different macroscopic states? The answer is
negative. The reason for this lies in the fact that {\bf the
statistical operator describing a statistical mixture, describes
at the same time infinitely many inequivalent statistical
mixtures} --- this is the weak point (for the problem we are
interested in) of the statistical operator formalism. In fact, let
us consider the following statistical mixture:
\begin{eqnarray} \label{sm2}
\lefteqn{\makebox{Half systems in state}\;
\frac{1}{\sqrt{2}}\,[|\makebox{Here}\rangle +
|\makebox{There}\rangle]\;} \\
& & \makebox{and half systems in state}\;
\frac{1}{\sqrt{2}}\,[|\makebox{There}\rangle -
|\makebox{There}\rangle].
\end{eqnarray}
Of course, (\ref{sm2}) describes a statistical ensemble which is
completely different from the one defined in (\ref{sm1}); however,
it is easy to check that the statistical operator describing it is
(\ref{sop}), i.e. the same one associated to the mixture
(\ref{sm1}).

The root of the problem should be clear: working only at the
statistical operator level, we cannot be sure that a dynamical
evolution like (\ref{deso}) transforms the pure state (\ref{smd})
into the statistical mixture (\ref{sm1}) --- a mixture whose
elements have definite macroscopic properties --- instead of
transforming it into a mixture like (\ref{sm2}), whose elements
are still superpositions of different macro--states. This means
that, in order to work out a fully consistent and unambiguous
theory of dynamical reductions, we have to assume that the
localizations affect directly the wavefunction, not only the
statistical operator.

\subsection{Linear and stochastic modifications of the Schr\"odinger
equation} \label{sec42}

The easiest way to implement a dynamical evolution like
(\ref{deso}), which suppresses the interference terms arising from
superpositions of different macro--states of a macroscopic system,
is to add a white noise stochastic potential $V(t)$ to the
standard Schr\"odinger equation for the wavefunction \cite{npfs}.
Here we give a simplified description of how this can be achieved,
considering the case of one particle in one dimension. We
disregard the Hamiltonian evolution and discretize the real axis
${\bf R}$ into intervals $\Delta_{i}$ of appropriate length. We
define the projection operators
\begin{equation}
P_{i}\, \psi(x) \quad = \quad \chi_{i}(x)\, \psi(x),
\end{equation}
where $\chi_{i}(x)$ is the characteristic function of the $i$--th
interval.

Let us now consider the following equation:
\begin{equation} \label{schsl}
i\hbar\,\frac{d}{dt}|\psi(t)\rangle \quad = \quad \sum_{i} P_{i}
\, V_{i}(t)\, |\psi(t)\rangle.
\end{equation}
$V_{i}(t)$ are white noise processes characterized by the
expectation values\footnote{We indicate with $\llangle \cdot
\rrangle$ the average value of the quantity contained within the
``brackets''.}:
\begin{equation}
\llangle V_{i}(t) \rrangle \; = \; 0 \qquad\quad \llangle
V_{i}(t)\, V_{j}(t') \rrangle \; = \;
\gamma\,\delta_{ij}\,\delta(t-t').
\end{equation}
The formal solution of equation (\ref{schsl}) is:
\begin{equation}
|\psi(t)\rangle \; = \; e^{\displaystyle -\frac{i}{\hbar}\,
\int_{0}^{t} d\tau \sum_{i} P_{i}\, V_{i}(\tau)} |\psi(0)\rangle
\; = \; \sum_{i}\, e^{\displaystyle -\frac{i}{\hbar}\,
\int_{0}^{t} d\tau V_{i}(\tau)} P_{i}\, |\psi(0)\rangle.
\end{equation}
If one defines the statistical operator
\begin{equation} \label{sjsfhj}
\rho(t) \quad = \quad \llangle |\psi(t)\rangle\langle \psi(t)|
\rrangle,
\end{equation}
the stochastic average is easily evaluated and one discovers that
$\rho(t)$ obeys the evolution equation:
\begin{equation} \label{efso9}
\frac{d}{dt}\, \rho(t) \quad = \quad \gamma \sum_{i} P_{i} \rho(t)
P_{i} \; - \; \frac{\gamma}{2} \sum_{i} \left\{ P^{2}_{i}, \rho(t)
\right\},
\end{equation}
which, as we shall see, is basically the same equation as the one
characterizing the dynamical reduction models we will discuss in
great detail in the following sections.

Let us now consider the vectors $|\Delta_{i}\rangle$ whose
position representation is $\langle x |\Delta_{i}\rangle =
\chi_{i}(x)$; then, equation (\ref{efso9}) leads to the following
equation for the matrix elements
$\langle\Delta_{i}|\rho(t)|\Delta_{j}\rangle$ of the statistical
operator:
\begin{equation} \label{radfd}
\frac{d}{dt}\, \langle\Delta_{i}|\rho(t)|\Delta_{j}\rangle \quad =
\quad \gamma\,(\delta_{ij} \; - \; 1)\,
\langle\Delta_{i}|\rho(t)|\Delta_{j}\rangle.
\end{equation}
We see that the diagonal elements do not change in time, while the
off--diagonal elements are exponentially damped, with a rate given
by $\gamma$; this means that equation (\ref{radfd}) embodies
precisely an evolution like (\ref{deso}). Does equation
(\ref{schsl}), then,  lead to the reduction of the statevector
into one of the states $|\Delta_{i}\rangle$, as it seems to follow
from equation (\ref{radfd})? The answer is no, for the following
reason.

Let us consider the average values $\langle \psi(t)| P_{i}
|\psi(t)\rangle$, measuring the portion of the wavefunction
$\psi(x, t) = \langle x|\psi(t)\rangle$ which is contained within
the interval $\Delta_{i}$. If equation (\ref{schsl}) induces, as
one could naively believe due to (\ref{radfd}), the reduction of
$\psi(x,t)$ into one interval, let us say $\Delta_{k}$, then the
following relation would necessarily hold:
\begin{equation}
\langle \psi(t)| P_{i} |\psi(t)\rangle \; \longrightarrow \;
\delta_{ik} \qquad \makebox{for $t \rightarrow \infty$}.
\end{equation}
On the contrary, for any realization of the stochastic potential
and for any time $t$, since $P_{i}$ commutes with the evolution operator:
\begin{equation}
\langle \psi(t)| P_{i} |\psi(t)\rangle \quad = \quad \langle
\psi(0)| P_{i} |\psi(0)\rangle.
\end{equation}
This means that, if $|\psi(0)\rangle$ corresponds to a non
localized function, the individual members of the ensemble are
always associated to non localized functions. The diagonalization
of $\rho$ arises from the random phases acquired by the states in
different intervals $\Delta_{i}$, not from a real reduction of the
statevector.

Stapp \cite{std} has considered this problem and has suggested to
accept a stochastic mechanism of this type, relating it to
fluctuations associated to the background radiation. We do not
like this proposal. At the macro--level we want that each
individual has actualities: those of being in a given region. If
this does not happen, how can one avoid the problem arising from
the many to one relation of ensembles with statistical operators,
discussed in the previous subsection? This shows that the fact
that equation (\ref{schsl}) leads to a statistical operator of
type (\ref{sm1}) is a necessary but not a sufficient condition in
order that the ensemble (\ref{sjsfhj}) can be considered a union
of pure cases corresponding to localized states. This is why, in
the following sections, we will confine our considerations to
models yielding Heisenberg reductions for the statevector.

To conclude: we have presented a stochastic equation for the
statevector, equation (\ref{schsl}), leading to equation
(\ref{efso9}) for the statistical operator, which induces
precisely an evolution like the one of equation (\ref{deso}), i.e.
an evolution which would be considered as transforming ensembles
of non localized states into ensembles of localized ones. Equation
(\ref{schsl}) is characterized by a hermitian coupling of the
stochastic noise to the operators $P_{i}$, and is linear; however,
it gives rise only to the diagonalization of the statistical
operator leaving the individual wavefunctions spatially extended.
This strongly suggests that the introduction of stochasticity into
the evolution equation (i.e. the possibility that a given state
evolves in different states according to its own story) combined
with the requirement that in the long run the statevector ends up
in one of the eigenmanifolds characterizing the preferred basis
(actual individual reductions) implies that the dynamics must be
nonlinear. As a matter of fact, the evolution laws of the
dynamical reduction models are stochastic and nonlinear.

As we will see in the next section, in a certain sense also the
converse is true, i.e. the consideration of nonlinear
modifications of the evolution equation requires, when some basic
relativistic constrains are added, the introduction of
stochasticity into the equation.

\subsection[Nonlinear and deterministic modifications]{Nonlinear
and deterministic modifications of the Schr\"o\-din\-ger equation}
\label{sec43}

Within standard quantum formalism the postulate of wavepacket
reduction is chosen in such a way that, even though the state of
the system can be instantaneously changed by a distant
measurement, such change cannot be used to send faster than light
signals between distant observers \cite{heb,grwfl,gwfl,shi1}. This
is a nice feature in absence of which an unacceptable violation of
relativistic requirements would occur. In fact, even though
quantum mechanics as considered here, and in particular the
process of wavepacket reduction, does not pretend to be a
relativistically invariant theoretical scheme, the fact that the
instantaneous changes induced by wavepacket reduction itself do
not depend in any way whatsoever from the distance between two
constituents one of which is subjected to a measurement, forbids
us to think that wavepacket reduction itself might represent some
non relativistic approximation of a relativistic process. If the
considered changes permit  one observer to become instantaneously
aware of the fact that the other (far away) constituent has been
subjected to a measurement, an explicit clash with basic
relativistic requirements would emerge, making the process
unacceptable.

Obviously the problem of possible instantaneous and detectable
effects at a distance must be faced when formulating a dynamical
reduction model, since, a priori, one cannot be sure that they do
not occur. In this respect it is appropriate to take into account
a quite general and interesting result obtained by N. Gisin
\cite{gisfl}. Let us consider a map from statistical ensembles to
statistical ensembles (we remember that we characterize a
statistical ensemble by specifying the pure states appearing in it
and the associated statistical weights):
\begin{equation}
M_{t}\, E(0) \quad \longrightarrow \quad E(t).
\end{equation}

In accordance with the analysis of section \ref{sec331}, we say
that two statistical ensembles $E$ and $E'$ are {\it equivalent}
if the corresponding statistical operators belong to the same {\it
equivalence class}:
\begin{equation}
E \; \sim \; E' \quad \makebox{iff: $\rho(E),\, \rho(E')\, \in \,
[\rho]$}.
\end{equation}
We can prove the following {\bf theorem}: a necessary condition in
order that the map $M_{t}$ describes an evolution which does not
conflict with relativity in the sense specified above (i.e. it
does not permit  faster than light signaling), is that the
equivalence relation be preserved by $M_{t}$, i.e.
\begin{equation}
E(0) \; \sim \; E'(0) \quad \Longrightarrow \quad E(t) \; \sim \;
E'(t)
\end{equation}
Technically one expresses this requirement by stating that the
evolution equation for the statistical operator is closed.

Let us sketch the proof: assume $E_{1}(0) \sim E_{2}(0)$ but
$E_{1}(t) \not\sim E_{2}(t)$, i.e. $\rho_{1}(t) \neq \rho_{2}(t)$;
this means that $E_{1}(t)$ and $E_{2}(t)$ do not belong to the
same equivalence class. $E_{1}(0)$ and $E_{2}(0)$ are ensembles
which are union of pure cases $|\psi_{i}\rangle$ with associated
weights $x_{i}$ and of pure cases $|\chi_{i}\rangle$ with weights
$y_{j}$, respectively. Then one can show\footnote{See, e.g.,
ref.\cite{qph} and references therein.} that it is possible to choose an
Hilbert space
${\mathcal K}$ and two orthonormal sets
$|\alpha_{i}\rangle$ and $|\beta_{j}\rangle$ in it such that, in
${\mathcal H} \otimes {\mathcal K}$ one has
\begin{equation}
\sum_{i} \sqrt{x_{i}}\,|\psi_{i}\rangle\otimes|\alpha_{i}\rangle
\; = \; \sum_{j}
\sqrt{y_{j}}\,|\chi_{j}\rangle\otimes|\beta_{j}\rangle \; = \;
|S+K\rangle.
\end{equation}
Moreover one can make the state $|S+K\rangle$ correspond to the
system $S$ and the auxiliary system $K$ (associated to the Hilbert
space ${\mathcal K}$) being located in two distant regions $R_{1}$
and $R_{2}$ respectively. The idea is then very simple: one
prepares an ensemble of systems $S+K$, all of which are in the
pure ``composite state" $|S+K\rangle$. Then, in region $R_{2}$ one
measures either the observable $A$ whose eigenstates are
$|\alpha_{i}\rangle$ or the observable $B$ whose eigenstates are
$|\beta_{j}\rangle$. Because of wavepacket reduction, the ensemble
of systems $S$ (in region $R_{1}$) becomes equivalent either to
the ensemble $E_{1}(0)$ or to the ensemble $E_{2}(0)$, according
to the measurement which has been performed on the auxiliary
system at the time $t=0$. These two ensembles are equivalent.
However, by hypothesis, they will no longer  be  equivalent at a
later time $t$, and  consequently the corresponding statistical
operators will be different: $\rho_{1}(t) \neq \rho_{2}(t)$. The
evolution of the two ensembles then yields, at subsequent times,
physically distinguishable situations in $R_{1}$. In this way the
observer in $R_{2}$ can let another observer in $R_{1}$ know what
measurement he has decided to perform on his (distant from
$R_{1}$) auxiliary system, and this allows faster than light
signaling. It is important to remark that the dynamics for the
statevector given by the dynamical reduction models which are the
subject of the present report actually leads to a closed evolution
equation for the statistical operator, a necessary condition,
according to Gisin's theorem, in order that they satisfy the no
faster than light constraint\footnote{A. Kent \cite{kenfl} has
proposed a dynamical reduction model which allows a simple
treatment of systems with identical constituents; however, this
can be easily shown to imply that equivalent ensembles can evolve
into inequivalent ones, with the possibility of faster than light
signaling, so that the proposal has to be disregarded.}.

The most interesting aspect of Gisin's result, from the point of
view we are interested in here is that, in a sense, it proves
``that nonlinearity requires stochasticity''. In fact, suppose we
consider a deterministic map of pure states into pure states
\begin{equation}
S_{t}\, |\psi(0)\rangle \quad \longrightarrow \quad
|\psi(t)\rangle
\end{equation}
Then a mixture of states $|\psi_{i}\rangle$ with weights $x_{i}$
evolves into a mixture of states $S_{t}\,|\psi_{i}\rangle$ with
the same weights. In particular
\begin{equation}
\sum_{i} x_{i}\, |\psi_{i}\rangle\langle\psi_{i}| \quad
\longrightarrow \quad \sum_{i} x_{i}\, S_{t}
|\psi_{i}\rangle\langle\psi_{i}| S_{t}^{\dagger}.
\end{equation}
Let us consider now, at the initial time $t=0$, two physically
different ensembles $E(0)$ with states $|\psi_{i}\rangle$ and
weights $x_{i}$ and $E'(0)$ with states $|\chi_{i}\rangle$ and
weights $y_{j}$, which are equivalent, i.e. $\rho(0) = \sum_{i}
x_{i}\, |\psi_{i}\rangle\langle\psi_{i}| = \sum_{j} y_{j}\,
|\chi_{i}\rangle\langle\chi_{i}|$. Two cases are then possible:
\begin{enumerate}
\item In at least one such case the evolved ensembles are
inequivalent. Then an unacceptable conflict with relativity
arises, as implied by Gisin's theorem.
\item The evolved ensembles are always equivalent. Then, by a
general theorem of Davies \cite{dav}, one can conclude that the
evolution given by $S_{t}$ must be linear and unitary.
\end{enumerate}
It is interesting to note that the above argument \cite{gisfl2}
shows that the attempt by Weinberg \cite{weifl} of introducing
nonlinear deterministic modifications of quantum mechanics turns
out to be unacceptable.

Taking the risk of being pedantic, we stress once more that from
our point of view the interest of Gisin's theorem lies in the fact
that it proves that if one wants to consider nonlinear
modifications of quantum mechanics one is forced to introduce
stochasticity and thus, in particular, the dynamics must allow the
transformation of ensembles corresponding to pure cases into
statistical mixtures.

\subsection{Brief history of dynamical reduction models}
\label{sec44}

We conclude the section with a brief review of the historical
development of dynamical reduction models. The history goes back
to the years 1970--1973, when G.C. Ghirardi, L. Fonda, A. Rimini and
T. Weber were working on quantum decay processes and in particular
on the possibility of deriving, within a quantum context, the
exponential decay law \cite{fons1,fons2}. Some features of their
approach have been extremely relevant for the subsequent elaboration of
the dynamical reduction program:
\begin{enumerate}
\item One deals with individual physical systems.
\item The statevector is supposed to suffer random processes
occurring at random times, leading to appropriate sudden changes
of it:
\[
|\psi\rangle \quad\longrightarrow\quad \frac{P_{u}|\psi
\rangle}{\|P_{u}|\psi\rangle\|};
\]
when $P_{u}$ is identified with the projection operator on the
unstable state manifold, one gets the desired result.
\item To make the treatment quite general (the apparatus does not
know which kind of unstable system it is testing) the authors have
been led to identify  the random processes with localization
processes of the relative coordinates of the decay fragments. Such
an assumption, combined with the peculiar ``resonant dynamics" of
an unstable system, yield completely in general the desired
result. The ``relative position basis" is the preferred basis of
this theory.
\item The authors have also applied their ideas to measurement
processes \cite{fons3}.
\item The final equation for the evolution at the ensemble level
is of the quantum dynamical semigroup type \cite{lin,gli} and has
a structure extremely similar to the final one of the GRW theory.
\end{enumerate}

In 1973 P. Pearle was the first to suggest to account for the
reduction process in terms of stochastic differential equations.
He pursued this line for various years. However, he did not
succeed in identifying the appropriate states to which the
dynamical equation should lead and consequently a mechanism whose
effectiveness could have been negligible for microsystems but
extremely relevant for the macroscopic ones. The lack of the
identification of the preferred basis, i.e. of what "is out
there", was the main obstacle for the success of the program.

The breakthrough is dated 1984. In that year the research program
suggested by Ghirardi, Rimini and Weber \cite{heid,grw,grw2}
started to be developed\footnote{Actually, even though
ref.\cite{heid} has been published in 1985, it appears among the
proceedings of a Conference held at Heidelberg in 1984.}. In these
papers, the first consistent and satisfactory model (QMSL) of
dynamical reductions, the one on which all subsequent attempts are
based, was presented and discussed in detail. The key assumption
is that each elementary constituent of any system is subjected, at
random times according to a Poisson distribution with mean
frequency $\lambda = 10^{-16}$ sec$^{-1}$, to random and
spontaneous localization processes around appropriate positions.
One assumes that the probability distribution of these processes
is such that hittings, i.e. spontaneous localizations around
specific points in space, occur with a higher probability at those
places where, in the standard quantum description, there is a
higher probability of finding the particle. As will be clear in
the following section, the above prescriptions can be satisfied by
introducing precise non--linear and stochastic elements in the
dynamics.

It is extremely easy also to convince oneself that the model
embodies the so--called {\bf trigger mechanism}, i.e. that the
spontaneous reductions become more and more frequent with
increasing the number of particles of the system under
consideration. The models thus ``has the property required ... of
having little impact for small systems but nevertheless
suppressing macroscopic superpositions'' \cite{bellqg}.

In the years 1989--90 the efforts of Ghirardi, Rimini, Weber on
the one side and of P. Pearle on the other, were joined together
and CSL (the Continuous Spontaneous Localizations model) was
developed \cite{csl0,csl}. CSL is based on a modified
Schr\"odinger equation containing new stochastic and  nonlinear
terms besides the standard hamiltonian. These new terms induce a
diffusion process for the statevector which acts like a continuous
hitting: it is precisely this diffusion process which is
responsible for the reduction of the statevector.

The next obvious step was to generalize dynamical reduction models
to relativistic quantum field theories. Various attempts have been
made \cite{p62, rel}, typically by considering models in which quantum
fields are locally coupled to scalar white noises. All the desired
properties of the non relativistic theory, the most important one
being the localization in space of macroscopic objects, hold also
in the relativistic case. However, the reduction mechanism induces
an infinite increase of the energy of physical systems per unit
time and unit volume: such models are then physically
unacceptable.

The work on relativistic dynamical reduction models is still in
progress; there are some hints that the difficulties so far
encountered can be solved by generalizing the coupling of the
fields to the stochastic noises \cite{ppo1, ppo2, bgrel}; such
generalizations however have still to be studied in detail.

\section{Quantum Mechanics with Spontaneous Localizations (QMSL)}
\label{sec5}

 The guiding lines which led Ghirardi, Rimini and
Weber to formulate the first consistent dynamical reduction model,
called Quantum Mechanics with Spontaneous Localizations (QMSL)
\cite{heid, grw}, are basically two:
\begin{enumerate}
\item The ``preferred basis'' --- the basis on which
reductions take place --- must be chosen in such a way to guarantee a
definite position in space to macroscopic objects.
\item The modified dynamics must have little impact on microscopic
objects, but at the same time must reduce the superposition of
different macroscopic states of macro--systems. There must then be
an ``amplification'' mechanism when moving from the micro to the
macro level.
\end{enumerate}
The section is devoted to the analysis of how these ideas have
been successfully implemented. We first list the axioms defining
the universal (i.e. valid both at the microscopic and at the
macroscopic level) dynamics of QMSL and we show, by resorting to a
simple example, how the reduction mechanism works; the general
discussion of statevector collapse is more easily accomplished
within the framework of the continuous model (CSL) analyzed in the
following sections, so it will be postponed. In subsection
\ref{sec52} we derive the dynamical evolution law of the
statistical operator, and we discuss the effect of the modified
dynamics on a free particle.

Subsections \ref{sec53} and \ref{sec54} are the core of the
section. In \ref{sec53} we specialize our analysis to the case of
a simple macroscopic system, a free macroscopic particle, proving
that the reducing terms of QMSL yield a classical description for
the macro--particle: this is how the macro--objectification
problem of Quantum Mechanics finds a natural solution within QMSL.
In subsection \ref{sec54} we deepen our analysis of macroscopic
systems, showing how their classical properties emerge from the
quantum properties of their constituents. This means that QMSL
embodies a single universal dynamics which takes into account both
the quantum properties of microscopic systems and the classical
properties of macroscopic objects.

In \ref{sec55} we discuss the possible numerical choices of the
two parameters appearing in the modified dynamics. They must be
chosen in such a way that all known properties of microscopic
quantum systems are not altered in any significant way, while the
classical properties of macroscopic systems must be guaranteed.
The final subsection is devoted to a mathematical review of
quantum dynamical semigroups, a class of evolution dynamics to
which the QMSL basic equation belongs.

\subsection{The assumptions of the model} \label{sec51}

Quantum Mechanics with Spontaneous Localizations \cite{heid,grw} is
based on the following assumptions:
\begin{enumerate}
\item Each particle of a system of $n$ distinguishable
particles experiences, with a mean rate $\lambda_{i}$, a sudden
spontaneous localization process.
\item In the time interval between two successive spontaneous
processes the system evolves according to the usual Schr\"odinger
equation.
\item The sudden spontaneous process is a localization described by:
\begin{equation} \label{rc}
|\psi\rangle \qquad \stackrel{\makebox{\tiny
localization}}{\longrightarrow} \qquad \frac{|\psi^{i}_{\bf
x}\rangle}{\| |\psi^{i}_{\bf x}\rangle\|},
\end{equation}
where $|\psi^{i}_{\bf x}\rangle = L^{i}_{\bf x}\, |\psi\rangle$.
$L^{i}_{\bf x}$ is a norm--reducing, positive, self--adjoint,
linear operator in the $n$--particle Hilbert space ${\mathcal H}$,
representing the localization of particle $i$ around the point
${\bf x}$.
\item The probability density for the occurrence of a localization
at point ${\bf x}$ is assumed to be
\begin{equation} \label{pd}
P_{i}({\bf x}) \quad = \quad \| |\psi^{i}_{\bf x}\rangle\|^{2}:
\end{equation}
This requires that
\begin{equation}
\int d^{3}x\, \left[ L^{i}_{\bf x}\right]^{2} \quad = \quad 1
\end{equation}
\item The localization operators $L^{i}_{\bf x}$ have been chosen
to have the form:
\begin{equation}
L^{i}_{\bf x} \quad = \quad\left(\frac{\alpha}{\pi}\right)^{3/4}\,
e^{\displaystyle -\frac{\alpha}{2}\, ({\bf q}_{i} - {\bf x})^{2}},
\end{equation}
${\bf q}_{i}$ being the position operator for particle $i$.
\end{enumerate}
Before going on, we want to make clear a fundamental conceptual
point. Here, and in the following, when we speak of ``particles''
we are are simply using the standard, somehow inappropriate,
quantum mechanical language. Within dynamical reduction models
particles are not point--like objects which move in space
following appropriate trajectories according to the forces they
are subjected to (as it is the case of, e.g., Bohmian mechanics).
In dynamical reduction models, like in standard quantum mechanics,
particles are represented just by the wavefunction which, in
general, is spread all over the space. As we will see, the basic
property of the models  analyzed here is that, when a large number
of ``particles'' interact with each other in appropriate ways,
they end up being always extremely well localized in space,
leading in this way to a situation which is perfectly adequate for
characterizing what we call a ``macroscopic object''. Thus,
strictly speaking \cite{bells} the are no particles in dynamical
reduction models at the fundamental level; there is simply a
microscopic, quantum, wave--like realm which gives rise to the
usual classical realm at the macroscopic level.

It is easy to see how the reduction mechanism works with the help
of the following simple example. Let us consider the superposition
of two gaussian functions, one centered around position $-a$ and
the other around position $a$ (for simplicity we deal with the
one--dimensional case):
\begin{equation}
\psi(z) \; = \; \frac{1}{\mathcal N}\, \left[ e^{\displaystyle
-\frac{\gamma}{2}\, (z + a)^{2}} \, + \; e^{\displaystyle
-\frac{\gamma}{2}\, (z - a)^{2}} \right];
\end{equation}
${\mathcal N}$ is a normalization constant. Let us assume that
$1/\sqrt{\gamma} \ll 1/\sqrt{\alpha}$ and $a \gg 1/\sqrt{\alpha}$:
the distance between the two gaussians is much greater than the
localization amplitude, while their width is much smaller than it.

Let us now consider a hitting centered around $a$; the
wavefunction changes as follows:
\begin{equation}
\psi(z) \; \longrightarrow \; \psi_{a}(z) = \frac{1}{{\mathcal
N}_{a}}\, \left[e^{\displaystyle -2\alpha a^{2}}\,
e^{\displaystyle -\frac{\gamma}{2}\, (z + a)^{2}} \, + \;
e^{\displaystyle -\frac{\gamma + \alpha}{2}\, (z - a)^{2}}
\right].
\end{equation}
We see that the gaussian function centered around position $-a$
has been exponentially suppressed with respect to the other term,
whose width is practically left unchanged: the new wavefunction
describes a particle very well localized around position $a$.
Moreover, it is easy to check that the probability for such a
hitting to occur, as given by (\ref{pd}), is almost equal to $1/2$
i.e. the quantum mechanical probability to find the particle in
$a$. Of course, a similar argument holds if the hitting takes
place around position $-a$.

Finally, let us consider the case in which a hitting takes place
in a region far from both gaussians, e.g. around the origin $0$.
In such a case it is easy to verify that the wavefunction does not
change in any appreciable way. The reduction to a localized state
does not occur. However, the probability for such an hitting to
occur is extremely small, about $e^{-\alpha a^{2}}$. Concluding,
reductions are more likely to occur where the probability to find
a particle, according to the standard interpretation of the
wavefunction, is greater.

\subsection{The equation for the statistical operator}
\label{sec52}

Within QMSL, the reduction mechanism transforms pure states into
statistical mixtures; we can then resort to the statistical
operator formalism to investigate specific physical consequences
of the theory, such as the time evolution of the mean values of
dynamical variables. However, it is important to stress once more that the
reduction mechanism must be effective at the wavefunction level,
as it should be clear according to the analysis of section
\ref{sec41}, and as we have already proved with the help of the
very simple example of the previous section (the general situation
will be discussed in the next section).

Let us consider a single particle. Suppose it suffers a hitting
process: its wavefunction $|\psi\rangle$ changes it into the new
wavefunction $|\psi_{\bf x}\rangle$. We do not know where the
hitting occurs, but only the probability for it to occur around
position ${\bf x}$. Accordingly, the pure state is transformed
into the following statistical mixture:
\begin{eqnarray}
|\psi\rangle\langle\psi| & \longrightarrow & \int d^{3} x\, P({\bf
x})\, \frac{|\psi_{\bf x}\rangle\langle\psi_{\bf x}|}{\|
|\psi_{\bf x}\rangle \|^{2}} \; = \nonumber \\ & = & \int d^{3}
x\, L^{i}_{\bf x} |\psi\rangle\langle\psi| L^{i}_{\bf x} \; \equiv
\; T[|\psi\rangle\langle\psi|].
\end{eqnarray}
Of course, if the initial state of the particle is not pure but a
statistical mixture given by the operator $\rho$, the effect of a
hitting process is the same as the one described above: $\rho$
changes into $T[\rho]$.

We  derive now the evolution equation for $\rho(t)$. In a time
interval $dt$, the statistical operator evolves in the following
way: since the localization mechanism is Poissonian, there is a
probability $\lambda\,dt$ for a hitting to occur during that time
interval, in which case $\rho$ changes to $T[\rho]$, and a
probability $1 - \lambda\,dt$ for no hittings to occur so that the
statistical operator evolves according to the usual Schr\"odinger
equation:
\[
\rho (t + dt) \quad = \quad (1 - \lambda\,dt) \left[ \rho(t) -
\frac{i}{\hbar}\, [H, \rho(t)] dt\right] \; + \; \lambda\, dt\,
T[\rho(t)],
\]
i.e.:
\begin{equation} \label{meqmsl}
\frac{d}{dt}\, \rho(t) \quad = \quad -\frac{i}{\hbar}\, [H,
\rho(t)] \; - \; \lambda\left(\rho(t)\, - \, T[\rho(t)] \right).
\end{equation}
This is the {\bf master equation} of QMSL, describing the quantum
evolution of a single particle which undergoes random localization
processes; it has a quantum--dynamical--semigroup structure, which
we will discuss at the end of the section.

In the coordinate representation one has, according to assumption
5:
\begin{equation} \label{maup}
\langle{\bf q}'|T[\rho]|{\bf q}''\rangle \; = \; e^{\displaystyle
-\frac{\alpha}{4}\, ({\bf q}' - {\bf q}'')^{2}} \langle{\bf
q}'|\rho|{\bf q}''\rangle.
\end{equation}
Since, owing to equation (\ref{maup}),
\begin{equation}
\langle{\bf q}|T[\rho]|{\bf q}\rangle \; = \; \langle{\bf
q}|\rho|{\bf q}\rangle,
\end{equation}
equation (\ref{meqmsl}) is obviously trace preserving. Moreover,
using equation (\ref{meqmsl}), it can be easily proved that
\begin{equation}
\frac{d}{dt}\, \makebox{Tr}[\rho^{2}] \; < \; 0.
\end{equation}
This implies that the dynamical evolution transforms pure states
into statistical mixtures.

Let us now consider equation (\ref{meqmsl}) in the case in which
$H$ is the Hamiltonian for a free particle; for simplicity we work
in one dimension. In the coordinate representation we get
\begin{eqnarray} \label{mepr}
\frac{\partial}{\partial t}\, \langle q'|\rho(t)| q''\rangle & = &
\frac{i\hbar}{2m}\, \left[\frac{\partial^{2}}{\partial {q'}^{2}} -
\frac{\partial^{2}}{\partial {q''}^{2}}\right] \langle q'|\rho(t)|
q''\rangle
\nonumber \\
& & \nonumber \\
& & -\,\lambda \left[ 1 - e^{\displaystyle - \frac{\alpha}{4}\,
(q' - q'')^{2}} \right] \langle q'|\rho(t)| q''\rangle.
\end{eqnarray}
One can express the solution of the above equation satisfying
given initial conditions in terms of the solution $\langle q' |
\rho_{\makebox{\tiny Sch}}(t) | q''\rangle$ of the pure
Schr\"odinger equation ($\lambda = 0)$ satisfying the same initial
conditions, according to \cite{grw}:
\begin{equation} \label{gfh}
\langle q'|\rho(t)| q''\rangle \; = \;
\frac{1}{2\pi\hbar}\int_{-\infty}^{+\infty} dk
\int_{-\infty}^{+\infty} dy\, e^{\displaystyle
-\frac{i}{\hbar}\,ky} F(k, q'-q'', t)\; \langle q' + y |
\rho_{\makebox{\tiny Sch}}(t) | q'' + y \rangle
\end{equation}
where
\begin{equation} \label{rcf}
F(k,q,t) \; = \; e^{\displaystyle -\lambda t + \lambda
\int_{0}^{t} d\tau\, e^{\displaystyle -\frac{\alpha}{4}\, \left( q
- \frac{k\tau}{m}\right)^{2}}}.
\end{equation}
The Hermitian symmetry of $\rho(t)$ follows from the property
$F(k,q,t) = F(-k, -q, t).$

To understand the dynamical evolution described by equation
(\ref{mepr}) we can evaluate the mean values, spreads, and
correlations for the position and momentum operators for all
times. In the considered case, it turns out \cite{grw} that these
variables are related to those of the pure Schr\"odinger evolution
by:
\begin{eqnarray}
\llangle q \rrangle & = & \llangle q \rrangle_{\makebox{\tiny
Sch}} \label{pedc}\\
\llangle p \rrangle & = & \llangle p \rrangle_{\makebox{\tiny
Sch}} \label{pedc2} \\
\{ q \} & \equiv & \llangle [ q - \llangle q \rrangle ]^{2}
\rrangle \; = \; \{ q \}_{\makebox{\tiny Sch}} \; + \;
\frac{\alpha\lambda\hbar^{2}}{6m^{2}}\, t^{3}
\label{pedc3}\\
\{ qp \} & \equiv & \llangle [(q - \llangle q \rrangle) (p -
\llangle p \rrangle)]_{\makebox{\tiny sym}} \rrangle \; = \; \{ qp
\}_{\makebox{\tiny Sch}} \;
+ \; \frac{\alpha\lambda\hbar^{2}}{4m}\, t^{2} \label{sym}\\
\{ p \} & \equiv & \llangle [ p - \llangle p \rrangle ]^{2}
\rrangle \; = \; \{ p \}_{\makebox{\tiny Sch}} \; + \;
\frac{\alpha\lambda\hbar^{2}}{2}\, t. \label{sedc}
\end{eqnarray}
In equation (\ref{sym}) we have denoted by
$[\,\cdot\,]_{\makebox{\tiny sym}}$ the Hermitian part of the
quantity in square brackets. The shorthands $\{ q \}$, $\{ qp \}$,
$\{ p \}$ have been introduced to simplify the notation of the
formal developments of the following sections. We note that the
mean values are not affected by the non Hermitian term in equation
(\ref{meqmsl}). For what concerns spreads and correlation, the
corrections depend only on the combination $\alpha\lambda$ of the
parameters $\alpha$ and $\lambda$.

\subsection[non Hamiltonian dynamics of a free macroscopic
particle]{Discussion of the non Hamiltonian dynamics for a free
macroscopic particle} \label{sec53}

In this subsection we begin the analysis of the effects of the
modified dynamics on macroscopic systems; for simplicity, we first
consider the case of a free macroscopic particle. Its time
evolution is embodied into equation (\ref{mepr}), where the mass
$m$ now has the order of magnitude of the mass of a macroscopic
object.

We remark that the standard quantum dynamics, in the case of a
free particle, induces for the mean values $\llangle q
\rrangle_{\makebox{\tiny Sch}}$ and $\llangle p
\rrangle_{\makebox{\tiny Sch}}$ exactly the classical evolution.
Moreover, for any reasonable choice of the initial spreads of the
position $\Delta\, q = \sqrt{\{ q\}}$ and of the momentum
$\Delta\, p = \sqrt{\{ p\}}$, the increase of $\Delta\, q$ when
time elapses, in virtue of the smallness of the Planck constant
and of the large value of the mass of a macroscopic object, can be
completely disregarded  for all interesting times. However, the
recognition of this fact does not exhaust the problem of the
derivation of the classical behaviour of a macroscopic object from
quantum principles, since problems remain open when linear
superpositions of macroscopically distinguishable states can
occur. In such cases, as already discussed, a satisfactory
classical description would require that the statistical ensemble
decomposes into a statistical mixture of macroscopically
distinguishable states. Let us discuss the above points within the
framework of the non Hamiltonian dynamics introduced in the
previous subsections.

First of all we can observe that Ehrenfest's theorem holds true
also for the modified dynamics. In fact, for any dynamical
variable $X$, which is a function of the operator $q$ only, it is
easily shown that
\begin{equation}
\makebox{Tr} \{ X(q)\,T[\rho]\} \quad = \quad \makebox{Tr}[X(q)\,
\rho].
\end{equation}
This in turns implies
\begin{equation}
\frac{d}{dt}\, \llangle X(q) \rrangle \; = \; \makebox{Tr} \left[
X(q)\, \frac{d\,\rho}{dt} \right] \; = \; - \frac{i}{\hbar} \,
\makebox{Tr} \{ X(q)\, [ H,\, \rho]\},
\end{equation}
as it happens for the Schr\"odinger evolution. It follows that
\begin{equation} \frac{d}{dt}\, \llangle q \rrangle \quad = \quad
\frac{1}{m}\, \llangle p \rrangle.
\end{equation}
For the operator $p$ one finds
\begin{equation}
\makebox{Tr} \{ p\, T[\rho]\} \quad = \quad \makebox{Tr} \{ p\,
\rho\}.
\end{equation}
Then, if $H = p^{2}/2m + V(q)$, we have
\begin{equation}
\frac{d}{dt}\, \llangle p \rrangle \quad = \quad - \LLangle
\frac{\partial V}{\partial q} \RRangle.
\end{equation}
In accordance with this property, equations (\ref{pedc}) and
(\ref{pedc2}) show that in the case of a free macroscopic
particle\footnote{Actually, of any free particle.} the mean values
of position and momentum are not affected by the non--Hamiltonian
terms. On the contrary, in the expression for the spreads
additional terms appear. These terms increase with time, so that
one can identify a characteristic time interval $T$ during which
they remain small with respect to those expressing the
Schr\"odinger evolution. $T$ turns out to be of the order of the
smaller of the two times $T_{1}$ and $T_{2}$ given by
\begin{equation} \label{nlhadun}
T_{1} \; = \; \left[\frac{6m^{2}(\Delta q_{\makebox{\tiny
Sch}})^{2}}{\alpha\lambda\hbar^{2}} \right]^{1/3} \qquad \quad
T_{2} \; = \; \frac{2(\Delta p_{\makebox{\tiny Sch}})^{2}}{\alpha
\lambda\hbar^{2}}.
\end{equation}
For the time interval $T$ the spreads given by equations
(\ref{pedc3})--(\ref{sedc}) coincide practically with the
Schr\"odinger values, which in turn are negligible for any
reasonable choice of their initial values. We shall discuss below
the values taken by $T$ when the parameters of the model are
appropriately chosen.

The fact that $\Delta \, q^{2}$ and $\Delta \, p^{2}$ are very
close to the Schr\"odinger values for an appropriate time interval
is strictly related to the small influence of the non--Hamiltonian
term on the matrix elements of the statistical operator $\langle
q'|\rho| q''\rangle$ when $| q' - q'' | \ll  1/\sqrt{\alpha}$. On
the contrary, the non--Hamiltonian dynamics has a drastic effect
on the off--diagonal elements when $| q' - q'' | \geq
1/\sqrt{\alpha}$. This can be easily understood by observing that
the properties of the function $F(k,q,t)$ are remarkably different
in the two cases $q=0$ and $q \neq 0$. In fact, when $q=0$ the
integral in the exponent in equation (\ref{rcf}) for sufficiently
small $t$ behaves like $t$, yielding the cancellation of the
factor $e^{-\lambda t}$ and making $F(k,0,t)$ very near to $1$.
Since $F=1$ implies $\langle q'|\rho(t)|q''\rangle = \langle
q'|\rho_{\makebox{\tiny Sch}}(t)|q''\rangle$, this shows that the
(almost) diagonal matrix elements of the statistical operator in
the coordinate representation are practically unaffected for an
appropriate time interval  by the non--Hamiltonian term in the
evolution equation. On the contrary, for $q\neq 0$, the integral
in equation (\ref{rcf}) cannot cancel, even for small times, the
damping factor $e^{-\lambda t}$, so that the off--diagonal
elements are rapidly suppressed.

To make these statements more precise we derive two inequalities
for the function $F(k,q,t)$ for the two cases $q=0$ and $q>0$.
\begin{description}
\item[a) $q=0$] Since
\begin{equation}
\frac{1}{t}\int_{0}^{t} d\tau\, e^{\displaystyle - \frac{\alpha
k^{2}\tau^{2}}{4m^{2}}} \; \geq \; e^{\displaystyle - \frac{\alpha
k^{2}t^{2}}{4m^{2}}},
\end{equation}
it follows that
\begin{equation} \label{hdgxTJ}
F(k,0,t) \; > \; e^{\displaystyle -\lambda t \left( 1 -
e^{\displaystyle - \alpha k^{2} t^{2}/4m^{2}} \right)} \; \geq \;
1 - \frac{\alpha\lambda k^{2} t^{3}}{4m^{2}},
\end{equation}
the last inequality being useful for $\alpha\lambda k^{2}
t^{2}/4m^{2} < 1$. We then have
\begin{equation} \label{lkjh}
1 - F(k,0,t) \quad \leq \quad \frac{\alpha\lambda k^{2}
t^{3}}{4m^{2}}.
\end{equation}
\item[b) $q>0$] The function $F$ can be written as
\begin{equation}
F(k,q,t) \; = \; e^{\displaystyle -\lambda t \left[ 1 - h\left(
(\sqrt{\alpha}/2)\frac{kt}{m}, (\sqrt{\alpha}/2)q\right) \right]},
\end{equation}
where
\begin{equation}
h(x,y) \quad = \quad \frac{1}{x}\int_{-y}^{x-y}dz\,
e^{\displaystyle - z^{2}}.
\end{equation}
The function $h(x,y)$ is the mean value of $e^{-z^{2}}$ on the
interval $(-y, x-y)$. Clearly one has
\begin{equation}
h(x,y) \; < \; h(y,y) \; = \; h(2y,y)
\end{equation}
for $x < y$, and
\begin{equation}
h(x,y) \; < \; h(2y,y)
\end{equation}
for $x > 2y$. For $y < x < 2y$ one finds
\begin{equation}
h(x,y) \; < \; \frac{1}{x}\int_{-y}^{y}dz\, e^{\displaystyle -
z^{2}} \; < \; \frac{1}{y}\int_{-y}^{y}dz\, e^{\displaystyle -
z^{2}} \; = \; 2\,h(2y,y),
\end{equation}
so that, on the whole,
\begin{equation}
h(x,y) \; < \; 2\, h(2y,y) \; = \; \frac{\sqrt{\pi}}{y}\,
\makebox{erf} (y).
\end{equation}
In turn the function $F$ obeys the inequality
\begin{equation} \label{ldem}
F(k,q,t) \quad < \quad e^{\displaystyle -\lambda\beta t},
\end{equation}
where
\begin{equation} \label{gupnld}
\beta \; =\; 1 \, - \, \frac{\sqrt{\pi}}{(\sqrt{\alpha}/2)q}\,
\makebox{erf}\,[(\sqrt{\alpha}/2)q].
\end{equation}
\end{description}
Inequalities (\ref{hdgxTJ}) and (\ref{ldem}) prove the correctness
of our previous statements about the behaviour of $F(k,q,t)$ and,
consequently, about the features of the dynamics concerning the
diagonal and off--diagonal elements of the density matrix in
configuration space.

We come back to the discussion of the diagonal elements of the
statistical operator, using the just derived inequalities
(\ref{hdgxTJ}) and (\ref{ldem}) for the function $F(k,q,t)$. From
equation (\ref{gfh}) we see that
\begin{eqnarray} \label{sgasd}
\langle q| \rho_{\makebox{\tiny Sch}}(t)| q\rangle - \langle q|
\rho(t)| q\rangle & = & \frac{1}{2\pi\hbar}
\int_{-\infty}^{+\infty} dk\, [1-F(k,0,t)]
\int_{-\infty}^{+\infty} dy\, e^{\displaystyle -\frac{i}{\hbar} \,
ky} \nonumber \\ & & \langle q+y|\rho_{\makebox{\tiny Sch}}(t)|
q+y\rangle.
\end{eqnarray}

To illustrate the implications of this equation, we discuss a
simple example. Suppose that $\langle q| \rho_{\makebox{\tiny
Sch}}(t)| q\rangle$ is a mixture of Gaussian terms whose spreads
are $\Delta_{i}$ with minimum $\Delta_{0}$. Then the Fourier
transform appearing in equation (\ref{sgasd}) yields terms
containing Gaussian factors $e^{-\Delta_{i}^{2}k^{2}/2\hbar^{2}}$,
whose maximum width is $\hbar/\Delta_{0}$, so that the integral in
$k$ is concentrated in a region $|k| < \hbar/\Delta_{0}$.
Inequality (\ref{lkjh}) shows then that the integrand in equation
(\ref{sgasd}) contains a factor smaller than
$(\alpha\lambda\hbar^{2}/4m\Delta_{0}^{2}) t^{3}$. The condition
$t \ll T_{1}$, the time $T_{1}$ being given by equation
(\ref{nlhadun}), implies
\begin{equation}
\langle q| \rho(t)| q\rangle \quad \simeq \quad \langle q|
\rho_{\makebox{\tiny Sch}}(t)| q\rangle.
\end{equation}
Obviously this result holds for those matrix elements which are
appreciably different from zero.

For the off--diagonal elements we consider the case $q' > q''$.
Obviously the same results are valid for $q' < q''$ due to the
Hermitian symmetry of $\rho(t)$. Inequality (\ref{ldem}) gives,
for $q' - q'' > 2\sqrt{\pi/\alpha}$, a significant bound on $F$
independent of $k$. This shows that expression (\ref{gfh}) for
$\langle q'|\rho(t)| q''\rangle$ contains an exponentially damping
factor whose lifetime is $\tau  = 1/\lambda\beta$, a consequence
of the fact that (in a time interval of the order of $\tau$)
linear superpositions of states separated by distances larger than
the characteristic localization distance $1/\sqrt{\alpha}$ are
transformed into one or the other of their terms.

As we shall see, one can choose the parameters $\lambda$ and
$\alpha$ in such a way that the time $T$ = min$\,(T_{1}, T_{2})$
is very large and $\tau$ extremely small so that we can conclude
that the modified dynamics leads to an evolution agreeing with the
classical one in the case of a macroscopic object and overcomes
the problems arising from linear superpositions of states
localized in far apart regions.

\subsection{Macroscopic dynamics from the microscopic one}
\label{sec54}

In the previous subsection we have introduced a
non--purely--Hamiltonian dynamics to describe the motion of a
macroscopic particle and we have outlined how this modification
can be used to overcome some of the difficulties  in the
description of such objects. However, macroscopic objects are
composite systems and standard quantum mechanics gives definite
prescriptions for their description. It is an important feature of
quantum mechanics that, under suitable conditions, the internal
and the center--of--mass  motions of the composite systems
decouple and, moreover, that the equation of motion for the center
of mass is formally identical to the equation prescribed by the
theory for the description of a single particle. Here we want to
investigate whether it is possible to obtain the
non--purely--Hamiltonian dynamics for macroscopic particles
described in the previous subsections from a modification of the
standard quantum dynamics for their microscopic constituents. If
such a modification leaves practically unaltered the behaviour of
microscopic systems as accounted for by quantum mechanics we can
say we have laid the foundations of a possible unified description
able to account for both the quantum and the classical behaviours
of  microscopic and macroscopic systems, respectively.

In section \ref{sec51} we have assumed that the localization
process $T[\,\cdot\,]$ occurs individually for each constituent of
a many--particle system. We consider now a system of $N$ particles
in one dimension. Assuming that the accuracy of the localizations
is the same for all constituents, the evolution equation for the
composite system is
\begin{equation} \label{pclm}
\frac{d}{dt}\, \rho(t) \quad = \quad -\frac{i}{\hbar}\, [H,
\rho(t)] \; - \; \sum_{i=1}^{N} \lambda_{i} \left( \rho(t) \, - \,
T_{i}[\rho(t)]\right),
\end{equation}
where
\begin{equation} \label{nhdn}
T_{i}[\rho(t)] \; = \; \sqrt{\frac{\alpha}{\pi}}\,
\int_{-\infty}^{+\infty} dx\, e^{\displaystyle -
\frac{\alpha}{2}\, (q_{i} - x)^{2}} \rho\, e^{\displaystyle -
\frac{\alpha}{2}\, (q_{i} - x)^{2}},
\end{equation}
$q_{i}$ being the position operator for the $i$th particle of the
system (throughout the subsection, we will keep working in 1
dimension).

It is worthwhile to illustrate the physical consequences of the
above equation for the important conceptual problem of the
possible occurrence of linear superpositions of states
corresponding to different locations of a macroscopic object. Such
a situation occurs, for instance, in the quantum theory of
measurement, in connection with possible macroscopically different
pointer positions. With reference to such a case we consider the
linear superposition $\psi = \psi_{1} + \psi_{2}$ of two states
corresponding to two different pointer positions. We remark that
in the case under discussion there is a macroscopic number $N$  of
particles which are located in macroscopically different positions
when the state is $\psi_{1}$ or $\psi_{2}$ (to be precise, in our
model this means located at a distance larger than
$1/\sqrt{\alpha}$). If a spontaneous localization process takes
place for one of such particles, this particle is constrained to
be either in the spatial region which it occupies when the state
is $\psi_{1}$, or in the one corresponding to $\psi_{2}$. The
linear superposition is consequently transformed into a
statistical mixture of states $\psi_{1}$ and $\psi_{2}$. Since the
number of differently located particles is $N$, the reduction of
states $\psi_{1}$ and $\psi_{2}$ occurs with a rate which is
amplified by a factor $N$ with respect to the one, $\lambda_{i}$,
which characterizes the elementary spontaneous localizations.

The model yields therefore a natural solution to the puzzling
situation originating from the occurrence of linear superpositions
of differently located states. These considerations, however, do
not exhaust the problems to be discussed. In fact, we must still
check that the modification of the dynamics for the microscopic
constituents does not imply physically unacceptable consequences
for the dynamics of the system as a whole. Actually, according to
the previous discussions, we would like to have for the
macroscopic object a dynamical equation of the type considered in
section \ref{sec53}. To discuss this point, let us introduce the
center  of mass and relative motion position operators $Q$ and
$r_{j}$ ($j=1,2,...,N-1$), related to the operators $q_{i}$ by
\begin{equation}
q_{i} \quad = \quad  Q \; + \; \sum_{j=1}^{N-1} c_{ij}\, r_{j}.
\end{equation}
Equation (\ref{pclm}), when the Hamiltonian $H$ can be split into
the sum of the center   of mass and internal motion parts $H_{Q}$
and $H_{r}$ acting in the respective state spaces, reads
\begin{equation} \label{pclm2}
\frac{d}{dt}\, \rho(t) \quad = \quad -\frac{i}{\hbar}\, [H_{Q},
\rho(t)] \; -\frac{i}{\hbar}\, [H_{r}, \rho(t)] \; - \; \sum_{i}
\lambda_{i} \left( \rho(t) \, - \, T_{i}[\rho(t)]\right),
\end{equation}
where the operator $T_{i}[\rho]$ can now be written as
\begin{equation}
T_{i}[\rho] \; = \; \sqrt{\frac{\alpha}{\pi}}
\int_{-\infty}^{+\infty} dx\, e^{\displaystyle -\frac{\alpha}{2}
\left[Q + \sum_{j=1}^{N-1} c_{ij} r_{j} - x \right]^{2}} \rho\,
 e^{\displaystyle -\frac{\alpha}{2}
\left[Q + \sum_{j=1}^{N-1} c_{ij} r_{j} - x \right]^{2}}.
\end{equation}
The dynamical evolution of the {\bf center of mass}  of the system
is described by the statistical operator
\begin{equation}
\rho_{Q} \quad = \quad \makebox{Tr}^{(r)}[\rho],
\end{equation}
obtained by taking the partial trace on the internal degrees of
freedom of the statistical operator $\rho$ for the complete
$N$--particle system. Taking the $r$ trace of the operation
$T_{i}[\rho]$ one gets
\begin{eqnarray}
\lefteqn{ \int dr_{1}\ \ldots \, dr_{N-1}\,
\sqrt{\frac{\alpha}{\pi}} \int_{-\infty}^{+\infty} dx \;
e^{\displaystyle -\frac{\alpha}{2} \left[Q + \sum_{j=1}^{N-1}
c_{ij} r_{j} - x \right]^{2}}\,
\cdot\qquad\qquad} \nonumber \\
& & \qquad\qquad \cdot\; \langle r_{1} \ldots r_{N-1}|\rho|r_{1}
\ldots r_{N-1}\rangle\; e^{\displaystyle -\frac{\alpha}{2} \left[Q
+ \sum_{j=1}^{N-1} c_{ij} r_{j} - x \right]^{2}},
\end{eqnarray}
so that, by shifting the integration variable $x$ by the amount
$\sum_{j} c_{ij} r_{j}$, one finds
\begin{equation} \label{dcfp}
\makebox{Tr}^{(r)}(T_{i}[\rho])\; = \; T_{Q}[
\makebox{Tr}^{(r)}(\rho)],
\end{equation}
where
\begin{equation}
T_{Q}[\,\cdot\,] \; = \; \sqrt{\frac{\alpha}{\pi}}\,
\int_{-\infty}^{+\infty} dx\, e^{\displaystyle -
\frac{\alpha}{2}\, (Q - x)^{2}} [\,\cdot\,]\; e^{\displaystyle -
\frac{\alpha}{2}\, (Q - x)^{2}}.
\end{equation}
If one takes the $r$ trace of equation (\ref{pclm2}) one then gets
\begin{equation} \label{nht}
\frac{d}{dt}\, \rho_{Q}(t) \quad = \quad -\frac{i}{\hbar}\,
[H_{Q}, \rho_{Q}] \; - \; \sum_{i} \lambda_{i} \left(\rho_{Q}\, -
\, T[\rho_{Q}]\right).
\end{equation}
We have thus shown that the equation describing the reduced
dynamics of the center  of mass has exactly the same form of
equation (\ref{meqmsl}), the parameter $\lambda$ being replaced by
the sum of the $\lambda_{i}$'s for the individual constituents of
the many--body system. This is a direct consequence of the formal
property (\ref{dcfp}).

It is worthwhile stressing that the non--Hamiltonian term in
equation (\ref{nht}) is directly generated by the analogous terms
of equation (\ref{pclm}) and is not due to the elimination of the
internal degrees of freedom. In fact, if within the standard
formalism one considers a composite system with an Hamiltonian $H
= H_{Q} + H_{r}$, the reduced dynamics for the center  of mass
motion is necessarily Hamiltonian, and therefore it allows for the
occurrence of linear superpositions of widely separated  states of
the center   of mass. To avoid this, one could couple the system
to some other system whose dynamics is then eliminated
\cite{krauqm}. This, however, gives rise to a chain process when
larger and larger external parts are included. If one wants to
reach a point where linear superpositions of far--away states
cannot actually occur, one has to break this chain in an arbitrary
way. In our approach the non--Hamiltonian dynamics for a
macroscopic object is induced by a basic non--Hami1tonian dynamics
for its microscopic constituents.

Let us now investigate briefly the effect of the modified dynamics
on the {\bf relative variables}. From a physical point of view it
is particularly simple and interesting to consider the case in
which the internal motion Hamiltonian gives rise to a sharp (with
respect to $1/\sqrt{\alpha}$) localization of the internal
coordinates, as it happens, for an appropriate choice of $\alpha$, in
an insulating solid. In such a case it is evident that localizing
with an accuracy $1/\sqrt{\alpha}$ any one of the points of the
almost rigid structure of the solid induces a corresponding
localization of the center  of mass. In this situation something
more can be proved, i.e., that the internal and the center  of
mass motion decouple almost exactly and the internal motion is not
affected by the non--Hamiltonian terms in (\ref{pclm}). To be
precise, we assume that the matrix elements $\langle
Q',r'|\rho|Q'',r''\rangle$ are non--negligible only when the
conditions
\begin{equation} \label{c11}
| \sum_{j=1}^{N-1} c_{ij}\, {r'}_{j} - a_{i} | \; \ll
\frac{1}{\sqrt{\alpha}} \qquad\quad | \sum_{j=1}^{N-1} c_{ij}\,
{r''}_{j} - a_{i} | \; \ll \frac{1}{\sqrt{\alpha}}, \qquad\quad i
= 1, \ldots, N
\end{equation}
are satisfied, $a_{i}$ being the equilibrium position of
constituent $i$ relative to the center  of mass. Since conditions
(\ref{c11}) imply
\begin{equation} \label{ac}
| \sum_{j=1}^{N-1} c_{ij} ({r'}_{j} - {r''}_{j}) | \; \ll
\frac{1}{\sqrt{\alpha}} \qquad \quad i = 1, \ldots, N,
\end{equation}
$\langle Q',r'|\rho|Q'',r''\rangle$ is negligibly small unless
condition (\ref{ac}) is satisfied. From the definition
(\ref{nhdn}) one gets
\begin{eqnarray} \label{vslp}
\langle Q',r'|T_{i}[\rho]|Q'',r''\rangle & = &
\sqrt{\frac{\alpha}{\pi}} \int_{-\infty}^{+\infty} dx\,
e^{\displaystyle -\frac{\alpha}{2} \left[Q' + \sum_{j=1}^{N-1}
c_{ij} {r'}_{j} - x \right]^{2}}\cdot \nonumber \\
& & \cdot\; \langle Q',r'|\rho|Q'',r''\rangle\; e^{\displaystyle
-\frac{\alpha}{2} \left[Q'' + \sum_{j=1}^{N-1}
c_{ij} {r''}_{j} - x \right]^{2}}\; = \nonumber \\
& & \nonumber \\ & = & e^{\displaystyle -\frac{\alpha}{4} \left[Q'
- Q'' + \sum_{j=1}^{N-1} c_{ij} ({r'}_{j} - {r''}_{j})
\right]^{2}}
\langle Q',r'|\rho|Q'',r''\rangle. \nonumber \\
& &
\end{eqnarray}
The exponential factor appearing in the last line of equation
(\ref{vslp}) is a Gaussian in the variable $Q' - Q''$ displaced by
the amount $\sum_{j} c_{ij} ({r'}_{j} - {r''}_{j})$. Because of
equation (\ref{ac}), the displacement of the Gaussian can be
neglected with respect to its width, so that in this approximation
\begin{equation} \label{cprc}
T_{i}[\rho] \quad = \quad T_{Q}[\rho].
\end{equation}
The physical meaning of equation (\ref{cprc}) is that, as
foreseen, a localization of a single constituent of a rigid system
is equivalent to a localization of the center  of mass. Equation
(\ref{pclm2}) shows that, if the initial statistical operator has
the form of a direct product $\rho_{Q}\, \rho_{r}$, it remains of
the same type, and the statistical operators $\rho_{r}$ and
$\rho_{Q}$ obey the equations
\begin{equation} \label{yjth}
\frac{d}{dt}\, \rho_{r} \quad = \quad -\frac{i}{\hbar}\, [H_{r},
\rho_{r}]
\end{equation}
and (\ref{nht}) respectively. We conclude that in the considered
case the internal and the center of mass motions decouple, the
internal motion of the solid being unaffected by the localization
process introduced in equation (\ref{pclm}) and the center of mass
motion being affected by such a process with a characteristic rate
equal to the sum of the rates for all single constituents.

The considerations which can be done in the case of an almost
rigid body ensure that the density operator retains the form $\rho
= \rho_{Q} \, \rho_{r}$, when it is initially of this form.
Therefore, the non Hamiltonian terms of equation (\ref{nht}),
which appear as a consequence of the localization mechanism,
expresses meaningfully the destruction of the long--distance
coherence, as they entail the suppression of the off--diagonal
elements of $\rho_{Q}$. The situation we have just discussed can
be considered, with some idealization, typical of the case in
which one is dealing with a macroscopic body.

To conclude this subsection we observe that if one assumes for
simplicity that the localization rates $\lambda_{i}$ of all
microscopic (e.g., atomic) constituents of a macroscopic body are
of the same magnitude ($\lambda_{i} = \lambda_{\makebox{\tiny
micro}}$), the center of mass is affected by the same process with
a rate $\lambda_{\makebox{\tiny macro}} = N\lambda_{\makebox{\tiny
micro}}$, where $N$ is of the order of Avogadro's number. As we
shall see in the next subsection, this will allow us to choose the
parameters $\lambda_{\makebox{\tiny micro}}$ and $\alpha$ in such
a way that standard quantum mechanics holds exactly for extremely
long times for microscopic systems, while for a macroscopic body
possible linear superpositions of far--away states are rapidly
suppressed, the dynamical evolution of the center--of--mass
position is the classical one and the internal structure remains
unaffected.

\subsection{Choice of the parameters and its consequences}
\label{sec55}

A crucial feature of the point of view which has been adopted in
QMSL, i.e., that of considering all elementary constituents of any
system as subjected to localizations, consists in the fact that
one can choose the parameters of the elementary processes in such
a way that (i) the quantum--mechanical predictions for microscopic
systems are valid for extremely long times, (ii) the dynamics of a
macroscopic object, when it is consistently derived from that of
its microscopic constituents, turns out to coincide with the
classical one for a sufficiently long time interval, (iii) the
suppression of long--distance coherence for macroscopic objects be
effective enough to imply that, after a microscopic system has
triggered a measuring apparatus, the dynamical evolution leads to
the reduction of the wavepacket with well--defined pointer
positions.

To give orientative indications on the numerical values of the
parameters appearing in our model, we start by remarking that, as
it is clear from the formulas of the previous subsections, all
physically significant effects of the modified dynamics for a
macroscopic object are governed (for a remarkably large range of
variability of these parameters) by the product
$\alpha\lambda_{\makebox{\tiny macro}}$. For the choice of the
parameter $\alpha\lambda_{\makebox{\tiny macro}}$ we have some
important criteria which must be taken into account. First of all we want
the mean time $1/\lambda_{\makebox{\tiny macro}}$ elapsing between
two successive localizations to be such that the transition to
statistical mixtures for states spreading over distances larger
than the localization distance $1/\sqrt{\alpha}$ takes place in a
very small fraction of a second. A further requirement which has
to be taken into account is that, when one is trying to identify
particle trajectories for a macroscopic system using the selective
form of our equation, the disagreement with the classical
predictions which, as shown in \cite{grw}, unavoidably arises for
large times, be unimportant for times which are long with respect
to those during which one can keep the macroscopic system
isolated. Finally, and more important, we want  the
modification of the dynamics for microscopic systems with respect
to the standard one to be totally irrelevant. The simplest way to
obtain this is to assume that the mean rate
$\lambda_{\makebox{\tiny micro}} = \lambda_{\makebox{\tiny
macro}}/N$ of the spontaneous localization processes for a
microscopic system be extremely small.

For what concerns the parameter $\alpha$ it is necessary to choose
the localization distance $1/\sqrt{\alpha}$ large with respect to
the atomic dimensions and to the mean spreads around the
equilibrium positions of the lattice points of a crystal. In this
way, even when one of the extremely infrequent localization
processes takes place for a constituent of an atomic system, the
localization itself does not modify the internal structure of that
system and the decoupling of the center of mass and relative
motions discussed in section \ref{sec54} still holds. On the other
hand $1/\sqrt{\alpha}$ represents the distance after which a
linear superposition is transformed into a statistical mixture.
This parameter must then be chosen in accordance with the
requirement of avoiding the embarrassing occurrence of linear
superpositions of appreciably different locations of a macroscopic
object.

These considerations lead us to discuss the following choice for
the order of magnitude of the parameters. For the localization
rate of the microscopic constituents of any system we choose
\begin{equation}
\lambda_{\makebox{\tiny micro}} \; \simeq \; 10^{-16}\,
\makebox{sec}^{-1}.
\end{equation}
This means that such systems are localized once every
$10^{8}$--$10^{9}$ years. For the parameter $1/\sqrt{\alpha}$ we
choose:
\begin{equation}
1/\sqrt{\alpha} \; \simeq \; 10^{-5}\,\makebox{cm}.
\end{equation}

The fact that a microscopic system is practically never localized,
entails that standard quantum mechanics remains fully valid for
this type of system.  Moreover, for a composite system for which
the relative coordinates are confined within a spatial range much
smaller than the localization distance $1/\sqrt{\alpha}$, as it
happens for atoms and molecules, the process $T[\,\cdot\,]$ is
almost ineffective even when it takes place, a fact that
strengthens the above conclusion.

For what concerns macroscopic objects (containing a number of
constituents of the order of Avogadro's number), according to the
considerations of section \ref{sec54} showing that the individual
tests on the constituents add for the center--of--mass dynamics,
we get as characteristic localization rate:
\begin{equation}
\lambda_{\makebox{\tiny macro}} \; \simeq \; 10^{7}\,
\makebox{sec}^{-1}.
\end{equation}
If we take, for the sake of definiteness, the mass of such an
object to be of the order of 1 g, and the initial spread of the
position $\Delta\,q_{0}$ again of the order of $10^{-5}$ cm, we
know that the quantum increase of the spread in the position is
negligible for extremely long times ($\sim 10^{10}$ yr), so that
the quantum evolution is practically the same as the classical
one. In such a case [compare equation (\ref{nlhadun})], the
additional term appearing in $\Delta\, q^{2}$ equals $\Delta\,
q_{0}^{2}$ at the time $T_{1}$, which is of the order of 100 yr.
This is a very long time for keeping isolated a macroscopic
object. A much longer time $T_{2}$ is required in order that the
additional term in $\Delta\, p^{2}$ has an appreciable effect for
any reasonably chosen initial spread of the momentum. As far as
the occurrence of linear superpositions of far away states is
concerned, as we have seen, the off--diagonal elements of the
statistical operator are exponentially suppressed with the
lifetime $\tau = 1/\lambda\beta$. For $|q - q'| = 4\times 10^{-5}$
cm we have $\tau = 10^{-6}$ sec [see equation (\ref{gupnld})].
Therefore after times of this order linear superpositions of
states separated by distances larger than $10^{-4}$ cm are
transformed into statistical mixtures.

Considerations of this type are important for the quantum theory
of measurement. In fact, at least in the case in which the
interaction leading to the triggering of the apparatus takes place
in a very short time, we can apply our treatment to the
macroscopic parts of the apparatus itself, obtaining in this way a
consistent solution of the difficulties related to the quantum
theory of measurement for what concerns wavepacket reduction and
the definite final position of the pointer.

It has to be remarked that the basic evolution equation
(\ref{meqmsl}), due to the appearance of the non--Hamiltonian
terms, implies a nonconservation of energy. Let us give an
estimate of this effect in the case of the free particle on the
basis of a choice for the parameters we have just made. From
equations (\ref{sedc}) we see that, in our case
\begin{equation} \label{ie}
\llangle E \rrangle \quad = \quad \llangle E
\rrangle_{\makebox{\tiny Sch}} \; + \;
\frac{\lambda\alpha\hbar^{2}}{4m}\, t.
\end{equation}
where $\llangle E \rrangle_{\makebox{\tiny Sch}}$ is the conserved
energy for free Schr\"odinger evolution\footnote{It is easy to
prove that the relation $d\llangle H \rrangle/dt = \lambda \alpha
\hbar^{2}/4m$, from which (\ref{ie}) can be derived, holds in
general even when a potential term $V(q)$ is present in the
Hamiltonian.}. Energy nonconservation is then expressed by the
term
\begin{equation}
\delta E \quad = \quad \frac{\lambda\alpha\hbar^{2}}{4m}\, t.
\end{equation}
Let us evaluate this term for the case of a microscopic system.
Since $\lambda_{\makebox{\tiny micro}} = 10^{-16}$ sec$^{-1}$, $m
\simeq 10^{-23}$ g,
\begin{equation}
\frac{\delta E}{t} \quad \simeq \quad 10^{-25} \;\; \makebox{eV
sec}^{-1},
\end{equation}
which means that to have an increase of 1 eV il takes a time of
$10^{18}$ yr. In the case of the center--of--mass equation for a
macroscopic system, since both the rate $\lambda$ and the mass
increase proportionally to the number of constituents, the energy
non conservation is of the same amount. However this argument
applies only to the increase of the energy of the center of mass.
There is also an increase of energy in the internal motion which,
as can be easily understood considering a system of free
particles, is the same for all constituents. When this fact is
taken into account one can conclude that the estimated energy
increase for a system of $N$  $[\simeq$ Avogadro's number] atoms is
\begin{equation}
\frac{\delta E}{t} \quad \simeq \quad 10^{-14} \;\; \makebox{erg
sec}^{-1},
\end{equation}
Referring to an ideal monoatomic gas the increase in temperature
with time is then of the order of $10^{-15}$ K per year.

We conclude that QMSL reproduces in a consistent way quantum
mechanics for microscopic objects and classical mechanics for
macroscopic objects, and provides the basis for a conceptually
appealing description of quantum measurement
process\footnote{Actually, this conclusion, to be taken seriously,
requires also the proof, which will be presented in what follows,
that the macroscopic outcomes which emerge in a measurement
process, occur with the probabilities attached to them by the
standard formalism.}, and of the behavior of macroscopic systems.

\subsection{Quantum dynamical Semigroups} \label{sec56}

Among the non--Hamiltonian evolution equations which have been
considered in the literature, there is a class which has been
studied in great detail and has proved to be useful in the
description of various physical processes, which is particularly
simple. This class of equations is usually referred to as {\bf
quantum dynamical semi--group} (QDS) equations \cite{lin}. Let as
give a precise definition of a QDS.

Consider the Banach space $T_{S}({\mathcal H})$ of the
self--adjoint trace--class operators (equipped with the trace
norm, denoted as usual by $\|\,\cdot\,\|_{\makebox{\tiny Tr}}$) on
the Hilbert space ${\mathcal H}$ of the considered physical
system. A QDS is a one parameter family of linear operators:
\[
\Sigma_{t}: \;\; T_{S}({\mathcal H}) \; \longrightarrow \;
T_{S}({\mathcal H}) \qquad \makebox{defined for $t \geq 0$},
\]
satisfying:
\[
\begin{array}{lll}
1) & \rho \; \geq \; 0 \quad \Longrightarrow \quad
\Sigma_{t}(\rho) \; \geq \; 0 & \forall\; t \, \geq \, 0 \\
& & \\
2) & \makebox{Tr} [\Sigma_{t}(\rho)] \; = \; \makebox{Tr} [\rho]
\quad & \forall\; \rho  \in
T_{S}({\mathcal H}), \;\;\; t \, \geq \, 0 \\
& & \\
3) & \Sigma_{t} \Sigma_{s} (\rho) \; = \; \Sigma_{t+s} (\rho), &
\forall\; \rho  \in T_{S}({\mathcal H}), \;\;\;
t, s \, \geq \, 0 \\
& & \\
4) & {\displaystyle \lim_{t\rightarrow 0}}\; \| \Sigma_{t}(\rho) -
\rho \|_{\makebox{\tiny Tr}} \; = \; 0 & \forall\; \rho \in
T_{S}({\mathcal H}).
\end{array}
\]
We note that, when $\rho$ is the statistical operator describing
the state of a quantum system, the first two conditions correspond
to the requirements which are necessary for probability
conservation and the third one expresses the Markovian nature of
the process (which implies the independence of the evolution law
from the time origin).

In terms of the map $\Sigma_{t}$ one can define the infinitesimal
generator $Z$ of the QDS by the equation
\begin{equation}
Z[\rho] \quad = \quad {\displaystyle \lim_{t\rightarrow 0}}\;
\left[ \frac{\Sigma_{t}(\rho) - \rho}{t} \right], \qquad
\makebox{in the norm $\| \cdot \|_{\makebox{\tiny Tr}}$.}
\end{equation}
Obviously, the simplest case of a QDS is represented by the
standard Hamiltonian evolution equation
\begin{equation}
\Sigma_{t}(\rho) \quad = \quad e^{\displaystyle -
\frac{i}{\hbar}\, H\, t} \rho\;  e^{\displaystyle
\frac{i}{\hbar}\, H\, t},
\end{equation}
where $H$ is self--adjoint; in such case
\begin{equation}
Z[\rho] \quad = \quad -\, \frac{i}{\hbar}\, [H, \,\rho].
\end{equation}
As we have already stated, QDS equations have been studied in
great detail \cite{dav} and many general results have been
obtained. Lindblad \cite{lin} has been able to identify the most
general form for the infinitesimal generator of a QDS when two
more conditions are added to those previously considered, i.e.
\[
\begin{array}{lll}
5) & Z \; \makebox{is bounded}, & \\
& & \\
6) & \Sigma_{t} \; \makebox{is completely positive definite}. &
\end{array}
\]
Complete positiveness has to be understood in the sense of
Stinespring \cite{sti}. In such a case, the evolution equation for
the statistical operator $\rho$ can be written as
\begin{equation} \label{vsqchd}
\frac{d}{dt}\, \rho(t) \quad = \quad -\, \frac{i}{\hbar} [H
,\rho(t)] \; + \; \lambda \left\{ T[\rho(t)] \; - \; \frac{1}{2}\,
\rho(t) J \; - \; \frac{1}{2}\, J \rho(t) \right\}.
\end{equation}
where
\begin{equation} \label{lhric}
T[\rho] \; = \; \sum_{i\in K} A_{i}\, \rho A^{\dagger}_{i},
\qquad\quad J \; = \; \sum_{i\in K} A^{\dagger}_{i} A_{i}.
\end{equation}
Here $K$ is a finite or countable set, $H$ is a bounded
self--adjoint operator, and $A_{i}$ are operators satisfying
\[
\sum_{i\in K_{0}} A^{\dagger}_{i} A_{i} \leq 1 \qquad \quad
\forall\; K_{0}\subset K.
\]
The series in equation (\ref{lhric}) converges in the trace norm
topology. Davies \cite{dav} has proved that equation
(\ref{vsqchd}) generates a QDS even when $H$ is not bounded. The
basic QMSL equation (\ref{meqmsl}) is a particular type of QDS
equation, where $J$ is the identity operator:
\begin{equation} \label{vsqchp}
\frac{d}{dt}\, \rho(t) \quad = \quad -\, \frac{i}{\hbar} [H
,\rho(t)] \; + \; \lambda \left\{ T[\rho(t)] \; - \; \rho(t)
\right\}.
\end{equation}
The map $T[\,\cdot\,]$ appearing in equations (\ref{vsqchd}) and
(\ref{lhric}) is a particular case of what is usually called an
operation. An operation $T[\,\cdot\,]$ is, in general, a map
\[
T: \;\; T({\mathcal H}) \; \longrightarrow \; T({\mathcal H}),
\]
of the set of trace class operators into itself which is linear,
positive and bounded with respect to the trace norm, with bound
less or equal to one.

Non--Hamiltonian equations of type (\ref{vsqchd}) have been proved
useful for the description of many interesting physical processes.
We recall here, in particular, the successful use of such
equations in the description of the Wigner--Weisskopf atom and of
 beam foil spectroscopy (see, e.g., reference \cite{dav},
section \ref{sec6} and references therein). The quantum
description of decay processes, and in particular the exponential
nature of the decay law, have obtained an important clarification
by the use of such an equation as describing the evolution of an
unstable quantum system in the presence of apparatuses devised to
detect the decay \cite{fons1}. A very interesting investigation
\cite{jz}, aimed to find a solution to the problems raised by the
quantum theory of measurement, has led Joos and Zeh to derive an
equation of the above type starting from the Hamiltonian dynamics
describing the unavoidable coupling of macroscopic systems to
their environment.

\section{Stochastic processes in Hilbert space} \label{sec6}

QMSL is the first consistent proposal to overcome the measurement
problem of Quantum Mechanics in which wavefunction collapse is
naturally induced by the unique dynamical principle governing the
evolution of all physical systems. QMSL exhibits all the desired
features one seeks in a theory of spontaneous reductions;
nontheless, it has to face two problems, one ``aesthetic'' and one
physical.

The aesthetic drawback is that the modified dynamical evolution of
QMSL, though perfectly definite, is not expressed in terms of a
compact mathematical equation for the statevector leading to
equation (\ref{meqmsl}) for the statistical operator. The physical
problem is that the dynamics does not preserve the symmetry character of
wavefunctions describing systems of identical particles. Both problems
have been solved by CSL, the Continuous Spontaneous Localization model
\cite{csl0,csl}.

In this section we review the formalism of stochastic processes in
Hilbert spaces, which is the mathematical background of CSL. In
subsection \ref{sec61} we resort to  It\^o's formalism to derive
a modified Schr\"odinger equation for the evolution of the
statevector. This equation is linear, but it does not conserve the
norm, so it needs to be supplemented by further formal
prescriptions which we will analyze. In \ref{sec62} we derive the
corresponding norm--preserving equation, which is non linear.

Subsection \ref{sec63} is devoted to the general discussion of
statevector reduction: we show that the modified Schr\"odinger
equation introduced previously leads to the spontaneous reduction
of the statevector into one, among a set, of appropriate manifolds
characterized by the equation itself.

In subsection \ref{sec64} we re--derive the results of subsections
\ref{sec61}--\ref{sec63}, resorting to the Stratonovich in place
of the It\^o formalism. This is an alternative, and physically
more intuitive formalism to deal with stochastic differential
equations.

In the final subsection we show that the modified Schr\"odinger
equation of CSL yields an evolution equation for the statistical
operator of the quantum dynamical semigroup type, analogous to the
equation of QMSL.

\subsection{Raw and physical processes: It\^o linear equation}
\label{sec61}

Within the Hilbert space, let us consider the Markov process
$|\psi_{B}(t)\rangle$ satisfying the It\^o stochastic differential
equation \cite{arn}:
\begin{equation} \label{edm1}
d\,|\psi\rangle \quad = \quad \left[C\, dt \; + \; {\bf A}\cdot
d{\bf B} \right]|\psi\rangle
\end{equation}
where $C$ in an operator, ${\bf A} = \{A_{i}\}$ is a set of
operators, and ${\bf B} = \{B_{i}\}$ is a set of real Wiener
processes such that
\begin{equation}
\llangle dB_{i} \rrangle \; = \; 0, \qquad \quad \llangle dB_{i}\,
dB_{j} \rrangle \; = \; \gamma\,\delta_{ij}\,dt,
\end{equation}
$\gamma$ being a real constant. The index $i$ can be continuous,
in which case the sum becomes an integral and the Kronecker
$\delta$ becomes a Dirac $\delta$. Given an initial state
$|\psi(0)\rangle$, equation (\ref{edm1}) generates at time $t$ an
ensemble of statevectors $|\psi_{B}(t)\rangle$, where $B$ denotes
a particular realization $B_{i}(t)$ of the Wiener processes. To
simplify the notation, the dependence of $|\psi(t)\rangle$ on $t$
and $B_{i}$ will be often dropped, as in equation (\ref{edm1}).
The process (\ref{edm1}) and the ensemble generated by it will be
called the {\it raw} process and ensemble. In the raw ensemble,
each statevector $|\psi_{B}(t)\rangle$ has the same probability as
the particular realizations $B_{i}(t)$ that originates it through
equation (\ref{edm1}).

The raw process (\ref{edm1}) does not conserve the norm of
vectors, in general. In fact, using It\^o calculus, one finds
\begin{eqnarray} \label{edm7}
d\, \| |\psi(t)\rangle\|^{2} & = & \langle d\,\psi |\psi\rangle \;
+ \langle \psi |d\,\psi\rangle \; + \llangle \langle d\,\psi
|d\,\psi\rangle\rrangle \;
\nonumber \\
& = & \langle\psi |({\bf A} + {\bf A}^{\dagger}) | \psi\rangle
\cdot d{\bf B} \; + \; \langle\psi| (C + C^{\dagger})
|\psi\rangle\, dt \; + \nonumber \\
& & \langle\psi | {\bf A}^{\dagger} \cdot{\bf A} | \psi\rangle \,
\gamma\, dt,
\end{eqnarray}
where we have used the notation $|d\,\psi\rangle =
d\,|\psi\rangle$. If the statevectors $|\psi_{B}(t)\rangle$ were
of norm $1$, the probabilities of occurrence for them, which are
characteristic of the raw ensemble, could naturally be interpreted
as the physical probabilities. Since this is not the case, we
consider the ensemble of the normalized vectors
\begin{equation} \label{edm4}
|\chi_{B}(t)\rangle \quad = \quad \frac{|\psi_{B}(t)\rangle}{\|
|\psi_{B}(t)\rangle\|},
\end{equation}
having the same probabilities as the corresponding vectors
$|\psi_{B}(t)\rangle$ [i.e., as the realizations $B_{i}(t)$ of the
Wiener processes] and the ensemble of the normalized vectors
\begin{equation} \label{edm5}
|\phi_{B}(t)\rangle \quad = \quad \frac{|\psi_{B}(t)\rangle}{\|
|\psi_{B}(t)\rangle\|},
\end{equation}
whose probabilities are those of the vectors $|\psi_{B}(t)\rangle$
times their squared norms $\| |\psi_{B}(t)\rangle\|^{2}$. We use
different symbols for the vector functions $|\chi_{B}(t)\rangle$
and $|\phi_{B}(t)\rangle$, in spite of the fact that the
right--hand sides of equations (\ref{edm4}) and (\ref{edm5})
coincide, because the associated probabilities are different, so
that as random vector functions they are different. In fact, as we
shall see and as it is obvious, they obey different stochastic
differential equations. We choose as the {\bf physical
probabilities} (which we shall often call ``cooked''
probabilities) those of the vectors (\ref{edm5}) rather than those
of the vectors (\ref{edm4}). The ensemble of vectors
$|\phi_{B}(t)\rangle$ and the stochastic process in the Hilbert
space that generates it will be called the {\it physical} ensemble
and process. The prescription leading to the physical ensemble is
the counterpart of the assumption $4$ of QMSL and of the postulate
of standard quantum mechanics on the probabilities of the outcomes
of measurement processes.

Let us now investigate the relation between the raw and the
physical processes. Indicating by $P_{\makebox{\tiny
Raw}}[B_{i}(t, t_{0})]$ the probability of the realizations
$B_{i}(t, t_{0})$ of the Wiener processes (or, equivalently, of
the statevector $|\psi_{B}(t)\rangle$) and by $P_{\makebox{\tiny
Cook}}[B_{i}(t,t_{0})]$ the probability of the statevector
$|\phi_{B}(t)\rangle$, one has by definition
\begin{equation} \label{edm10}
P_{\makebox{\tiny Cook}}[B_{i}(t,t_{0})] \quad = \quad
P_{\makebox{\tiny Raw}}[B_{i}(t, t_{0})]\, \| |\psi_{B}(t,
t_{0})\rangle\|^{2}.
\end{equation}
It is easily shown that, because of linearity of equation
(\ref{edm1}) together with the Markov nature of the Wiener process
$B_{i}$, the procedure leading from the raw to the physical
ensemble can be performed just at the considered final time or, in
addition, any number of times between the initial and the final
times. It follows that equation (\ref{edm10}) can be replaced by
its specialization to the infinitesimal time interval $(t_{0},
t_{0} + dt)$, i.e.,
\begin{equation} \label{edm8}
P_{\makebox{\tiny Cook}}[dB_{i}] \quad = \quad P_{\makebox{\tiny
Raw}}[dB_{i}]\, \left[ 1 + d\, \| |\psi_{B}\rangle\|^{2} \right].
\end{equation}
The possibility of considering the physical ensemble depends on
the fulfillment of the condition that the total probability
associated with the distribution $P_{\makebox{\tiny Cook}}$ is
$1$. This amounts to requiring that, for any $|\psi\rangle$, the
average relative to the distribution $P_{\makebox{\tiny Raw}}$ of
the weighting factor $\| |\psi\rangle \|^2$ is $1$, i.e. $d \,
\llangle\| |\psi\rangle \|^2\rrangle = \llangle d \, \|
|\psi\rangle \|^2\rrangle = 0$. From equation (\ref{edm7}), one
finds
\begin{equation} \label{aqnld}
C\; + \; C^{\dagger} \quad = \quad -\, \gamma {\bf
A}^{\dagger}\cdot {\bf A}.
\end{equation}
When this condition is taken into account, denoting by
$-(i/\hbar)H$ the anti--Hermitian part of $C$, equation
(\ref{edm1}) becomes:
\begin{equation} \label{imef}
d\, |\psi(t)\rangle \quad = \quad \left[ -\frac{i}{\hbar}\, H\,dt
\; + \; {\bf A}\cdot d{\bf B} \; - \; \frac{\gamma}{2} {\bf
A}^{\dagger}\cdot {\bf A}\,dt \right] |\psi(t)\rangle.
\end{equation}

\subsection{It\^o non linear equation} \label{sec62}

The linear It\^o equation (\ref{imef}) and the cooking
prescription (\ref{edm10}) can be joined into a single {\it non
linear} stochastic differential equation for the physical vectors
$|\phi(t)\rangle$. Let us see how this can be accomplished.

Because of relation (\ref{aqnld}), equation (\ref{edm7})
simplifies to
\begin{equation} \label{mdc}
d\,\| |\psi(t)\rangle\|^{2} \quad = \quad \langle\psi(t)|({\bf A}
+ {\bf A}^{\dagger})|\psi(t) \rangle\cdot d{\bf B}.
\end{equation}
Then equation (\ref{edm8}) becomes
\begin{equation}
P_{\makebox{\tiny Cook}}[dB_{i}] \quad = \quad \left[ 1\; + \;
2\,{\bf R}\cdot d{\bf B} \right]\, P_{\makebox{\tiny
Raw}}[dB_{i}],
\end{equation}
where
\begin{equation}
{\bf R}\quad = \quad \frac{1}{2}\, \langle\psi|({\bf A} + {\bf
A}^{\dagger})|\psi \rangle
\end{equation}
and the probability distribution $P_{\makebox{\tiny Cook}}$ is
normalized. Indicating by $dB_{i}'$ the random variable whose
distribution is $P_{\makebox{\tiny Cook}}$, one has
\begin{equation}
\llangle dB_{i}' \rrangle \; = \; 2\,\gamma\,R_{i}\,dt, \qquad
\quad \llangle dB_{i}'\, dB_{j}' \rrangle \; = \;
\gamma\,\delta_{ij}\,dt,
\end{equation}
so that
\begin{equation}
d{\bf B}' \quad = \quad d{\bf B} + 2\gamma {\bf R} dt
\end{equation}
and $B_{i}'$ is a diffusion process having the same diffusion as
$B_{i}$ and drift  $2 R_{i} \gamma$. The meaning of the process
$B_{i}'$ and of its differential $d B_{i}'$ follows from the one
of the probability distribution $P_{\makebox{\tiny Cook}}$ which
defines them. The set of all realizations $B_{i}'(t)$ coincides
with that of all realizations $B_{i}(t)$ (in fact both sets
coincide with the set of all functions satisfying a given initial
condition), but their probabilities, according to the definition
(\ref{edm10}) of $P_{\makebox{\tiny Cook}}$, are those of the
physical ensemble instead of those of the raw ensemble. The
stochastic differential equation for the physical process can now
easily be written. We first write down the equation for the
process generating the normalized vectors $| \chi\rangle$. From
equation (\ref{imef}) and (\ref{mdc}), by direct evaluation, one
gets:
\begin{eqnarray} \label{ime2}
d\, |\chi(t)\rangle & = & \left[ -\frac{i}{\hbar} H\, dt \; + \;
\left( - \; \frac{1}{2}\gamma {\bf A}^{\dagger}\cdot {\bf A} \; -
\; \gamma {\bf A}\cdot {\bf R} \; + \; \frac{3}{2} \gamma {\bf R}
\cdot {\bf R} \right) dt \right. \; + \nonumber \\
&  & \left. ( {\bf A} - {\bf R})\cdot d{\bf B} \right]
|\chi(t)\rangle, \\
{\bf R} & = & \frac{1}{2}\, \langle \chi|({\bf A}^{\dagger} + {\bf
A}) |\chi \rangle. \nonumber
\end{eqnarray}
It is easily checked that equation (\ref{ime2}) conserves the norm
and that this feature does not depend on $B_{i}$ having drift
zero. The physical process is obtained by replacing each
realization $B_{i}(t)$ of the random function $B_{i}(t)$ by an
equivalent realization having the appropriate different
probability, i.e. an equivalent realization $B'_{i}(t)$ of the
random function $B'_{i}(t)$. This amounts to replace $dB_{i}$ by
$dB'_{i}$ in equation (\ref{ime2}), so that we get
\begin{eqnarray} \label{ime3}
d\, |\phi(t)\rangle & = & \left[ -\frac{i}{\hbar} H\, dt \; + \;
\left( - \; \frac{1}{2}\gamma {\bf A}^{\dagger}\cdot {\bf A} \; -
\; \gamma {\bf A}\cdot {\bf R} \; + \; \frac{3}{2} \gamma {\bf R}
\cdot {\bf R} \right) dt \right. \; + \nonumber \\
&  & \left. ( {\bf A} - {\bf R})\cdot d{\bf B}' \right]
|\phi(t)\rangle, \\
{\bf R} & = & \frac{1}{2}\, \langle \phi|({\bf A}^{\dagger} + {\bf
A}) |\phi \rangle. \nonumber
\end{eqnarray}
It is convenient to rewrite the above equation in terms of the
original Wiener processes $B_{i}(t)$. One gets the final equation:
\begin{eqnarray} \label{ime4}
d\, |\phi(t)\rangle & = & \left[ -\frac{i}{\hbar} H\, dt \; + \;
\left( - \; \frac{1}{2}\gamma ({\bf A}^{\dagger} \, - \, {\bf
R})\cdot {\bf A} \; + \; \frac{1}{2}\gamma ({\bf A} \, - \,
{\bf R})\cdot {\bf R} \right) dt \right. \; + \nonumber \\
&  & \left. ( {\bf A} - {\bf R})\cdot d{\bf B} \right]
|\phi(t)\rangle, \\
{\bf R} & = & \frac{1}{2}\, \langle \phi|({\bf A}^{\dagger} + {\bf
A}) |\phi \rangle. \nonumber
\end{eqnarray}
We note that the equations for the norm conserving processes
(\ref{ime2}) and (\ref{ime3}) or (\ref{ime4}), contrary to
equations (\ref{edm1}) or (\ref{imef}), are nonlinear.

The case in which $A_{i}$ is a set of self--adjoint operators is
of particular interest. In this case equation (\ref{ime4})
becomes\footnote{Stochastic equations having a formal structure of
the type (\ref{ime4}) have been considered in previous works
\cite{gis1,pea1}, but there the random terms appearing at the
right--hand side had a specific form devised to describe a
specific measurement that was supposed to be performed. The
considered equations, therefore, did not have the universal
character of the CSL equations. Other investigations
\cite{gisfl,di1,di2} deal with dynamical reduction models similar
to the one considered in reference \cite{csl0} and here. In
reference \cite{di1} an equation very close to equation
(\ref{ime4}) is introduced (without deriving it from a linear
process), but it is not specialized to the use of densities around
space points to discriminate among different configurations. The
idea of using densities is considered in reference \cite{di2},
where, however, the dynamical equation has a more complicated
structure than CSL.}
\begin{eqnarray} \label{ime5}
d\, |\phi(t)\rangle & = & \left[-\frac{i}{\hbar} H\, dt \, - \,
\frac{1}{2}\gamma ({\bf A} - {\bf R})^{2}  dt \, + \, ( {\bf A} -
{\bf R})\cdot d{\bf B}
\right] |\phi(t)\rangle \nonumber \\
& & \nonumber \\
{\bf R} & = &  \langle \phi|{\bf A}|\phi \rangle.
\end{eqnarray}
The analysis of this and of the previous subsection have shown
that one can take two different attitudes to describe the
diffusion process: either one solves equation (\ref{imef}), taking
as physical vectors the normalized ones and taking
$P_{\makebox{\tiny Cook}}$ as the physical probability
distribution; or one considers equation (\ref{ime4}), without the
need to normalize vectors and without the cooking prescription. At
the non relativistic level, these two attitudes are equivalent.
However, relativistic considerations we will discuss in section
\ref{sec011} will indicate that the first attitude
--- based on the linear equation + the cooking prescription --- is
more suited to describe the physics of the stochastic process.

\subsection{Reduction of the statevector} \label{sec63}

We shall now show that, when $\{A_{i}\}$ is a set of commuting
self--adjoint operators, the new terms in equation (\ref{ime5})
induce, for large times, the reduction of the statevector on the
common eigenspaces of the operators $A_{i}$ \cite{csl}.

Since here we are interested in discussing the physical effects of
the new terms, we disregard for the moment the Schr\"odinger part
of the dynamical equation. Then equation (\ref{ime5}) becomes
simply
\begin{eqnarray} \label{ime6}
d\, |\phi(t)\rangle & = & \left[ -\,\frac{1}{2}\gamma ({\bf
A}\,-\, {\bf R})^{2}\, dt  \, + \, ( {\bf A} - {\bf R})\cdot d{\bf
B} \right]
|\phi(t)\rangle  \\
& & \nonumber \\
{\bf R} & = &  \langle \phi|{\bf A}|\phi \rangle. \nonumber
\end{eqnarray}
Let us write
\begin{equation} \label{sra}
{\bf A} \quad = \quad \sum_{\sigma} {\bf a}_{\sigma}\,P_{\sigma},
\end{equation}
where the orthogonal projection operators $P_{\sigma}$ sum up to
the identity and it is understood that $\sigma \neq \tau
\Rightarrow {\bf a}_{\sigma} \neq {\bf a}_{\tau}$ (i.e.
$a_{i\sigma} \neq a_{i\tau}$ for at least one value of $i$). We
consider the real non--negative variables
\begin{equation}
\langle \phi|P_{\sigma}|\phi \rangle \quad = \quad z_{\sigma},
\end{equation}
having the property
\begin{equation}
\sum_{\sigma} z_{\sigma} \quad = \quad 1.
\end{equation}
In terms of such variables, one finds:
\begin{eqnarray}
{\bf R} & = & \sum_{\sigma} {\bf a}_{\sigma}\, z_{\sigma}, \\
({\bf A} - {\bf R})|\phi\rangle & = & \sum_{\sigma} \sum_{\tau}
z_{\tau}\, ({\bf a}_{\sigma} - {\bf a}_{\tau})\, P_{\sigma}\,
|\phi\rangle,
\\
({\bf A} - {\bf R})^{2}|\phi\rangle & = & \sum_{\sigma} \left[
\sum_{\tau} z_{\tau}\, ({\bf a}_{\sigma} - {\bf
a}_{\tau})\right]^{2} P_{\sigma}\, |\phi\rangle.
\end{eqnarray}
It follows that the stochastic differential equation (\ref{ime6})
can be written
\begin{eqnarray} \label{ime7}
d P_{\sigma}|\phi(t)\rangle & = & \left[- \frac{\gamma}{2} \left(
\sum_{\tau} z_{\tau}({\bf a}_{\sigma} - {\bf a}_{\tau})\right)^{2}
dt + \right.
\nonumber \\
& & \qquad\quad \left. \sum_{\tau}z_{\tau}({\bf a}_{\sigma} - {\bf
a}_{\tau})\cdot d{\bf B} \right] P_{\sigma} |\phi(t)\rangle.
\end{eqnarray}
Using this equation in the relation:
\begin{eqnarray*}
d\, \langle \phi|P_{\sigma}|\phi \rangle & = & [d\,\langle
\phi|P_{\sigma}]\,P_{\sigma}|
\phi \rangle + \\
& & \langle \phi|P_{\sigma} \,[d\,P_{\sigma}|\phi \rangle] \; + \;
\llangle [\langle \phi|d\,P_{\sigma}] \,[d\,P_{\sigma}|\phi
\rangle] \rrangle,
\end{eqnarray*}
gives for the variables $z_{\sigma}$ the set of stochastic
differential equations
\begin{equation} \label{efz}
d\,z_{\sigma} \quad = \quad 2\,z_{\sigma}\, \sum_{\tau} z_{\tau}\,
({\bf a}_{\sigma} \, - \, {\bf a}_{\tau})\cdot d{\bf B}.
\end{equation}
Qualitatively, equations (\ref{efz}) shows that the diffusion of the
$\{z_{\sigma}\}$ vanishes when they approach the
solution of the set of equations:
\begin{equation} \label{www}
z_{\sigma}\, \sum_{\tau} z_{\tau} \, ({\bf a}_{\sigma} \, - \,
{\bf a}_{\tau}) \quad = \quad 0,
\end{equation}
so that the values of $\{z_{\sigma}\}$ eventually accumulate
towards such solutions. A formal proof of the fact that
$\{z_{\sigma}\}$ asymptotically reduce to one of the solutions of
equation (\ref{www}) is easily obtained. From equation (\ref{efz})
one finds:
\begin{equation}
d\, z^{2}_{\sigma} \; = \; 2\,z_{\sigma}\, dz_{\sigma} \; + \;
\left[ 2\,z_{\sigma}\, \sum_{\sigma} z_{\tau}\, ({\bf a}_{\sigma}
\, - \, {\bf a}_{\tau}) \right]^{2} \gamma\, dt
\end{equation}
and in turn
\begin{equation} \label{vmz}
d\, \llangle z^{2}_{\sigma} \rrangle \; = \; \llangle d\,
z^{2}_{\sigma} \rrangle \; = \; \gamma \left[2\,z_{\sigma}\,
\sum_{\tau} z_{\tau}\, ({\bf a}_{\sigma} \, - \, {\bf a}_{\tau})
\right]^{2} dt.
\end{equation}
It follows that
\begin{equation}
\frac{d}{dt}\, \llangle z^{2}_{\sigma}(t) \rrangle \; \geq \; 0.
\end{equation}
This result, together with the boundedness property
\begin{equation}
\llangle z^{2}_{\sigma}(t) \rrangle \; \leq \; 1,
\end{equation}
entails that for $t \rightarrow \infty$
\begin{equation}
\frac{d}{dt}\, \llangle z^{2}_{\sigma}(t) \rrangle \;
\longrightarrow \; 0.
\end{equation}
Using again equation (\ref{vmz}), we get
\begin{equation} \label{soe}
z_{\sigma}\, \sum_{\tau} z_{\tau}\, ({\bf a}_{\sigma} \, - \, {\bf
a}_{\tau}) \quad \longrightarrow \quad 0.
\end{equation}
In reference \cite{csl} it is shown that the only solutions to the
set of equations (\ref{www}) are of the form
\[
z_{1} = 0 \qquad z_{2} = 0 \quad \ldots \quad z_{\sigma} = 1 \quad
\ldots,
\]
corresponding to $|\phi\rangle$ lying in one of the common
eigenspaces of the operators $A_{i}$.  Since equations
(\ref{ime7}) do not change  the Hilbert space ray to which each
component $P_{\sigma}|\phi(t)\rangle$ belongs, we conclude that
$|\phi(t)\rangle$ asymptotically reduces to one of its initial
components $P_{\sigma}|\phi(0)\rangle$ times a normalization
factor.

The probabilities for the various possible issues are also easily
calculated. In fact, since $d\, \llangle z_{\sigma} \rrangle  =
\llangle d\, z_{\sigma} \rrangle = 0$, one has
\begin{equation}
\llangle z_{\sigma} \rrangle = z_{\sigma}(0).
\end{equation}
On the other hand,
\begin{equation}
\llangle z_{\sigma} \rrangle \; \longrightarrow \;
\makebox{Prob}[z_{\sigma}(\infty) = 1],
\end{equation}
so that one finds
\begin{equation}
{\makebox{Prob}}[z_{\sigma}(\infty) = 1] \; = \; z_{\sigma}(0),
\end{equation}
i.e.,
\begin{equation} \label{ixfgjb}
\makebox{Prob}[|\phi(\infty)\rangle \propto
P_{\sigma}|\phi(0)\rangle] \quad = \quad \langle\phi(0)|P_{\sigma}
|\phi(0)\rangle.
\end{equation}
As one can see, this result is a direct consequence of the
martingale property $\llangle d\, z_{\sigma} \rrangle  = 0$
\cite{pea3,pea4,csl0}. Physically, the above equation plays a
fundamental role since it guarantees that the dynamical reduction
models reproduce the quantum predictions about measurement
outcomes.

\subsection{Linear and non linear equations: the Stratonovich
formalism} \label{sec64}

The analysis of the previous subsections was based on the It\^o
formalism for stochastic differential equations. It is not
difficult to re--write all the above equations, resorting to the
Stratonovich formalism \cite{arn}, which is easier to handle. The
linear equation, corresponding to (\ref{imef}), is:
\begin{equation} \label{smef}
\frac{d}{dt}\, |\psi(t)\rangle \quad = \quad \left[
-\frac{i}{\hbar}\, H \; + \; {\bf A}\cdot {\bf V}(t) \; - \;
\frac{\gamma}{2} \left( {\bf A}^{\dagger}\cdot {\bf A} \, + \,
{\bf A}^{2}\right) \right] |\psi(t)\rangle,
\end{equation}
where $V_{i}(t)$ are $c$--number stochastic processes with
probability of occurrence given by:
\begin{equation} \label{edm10g}
P_{\makebox{\tiny Cook}}[V_{i}(t)] \quad = \quad P_{\makebox{\tiny
Raw}}[V_{i}(t)]\, \||\psi(t)\rangle\|^{2}.
\end{equation}
$P_{\makebox{\tiny Raw}}[V_{i}(t)]$ is the probability
distribution of gaussian white noises satisfying:
\begin{equation} \label{probwn}
\llangle V_{i}(t) \rrangle \; = \; 0, \qquad \quad \llangle
V_{i}(t_{1})\, V_{j}(t_{2}) \rrangle \; = \; \gamma\,\delta_{ij}\,
\delta(t_{1} - t_{2}).
\end{equation}
In the particular but very important case in which the operators
$A_{i}$ are self--adjoint, (\ref{smef}) reduces to:
\begin{equation} \label{smefsa}
\frac{d}{dt}\, |\psi(t)\rangle \quad = \quad \left[
-\frac{i}{\hbar}\, H \; + \; {\bf A}\cdot {\bf V}(t) \; - \;
\gamma {\bf A}^{2}\right] |\psi(t)\rangle.
\end{equation}
The physical meaning of equations (\ref{smef}) is the same of the
corresponding It\^o equation: if a homogeneous ensemble (pure
case) at the initial time $t_{0}$ is associated with the
statevector $|\psi(t_{0})\rangle$, then the ensemble at a
subsequent time $t$ is the union of homogeneous ensembles
associated with the normalized vectors $|\psi(t)\rangle/\|
|\psi(t)\rangle \|$, where $|\psi(t)\rangle$ is the solution of
equation (\ref{smef}) with the assigned initial conditions and for
a specific stochastic process $V(t) = \{ V_{i}(t) \}$ which has
occurred in the interval $(0,t)$. The probability associated to
any such homogeneous ensemble is given by (\ref{edm10g}).

It is not difficult to write down the non linear equation for the
{\it physical} vectors $|\phi(t)\rangle = |\psi(t)\rangle/\|
|\psi(t)\rangle \|$; here we limit ourselves only to the case in
which $A_{i}$ are self--adjoint:
\begin{eqnarray} \label{qelaoj}
\frac{d}{dt}\, |\phi(t)\rangle & = & \left[ -\frac{i}{\hbar}\, H\;
+ \; \left( {\bf A} - {\bf R} \right)\cdot {\bf V}(t) \; - \;
\gamma \left( {\bf A} - {\bf R} \right)^{2} \right. \nonumber
\\
& + & \left. \gamma \left( {\bf Q}^{2} - {\bf R}^{2} \right)
\right] |\phi(t)\rangle,
\\
{\bf R} & = & \langle\phi| {\bf A} |\phi\rangle \qquad {\bf Q}^{2}
\; = \; \langle\phi| {\bf A}^{2} |\phi\rangle \nonumber.
\end{eqnarray}

It is instructive to reconsider the reduction mechanism discussed
in the previous subsection, using the Stratonovich equation
(\ref{smef}); for simplicity we will limit ourselves to study the
case of a single self--adjoint operator $A$, so that only one
stochastic field $V(t)$ appears. Suppose the initial
statevector $|\psi(0)\rangle$ (for simplicity we take $t_{0} = 0$)
has non vanishing projections on two distinct eigenmanifolds of
$A$, corresponding to the eigenvalues $\alpha$ and $\beta$
respectively:
\begin{equation} \label{ispe}
|\psi(0)\rangle \quad = \quad P_{\alpha}\,|\psi(0)\rangle \; + \;
P_{\beta}\,|\psi(0)\rangle.
\end{equation}
When the hamiltonian is disregarded, the solution of equation
(\ref{smef}) is\footnote{In equation (\ref{ccv}) and following the
statevector is labeled by the Brownian motion symbol $B$, to
stress the fact that, under our assumptions, the state at time $t$
does not depend on the specific sample function $V(t)$ in the
interval $(0,t)$ but only on its integral given by equation
(\ref{bmp}).}:
\begin{equation} \label{ccv}
|\psi_{B}(t)\rangle \quad = \quad e^{\displaystyle \;\alpha B(t) -
\alpha^{2}\gamma\, t} P_{\alpha}\,|\psi(0)\rangle \; + \;
e^{\displaystyle \;\beta B(t) - \beta^{2}\gamma\, t}
P_{\beta}\,|\psi(0)\rangle.
\end{equation}
Here $B(t)$ is the Brownian process:
\begin{equation} \label{bmp}
B(t) \quad = \quad \int_{0}^{t} d\tau\, V(\tau).
\end{equation}

Taking into account equation (\ref{ccv}) and the cooking
prescription, one gets the cooked probability density for the
value $B(t)$ of the Brownian process at time $t$:
\begin{eqnarray} \label{waxmi}
P_{\makebox{\tiny Cook}}[w(t)] & = & \|
P_{\alpha}\,|\psi(0)\rangle \|^{2} \frac{1}{\sqrt{2\pi\gamma t}}\,
\; e^{\displaystyle\; -\frac{1}{2\gamma t}\,
[B(t) - 2 \gamma \alpha t]^{2}} \quad + \nonumber \\
& & \| P_{\beta}\,|\psi(0)\rangle \|^{2} \frac{1}{\sqrt{2\pi\gamma
t}}\, \; e^{\displaystyle\; -\frac{1}{2\gamma t}\, [B(t) - 2
\gamma \beta t]^{2}}.
\end{eqnarray}
>From the above equation it is clear that, for $t \rightarrow
\infty$, the stochastic process $B(t)$ can assume only values
belonging to an interval of width $\sqrt{\gamma t}$
around\footnote{Note that even though the spread $\sqrt{\gamma t}$
tends to $\infty$ for $t \rightarrow \infty$, its ratio to the
distance $2(\alpha -\beta)\gamma t$ between the two considered
peaks of the distribution tends to zero.} either the value
$2\gamma \alpha t$ or the value $2\gamma \beta t$. The
corresponding probabilities are $\| P_{\alpha}\,|\psi(0)\rangle
\|^{2}$ and $\| P_{\beta}\,|\psi(0)\rangle \|^{2}$ respectively.
The occurrence of a value ``near'' to $2 \alpha\gamma t$ for the
random variable $B(t)$ leads, according to (\ref{ccv}), to a
statevector which for $t \rightarrow \infty$ is driven into the
eigenmanifold corresponding to the eigenvalue $\alpha$ of $A$. In
fact, in such a case one gets:
\begin{equation}
\frac{\| P_{\beta}\,|\psi_{B}(t)\rangle \|^{2}}{\| P_{\alpha}\,
|\psi_{B}(t)\rangle \|^{2}} \quad \simeq \quad e^{\displaystyle
-2\gamma (\alpha -\beta)^{2}t} \; \frac{\|
P_{\beta}\,|\psi(0)\rangle \|^{2}}{\| P_{\alpha}\,|\psi(0)\rangle
\|^{2}} \quad \stackrel{t \rightarrow \infty}{\longrightarrow}
\quad 0.
\end{equation}
Analogously, when the random variable $B(t)$ takes a value
``near'' $2 \beta\gamma t$, for $t \rightarrow \infty$ the
statevector is driven into the eigenmanifold corresponding to the
eigenvalue $\beta$ of $A$.

It is then clear that the model establishes a one--to--one
correspondence between the ``outcome'' (the final ``preferred''
eigenmanifold into which an individual statevector is driven) and
the specific value (among the only ones having an appreciable
probability) taken by $B(t)$ for $t \rightarrow \infty$, a
correspondence irrespective of what $|\psi(0)\rangle$
is\footnote{Obviously, $|\psi(0)\rangle$ enters in a crucial way
in determining the probability of occurrence of the Brownian
processes $B(t)$.}. In the general case of several operators
$A_{i}$, a similar conclusion holds for the ``outcomes''
$\alpha_{i}$ of $A_{i}$ and the corresponding noises $B_{i}(t)$.

\subsection{The statistical operator} \label{sec65}

The statistical operator corresponding to the physical ensemble
and its evolution equation are easily obtained from the
definition:
\begin{equation} \label{hjkg}
\rho \; = \;  \int {\mathcal D}[B_{i}(t)]\, \frac{|\psi\rangle}{\|
|\psi\rangle \|}\, \frac{\langle \psi|}{\|\langle\psi|\|}\,
P_{\makebox{\tiny Raw}}[B_{i}(t)]\| | \psi\rangle \|^{2}\, \; = \;
\llangle |\psi \rangle \langle \psi| \rrangle_{P_{\makebox{\tiny
Raw}}}
\end{equation}
and equation (\ref{imef}), or from
\begin{equation}
\rho \; = \; \int {\mathcal D}[B_{i}(t)]\, |\phi \rangle\langle
\phi|\, P_{\makebox{\tiny Cook}}[B_{i}(t)]
\end{equation}
and equation (\ref{ime4}). $\int {\mathcal D}[B_{i}(t)]$ is the
functional integral with respect to all the possible realizations
of the stochastic processes $B_{i}(t)$.

Using once more It\^o calculus in evaluating $d\rho$, one
gets\footnote{Of course, the same equation is obtained starting
from the Stratonovich equations (\ref{smef}) or (\ref{qelaoj}),
which correspond to the It\^o equation (\ref{imef}) and
(\ref{ime4}), respectively.}:
\begin{equation} \label{efso}
\frac{d}{dt}\, \rho(t) \quad = \quad -\frac{i}{\hbar}\, [H,
\rho(t)] \; + \; \gamma {\bf A} \rho(t)\cdot {\bf A}^{\dagger} \;
- \; \frac{\gamma}{2} \left\{ {\bf A}^{\dagger}\cdot {\bf A},
\rho(t) \right\}.
\end{equation}
where $\{ \, \cdot \, ,\cdot \, \}$ denotes the the
anticommutator. This is the Lindblad  form for the generator of a
quantum dynamical semigroup, as already discussed. It is
remarkable that the general Lindblad generator can be obtained
from a stochastic process in Hilbert space. Note that the way we
have followed to get equation (\ref{efso}) describing an ensemble
associated to the statistical operator $\rho \! \left( t \right)$
makes clear that each member of the ensemble has a definite
statevector at any time, a statevector which, eventually, ends up
in one of the eigenmanifolds of the preferred basis.

\section{Continuous Spontaneous Localizations (CSL)} \label{sec7}

All the necessary mathematical tools to work with stochastic
differential equations in Hilbert space have been developed; we
can now apply this formalism to work out a model of dynamical
reductions which has all the desired features of QMSL, but, at the
same time, overcomes the difficulties we have mentioned at the
beginning of the previous section.

It should be clear from the above analysis that what we have to do
is to choose the ``preferred basis'', i.e. the operators $A_{i}$
whose common eigenmanifolds are the manifolds in which the
statevector is driven by the diffusion process. These operators
have to be chosen is such a way that:
\begin{enumerate}
\item Macroscopic objects are always localized in {\it space}.
\item Microscopic dynamics is not altered in an appreciable way
with respect to the standard quantum evolution.
\item In particular, the energy increase of the system --- due to
space localizations --- must not be detectable.
\item The symmetry properties of systems containing identical
particles must be preserved.
\end{enumerate}
We now see how all these requirements are met by CSL. In the first
subsection we set out the preferred basis: the operators $A_{i}$
will be chosen to be appropriate functions of the creation and
annihilation operators of particles in space. In subsection
\ref{sec72} we discuss the implications of such a choice in the
case of macroscopic systems, showing how their classical
properties stem from the quantum properties of their microscopic
constituents; this subsection parallels the analysis of subsection
\ref{sec54}, which was performed only at the statistical operator
level, not at the wavefunction level like in the present section.

In subsection \ref{sec73} we determine the reduction rates induced
by CSL, proving that, with an appropriate choices of the
parameters, they are compatible with those of QMSL. In subsection
\ref{sec74} we discuss how the average value of physical
observables are affected by the new non--Hamiltonian terms,
showing once more that there are no appreciable differences with
respect to QMSL. In subsection \ref{sec75} we put forward a
simple, pedagogical, CSL model, which is illuminating in order to
understand how the reduction process amplifies when moving from
the micro to the macro level.

In the last two subsections we discuss two new CSL models; in the
first one the collapse mechanism is related to the mass density
distribution of the object, rather than to the density--number
operator, while in the second one the reduction mechanism is
related to gravity.

\subsection{The choice of the ``preferred basis''} \label{sec71}

Let us consider the creation and annihilation operators
$a^{\dagger}({\bf y}, s)$ and $a({\bf y}, s)$  of a particle at
point ${\bf x}$ with spin component $s$ satisfying canonical
commutation or anticommutation relations. We define a locally
averaged density operator
\begin{equation} \label{ando}
N({\bf x}) \quad = \quad \sum_{s}\int d^{3}y\, g({\bf y} - {\bf
x})\, a^{\dagger}({\bf y}, s)a({\bf y}, s),
\end{equation}
where $g({\bf x})$ is a spherically symmetric, positive real
function peaked around ${\bf x} = 0$, normalized in such a way
that:
\[
\int d^{3}x\, g({\bf x}) \quad = \quad 1,
\]
so that
\[
\int d^{3}x\, N({\bf x}) \quad = \quad N,
\]
$N$ being the total number operator. The operators $N({\bf x})$
are self--adjoint and commute with each other. In what follows we
choose
\begin{equation} \label{gggg}
g({\bf x}) \quad = \quad \left(\frac{\alpha}{2\pi}\right)^{3/2}
e^{\displaystyle -\frac{\alpha}{2}\, ({\bf x})^{2}},
\end{equation}
where $\alpha$ is a parameter such that $\alpha^{-3/2}$ represents
essentially the volume over which the average is taken in the
definition of $N({\bf x})$. The improper vectors
\begin{equation} \label{ces}
|q, s\rangle \quad = \quad {\mathcal N}\, a^{\dagger}({\bf q}_{1},
s_{1}) a^{\dagger}({\bf q}_{2}, s_{2}) \ldots a^{\dagger}({\bf
q}_{n}, s_{n})|0\rangle
\end{equation}
are the normalized common eigenstates of the operators $N({\bf
x})$ belonging to the eigenvalues
\[
n({\bf x}) \quad = \quad \sum_{i=1}^{n}\, g({\bf q}_{i} - {\bf
x}).
\]

We identify now, with reference to the previous section, the index
$i$ which labels the operators $A_i$ with the space point ${\bf
x}$ and the operators $A_{i}$ with the density operators $N({\bf
x})$. {\bf It\^o equation} (\ref{imef}) then becomes:
\begin{equation} \label{iecsl}
d |\psi(t) \rangle = \left[ -\frac{i}{\hbar}\, H\, dt \; + \; \int
d^3 x \, N({\bf x}) dB({\bf x}) - \frac{\gamma}{2} \int d^3 x \,
N^2({\bf x}) dt \right] | \psi(t) \rangle,
\end{equation}
where:
\begin{equation}
\llangle dB({\bf x})\rrangle \; = \; 0 \qquad\quad \llangle
dB({\bf x})\,dB({\bf y})\rrangle \; = \;  \gamma \, \delta^3({\bf
x} - {\bf y}) dt.
\end{equation}
This is, in a different notation, the process considered in
references \cite{csl0,csl}  for identical particles. The
generalization to several kinds of particles is immediate. For
completeness, we write down also the corresponding {\bf
Stratonovich equation} (\ref{smef}):
\begin{equation} \label{gtg}
\frac{d}{dt}\, |\psi(t) \rangle = \left[ -\frac{i}{\hbar}\, H\, \;
+ \; \int d^3 x \, N({\bf x}) V({\bf x}, t) \; - \; \gamma \int
d^3  x \, N^2({\bf x}) dt \right] | \psi(t) \rangle;
\end{equation}
the first two moments of the white noise $V({\bf x}, t)$ are:
\begin{equation} \label{sch1d2}
\llangle V({\bf x}, t)\rrangle \; = \; 0 \qquad\quad \llangle
V({\bf x}, t_{1})\, V({\bf y}, t_{2})\rrangle \; = \;  \gamma \,
\delta^3({\bf x} - {\bf y})\, \delta (t_{1} - t_{2}).
\end{equation}
Note that equation (\ref{gtg}) can be rewritten in the following
way:
\begin{eqnarray} \label{sch1d}
\frac{d|\psi(t)\rangle}{dt} & = & \left[ -\frac{i}{\hbar} H\; +\;
\int d^{\,3}x\; {\mathcal N}({\bf x}) V'({\bf x},t) \right.
\nonumber \\
& & - \; \left.\gamma \int d^{\,3}x d^{\,3}y \; {\mathcal N}({\bf
x}) D({\bf x} - {\bf y}) {\mathcal N}({\bf y}) \right]
|\psi(t)\rangle,
\end{eqnarray}
where:
\begin{equation}
{\mathcal N}({\bf x}) \quad = \quad \sum_{s} a^{\dagger}({\bf x},
s)\,a({\bf x}, s)
\end{equation}
is the number--density operator, and $V'({\bf x},t)$ is a new
Gaussian stochastic process defined as:
\begin{equation} \label{nsf}
V'({\bf x}, t) \quad = \quad
\left(\frac{\alpha}{2\pi}\right)^{\frac{3}{2}} \int d^{3}y\,
e^{\displaystyle -\frac{\alpha}{2}({\bf x} - {\bf y})^{2}} V({\bf
y}, t).
\end{equation}
It is easy to check that its average value is zero, while the
correlation function is:
\begin{eqnarray} \label{fcrov}
\llangle V'({\bf x}, t_{1})\, V'({\bf y}, t_{2}) \rrangle & = &
\gamma\, D({\bf x} - {\bf y}) \, \delta(t_{1} - t_{2})
\; = \nonumber \\
& = & \gamma\, \left(\frac{\alpha}{4\pi}\right)^{\frac{3}{2}}
e^{\makebox{$-\frac{\alpha}{4}({\bf x} - {\bf y})^{2}$}}
\delta(t_{1} - t_{2}).
\end{eqnarray}

The equation  for the {\bf statistical operator} (\ref{efso})
reads:
\begin{equation} \label{cslso}
\frac{d}{dt}\, \rho(t) \; = \; - \frac{i}{\hbar}\, \left[ H,
\rho(t) \right] \; + \; \gamma \int d^3 x \, N({\bf x}) \rho(t)
N({\bf x}) \; - \; \frac{\gamma}{2}  \int d^3 x \, \{ N^{2}({\bf
x}) , \rho(t) \}.
\end{equation}
In the representation given by the improper vectors (\ref{ces}),
equation (\ref{cslso}) becomes:
\begin{eqnarray} \label{sclg}
\frac{\partial}{\partial t} \langle q', s'| \rho(t) | q'', s''
\rangle & = & - \frac{i}{\hbar} \langle q', s' | \left[ H, \rho(t)
\right]  | q'', s'' \rangle \nonumber + \frac{\gamma}{2} \sum
_{ij} \left[ 2\,
G({\bf q}_{i}' - {\bf q}_{j}'') - \right. \nonumber \\
& & \left. G({\bf q}_{i}' - {\bf q}_{j}') - G({\bf q}_{i}'' - {\bf
q}_{j}'') \right] \langle q' ,s' | \rho(t) | q'', s'' \rangle,
\end{eqnarray}
where
\begin{eqnarray}
G({\bf y}' - {\bf y}'') & = & \int d^{3} x \, g({\bf y}' - {\bf
x})\, g({\bf y}'' - {\bf x}) \nonumber \\ & = & \left(
\frac{\alpha}{4\pi} \right)^{3/2}\, e^{\displaystyle -
\frac{\alpha}{4}\, ({\bf y}' - {\bf y}'')^{2}}.
\end{eqnarray}
For a single particle, equation (\ref{sclg}) reduces to:
\begin{eqnarray} \label{mepr2}
\frac{\partial}{\partial t}\, \langle{\bf q}'|\rho(t)|{\bf
q}''\rangle & = & - \frac{i}{\hbar}\, \langle{\bf q}'|[H, \rho(t)
]|{\bf q}''\rangle \, - \, \nonumber \\ & &
\gamma\left(\frac{\alpha}{4\pi}\right)^{\frac{3}{2}} \left[ 1 -
e^{\displaystyle - \frac{\alpha}{4}\, ({\bf q}' - {\bf q}'')^{2}}
\right] \langle{\bf q}'|\rho(t)|{\bf q}''\rangle.
\end{eqnarray}
We note that, taking
\begin{equation} \label{rrde}
\lambda \quad = \quad \gamma
\left(\frac{\alpha}{4\pi}\right)^{3/2},
\end{equation}
equation (\ref{mepr2}) coincides with the QMSL equation
(\ref{mepr}).

\subsection{Dynamical reductions for macroscopic rigid bodies}
\label{sec72}

We now discuss the physical implications of the modified dynamical
equation (\ref{iecsl}) for a macroscopic system, under the
assumption that the order of magnitude of the length parameter
$1/\sqrt{\alpha}$ is such that it can reasonably be admitted that
the wavefunction of the internal variables of a macroscopic body
is sharply localized with respect to $1/\sqrt{\alpha}$: the
conclusions we will reach will be analogous to those derived in
section \ref{sec54}, when macroscopic objects where analyzed
within QMSL.

Let ${\bf Q}$ be the center  of mass coordinate of the system of
identical particles which constitutes the considered macroscopic
body,
\[
{\bf Q} \quad = \quad \frac{1}{N} \sum_{i=1}^{N} {\bf q}_{i},
\]
and write
\[
{\bf q}_{i} \quad = \quad {\bf Q} + \tilde{\bf q}_{i}.
\]
The coordinates $\tilde{\bf q}_{i}$ with respect to the center of
mass sum up to zero, so that they are functions of $3N - 3$
independent internal variables\footnote{The internal variables, as
defined here, describe also rotations of the $N$--particle system.
We assume that the the wavefunction is such that the orientation
of the system (and consequently of its internal structure) is
sharply defined. In the general case, one could consider $3$
orientation variables, to be treated along the same lines as the
center  of mass coordinate, and $3N - 6$ truly internal variables,
to be assumed sharply localized in the wave function. However, in
this case the problem would be considerably more complicated
without gaining very much as regards to physical insight.}, which
we indicate by $r$. The internal variables $r$, together with the
center of mass coordinates ${\bf Q}$, are functions of the
coordinates ${\bf q}_{i}$.  So, we consider the wavefunction
\begin{equation} \label{wfrb}
\psi (q, s) \; = \; \Psi ({\bf Q})\, \chi (r, s), \qquad \quad
\chi (r, s) \; = \; \left[
\begin{array}{c}
\makebox{S} \\ \makebox{A}
\end{array}
\right] \Delta (r, s),
\end{equation}
where ``S'' and ``A'' mean symmetrization or antisymmetrization
with respect to interchanges of the arguments $({\bf q}_{i},
s_{i})$. The wavefunctions $\Psi$ and $\chi$ are understood to be
separately normalized. The function $\Delta (r, s)$ is assumed to
be sharply (with respect to $1/\sqrt{\alpha}$) peaked around the
value $r_{0}$ of $r$.

The action of the operator $N({\bf x})$ on the wavefunction
(\ref{wfrb}) is easily worked out. One finds:
\begin{equation}
N({\bf x})\, \Psi ({\bf Q})\, \chi (r, s) \; = \; \Psi ({\bf Q})
\left[
\begin{array}{c}
\makebox{S} \\ \makebox{A}
\end{array}
\right]
 \sum_{i} \left(\frac{\alpha}{2\pi}\right)^{3/2}\,
e^{\displaystyle  - \frac{\alpha}{2} [{\bf Q} + \tilde{\bf
q}_{i}(r) - {\bf x} ]^{2}}\, \Delta (r,s).
\end{equation}
According to our assumptions, the factor in front of the function
$\Delta$ varies much more slowly than $\Delta$ itself, so that we
can take $r = r_{0}$ in the factor. In other words, we treat the
factor as if $\Delta (r, s)$ were of the form $\delta^{3n-3} (r -
r_{0}) \xi (s)$. Then:
\begin{equation} \label{efn}
N({\bf x})\, \Psi ({\bf Q})\, \chi (r, s) \quad = \quad F({\bf Q}
- {\bf x})\, \Psi ({\bf Q})\, \chi (r, s) ,
\end{equation}
where:
\begin{equation} \label{ecslw}
F({\bf Q} - {\bf x})\; = \; \sum_{i}
\left(\frac{\alpha}{2\pi}\right)^{3/2}\, e^{\displaystyle  -
\frac{\alpha}{2} [{\bf Q} + \tilde{\bf q}_{i}(r_{0}) - {\bf x}
]^{2}}.
\end{equation}

According to equation (\ref{efn}) the operator $N({\bf x})$ acts
only on the factor $\Psi$ of $\psi$. As a consequence, under the
assumption that:
\[
H \quad = \quad  H_{Q}\; + \; H_{r},
\]
if $\Psi$ and $\chi$ satisfy the equations
\begin{eqnarray}
d |\Psi \rangle & = & \left[ - \frac{i}{\hbar} H_{Q}\, dt  \; + \;
\int d^{3} x\, F({\bf Q} - {\bf x})\, dB({\bf x}) \; - \;
\frac{\gamma}{2} \int d^{3} x\, F^{2}({\bf Q} - {\bf x})\, dt
\right] |\Psi\rangle \nonumber \\
& & \label{cmie} \\
d |\chi\rangle & = &  \left[ - \frac{i}{\hbar}\, H_{ r}\, dt
\right]\, |\chi\rangle \label{odghj}
\end{eqnarray}
respectively, the wavefunction (\ref{wfrb}) satisfies equation
(\ref{iecsl}). We can conclude that, under our assumptions, the
center of mass and the internal motions decouple as in the absence
of the stochastic terms in equation (\ref{iecsl}). Furthermore,
the stochastic terms do not affect the internal structure, while
the center of mass wavefunction obeys a stochastic differential
equation, again of the type (\ref{imef}), whose consequences will
be discussed below.

Note that equations (\ref{cmie}) and (\ref{odghj}) are exactly the
counterpart of equations (\ref{nht}) and (\ref{yjth}) of QMSL,
respectively. This proves that the separation of the center of
mass and internal motion takes place also at the wavefunction
level, not only at the statistical operator level, as seen in
section \ref{sec5}.

\subsection{Reduction rates} \label{sec73}

The operators $F({\bf Q}-{\bf x})$ appearing in equation
(\ref{cmie}), which correspond to the operators $A_{i}$ of
equation (\ref{imef}), are real functions of the center  of mass
position operator ${\bf Q}$. They are a set of commuting
self--adjoint operators, so that, as we know from the results of
section \ref{sec6}, the non--Schr\"odinger terms in equation
(\ref{cmie}) induce the reduction of the statevector on the
eigenvectors of the position ${\bf Q}$. Of course, such a process
requires an infinitely long time, while, in finite times, only the
reduction on approximate eigenstates of ${\bf Q}$ takes
place\footnote{Of course, for very large times the Hamiltonian $H$
cannot be any more ignored: a sort of balance between the
Hamiltonian spreading of the wavefunction and the reduction
mechanism is established, which keeps constant the spread of the
wavefunction.}. We discuss here the time rate of the localization
process by studying the time dependence of the off--diagonal
elements of the statistical matrix $\langle {\bf Q}'| \rho |{\bf
Q}'' \rangle$. Again, we disregard the effect of the Schr\"odinger
term, this approximation being justified by the fact that, for the
values of $|{\bf Q}' - {\bf Q}''|$ in which we are interested, the
reduction process will turn out to be very fast.

Equation (\ref{efso}) becomes in the present case:
\begin{equation} \label{ecsl1}
\frac{\partial}{\partial t} \langle {\bf Q}'| \rho |{\bf Q}''
\rangle \quad = \quad - \Gamma ({\bf Q}',{\bf Q}'')\, \langle {\bf
Q}'| \rho |{\bf Q}'' \rangle,
\end{equation}
where:
\begin{equation} \label{Paaaqvl}
\Gamma ({\bf Q}',{\bf Q}'') \; = \; \gamma \int d^{3} x \left[
\frac{1}{2} F^{2} ({\bf Q}' - {\bf x}) +  \frac{1}{2} F^{2} ({\bf
Q}'' - {\bf x})  - F({\bf Q}' - {\bf x}) F({\bf Q}'' - {\bf x})
\right].
\end{equation}
Equation (\ref{ecsl1}) gives:
\begin{equation} \label{qwer}
\langle {\bf Q}'| \rho(t) |{\bf Q}'' \rangle \quad = \quad
e^{\displaystyle - \Gamma t}\, \langle {\bf Q}'| \rho(0) |{\bf
Q}'' \rangle.
\end{equation}
It is easily found that $\Gamma$ is an even function of ${\bf Q}'
- {\bf Q}''$. Since it is assumed that very many constituents of
the considered body are contained in a volume $\alpha^{-3/2}$, we
can use the macroscopic density approximation, consisting in
replacing the sum by an integral in equation (\ref{ecslw}). Then
one writes:
\begin{equation} \label{escl3}
F({\bf Q} - {\bf x}) \; = \; \int d^{3} \tilde{y}\, D(\tilde{\bf
y}) \left(\frac{\alpha}{2\pi}\right)^{3/2} e^{\displaystyle -
\frac{\alpha}{2}\, ({\bf Q} + \tilde{\bf y} - {\bf x})^{2}} ,
\end{equation}
where $D(\tilde{\bf y})$ is the number of particles per unit
volume in the neighborhood of the point ${\bf y} = {\bf Q} +
\tilde{\bf y}$.

A further approximation, which we call the sharp scanning
approximation, can be used, since we are not interested here in
the details of the function $\Gamma$ for ${\bf Q}' - {\bf Q}''
\rightarrow 0$. The sharp scanning approximation consists in
replacing the normalized Gaussian function appearing in equation
(\ref{escl3}) by the corresponding delta function. Then one has:
\begin{equation}
F({\bf Q} - {\bf x}) \quad = \quad D({\bf x} - {\bf Q}),
\end{equation}
so that one gets:
\begin{equation} \label{gss}
\Gamma ({\bf Q}' - {\bf Q}'') \; = \; \gamma \int d^{3} x\, \left[
D^{2}({\bf x}) - D({\bf x}) D({\bf x} + {\bf Q}' - {\bf Q}'')
\right],
\end{equation}
where suitable changes of the integration variable have also been
made. The physical meaning of $\Gamma$ is easily understood by
making reference to a homogeneous macroscopic body of density
$D_{0}$. Then:
\begin{equation} \label{ecsl5}
\Gamma \quad = \quad \gamma\, D_{0}\, n_{\makebox{\tiny out}},
\end{equation}
$n_{\makebox{\tiny out}}$ being the number of particles of the
body when the center of mass position is ${\bf Q}'$, which do not
lie in the volume occupied by the body when the center of mass
position is ${\bf Q}''$. The ratio between the macroscopic rate
(\ref{ecsl5}) and the microscopic rate (\ref{rrde}) is
$n_{\makebox{\tiny out}}\, D_{0} ( 4\pi /\alpha)^{3/2}$.

The results  (\ref{qwer}) and (\ref{ecsl5}) have to be compared
with the result:
\begin{eqnarray}
\langle {\bf Q}'| \rho(t) |{\bf Q}'' \rangle & = &
e^{\displaystyle - \lambda_{\makebox{\tiny macro}} t}\,
\langle {\bf Q}'| \rho(0) |{\bf Q}''\rangle, \\
\lambda_{\makebox{\tiny macro}} & = & N\, \lambda,
\end{eqnarray}
valid for $|{\bf Q}' - {\bf Q}''| \gg 1/\sqrt{\alpha}$, obtained
in section \ref{sec55} for the case of distinguishable particles.
We note that in the present case an additional factor $D_{0} (4\pi
/\alpha)^{3/2}$ appears in the macro--to--micro ratio, but such a
factor is multiplied by the number of uncovered particles
$n_{\makebox{\tiny out}}$ rather than by the total number $N$.
Clearly, this is a consequence of the indistinguishability of
particles and of the choice of the density as the dynamical
variable governing the process. In section \ref{sec55}, the length
parameter $1/\sqrt{\alpha}$ was chosen to be of the order of
$10^{-5}$ cm and the microscopic rate $\lambda$ was suggested to
be of the order of $10^{-16}$ sec$^{-1}$ with the aim of obtaining
$\lambda_{\makebox{\tiny macro}} \approx 10^7$ sec$^{-1}$ for a
typical macroscopic number $N \approx 10^{23}$. We repeat here the
same choice,
\begin{equation} \label{fn1}
\frac{1}{\sqrt{\alpha}} \quad \approx \quad 10^{-5}\; \makebox{cm}
\end{equation}
and look for a value of $\gamma$ such that the macroscopic rate
$\Gamma$ is again of the order of $10^7$ sec$^{-1}$ for
$n_{\makebox{\tiny out}} \approx 10^{13}$. Since $D_{0} \approx
10^{24}$ cm$^{-3}$, we get
\begin{equation} \label{fn2}
\gamma \quad \approx \quad 10^{-30}\; \makebox{cm$^3$ sec$^{-1}$},
\end{equation}
corresponding (according to the relation (\ref{rrde}) between
$\lambda$, $\gamma$ and $\alpha$) to $\lambda \approx 10^{-17}$
sec$^{-1}$. This value is such that nothing changes in the
dynamics of a microscopic particle even in the case in which it
has an extended wavefunction\footnote{Note that, in the case of a
macroscopic object, QMSL requires a displacement of $\simeq
10^{24}$ particles for the reduction rate to be equal to $10^{7}$
sec${}^{-1}$. CSL, on the other hand -- in the case of normal
density -- requires a displacement only of about $10^{13}$ particles in
order to have the same localization rate. This improvement of CSL with
respect to QMSL is due to the indistinguishability of identical
particles, whose effects are explicitly taken into account in CSL.}.

\subsection{Position and momentum spreads} \label{sec74}

According to equation (\ref{gss}) or to the original expression
(\ref{Paaaqvl}), the diagonal elements $\langle {\bf Q}| \rho
|{\bf Q} \rangle$ of the statistical operator in the position
representation are not affected by the reduction process, as a
consequence of the process being a localization. Of course, this
does not mean that the time evolution of $\langle {\bf Q}| \rho
|{\bf Q} \rangle$ is the same as the one given by the pure
Schr\"odinger dynamics: some changes are expected in the time
dependence of both position and momentum spreads, as a consequence
of the presence of the localization process. An explicit
evaluation of these effects is necessary in order to check that no
unacceptable behaviour arises.
\par The equation for the statistical operator, in operator form,
is written
\begin{equation} \label{vfc}
\frac{d}{dt} \rho(t) =  - \frac{i}{\hbar} [H, \rho(t)] + \gamma
\int d^3 x \left[ F({\bf Q} - {\bf x}) \rho(t) F({\bf Q} - {\bf
x}) - \frac{1}{2} \left\{ F^2({\bf Q} - {\bf x}), \rho(t) \right\}
\right]
\end{equation}
where we have retained also the Schr\"odinger term. We consider the case
of a free macroscopic body, so that, in our notation, $H = P^2 /
(2M)$, $M$ being the total mass. For a dynamical variable $S$, we
define the mean value
\begin{equation}
\llangle S \rrangle \quad = \quad \makebox{Tr} \left[ S \rho
\right].
\end{equation}
The time derivative of $\llangle S \rrangle$, according to
equation (\ref{vfc}), is given by
\begin{eqnarray} \label{tdc}
\frac{d}{dt} \llangle S \rrangle & = &  -\frac{i}{\hbar}
\makebox{Tr} \left( [S,H] \rho \right) \; + \\
& & \;\; \gamma \int d^3 x \;\makebox{Tr} \left[ \left( F({\bf Q}
- {\bf x}) S F({\bf Q} - {\bf x}) - \frac{1}{2} \left\{ S,
F^2({\bf Q} - {\bf x}) \right\} \right) \rho \right]. \nonumber
\end{eqnarray}
>From equation (\ref{tdc}), through tedious but elementary
calculations, one gets expressions for the average values of
position, momentum and their combinations which are analogous to
the QMSL equations (\ref{pedc})--(\ref{sedc}). In particular we
have:
\begin{eqnarray}
& \displaystyle\{ Q_i \} & = \quad \{ Q_i \}_{\makebox{\tiny Sch}}
\; + \; \gamma \delta_i \frac{\hbar^2}{6M^2} t^3,
\label{dfn1} \\
& \displaystyle\{ P_i \} & = \quad \{ P_i \}_{\makebox{\tiny Sch}}
\; + \; \gamma \delta_i \frac{\hbar^2}{2} t, \label{dfn3}
\end{eqnarray}
where
\begin{equation}
\delta_i \quad = \quad \int d^3 y \left[ \frac{\partial F ({\bf
y})}{\partial y_i} \right]^2.
\end{equation}
To evaluate the quantities $\delta_i$ the sharp scanning
approximation is no longer sufficient, because here the derivative
of the function $F$ is required. We then use the macroscopic
density approximation (\ref{escl3}). For definiteness and
simplicity, we make reference to a homogeneous macroscopic
parallelepiped of density $D_0$ having edges of lengths $L_i$
parallel to the coordinate axes. Then, as shown in \cite{csl}, one
has with high accuracy
\begin{equation} \label{meps}
\delta_i \quad = \quad \sqrt{\frac{\alpha}{\pi}}\, D_0^2\, S_i,
\end{equation}
where $S_i = L_1 L_2 L_3 / L_i$ is the transverse section of the
macroscopic parallelepiped.

If the choice (\ref{fn1}) and (\ref{fn2}) is used for $\alpha$ and
$\gamma$ together with $D_0 \approx 10^{24}$ cm$^{-3}$, one gets
from equation (\ref{meps}) for the momentum diffusion coefficient,
\begin{equation}
\frac{1}{2}\, \gamma\, \delta_i \, \hbar^2 \quad \approx \quad
10^{-32} (\makebox{g cm  sec$^{-1}$})^{2}\, \makebox{sec}^{-1}\,
S_i\, \makebox{cm}^{-2} .
\end{equation}
For an ordinary macroscopic body, this value appears too small to
give detectable effects. For a very small macroscopic particle,
due to the $1 / M^2$ factor in the extra term of equation
(\ref{dfn1}), a non--negligible stochasticity could appear. For
$L_j \approx 10^{-4}$ cm (this is the smallest order of magnitude
for which the approximations leading to equation (\ref{meps})
remain valid), a time of the order of $10^2$ sec is required to
make the extra term of the order of $10^{-10}$ cm$^2$. We do not
know whether this kind of effect could be used to provide a
significant experimental bound on the product $\gamma
\sqrt{\alpha}$ contained in the momentum diffusion coefficient. We
note, however, that the value (\ref{meps}) could overestimate
$\delta_i$, because of the assumption of a rectangular profile for
the object under consideration.

\subsection{Reduction mechanism: a simple model} \label{sec75}

After this long and detailed analysis of the reduction mechanism
within CSL, we propose here a second simpler way of looking at
the localization process \cite{cc}; for simplicity's  sake we
restrict ourselves to a simplified version of CSL obtained by
disregarding the hamiltonian term and discretizing the space.

We divide the space into cells of volume $(\alpha/2\pi)^{-3/2}$
and we denote by $N_{i}^{(k)}$ the number operator counting the
particles of type $k$ in the $i$--th cell. As follows from the
discussion of the preceding subsections in the considered case,
the dynamical evolution drives the statevector into a manifold
such that the number of particles present in any cell is definite.
The simplified equation for the statistical operator (\ref{cslso})
reads:
\begin{equation} \label{cslsod}
\frac{d}{dt}\, \rho(t) \; = \;
\gamma\left(\frac{\alpha}{4\pi}\right)^{3/2} \sum_{k} \left[
\sum_{i} N_{i}^{(k)}\, \rho(t)\, N_{i}^{(k)} \; - \;
\frac{1}{2}\sum_{i} \{ N_{i}^{(k)\,2} , \rho(t) \} \right].
\end{equation}
In accordance with relation (\ref{rrde}), we will often use the
QMSL rate parameter $\lambda$ in place of the expression $\gamma
(\alpha/4\pi)^{3/2}$. If we denote by $|n_{1}^{(k)}, n_{2}^{(k)},
\ldots n_{i}^{(k)}, \ldots \rangle$ the state with the indicated
occupation numbers for the various types of particles and for the
various cells, the solution of equation (\ref{cslsod}) reads, in
the considered basis:
\begin{eqnarray} \label{dsct}
\lefteqn{ \langle n_{1}^{(k)}, n_{2}^{(k)},\ldots |\rho(t)|
m_{1}^{(k)} m_{2}^{(k)},\ldots\rangle \; = \;} \nonumber \\ & &
=\; e^{\displaystyle -\, \frac{\lambda}{2}\sum_{k,i} (n_{i}^{(k)}
- m_{i}^{(k)})^{2}t} \langle n_{1}^{(k)}, n_{2}^{(k)},\ldots
|\rho(0)| m_{1}^{(k)} m_{2}^{(k)},\ldots\rangle.
\end{eqnarray}
Equation (\ref{dsct}) is an indirect proof\footnote{The direct
proof, as repeatedly stated, comes from the study of the
localization mechanism at the wavefunction level.} that linear
superpositions of states containing different number of particles
in the various cells are dynamically reduced to one of the
superposed states with an exponential time rate depending on the
expression
\[
\frac{\lambda}{2}\, \sum_{k} \sum_{i}\, (n_{i}^{(k)} -
m_{i}^{(k)})^{2}.
\]

The amplification process in going from the micro to the
macroscopic case and the preferred role assigned to position make
it clear how such models overcome the difficulties of quantum
measurement theory. In fact in measurement processes one usually
assumes that different eigenstates of the measured
micro-observable trigger (through the system--apparatus
interaction) different displacements of a macroscopic pointer from
its ``ready'' position. The unique dynamical principle of QMSL or
CSL leads then, in extremely short times, to the dynamical
suppression, with the appropriate probability, of all but one of
the terms in the superposition, i.e., to the emergence of an
outcome.

\subsection{Relating reductions to the mass density} \label{sec76}

In this subsection we consider a CSL type dynamics in which, in
place of the operators $N({\bf x})$ previously considered,  we
introduce the mass density operators \cite{cc}:
\begin{equation}
M({\bf x}) \quad = \quad \sum_{k}\, m_{k} N_{k}({\bf x}),
\end{equation}
where $m_{k}$  and $N_{k}$ are the mass and the average mass
density operator for a particle of type $k$, respectively. The
Stratonovich stochastic evolution equation for the statevector is:
\begin{equation}
\frac{d}{dt} |\psi(t)\rangle \; = \; \left[ - \frac{i}{\hbar}\, H
\; + \; \int d^{3} x\, M({\bf x}) V({\bf x}, t) \; - \;
\frac{\gamma}{m_{0}^{2}} \int d^{3} x\, M^{2}({\bf x}) \right]
|\psi(t)\rangle,
\end{equation}
where $m_{0}$ is a reference mass and $\gamma$ is the parameter
appearing in standard CSL of section \ref{sec71}. We identify the
mass $m_{0}$ with the nucleon mass; in this way the reduction
rates for macroscopic objects are practically equal to those of
the standard CSL model. With this choice the decoherence is
governed by the mass of the nucleons in ordinary matter, the
contribution due to electrons being negligible.

As usual the corresponding equation for the statistical operator
is easily obtained:
\begin{equation} \label{cslsob}
\frac{d}{dt}\, \rho(t) \; = \; - \frac{i}{\hbar}\, \left[ H,
\rho(t) \right]  +  \frac{\gamma}{m^{2}_{0}} \int d^3 x \, M({\bf
x}) \rho(t) M({\bf x})  - \frac{\gamma}{2m^{2}_{0}}  \int d^3 x \,
\{ M^{2}({\bf x}) , \rho(t) \}.
\end{equation}

One of main motivations to replace the number density operators
$N^{(k)}({\bf x})$ in the CSL dynamics with the mass density
operators $M({\bf x})$ derives from the desire to relate
reductions to gravity as suggested by various authors
\cite{kar,kom,pen1,emn,di2,fre} (a model with analogous
characteristics will be presented in the next section). Another
important feature of the above choice has been pointed out by P.
Pearle and E. Squires \cite{pes}: while the reduction rates for
macro--objects are practically the same as those of the standard
CSL model, the probabilities of excitation or dissociation of
microscopic bound systems turn out to be depressed by large
factors \cite{pes,bggg}, thus leading to a smaller disagreement
with the predictions of quantum mechanics for such systems. In
particular, a simple computation within the quark model for
nucleons (disregarding all relativistic effects which however
could turn out to be very important at this level) gives a
dissociation rate for the proton well below the experimental
bounds, while  the standard CSL model would lead to the violation of
such  bounds for the proton life-time. The advantages of taking the above
position have also been discussed by A. Rimini \cite{rimg}.

\subsection{A reduction model involving gravity} \label{sec8}

In 1989 L. Di\`osi \cite{di2} proposed a modification of QMSL,
different from CSL, with the explicit aim of eliminating the new
constants of nature $\alpha$ and $\gamma$ and of relating the
process to gravity. Such model has been called {\it quantum
mechanics with universal density localization} (QMUDL) by the
author.

Di\`osi considers an extended macroscopic object, such as e.g. a
homogeneous sphere, and introduces a mass density operator $f({\bf
x})$ at point ${\bf x}$. For the case of a homogeneous sphere of
mass $M$ and radius $R$:
\begin{equation} \label{jdt}
f({\bf x}) \quad = \quad \frac{M}{V}\, \theta(R - |{\bf q} - {\bf
x}|),
\end{equation}
where $V$ is the volume of the sphere and ${\bf q}$ is the center
of mass position operator. The stochastic dynamical equation is
assumed to be:
\begin{eqnarray} \label{gcsl1}
d\,|\psi(t)\rangle & = & \left[ -\,\frac{i}{\hbar}\,H\, dt \; - \;
\frac{G}{2\hbar} \int\int d^{3} x_{1}\, d^{3} x_{2}
\frac{1}{x_{12}} [f({\bf x}_{1}) - f_{\psi}({\bf x}_{1})] \times
\right. \nonumber \\
& & \; \times\left. [f({\bf x}_{2}) - f_{\psi}({\bf x}_{2})] dt  +
\int d^{3} x [f({\bf x}) - f_{\psi}({\bf x})] d\xi ({\bf x})
\right] |\psi(t)\rangle, \qquad
\end{eqnarray}
where $f_{\psi}({\bf x}) = \langle\psi|f({\bf x})| \psi\rangle$;
$\xi({\bf x})$ is a real Wiener process satisfying:
\begin{equation}
\llangle d\xi({\bf x}) \rrangle \; = \; 0 \qquad \quad \llangle
d\xi({\bf x}_{1}) d\xi({\bf x}_{2}) \rrangle \; = \;
\frac{G}{\hbar}\, \frac{1}{{\bf x}_{12}}\, dt.
\end{equation}
In the above equation ${\bf x}_{12} = |{\bf x}_{1} - {\bf x}_{2}|$
and $G$ is Newton's gravitational constant.

Equation (\ref{gcsl1}) can be put in the form (\ref{ime5}), as
shown in \cite{ggrg}, the corresponding operators $A_{i}$ being
functions of the center of mass position operator ${\bf q}$: the
non Hamiltonian terms then induce the reduction onto states in
which the position of the center of mass is more and more
definite.

It is easy to prove by It\^o stochastic calculus that equation
(\ref{gcsl1}) implies, for the statistical operator, the dynamical
equation
\begin{equation} \label{gcsl2}
\frac{d}{dt}\, \rho(t) \quad = \quad -\, \frac{i}{\hbar}\, [H,
\rho(t)] \; - \; \frac{G}{2\hbar}\, \int\int d^{3} x_{1}\, d^{3}
x_{2}\, \frac{1}{x_{12}}\, [f({\bf x}_{1}), [ f({\bf x}_{2}),
\rho(t)]].
\end{equation}

>From equation (\ref{gcsl2}), disregarding the Hamiltonian term, in
the considered case of a homogeneous sphere, one gets
\begin{equation} \label{gcsl3}
\langle {\bf q}'| \rho(t) |{\bf q}''\rangle \quad = \quad e^{\;
\displaystyle \Gamma(\Delta)\, t}\, \langle {\bf q}'| \rho(0)
|{\bf q}''\rangle,
\end{equation}
where
\begin{equation} \label{gcsl4}
\Gamma(\Delta) \; = \; -\, \frac{1}{\hbar}\, [U(0) - U(\Delta)],
\qquad\quad \Delta \; = \; |{\bf q}' - {\bf q}''|.
\end{equation}
In equation (\ref{gcsl4}) $U(\Delta)$ is the gravitational
interaction energy of two homogeneous spheres of the considered
mass and radius, with the centers at distance $\Delta$. For
instance, for $R = 1$ cm, $M = 1$ g and $\Delta = 10^{-5}$ cm,
$\Gamma(\Delta) \simeq 10^{9}$ sec$^{-1}$, so that a linear
superposition of two such states evolves into one of them in about
$10^{-9}$ sec.  This is comparable to the reduction rate implied by QMSL
in such a case.

This model has many appealing features, e.g. the reduction
mechanism is related to the gravitational potential without
resorting to any physical constant, besides Newton's constant $G$.
However, QMUDL, when taken literally, meets some serious
difficulties: in fact, it has been proven \cite{ggrg} that the
localization mechanism of QMUDL induces, in the case of
macroscopic systems (with a number of constituents of the order of
Avogadro's number), a total energy increase of about $10^{3}$ erg
sec${}^{-1}$, which clearly is unacceptable.

Ghirardi, Grassi and Rimini \cite{ggrg} have shown that it is
possible to remove such a difficulty by making the following
choice for the operators $f({\bf x})$ (which replaces
(\ref{jdt})):
\begin{equation} \label{mig1}
f({\bf x}) \quad = \quad m\, N({\bf x}),
\end{equation}
where $N({\bf x})$ is given by equation (\ref{ando}) and $m$ is
the mass of the particles created by $a^{\dagger}({\bf y},s)$.
Obviously if the system contains different types of particles, in
(\ref{mig1}) a sum over all types is understood. They chose for
the parameter $\alpha$ again the value given by CSL.

It is not difficult to show that  the model possesses all the
appealing features of CSL, in particular it induces a fast
suppression of the linear superpositions of states containing a
macroscopic number of particles which are differently located, it
does not alter in any appreciable way the dynamics of
microsystems, and, in the case of a body with the internal
coordinate sharply localized with respect to $1/\sqrt{\alpha}$, it
allows the decoupling of the internal and center of mass motions,
the internal motion being governed with high accuracy by the
standard quantum dynamics. The price which has been paid is, with
respect to Di\'osi's proposal which aimed to get rid of any new
constant, the introduction of a new physical constant, namely the
localization width $1/\sqrt{\alpha}$. In our opinion, however,
this is not a serious drawback. Actually, as we have stressed many
times, in order to have a precise theory one needs to identify the
borderline between quantum and classical, to get rid of the shifty
character of the standard theory and of any proposed
interpretation of it. The new parameter plays such a role in the
just discussed model.

\section{Dynamical reduction models with general gaussian noises}
\label{nsec}

The dynamical reduction models analyzed in the previous sections
have proven to yield (at the non relativistic level) a perfectly
consistent solution to the macro--objectification problem of
Quantum Mechanics. They are based on a stochastic modification of
Schr\"odinger equation: besides the standard Hamiltonian, new
terms are added, which contain Gaussian white noises.

It is an interesting question to analyze whether the most
important features of CSL (and consequently of QMSL), in
particular the localization mechanism, depend in any essential way
on the white noise character of the stochastic processes
considered \cite{pero, p1wq, gi2wq, gi2}. This is the subject of
the present section.

There is a second, more important, reason to consider dynamical
reduction models governed by more general noises. As we shall see
in the fourth part of the report, relativistic CSL meets serious
difficulties, since the reduction process yields an infinite
increase of the energy (per unit time and unit volume). This
divergent increase is due to the local coupling between quantum
fields and the {\it white noise} stochastic filed: it is still an
open problem to understand whether a non--white stochastic field
can remove these divergences, and lead to a fully consistent
relativistic model of dynamical reductions.

The content of the present section is the following. In subsection
\ref{nsec1} we derive a modified Schr\"odinger equation, in which
the new stochastic terms contain a general Gaussian noise. In
subsection \ref{nsec2} we analyze two important cases in which the
modified Schr\"odinger equation can be studied in detail. In
subsection \ref{nsec3} we analyze the reduction properties of such
an equation, while in subsection \ref{nsec4} we study the time
evolution of the average value of physical quantities. In this way
we prove that the considered model shares all the essential
features of standard CSL.

The section ends with an explicit application of the above results
to a specific model of dynamical reductions in {\it space}, like
we did in section \ref{sec7} for white noise models.

\subsection{The modified Schr\"odinger equation} \label{nsec1}

In this subsection we begin the analysis of dynamical reduction
models in which the reduction mechanism is controlled by general
Gaussian noises. The first task is to derive a modified
Schr\"odinger equation generalizing equation (\ref{smefsa}), and
preserving the average value of the square norm of vectors, so
that the cooking prescription can be applied to it.

Let us then consider the following equation:
\begin{equation} \label{nm1}
\frac{d|\psi(t)\rangle}{dt} \quad = \quad \left[ -\frac{i}{\hbar}
H_{0}\; +\; \sum_{i} A_{i}w_{i}(t) \right] |\psi(t)\rangle,
\end{equation}
where, as before, $H_{0}$ is the Hamiltonian of the system, $\{
A_{i} \}$ is a set of commuting self--adjoint operators, and
$w_{i}(t)$ are $c$--number gaussian stochastic processes whose
first two moments are\footnote{There is no loss of generality in
considering gaussian processes with zero mean. In fact, if
$\llangle w_{i}(t) \rrangle = m_{i}(t) \neq 0$, we can always
define new processes $z_{i}(t) = w_{i}(t) - m_{i}(t)$, which have
zero mean, and rewrite the modified Schr\"odinger equation (\ref{nm1}) in
terms of the processes $z_{i}(t)$.}:
\begin{equation} \label{cn}
\llangle w_{i}(t) \rrangle = 0, \qquad \quad \llangle w_{i}(t_{1})
w_{j}(t_{2})\rrangle = \gamma\,D_{ij}(t_{1}, t_{2}).
\end{equation}
We already know that the evolution described by equation
(\ref{nm1}) is not unitary and it does not preserve the norm of
the statevector; we then follow the same prescription outlined in
section \ref{sec6}. We consider as {\it physical} vectors the
normalized ones:
\begin{equation} \label{co2}
|\phi(t)\rangle \quad = \quad \frac{|\psi(t)\rangle}{\|
|\psi(t)\rangle \|},
\end{equation}
and we assume that any particular realization of the stochastic
processes $w_{i}(t)$  has a probability of occurrence
$P_{\makebox{\tiny Cook}}[w(t)]$ equal to:
\begin{equation} \label{prob2}
P_{\makebox{\tiny Cook}}[w(t)] \quad = \quad P_{\makebox{\tiny
Raw}}[w(t)]\, \||\psi(t)\rangle\|^{2},
\end{equation}
where $P_{\makebox{\tiny Raw}}[w(t)]$ is now the gaussian
probability distribution defined by (\ref{cn}).  The above
assumptions guarantee that the reduction probabilities reproduce
the standard quantum mechanical probabilities.

Of course, we have to check that equation (\ref{prob2}) correctly
defines a probability distribution, i.e. that it sums to 1.
Following the discussion of section \ref{sec6}, we know that this
is equivalent to requiring that the time derivative of $\llangle
\langle\psi(t)|\psi(t)\rangle \rrangle$ is zero. Let us evaluate
it:
\begin{eqnarray*}
\frac{d}{dt}\llangle\langle\psi(t)|\psi(t)\rangle\rrangle & = &
\LLangle\left[\frac{d\langle\psi(t)|}{dt}\right]|\psi(t)
\rangle\RRangle\; + \;
\LLangle\langle\psi(t)|\left[\frac{d|\psi(t)\rangle}{dt}\right]
\RRangle\quad = \\
& = & \LLangle\langle\psi(t)|\left[ +\;\frac{i}{\hbar} H_{0}\; +
\;\sum_{i}
A_{i}w_{i}(t)\right]|\psi(t)\rangle\RRangle \; + \\
& &  \LLangle\langle\psi(t)|\left[ -\;\frac{i}{\hbar}H_{0} \; +
\;\sum_{i} A_{i}w_{i}(t)\right]|\psi(t)\rangle\RRangle.
\end{eqnarray*}
The two terms involving the Hamiltonian $H_{0}$ cancel out (in
fact they describe the unitary part of the evolution); the noises
$w_{i}(t)$, being $c$--numbers, can be taken out of the scalar
product, so that:
\begin{equation} \label{ie1}
\frac{d}{dt}\llangle\langle\psi(t)|\psi(t)\rangle\rrangle \quad =
\quad 2 \sum_{i} \llangle\langle\psi(t)|A_{i}|\psi(t)\rangle
w_{i}(t)\rrangle.
\end{equation}
The right hand side of (\ref{ie1}) can be rewritten with the help
of the {\it Furutsu--Novikov formula} \cite{ksv, gi2}:
\begin{equation} \label{fn}
\llangle F[w(t)]w_{i}(t) \rrangle \quad = \quad \gamma \sum_{j}
\int_{0}^{+\infty} D_{ij}(t,s) \LLangle \frac{\delta
F[w(t)]}{\delta w_{j}(s)} \RRangle\, ds
\end{equation}
(for simplicity, throughout this subsection we take $t_{0} = 0$ as
the initial time). $F[w(t)]$ is any functional of the stochastic
fields $w_{i}(t)$; in the present case case, $F[w(t)] =
\langle\psi(t)|A_{i}|\psi(t)\rangle$.

The formal solution of equation (\ref{nm1}) is:
\begin{equation} \label{fs}
|\psi(t)\rangle = T\, e^{\displaystyle -\, \frac{i}{\hbar}H_{0}
t\, + \sum_{i} A_{i} \int_{0}^{t} w_{i}(s)\, ds }|\psi(0)\rangle.
\end{equation}
Note that, since $|\psi(t)\rangle$ depends on the stochastic
processes $w_{i}(s)$ only within the time--interval $[0,t]$, the
functional derivative of $|\psi(t)\rangle$ with respect to
$w_{j}(s)$ is zero if $s \not\in [0,t]$. We then have:
\begin{eqnarray} \label{ie2}
\frac{d}{dt}\llangle\langle\psi(t)|\psi(t)\rangle\rrangle & = &
2\gamma \sum_{i,j} \int_{0}^{t} D_{ij}(t,s) \LLangle \left[
\frac{\delta \langle\psi(t)|}{\delta w_{j}(s)}\right]
A_{i}|\psi(t)\rangle\RRangle \,ds \nonumber \\ &  &  \\ & + &
2\gamma \sum_{i,j} \int_{0}^{t} D_{ij}(t,s) \LLangle
\langle\psi(t)|A_{i} \left[ \frac{\delta |\psi(t)\rangle}{\delta
w_{j}(s)} \right]\RRangle \, ds \; \neq \; 0. \nonumber
\end{eqnarray}
Since the time derivative of the average value of the square norm
of the statevector is not zero, we have to add an extra term to
equation (\ref{nm1}), as expected and as it happens also in the
case of white noise. Relation (\ref{ie2}) tells us which kind of
term must be added. The conclusion follows: with reference to our
procedure, the request that $P_{\makebox{\tiny Cook}}[w(t)]$
correctly defines a probability distribution, i.e. that the
average value of the square norm of the statevector
$|\psi(t)\rangle$ is conserved, leads to the stochastic
Schr\"odinger equation:
\begin{equation}\label{sch2}
\frac{d|\psi(t)\rangle}{dt} =  \left[ -\frac{i}{\hbar} H_{0} +
\sum_{i} A_{i} w_{i}(t) - 2\gamma \sum_{i,j}A_{i} \int_{0}^{t}
ds\, D_{ij}(t,s) \frac{\delta}{\delta w_{j}(s)} \right]
|\psi(t)\rangle.
\end{equation}
This is the main result of this subsection. Note that an equation
like (\ref{sch2}) has been derived also in \cite{gi2wq, bud} by
following a different line of thought.

Some comments are appropriate:
\begin{itemize}
\item Equation (\ref{sch2}) no longer describes a Markovian
evolution for the sta\-te\-vector unless the correlation functions
$D_{ij}(t,s)$ are Dirac--$\delta$'s in the time variable --- i.e.
the stochastic processes $w_{i}(t)$ are white in time. As a
consequence, the corresponding equation for the statistical
operator is not of the quantum--dynamical--semigroup
type\footnote{See section \ref{sec56}.}, contrary to what happen
for the case of CSL (see equation (\ref{efso})).

\item In general,  the explicit form of the functional derivatives
of $|\psi(t)\rangle$ with respect to the noise $w_{i}(t)$ cannot
be evaluated exactly, except for few special cases, two of which
will be considered in the next subsection. Therefore, in the
general case it is difficult to analyze the time evolution of the
statevector and the statistical properties of the ensemble of
states generated by the stochastic processes. In particular, one
cannot write a closed equation for the evolution of the
statistical operator.
\end{itemize}

\subsection{Two special cases} \label{nsec2}

In order to understand the kind of difficulties one encounters
when working with non--white stochastic processes, and in
particular the reasons for which the functional derivative of the
statevector $|\psi(t)\rangle$ in general cannot be computed
exactly, let us reconsider equation (\ref{fs}), writing explicitly
its perturbative expansion:
\begin{eqnarray} \label{fs2}
\lefteqn{ T\, e^{\displaystyle -\, \frac{i}{\hbar}H_{0}t\, +
\sum_{i} A_{i} \int_{0}^{t} w_{i}(s)\, ds } \; =} \qquad\qquad
\nonumber \\
& = & \sum_{n = 0}^{\infty} \left[-\frac{i}{\hbar}\right]^{n}\,
\frac{1}{n!} \, \int_{0}^{t} dt_{1} \cdots \int_{0}^{t} dt_{n}\,
T\left\{H(t_{1}) \ldots H(t_{n}) \right\}, \quad
\end{eqnarray}
where we have defined the operator:
\begin{equation} \label{yhj}
H(t) \quad = \quad H_{0} \; + \; i\hbar\sum_{i}A_{i}\,w_{i}(t).
\end{equation}

The functional derivative of $|\psi(t)\rangle$ with respect to
$w_{j}(s)$ can be obtained deriving  term by term the
series\footnote{We assume that the initial state $|\psi(0)\rangle$
does not depend on the stochastic processes $w_{i}(t)$.}
(\ref{fs2}). The derivative of the term $n=0$ is zero; the
derivative of the term $n=1$ is:
\begin{equation} \label{topd}
\frac{\delta}{\delta w_{j}(s)} \left[ -\frac{i}{\hbar}\,
\int_{0}^{t} dt_{1}\, H(t_{1}) \right] \; = \; -\frac{i}{\hbar}\,
\int_{0}^{t} dt_{1}\, \left[ i\hbar\, \delta(s - t_{1})\, A_{j}
\right] \; = \; A_{j}.
\end{equation}
The next ($n=2$) term is:
\begin{equation} \label{nt314}
\left[-\frac{i}{\hbar}\right]^{2}\, \frac{1}{2}\, \int_{0}^{t}
dt_{1} \int_{0}^{t} dt_{2}\, T\left\{H(t_{1}) \, H(t_{2})
\right\}.
\end{equation}
The functional derivative of the time--ordered product $T \{
H(t_{1})\, H(t_{2}) \} = \theta(t_{1}-t_{2})H(t_{1})H(t_{2}) +
\theta(t_{2} - t_{1}) H(t_{2}) H(t_{1})$ is:
\begin{eqnarray} \label{dftt}
\lefteqn{\frac{\delta}{\delta w_{j}(s)} T\left\{ H(t_{1}) \,
H(t_{2}) \right\} \; = \qquad\quad} \nonumber \\
& = & i\hbar\,\theta(t_{1}-t_{2})\left[ \delta(t_{1} - s)\,
A_{j}\, H(t_{2}) \; + \; \delta(t_{2} - s)\, H(t_{1})\, A_{j}
\right] \nonumber \\
& + & i\hbar\,\theta(t_{2}-t_{1})\left[ \delta(t_{2} - s)\,
A_{j}\, H(t_{1}) \; + \; \delta(t_{1} - s)\, H(t_{2})\, A_{j}
\right].
\end{eqnarray}
We note that the first and third terms at the right hand side of
(\ref{dftt}) differ only for the exchange of the dummy variables
$t_{1} \leftrightarrow t_{2}$; the same is true for the second and
the fourth term. The derivative of the $n=2$ term (i.e. of Eq.
(\ref{nt314}))  is then:
\begin{equation} \label{rftt}
A_{j} \left[ -\frac{i}{\hbar}\, \int_{0}^{s} dt_{1}\, H(t_{1})
\right] \; + \; \left[ -\frac{i}{\hbar}\, \int_{s}^{t} dt_{1}\,
H(t_{1}) \right] A_{j}.
\end{equation}
Equation (\ref{rftt}) does not have a simple form, contrary to
(\ref{topd}), and derivatives of higher terms are more and more
complicated, due to the fact that the operators $A_{j}$ in general
do not commute with the Hamiltonian $H_{0}$. In fact, would they
commute, equation (\ref{rftt}) would simplify to:
\begin{equation}
A_{j} \left[ -\frac{i}{\hbar}\, \int_{0}^{t} dt_{1}\, H(t_{1})
\right],
\end{equation}
i.e. the derivative of the second term would give $A_{j}$ times
the first term. Moreover, if $[A_{j}, H_{0}] = 0$, the functional
derivative of the term $n+1$ gives $A_{j}$ times the $n$--th term
so that :
\begin{equation} \label{nhlqr}
\frac{\delta}{\delta w_{j}(s)}\, |\psi(t)\rangle \quad = \quad
A_{j}\, |\psi(t)\rangle,
\end{equation}
as we are going to prove. In fact, the hypothesis that the
operators $A_{i}$ commute with the Hamiltonian $H_{0}$ is
equivalent to the (more elegant) requirement that the operators
$H(t)$ defined in (\ref{yhj}) commute at different times. In this
case, the time--ordered product in the exponential series
(\ref{fs2}) can be omitted, and the functional derivative of the
$n$--th term is:
\begin{eqnarray}
\lefteqn{\frac{\delta}{\delta\, w_{j}(s)} \left[-\frac{i}{\hbar}
\right]^{n}\, \frac{1}{n!} \, \int_{0}^{t} dt_{1} \cdots
\int_{0}^{t} dt_{n}\, \left\{H(t_{1}) \ldots H(t_{n}) \right\} \;
=}
\nonumber \\
& = & \left[-\frac{i}{\hbar}\right]^{n}\, \frac{1}{n!}
\sum_{i=1}^{n} \int_{0}^{t} dt_{1} \cdots  \int_{0}^{t} dt_{n}\,
\left\{H(t_{1}) \ldots \frac{\delta\, H(t_{i})}{\delta\ w_{j}(s)}
\ldots H(t_{n})
\right\}\; = \nonumber \\
& = & \left[-\frac{i}{\hbar}\right]^{n}\, \frac{1}{(n-1)!}
\int_{0}^{t} dt_{1} \cdots  \int_{0}^{t} dt_{n}\, \left\{
\frac{\delta\, H(t_{1})}{\delta\ w_{j}(s)} \ldots H(t_{n})
\right\}\; = \nonumber \\
& = & A_{j} \left[-\frac{i}{\hbar}\right]^{n-1}\, \frac{1}{(n-1)!}
\int_{0}^{t} dt_{1} \cdots  \int_{0}^{t} dt_{n-1}\, \left\{
H(t_{1}) \ldots H(t_{n-1}) \right\}.
\end{eqnarray}
This completes the proof. Note also that, when $s=t$, an extra
factor $1/2$ appears in (\ref{nhlqr}), because in this case the
Dirac delta function arising from the functional derivative of
$H(t)$ is centered in one of the two extreme points of the
interval of integration.

Recently, L. Hughston \cite{hugh}, S. Adler and P. Horwitz
\cite{ad1, ad2} have proposed a white--noise model of dynamical
reductions in which the operators $A_{i}$ are taken to be
functions of the Hamiltonian $H_{0}$; this implies that the
stochastic terms of equation (\ref{smefsa}) drive the statevector
into the {\it energy} eigenmanifolds of the physical system.
Making such a choice in the non--white equation (\ref {sch2}), the
operators $H(t)$ at different times commute among themselves, the
functional derivatives of the statevector $|\psi(t)\rangle$ can be
computed, and equation (\ref{sch2}) becomes:
\begin{equation}\label{sch2a}
\frac{d|\psi(t)\rangle}{dt} =  \left[ -\frac{i}{\hbar} H_{0} +
\sum_{i} A_{i} w_{i}(t) - 2\gamma \sum_{i,j}A_{i} A_{j}
\int_{0}^{t} D_{ij}(t,s)\,ds \right] |\psi(t)\rangle,
\end{equation}
with $A_{i} = A_{i}(H_{0})$. Equation (\ref{sch2a}) is exact and,
correspondingly, one can easily derive a closed equation for the
time evolution of the statistical operator. All the statistical
properties concerning the physical system can be evaluated
exactly.

We conclude the subsection showing that the functional derivatives
of $|\psi(t)\rangle$ can be explicitly evaluated also in the case
of general white noise stochastic processes,  without having to
require that $H_{0}$ commutes with $A_{i}$. Moreover, we will
prove that in this case equation (\ref{sch2}) reduces to
(\ref{smefsa}), as expected.

Under the assumption of white--noise stochastic processes
($D_{ij}(t_{1}, t_{2}) \quad = \quad \delta_{ij}\,\delta(t_{1} -
t_{2})$), the Furutsu--Novikov relation
\begin{equation} \label{fnwn}
\llangle F[w(t)]w_{i}(t) \rrangle \quad = \quad \gamma \LLangle
\frac{\delta F[w(t)]}{\delta w_{i}(t)} \RRangle
\end{equation}
leads to the following expression for the time derivative of the
average value of the square norm of the statevector
$|\psi(t)\rangle$ satisfying equation (\ref{nm1}):
\begin{eqnarray} \label{ie2wn}
\frac{d}{dt}\llangle\langle\psi(t)|\psi(t)\rangle\rrangle & = &
2\gamma \sum_{i} \LLangle \left[ \frac{\delta
\langle\psi(t)|}{\delta w_{i}(t)}\right]
A_{i}|\psi(t)\rangle\RRangle  \; + \nonumber \\  &  & 2\gamma
\sum_{i} \LLangle \langle\psi(t)|A_{i} \left[ \frac{\delta
|\psi(t)\rangle}{\delta w_{i}(t)} \right]\RRangle.
\end{eqnarray}
We now have to evaluate the functional derivatives of the
statevector, taking into account that the noises $w_{i}$
(appearing in the derivatives) are taken at time $t$.

The derivative of the term $n=1$ is equal to $(1/2)A_{j}$ (see
equation (\ref{topd})), the factor $(1/2)$ deriving from the Dirac
delta function $\delta(t - t_{1})$ which is integrated between $0$
and $t$. For the derivative of the $n=2$ term, let us look at
expression (\ref{rftt}). If we take $s=t$, the second term goes to
zero, while the first one gives\footnote{The factor $(1/2)$
appears for the same reason as before.}:
\begin{equation}
\frac{1}{2}\, A_{j} \left[ -\frac{i}{\hbar} \int_{0}^{t} dt_{1}\,
H(t_{1}) \right].
\end{equation}
In general, the functional derivative of any term of the
exponential series (\ref{fs2}) gives $(1/2)A_{j}$ times the
previous term, so that, in general:
\begin{equation} \label{rsncwn}
\frac{\delta}{\delta w_{j}(t)}\, |\psi(t)\rangle \quad = \quad
\frac{1}{2}\,A_{j}\, |\psi(t)\rangle.
\end{equation}
This means that the square--norm--preserving Schr\"odinger
equation is:
\begin{equation}\label{sch3}
\frac{d|\psi(t)\rangle}{dt} =  \left[ -\frac{i}{\hbar} H_{0} +
\sum_{i} A_{i}w_{i}(t) - \gamma \sum_{i} \,A_{i}\, A_{j} \right]
|\psi(t)\rangle,
\end{equation}
which coincides with the original CSL equation (\ref{smefsa}). An
alternative and quicker way to derive the white--noise limit is to
replace $D_{ij}(t,s)$ with $\delta_{ij} \delta(t-s)$ in equation
(\ref{sch2}) and to show that (\ref{rsncwn}) is a consistent
solution.

\subsection{The reduction mechanism} \label{nsec3}

Here, we will analyze under which conditions the new terms in the
modified Schr\"odinger equation (\ref{sch2}) induce, for large
times, the reduction of the statevector to one of the common
eigenstates of the commuting operators $A_{i}$.

For this purpose, let us disregard the Hamiltonian $H_{0}$; under
this assumption the operators $H(t)$ commute at different times
and (as discussed in the previous subsection) the functional
derivatives of the statevector $|\psi(t)\rangle$ give the
operators $A_{i}$ times $|\psi(t)\rangle$. Equation (\ref{sch2})
becomes then\footnote{Here and in what follows, we consider a
generic initial time $t_{0}$.}:
\begin{equation}\label{schred}
\frac{d|\psi(t)\rangle}{dt} =  \left[ \sum_{i} A_{i}\, w_{i}(t) -
2\gamma \sum_{i,j}A_{i} A_{j} \int_{t_{0}}^{t} D_{ij}(t,s)\,ds
\right] |\psi(t)\rangle.
\end{equation}
The equation for the statistical operator can now be easily
derived; using the definition (\ref{hjkg}), we get:
\begin{equation} \label{erops}
\frac{d \rho(t)}{dt} \quad = \quad -\gamma \sum_{i,j} \left[
A_{i}, \left[ A_{j}, \rho(t)\right]\right] \int_{t_{0}}^{t}
D_{ij}(t,s)\, ds,
\end{equation}
which is a consistent generalization of the CSL equation
(\ref{efso}) when the Hamiltonian $H_{0}$ is omitted: in fact, if
the stochastic processes $w_{i}(t)$ are independent and white
($D_{ij}(t_{1}, t_{2}) = \delta_{ij}\, \delta(t_{1} - t_{2})$),
then (\ref{erops}) reduces exactly to (\ref{efso}).

In order to test the reduction properties, we will show first of
all how the reduction mechanism works for the statistical
operator. As in section \ref{sec6}, let us suppose that the common
eigenmanifolds of the operators $A_{i}$, which we assume to have a
purely discrete spectrum, are one--dimensional; let
$|\alpha\rangle$ be the vector spanning the
$\alpha$--eigenmanifold. The equation for the matrix elements
$\langle\alpha|\rho(t)|\beta\rangle$ is:
\begin{equation} \label{dada}
\frac{d \langle\alpha|\rho(t)|\beta\rangle}{dt} \; = \; - \,
\gamma \sum_{i,j} (a_{i\alpha} - a_{i\beta})(a_{j\alpha} -
a_{j\beta}) \int_{t_{0}}^{t} D_{ij}(t,s)\, ds \,
\langle\alpha|\rho(t)|\beta\rangle.
\end{equation}
Making use of the symmetry property of the correlation functions:
\begin{equation}
D_{ij}(t_{1}, t_{2}) \quad = \quad D_{ji}(t_{2}, t_{1}),
\end{equation}
we can write the solution of equation (\ref{dada}) in the
following form:
\begin{equation} \label{sesr}
\langle\alpha|\rho(t)|\beta\rangle  =  e^{\displaystyle - \,
\frac{\gamma}{2} \sum_{i,j} (a_{i\alpha} - a_{i\beta})(a_{j\alpha}
- a_{j\beta}) \int_{t_{0}}^{t} dt_{1} \int_{t_{0}}^{t} dt_{2}\,
D_{ij}(t_{1},t_{2})} \langle\alpha|\rho(t_{0})|\beta\rangle.
\end{equation}
>From equation (\ref{sesr}), we sees that if $|\alpha\rangle =
|\beta\rangle$, the exponent is zero: as in CSL, the diagonal
elements of the density matrix do not change in time. If, on other
other hand $|\alpha\rangle \neq |\beta\rangle$, the evolution of
the matrix element depends on the time behavior of the correlation
functions $D_{ij}(t_{1},t_{2})$.

If we want the off--diagonal elements to be damped at large times,
two conditions must be satisfied. The first one is that the {\bf
exponent} in (\ref{sesr}) must be {\bf negative}: this is always
true, since the correlation function of a Gaussian process is
positive definite\footnote{We assume that the correlation function
is non degenerate.}.

The second condition is that the {\bf double integral} of the
correlation function must {\bf diverge} for large times:
\begin{equation} \label{cond2}
\int_{t_{0}}^{t} dt_{1} \int_{t_{0}}^{t} dt_{2}\,
D_{ij}(t_{1},t_{2}) \; \longrightarrow \; +\infty \qquad
\makebox{for $t \rightarrow +\infty$},
\end{equation}
so that the off--diagonal elements of the density matrix go to
zero. This condition is not {\it a priori} satisfied by a generic
Gaussian stochastic field. At any rate, physically reasonable
stochastic fields always satisfy it: here we present just a couple
of meaningful examples.

Suppose the stochastic fields $w_{i}(t)$ are equal and
independent, with a (normalized) Gaussian correlation function:
\begin{equation} \label{gcf}
D_{ij}(t_{1},t_{2}) \quad = \quad \delta_{ij}\,
\frac{1}{\sqrt{2\pi}\tau}\,e^{\displaystyle - \frac{(t_{1} -
t_{2})^{2}}{2\tau^{2}}}.
\end{equation}
Let us also take $t_{0} = -\infty$. Equation (\ref{dada}) then
becomes:
\begin{equation} \label{dadae1}
\frac{d \langle\alpha|\rho(t)|\beta\rangle}{dt} \; = \; - \,
\frac{\gamma}{2} \sum_{i} (a_{i\alpha} - a_{i\beta})^{2} \,
\langle\alpha|\rho(t)|\beta\rangle,
\end{equation}
which is independent from the correlation time $\tau$, and
moreover it corresponds exactly to the CSL evolution. Note that if
we take the limit $\tau \rightarrow 0$, the gaussian process
becomes a white noise process with a Dirac--$\delta$ correlation
function and we recover, again, the CSL theory.

As a second example, suppose the correlation function is:
\begin{equation} \label{gcf2}
D_{ij}(t_{1},t_{2}) \quad = \quad \delta_{ij}\,
\frac{1}{2\tau}\,e^{\displaystyle - \frac{|t_{1} - t_{2}|}{\tau}}.
\end{equation}
Equation (\ref{dada}) becomes:
\begin{equation} \label{dadae2}
\frac{d \langle\alpha|\rho(t)|\beta\rangle}{dt} \; = \; - \,
\frac{\gamma}{2} \left[1- e^{\displaystyle -\frac{(t -
t_{0})}{\tau}}\right] \sum_{i} (a_{i\alpha} - a_{i\beta})^{2} \,
\langle\alpha|\rho(t)|\beta\rangle.
\end{equation}
As before, the off--diagonal elements are exponentially damped
and, in the limit $t \rightarrow +\infty$ we recover the behaviour
of CSL. Note that the effect of a non--white correlation function
is that of decreasing the reduction rate of the localization
mechanism.

We now analyze how the reduction mechanism works at the
wavefunction level. As in section \ref{sec64}, we consider a
simplified dynamics in which only one operator $A$ appears in
equation (\ref{schred}). This operator is coupled to a single
stochastic process $w(t)$, whose correlation function is $D(t_{1},
t_{2})$. Finally, we assume that at the initial time $t_{0}$ the
statevector is:
\begin{equation}
|\psi(t_{0})\rangle \quad = \quad P_{\alpha} |\psi(t_{0})\rangle
\; + \; P_{\beta} |\psi(t_{0})\rangle,
\end{equation}
where $P_{\alpha}$ and $P_{\beta}$ are projection operators onto
the eigenmanifolds of $A$ corresponding to two different
eigenvalues $\alpha$ and $\beta$, respectively. The solution of
equation (\ref{schred}) is:
\begin{equation}
|\psi(t)\rangle \; = \; e^{\displaystyle \alpha x(t) - \alpha^{2}
\gamma f(t)}P_{\alpha} |\psi(t_{0})\rangle \; + \;
e^{\displaystyle \beta x(t) - \beta^{2} \gamma f(t)}P_{\alpha}
|\psi(t_{0})\rangle,
\end{equation}
where
\begin{equation}
x(t) \; = \; \int_{t_{0}}^{t} w(s)\, ds, \qquad\quad f(t) \; = \;
\int_{t_{0}}^{t} ds_{1} \int_{t_{0}}^{t} ds_{2}\, D(s_{1}, s_{2}).
\end{equation}
Note that $\gamma f(t) = \llangle x^{2}(t) \rrangle$, i.e. such a
quantity is the variance of the stochastic process $x(t)$.

Since the ``raw'' probability distribution of the process $x(t)$
is:
\begin{equation}
P_{\makebox{\tiny Raw}}[x(t)] \quad  = \quad \frac{1}{\sqrt{2\pi
\gamma f(t)}}\, e^{\displaystyle -\frac{1}{2\gamma f(t)}\,
x^{2}(t)},
\end{equation}
taking into account the cooking prescription (\ref{prob2}) we
obtain:
\begin{eqnarray} \label{nmvd}
P_{\makebox{\tiny Cook}}[x(t)] &  = & \|P_{\alpha}
|\psi(t_{0})\rangle \|^{2}\, \frac{1}{\sqrt{2\pi\gamma f(t)}}\,
e^{\displaystyle -\frac{1}{2\gamma f(t)}\, [x(t) - 2\alpha\gamma
f(t)]^{2}} \nonumber \\
&  + & \|P_{\beta} |\psi(t_{0})\rangle \|^{2}\,
\frac{1}{\sqrt{2\pi\gamma f(t)}}\, e^{\displaystyle
-\frac{1}{2\gamma
f(t)}\, [x(t) - 2\beta\gamma f(t)]^{2}}. \nonumber \\
\end{eqnarray}
Equation (\ref{nmvd}) implies that, if $f(t) \rightarrow +\infty$
when $t \rightarrow +\infty$, the stochastic process $x(t)$ will
take either a value close to $2\alpha\gamma f(t)$ --- within an
interval of width $\sqrt{\gamma f(t)}$ --- or a value close to
$2\beta\gamma f(t)$, within the same interval\footnote{As noted in
section \ref{sec64}, even though the interval $\sqrt{\gamma f(t)}$
tends to infinity as time increases, the ratio $\sqrt{\gamma
f(t)}/ 2(\alpha - \beta)\gamma f(t)$ goes to zero.}. Of course,
the requirement that $f(t) \rightarrow +\infty$ as time increases
is exactly the same as requirement (\ref{cond2}) which guarantees
the damping of the off--diagonal elements of the density matrix.

Suppose now that the actual realization of the stochastic process
$x(t)$ occurs around $2\alpha\gamma f(t)$; the corresponding
probability is $\|P_{\alpha} |\psi(0)\rangle \|^{2}$. We then
have:
\begin{equation}
\frac{\|P_{\beta} |\psi(t)\rangle \|^{2}}{\|P_{\alpha}
|\psi(t)\rangle \|^{2}} \; \simeq \; e^{\displaystyle
-2\gamma(\alpha - \beta)^{2} f(t)} \frac{\|P_{\beta}
|\psi(0)\rangle \|^{2}}{\|P_{\alpha} |\psi(0)\rangle \|^{2}} \;
\rightarrow \; 0 \qquad \makebox{as $t \rightarrow \infty$},
\end{equation}
which means that the statevector $|\psi(t)\rangle$ is driven into
the eigenmanifold of the operator $A$ corresponding to the
eigenvalue $\alpha$. By the same reasoning, it is immediate to see
that, with a probability equal to $\|P_{\beta} |\psi(0)\rangle
\|^{2}$, the statevector is driven into the eigenmanifold
associated to the eigenvalue $\beta$. We have thus proved that the
statevector $|\psi(t)\rangle$  undergoes a random spontaneous
reduction to one of the two eigenmanifolds of the operator
$A$, with a probability which coincides with the one assigned by
standard Quantum Mechanics to the outcomes of an experiment aimed
to measure the observable $A$.

\subsection{The average value of observables} \label{nsec4}

When one disregards the Hamiltonian term $H_{0}$, it is not
difficult to see how the stochastic terms affect the average value
of physical quantities. Its time derivative can be calculated
following almost the same steps which, in the previous subsection,
have led to equation (\ref{erops}) for the statistical operator; the
final equation is:
\begin{equation} \label{av}
\frac{d\langle O\rangle}{dt} \quad  = \quad -\gamma \sum_{i,j}
\llangle \langle\psi(t)| \left[ A_{i},\left[ A_{j}, O
\right]\right] |\psi(t)\rangle \rrangle \int_{t_{0}}^{t}
D_{ij}(t,s)\, ds,
\end{equation}
to be compared with the corresponding CSL--white noise equation:
\begin{equation} \label{avcsl}
\frac{d\langle O\rangle}{dt} \quad  = \quad -\,\frac{\gamma}{2}
\sum_{i} \llangle \langle\psi(t)| \left[ A_{i},\left[ A_{i}, O
\right]\right] |\psi(t)\rangle \rrangle.
\end{equation}
The analysis of the previous subsection should have made clear how
(\ref{av}) differs from (\ref{avcsl}), so we will not repeat it
here.

\subsection{Connection with CSL} \label{nsec5}

We now apply the formalism introduced in the previous subsections
to derive an equation with the property of localizing macroscopic
systems in {\it space}, as it happens for CSL. In other words, we specify
the choice of the ``preferred basis'' $\{ A_{i} \}$ in such a way
to have a physically meaningful theory for our purposes.

The most natural choice for the operators $A_{i}$ is the number
density operator for a system of identical particles:
\begin{equation}
A_{i} \quad \longrightarrow \quad {\mathcal N}({\bf x}) \; = \;
\sum_{s} a^{\dagger}({\bf x}, s)\,a({\bf x}, s).
\end{equation}
Correspondingly, the noises $w_{i}(t)$ are replaced by a
stochastic field $w({\bf x}, t)$, whose correlation function is
$D({\bf x}, t_{1}; {\bf y}, t_{2})$.

In section \ref{sec9}, the transformation and invariance
properties of dynamical reduction models will be discussed in
detail. In particular, it will be proved that, in order for the
physics of the model to be invariant under Galilean
transformations (we speak of {\it stochastic Galilean
invariance}), the correlation function $D({\bf x}, t_{1}; {\bf y},
t_{2})$ itself must be invariant under the considered group of
transformations, i.e.
\begin{equation} \label{fhf}
D({\bf x}, t_{1}; {\bf y}, t_{2}) \quad = \quad D(|{\bf x} - {\bf
y}|, t_{1}- t_{2});
\end{equation}
the easiest way to construct a function like (\ref{fhf}) is to
take the product of two functions of the space and time variables,
respectively:
\begin{equation}
D({\bf x}, t_{1}; {\bf y}, t_{2}) \quad = \quad g(|{\bf x} - {\bf
y}|)\, h(t_{1} - t_{2}).
\end{equation}
As regards $g(|{\bf x} - {\bf y}|)$, a reasonable choice is a
gaussian function, like in CSL:
\begin{equation}
g(|{\bf x} - {\bf y}|) \quad = \quad \gamma
\left(\frac{\alpha}{4\pi}\right)^{\frac{3}{2}}\, e^{\displaystyle
- \frac{\alpha}{4} ({\bf x} - {\bf y})^{2}},
\end{equation}
with $1/\sqrt{\alpha} \simeq 10^{-5}$ cm.

It is  natural to choose a gaussian function also for $h(t_{1} -
t_{2})$:
\begin{equation}
h(t_{1} - t_{2}) \quad = \quad
\left(\frac{\beta}{4\pi}\right)^{\frac{1}{2}}\, e^{\displaystyle -
\frac{\beta}{4} (t_{1} - t_{2})^{2}}.
\end{equation}
In making the above choice, we have introduced a new parameter
($\beta$); this can be considered as a drawback of the model.
However, we note that it always is possible to define $\beta$ in
terms of $\alpha$, $\gamma$ and fundamental constants of nature,
so that, essentially,  no new arbitrary parameter is introduced
into the model. As an example, we can choose $\beta = c^{2}\alpha
\simeq 10^{30}$ sec${}^{-2}$, where $c$ is the speed of light.
This choice is particularly appropriate from the point of view of
a possible relativistic generalization of the theory, which we
will discuss in the fourth part of this report. Moreover, such a
choice corresponds to an extremely small correlation time, so that
for ordinary systems (moving slower than the speed of light) the
behaviour of the model is similar to the one arising from the
white--noise CSL.

The modified equation (\ref{sch2}) for the statevector evolution
becomes now:
\begin{eqnarray} \label{ecnmr}
\frac{d |\psi(t)\rangle}{dt} & = & \left[ -\frac{i}{\hbar} H_{0}
\; + \; \int d^{3}x\, {\mathcal N}{(\bf x}) w({\bf x}, t) \; -
\right.
\\
& & \left. - \; 2\gamma \int d^{3}x d^{3}y\, {\mathcal N}({\bf x})
g(|{\bf x} - {\bf y}|) \int_{t_{0}}^{t} ds\, h(t - s)
\frac{\delta}{\delta w({\bf y}, s)} \right]|\psi(t)\rangle.
\nonumber
\end{eqnarray}
If we ignore the free Hamiltonian $H_{0}$, i.e. if we confine our
considerations to the reduction mechanism\footnote{For the
physically interesting cases, e.g. for the dynamical evolution of
macrosystems, such an assumption is justified by the fact that the
effect of the reduction is much faster than the typical times in
which the Hamiltonian can induce appreciable dynamical changes of
the statevector.}, equation (\ref{ecnmr}) becomes:
\begin{equation} \label{enmrsh}
\frac{d |\psi(t)\rangle}{dt} \; = \; \left[ \int d^{3}x\,
{\mathcal N}{(\bf x}) w({\bf x}, t) \; - \; \gamma(t) \int d^{3}x
d^{3}y\, {\mathcal N}({\bf x})g(|{\bf x} - {\bf y}|) {\mathcal
N}({\bf y}) \right] |\psi(t)\rangle
\end{equation}
with:
\begin{equation}
\gamma(t) \quad = \quad 2\gamma \int_{t_{0}}^{t} ds\, h(t - s).
\end{equation}
The corresponding equation for the statistical operator is:
\begin{equation} \label{eosnsh}
\frac{d}{dt}\, \rho(t) \; = \; -\frac{\gamma(t)}{2} \int d^{3}x\,
d^{3}y \left[ {\mathcal N}({\bf x}), \left[ {\mathcal N}({\bf y}),
\rho(t) \right] \right] g(|{\bf x} - {\bf y}|).
\end{equation}

Equation (\ref{ecnmr}) can be rewritten in a form closer to
equation (\ref{gtg}), a fact which will be useful for the subsequent
discussion. Let us define a new Gaussian stochastic process
$\overline{w}({\bf x}, t)$, which is connected to $w({\bf x}, t)$
by the relation:
\begin{equation}
w({\bf x}, t) \quad = \quad
\left(\frac{\alpha}{2\pi}\right)^{\frac{3}{2}} \int d^{3}x\,
e^{\displaystyle -\frac{\alpha}{2} ({\bf x} - {\bf y})^{2}}
\overline{w}({\bf y}, t).
\end{equation}
The process $\overline{w}({\bf x}, t)$ has zero mean and
correlation function
\begin{equation}
\llangle \overline{w}({\bf x}, t_{1})\, \overline{w}({\bf y},
t_{2}) \rrangle \quad = \quad \gamma\, \delta^{(3)}({\bf x} - {\bf
y})\, h(t_{1} - t_{2}).
\end{equation}
Using the following relation:
\begin{eqnarray}
\frac{\delta}{\delta \overline{w}({\bf x}, s)}\, |\psi(t)\rangle &
= & \int d^{3}y\; \frac{\delta w({\bf y}, s)}{\delta
\overline{w}({\bf x}, s)}\, \frac{\delta}{\delta w({\bf y}, s)}\,
|\psi(t)\rangle
\quad = \nonumber \\
& = & \left(\frac{\alpha}{2\pi}\right)^{\frac{3}{2}} \int d^{3}y\;
e^{\displaystyle -\frac{\alpha}{2} ({\bf x} - {\bf y})^{2}}
\frac{\delta}{\delta w({\bf y}, s)}\, |\psi(t)\rangle,
\end{eqnarray}
it can be easily seen that (\ref{ecnmr}) is equivalent to the
 equation:
\begin{eqnarray} \label{ecnmr2}
\frac{d |\psi(t)\rangle}{dt} & = & \left[ -\frac{i}{\hbar} H_{0}
\; + \; \int d^{3}x\, N{(\bf x}) w({\bf x}, t) \; - \right.
\\ & &
\left. - \; 2\gamma \int d^{3}x\, N({\bf x}) \int_{t_{0}}^{t} ds\,
h(t - s) \frac{\delta}{\delta \overline{w}({\bf x}, s)}
\right]|\psi(t)\rangle, \nonumber
\end{eqnarray}
with $N{(\bf x})$ defined by (\ref{ando}).

\subsubsection{Dynamics for macroscopic rigid bodies} \label{nsec51}

As for CSL, it is not difficult to discuss the physical
implications of equation (\ref{ecnmr}) --- or equation
(\ref{ecnmr2}) --- for the case of an almost rigid macroscopic  body,
i.e. a body such that the wavefunctions of its constituents can be
considered very well localized with respect to the localization
length $1/\sqrt{\alpha}$. Under the same assumptions of section
\ref{sec72}, we obtain that if $|\Psi\rangle$ and $|\chi\rangle$
satisfy the equations
\begin{eqnarray} \label{ecnmr3}
\frac{d |\Psi(t)\rangle}{dt} & = & \left[ -\frac{i}{\hbar} H_{\bf
Q} \; + \; \int d^{3}x\, {\mathcal N}{(\bf x}) \overline{w}({\bf
x}, t) \; - \right.
\\ & &
\left. - \; 2\gamma \int d^{3}x\, {\mathcal N}({\bf x})
\int_{t_{0}}^{t} ds\, h(t - s) \frac{\delta}{\delta
\overline{w}({\bf x}, s)} \right]|\Psi(t)\rangle, \nonumber
\\
& & \nonumber
\\ \label{tyj} \frac{d |\chi (t)\rangle}{dt}
& = & \left[ -\frac{i}{\hbar} H_{r} \right] |\chi (t)\rangle,
\end{eqnarray}
then $|\psi(t)\rangle = |\Psi(t)\rangle\, |\chi (t)\rangle$
satisfies equation (\ref{ecnmr2}) or, equivalently, equation
(\ref{ecnmr}).

Equations (\ref{ecnmr3}) and (\ref{tyj}) imply that the center of
mass and internal motion decouple, and that the stochastic terms
affect only the center of mass and not the internal structure, as
it happens for CSL.

Following the same arguments of section \ref{sec73}, it can also
be proven that the localization rate of the center  of mass
wavefunction grows linearly with the number of particles of the
rigid body. In fact, by disregarding the Hamiltonian $H_{0}$, one
derives an evolution equation for the matrix elements $\langle
{\bf Q}'|\rho_{\bf Q}(t)|{\bf Q}''\rangle$ of the statistical
operator  of the centre of mass, which is similar to
(\ref{ecsl1}), with
$\gamma(t)$ replacing
$\gamma$:
\begin{equation}
\frac{\partial \langle {\bf Q}'|\rho_{\bf Q}(t)|{\bf
Q}''\rangle}{\partial t} \quad = \quad - \tilde{\Gamma}({\bf Q}',
{\bf Q}'', t) \, \langle {\bf Q}'|\rho_{\bf Q}(t)|{\bf Q}''\rangle
\end{equation}
with
\begin{eqnarray}
\tilde{\Gamma}({\bf Q}', {\bf Q}'', t) & = & \gamma(t) \int
d^{3}x\, \left[ \frac{1}{2} F^{2}({\bf Q}' - {\bf x}) +
\frac{1}{2}
F^{2}({\bf Q}'' - {\bf x}) - \right. \nonumber \\
& & \left. \frac{}{} F({\bf Q}' - {\bf x}) F({\bf Q}'' - {\bf x})
\right].
\end{eqnarray}
This proves that also in the present model the reduction rate of
the center of mass of the system grows linearly with the number of
its constituents. Moreover, taking a large value for $\beta$, as
it has been suggested previously, $\gamma(t) \rightarrow \gamma$
in very short times, so that the reducing dynamics is practically
the same as the one of CSL.

\part{The Interpretation of GRW and CSL Models}

\section{The mass density function} \label{sec012}

    In the second part of this report we have made plausibe that the
Dynamical Reduction Models allow to overcome the macro-objectification
problem. However, in accordance with the appropriate and strict requests
by J.S. Bell (see below) about the fact that a theory must first of
all make  perfectly clear what it is actually about, it is necessary to
supplement the formal apparatus with a precise interpretation which
specifies how the mathematical entities entering into play are related to
the physical aspects of natural processes we experience.
Accordingly, in this section we tackle the subtle problem of working out a
consistent and unambiguous interpretation of the theoretical  models
under study, in the non relativistic case. We will show how, by taking
advantage of their specific features, one can give a description of the
world in terms of the mean values ${\mathcal M}({\bf r},t)$, at different
places and at different times, of appropriately defined mass density
operators. The presentation is organized as follows.

We start with a historical account of J.S. Bell's contribution to
the elaboration of a sensible interpretation of QMSL (subsection
\ref{sec121}). Next, we introduce the mass--density function
${\mathcal M}({\bf r},t)$ and we show that, within standard
Quantum Mechanics, i.e. in the absence of a mechanism restricting
the possible states of the Hilbert space of ``our universe'', one
unavoidably meets situations which cannot be consistently
described in terms of ${\mathcal M}({\bf r},t)$ (subsection
\ref{sec122}).

Fortunately, since the universal dynamics of the reduction models
does not permit  the persistence for \cite{bells} {\it more than a
split second} of the just mentioned unacceptable states, it allows  to
identify the function ${\mathcal M}({\bf r},t)$ as the basic element for
the description of the world (subsection \ref{sec123}). In terms of it
one can  define an appropriate ``topology" (subsection
\ref{sec124}) which is the natural candidate for establishing a
satisfactory psycho--physical correspondence. We conclude the
section with a general discussion of how a theory should describe
the physical world, and how this is accomplished by dynamical
reduction models (subsection \ref{sec125}).

\subsection{The position of J. Bell about dynamical reduction
models} \label{sec121}

J. S. Bell has always been in the forefront of the struggle for
clarifying the conceptual status of quantum theory. He has
repeatedly stressed the points he considered as essential to have
``an exact theory'' i.e., in his words \cite{bellqg}, one which
{\it neither needs nor is embarrassed by an observer}. We may
appropriately recall some of his more passionate statements \cite{bell89}:
\begin{quotation}
A good word is `beable' from the verb `to be', `to exist'. In your
theory you should identify some things as being really there, as
distinct from the many mathematical concepts you can easily devise
--- like the projection of the side of a triangle to infinity and
so on. We must decide that some things are really there and that
you are going to take them seriously. These are the beables, and
if you are going eventually to have `observers', for example, they
must be made out of beables. ... Another good word is
`kinematics'. Many accounts of quantum mechanics start by telling
you how to calculate probabilities; and I consider them to be
dynamics. The kinematics should list the possibilities that you
are envisaging, then afterwords you can attach probabilities to
the different possibilities. I won't accept as a list of
possibilities the `possible results of experiments', because that
is to try again to begin with these vague concepts. I would want
the kinematics of your theory tell me what it is you are talking
about before you tell me what about it.
\end{quotation}
And it is just in the spirit of the above sentences that he has
analyzed the GRW theory, a theory that \cite{bell89}:
\begin{quotation}
looks like a rather neat resolution of the problem of quantum
mechanics. It is very close to what one does in practice, but
instead of having this funny jump at an arbitrarily defined act of
`measurement', it has it as something which happens all the time
and more often in systems which are big --- big in a way which is
controlled by the parameters of the theory ... .
\end{quotation}

Since his first writing on this theory \cite{bells}, Bell has
proposed an interesting interpretation for it in terms of beables:
\begin{quotation}
There is nothing in this theory but the wavefunction. It is in the
wavefunction that we must find an image of the physical world, and
in particular of the arrangement of things in ordinary
3--dimensional space. But the wavefunction as a whole lives in a
much bigger space, of $3N-$dimensions. It makes no sense to ask
for the amplitude or phase or whatever of the wavefunction at a
point in ordinary space. It has neither amplitude nor phase nor
anything else until a multitude of points in ordinary 3--space are
specified. However, the GRW jumps (which are part of the
wavefunction, not something else) are well localized in ordinary
space. Indeed each is centered on a particular spacetime point
$({\bf x},t)$. So we can propose these events as the basis of the
`local beables' of the theory. These are the mathematical
counterparts in the theory to real events at definite places and
times in the real world (as distinct from the many purely
mathematical constructions that occur in the working out of
physical theories, as distinct from things which may be real but
not localized, and as distinct from the `observables' of other
formulations of quantum mechanics, for which we have no use here).
A piece of matter then is a galaxy of such events. As a schematic
psychophysical parallelism we can suppose that our personal
experience is more or less directly of events in particular pieces
of matter, our brains, which events are in turn correlated with
events in our bodies as a whole, and they in turn with events in
the outer world.
\end{quotation}

The suggestion about the possibility of establishing
an appropriate psycho--physical parallelism has been subsequently
proved  \cite{abp}  to be perfectly appropriate (see also the discussion of
section
\ref{sec014}): a perception process
requires the displacement of a certain number of particles within
our brain. The definite perception corresponds therefore
unambiguously to a mini--galaxy of localizations in the axons or
the cerebral cortex, such mini--galaxies referring to different
`brain regions' according to the precise situation which triggers the
perception\footnote{Actually, even triggering processes involving
superpositions of states of microsystems which are such to induce
precise perceptions (due to the extreme sensitivity, e.g. of the
visual process) lead to definite perceptions and not to a confused
state of mind, just because they imply different modalities of the
localization processes in the brain (see reference \cite{abp}).}.
This is the precise sense in which the GRW theory allows  to
`close the circle', i.e., it yields a picture of natural processes
which agrees with quantum predictions at the micro--level but also
with our definite perceptions and conceptualizations at the
macroscopic one. We do not see what more, on an ontological basis,
can be required from a theory of natural processes.

Starting from
these remarks we can now pass to discuss what we have denoted as
the mass density interpretation and raise the question: why can't
one take the mass density function seriously as the ``local beable" of
the theory and, in particular, replace the `galaxy' of definite
localizations occurring in definite space regions with the existence of
precise regions in which the mass density is ``accessible'' (the exact
meaning of this expression will be discussed later)  and the
`mini--galaxies' of localizations in our bodies and our brains with the
locations of the macroscopic number of ions which are involved in the
transmission of a  nervous signal and in triggering the conscious
perception?

Before coming to deepen this point we have to mention that subsequently
J. Bell himself has slightly changed his mind. In reference \cite{bellam}
he wrote:
\begin{quotation}
The GRW type theories have nothing in their kinematics but the
wavefunction. It gives the density (in a multidimensional
configuration space!) of stuff. To account for the narrowness of
that stuff in macroscopic dimensions, the linear Schr\"{o}dinger
equation has to be modified, in the GRW picture by a
mathematically prescribed spontaneous collapse mechanism.
\end{quotation}
One of us (G.C.G.) has exchanged with him various letters devoted to
deepening this point. In a letter of October 3, 1989, J.S. Bell
wrote:
\begin{quotation}
As regards $\Psi $ and the density of stuff, I think it is
important that this density is in the $3N-$dimensional
configuration space. So I have not thought of relating it to
ordinary matter or charge density in $3-$space. Even for one
particle I think one would have problems with the latter. So I am
inclined to the view you mention `as it is sufficient for an
objective interpretation' ... And it has to be stressed that the
`stuff' is in $3N-$space --- or whatever corresponds in field
theory.
\end{quotation}

This concludes our analysis of the stimulating suggestions of J.S.
Bell about the interpretation of dynamical reduction theories.

As already anticipated, in
the next subsection we will analyze a different and more precise
proposal, based on the consideration of an appropriately averaged
mass density distribution in ordinary space, which has been put
forward in ref. \cite{cc}. Such a proposal, among other interesting
features, will make perfectly clear the links, at the macroscopic
level, between the formal elements which characterize the states of
macroscopic systems according to the theory, the properties which
can be considered as objectively possessed by them and the
practical way of testing such properties. In fact one could remark
that J.S. Bell has been not wholly explicit about the problem of
establishing precise relations between the specific formal features
characterizing macro--systems (\textit{the narrowness of the stuff}) and
the experiments aimed to ascertain the associated properties, and has
simply suggested (\textit{as it is sufficient for an objective
interpretation}) that it could be easily settled within the GRW theory.

\subsection{Mass density function} \label{sec122}

In order to prepare the grounds for the analysis we are going to
perform it is useful to recall that the universal dynamics of
dynamical reduction models strives to make some precise properties
as objectively possessed by individual physical systems. The very
fact that no analogous mechanism is at work within the standard
theory forbids to adopt, within such a theory, the natural
interpretation we are going to propose as the basic ontology for
CSL. As we have discussed in great details, the Dynamical
Reduction Theories make almost definite the positions of massive
particles. Accordingly, the natural quantity which will end up
having an objective (i.e. independent of any measurement process)
value is the
 locally averaged (over the characteristic volume of the
theory, i.e., $10^{-15} cm^{3}$) mass distribution of the universe.

In this subsection we characterize in a mathematically precise way
the $c$-number function representing the just mentioned average
mass density in ordinary 3-dimensional space, the quantity to
which we will attach an absolutely prominent role for the
interpretation of the theory. We will also make clear that, as
already mentioned, within standard quantum mechanics such a
quantity exhibits problematic features (which parallel the ones
connected with the macro-objectification problem) implying that
an ontology based on the mass density function cannot be
consistently adopted within such a theory. The clarification of
this important point will pave the road for the proof that, on the
contrary, such an ontology leads, within the dynamical reduction
formalism, to a clear, precise and fully consistent worldview
which fits perfectly our experience with the reality around us.

To begin with, let us then consider a physical system $S$ which
will constitute ``our universe'' and let us denote by ${\mathcal
H}(S)$ the associated Hilbert space. Let $|\psi(t)\rangle$ be the
normalized statevector describing our individual system at time
$t$; in terms of it we define an average mass density $c$--number
function ${\mathcal M}({\bf r},t)$ in ordinary space as
\begin{equation} \label{mdf}
{\mathcal M}({\bf r},t) \quad = \quad \langle \psi(t)| M({\bf
r})|\psi(t)\rangle,
\end{equation}
where $M({\bf r},t)$ is the mass density operator defined in
section \ref{sec76}. Equation
(\ref{mdf}) establishes, for a given t, a mapping of ${\mathcal
H}(S)$ into the space of positive and bounded functions of ${\bf
r}$.

Obviously this map is many to one; in particular, to better focus
on this point as well as for making clear the difficulties of the
standard theory with such a function, it turns out to be useful to
compare two statevectors $|\psi^{\oplus}\rangle$ and
$|\psi^{\otimes}\rangle$  defined as follows. Let us consider a
very large number $N$ of particles and two space regions $A$ and
$B$ with spherical shape and radius $R$. The state
$|\psi^{\oplus}\rangle$  is the linear superposition, with equal
amplitudes, of two states $|\psi^{A}_{N}\rangle$ and
$|\psi^{B}_{N}\rangle$  in which the $N$ particles are well
localized with respect to the characteristic length ($10^{-5}$ cm)
of the model and uniformly distributed in regions $A$ and $B$,
respectively, in such a way that the density turns out to be of
the order of 1 gr/cm$^{3}$. On the other hand,
$|\psi^{\otimes}\rangle$ is the tensor product of two states
$|\phi^{A}_{N/2}\rangle$ and  $|\phi^{B}_{N/2}\rangle$
corresponding to $N/2$ particles being distributed in region $A$
and $N/2$ in region $B$, respectively:
\begin{equation}
|\psi^{\oplus}\rangle \; = \; \frac{1}{\sqrt{2}}\left[
|\psi^{A}_{N}\rangle + |\psi^{B}_{N}\rangle \right] \qquad\quad
|\psi^{\otimes}\rangle \; = \; |\phi^{A}_{N/2}\rangle\otimes
|\phi^{B}_{N/2}\rangle
\end{equation}
It is trivial to see that the two considered states give rise to
the same function ${\mathcal M}({\bf r})$ and it is clear that if
one attempts to attach some meaning to it one has to be very
careful in keeping in mind from which state ${\mathcal M}({\bf
r})$ originates.

In particular, it is quite obvious that in the case of
$|\psi^{\oplus}\rangle$, ${\mathcal M}({\bf r})$ cannot be
considered as describing an ``actual" mass density distribution.
To see this, let us suppose that one can use standard quantum
mechanics to describe the gravitational interaction between
massive bodies and let us consider the following gedanken
experiment: a test mass is sent through the middle point of the
line joining the centers of regions $A$ and $B$ with its momentum
orthogonal to it (see figures 2a and 2b).
\begin{center}
\begin{picture}(350,150)(0,0)
\put(0,0){\line(1,0){170}} \put(0,150){\line(1,0){170}}
\put(0,0){\line(0,1){150}} \put(170,0){\line(0,1){150}}
\put(180,0){\line(1,0){170}} \put(180,150){\line(1,0){170}}
\put(180,0){\line(0,1){150}} \put(350,0){\line(0,1){150}}
\put(85,110){\circle{20}} \put(85,110){\circle{16}}
\put(85,110){\circle{12}} \put(85,110){\circle{8}}
\put(85,110){\circle{4}} \put(85,110){\circle{1}}
\put(30,75){\circle*{5}} \put(40,75){\vector(1,0){90}}
\put(140,75){\circle*{5}} \put(85,40){\circle{20}}
\put(85,40){\circle{16}} \put(85,40){\circle{12}}
\put(85,40){\circle{8}} \put(85,40){\circle{4}}
\put(85,40){\circle{1}}
\put(10,115){\makebox(150,40){\tiny $|\psi^{\otimes}\rangle \; =
\; |\phi^{A}_{N/2}\rangle\otimes |\phi^{B}_{N/2}\rangle$}}
\put(105,107){$A$} \put(7,85){\footnotesize test particle}
\put(82,80){\footnotesize $\otimes$} \put(25,60){\tiny
$|\phi\rangle$} \put(105,37){$B$}
\put(10,3){\makebox(150,20){\tiny $|\psi^{\otimes}\rangle \otimes
|\phi\rangle \rightarrow \{ |\phi^{A}_{N/2}\rangle\otimes
|\phi^{B}_{N/2}\rangle \} |\phi\rangle$}}
\put(3,137){\framebox(15,10){2a}}
\put(265,110){\circle{20}} \put(265,110){\circle{16}}
\put(265,110){\circle{12}} \put(265,110){\circle{8}}
\put(265,110){\circle{4}} \put(265,110){\circle{1}}
\put(210,75){\circle*{5}} \put(220,75){\line(1,0){20}}
\put(240,75){\vector(3,1){70}} \put(240,75){\vector(3,-1){70}}
\put(320,102){\circle*{5}} \put(320,48){\circle*{5}}
\put(265,40){\circle{20}} \put(265,40){\circle{16}}
\put(265,40){\circle{12}} \put(265,40){\circle{8}}
\put(265,40){\circle{4}} \put(265,40){\circle{1}}
\put(190,115){\makebox(150,40){\tiny $|\psi^{\oplus}\rangle \; =
\; \frac{1}{\sqrt{2}}\left[ |\psi^{A}_{N}\rangle +
|\psi^{B}_{N}\rangle \right]$}} \put(285,107){$A$}
\put(187,85){\footnotesize test particle}
\put(262,73){\footnotesize $+$} \put(318,73){\footnotesize $+$}
\put(205,60){\tiny $|\phi\rangle$} \put(285,37){$B$}
\put(190,3){\makebox(150,20){\tiny $|\psi^{\oplus}\rangle \otimes
|\phi\rangle \rightarrow
\frac{1}{\sqrt{2}}\left[|\psi^{A}_{N}\rangle |\phi^{\makebox{\tiny
up}}\rangle + |\psi^{B}_{N}\rangle |\phi^{\makebox{\tiny
down}}\rangle \right]$}}
\put(183,137){\framebox(15,10){2b}}
\end{picture}

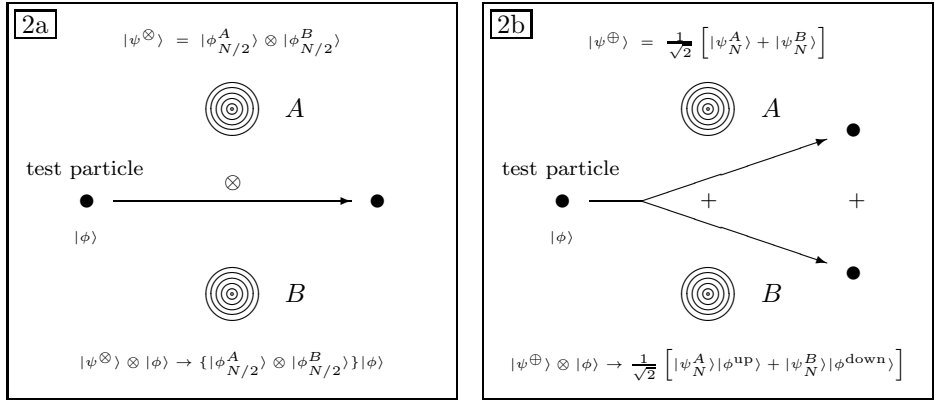
\vspace{0.2cm} \footnotesize \parbox{4.8in}{Figure 2: Accessible
and non--accessible mass density distribution ${\mathcal M}({\bf
r})$. In case 7a, corresponding to the factorized state
$|\psi^{\otimes}\rangle$, the mass density in regions $A$ and $B$
is accessible and the test particle, interacting with
$|\psi^{\otimes}\rangle$, behaves in such a way as to give rise to
the appropriate density along its natural trajectory. In case 7b,
corresponding to the superposition $|\psi^{\oplus}\rangle$, the
densities in $A$ and $B$ are non--accessible and the same holds
for the density distribution generated by the interaction of the
test particle with $|\psi^{\oplus}\rangle$.} \normalsize
\end{center} \vspace{0.5cm}
In the case of the state $|\psi^{\otimes}\rangle$  for the system
of the $N$ particles  standard quantum mechanics predicts that the test
particle will not be deflected. On the other hand, if the same
test is performed when the state is $|\psi^{A}_{N}\rangle$
($|\psi^{B}_{N}\rangle$), quantum mechanics predicts an upward
(downward) deviation of the test particle. Due to the linear
nature of the theory this implies that if one would be able to
prepare the state  $|\psi^{\oplus}\rangle$ the final state would
be
\begin{equation} \label{edcvf}
|\psi\rangle \quad = \quad \frac{1}{\sqrt{2}} \left[
|\psi^{A}_{N}\rangle \otimes |\psi^{\makebox{\tiny UP}}\rangle \;
+ \; |\psi^{B}_{N}\rangle \otimes |\psi^{\makebox{\tiny
DOWN}}\rangle \right]
\end{equation}
with obvious meaning of the symbols. If one includes the test
particle into the ``universe" and considers the mass density
operator in regions corresponding to the wavepackets
$|\psi^{\makebox{\tiny UP}}\rangle$  and $|\psi^{\makebox{\tiny
DOWN}}\rangle$, one discovers once more that nowhere in the
universe one can ``detect" or ``perceive" a density corresponding
to the density of the test particle. In a sense, if one would
insist in giving a meaning to the density function he would be led
to conclude that the particle has been split by the interaction
into two pieces of half its density. This analysis shows that
great attention should be paid in assuming that the function
${\mathcal M}({\bf r})$ describes the actual state of affairs.

Before going on we consider also another quantity which will be
useful in what follows. It is the mass density variance at ${\bf
r}$ at time $t$ defined by the following map from ${\mathcal
H}(S)$ into ${\bf R}^{3}$:
\begin{equation}
{\mathcal V}({\bf r}, t) \quad = \quad \langle \psi(t)|[ M({\bf
r}) - \langle \psi(t)| M({\bf r}) | \psi(t) \rangle]^{2}
|\psi(t)\rangle
\end{equation}
$|\psi(t)\rangle$ being a normalized statevector.

With these premises we have all the elements which are necessary
to discuss the problems one meets when dealing with ${\mathcal
M}({\bf r})$ and the way to overcome them. We will do this in the
next subsection. Before doing that, we consider it appropriate to
simply mention the obvious fact that the states giving rise to
puzzling, non objective, density functions are those corresponding
to superpositions of differently located macroscopic bodies, i.e.
the infamous states which are at the centre of the long debated
problems about the meaning of quantum mechanics at the
macro--level.

For future purposes it is useful to introduce a mathematical
criterion which permits  to clarify  the different status of
the mass densities in the two above considered cases
(corresponding to the states $|\psi^{\oplus}\rangle$  and
$|\psi^{\otimes}\rangle$, respectively). This is more easily
expressed by resorting to a discretization of space in analogy
with what has been done in section \ref{sec75}. Obviously, in
place of the space functions ${\mathcal M}({\bf r}, t)$ and
${\mathcal V}({\bf r}, t)$ we will consider the mean value
${\mathcal M}_{i}(t)$ and the variance ${\mathcal V}_{i}(t)$ of
the mass operator in the $i$--th cell. For any cell $i$ we define
the ratio:
\begin{equation}
{\mathcal R}_{i}^{2} \quad = \quad {\mathcal V}_{i} / {\mathcal
M}_{i}^{2}
\end{equation}
We then state that the mass ${\mathcal M}_{i}$ is ``accessible" if
${\mathcal R}_{i}$ turns out to be much smaller than one, that is:
\begin{equation} \label{acc}
{\mathcal R}_{i} \quad \ll \quad 1
\end{equation}
This criterion is clearly reminiscent of the probabilistic
interpretation of the statevector in standard quantum mechanics.
Actually, within such a theory equation (\ref{acc}) corresponds to
the fact that the spread of the mass operator $M_{i}$ is much
smaller than its mean value. Even though in this paper we take a
completely different attitude with respect to the mean value
${\mathcal M}_{i}$, it turns out to be useful to adopt the above
criterion also within the new context. In fact, as we will discuss
in what follows, when one has a space region such that for all
cells contained in it (\ref{acc}) holds, it behaves as if it would
have the ``classical" mass corresponding to ${\mathcal M}_{i}$. This
remark should clarify the reason for having characterized as
``accessible" the mass (or equivalently the mass density) when the above
conditions are satisfied.

With reference to the previous example we stress that in the case
of $|\psi^{\otimes}\rangle$ all cells within regions $A$ and $B$
are such that criterion (\ref{acc}) is very well satisfied. In the
case of $|\psi^{\oplus}\rangle$ one has for the same cells:
\begin{equation}
{\mathcal M}_{i} \; \simeq \; \frac{n}{2}\,M_{0}, \quad\qquad
{\mathcal V}_{i} \; \simeq \; \frac{n^{2}}{4}\, M_{0}^{2}
\end{equation}
where $n$ is the number of particles per cell. It follows that
\begin{equation} {\mathcal R}_{i} \quad \simeq \quad 1.
\end{equation}

\subsection{The mass density function within dynamical reduction
models} \label{sec123}

In the previous subsection we have presented a meaningful example
of the difficulties one meets when one keeps the standard quantum
dynamics and tries to base a description of the world and an
acceptable psycho--physical correspondence on the mass density
function ${\mathcal M}({\bf r})$. The unacceptable features find
their origin in the fact that, when the macrostate is
$|\psi^{\oplus}\rangle$, while the density function takes the
value of about 1/2 gr/cm$^{3}$ within regions $A$ and $B$, if one
performs a measurement of the density in the considered regions,
or if a measurement like process (such as the passage of the test
particle in between $A$ and $B$) occurs, things proceed in such a
way that is incompatible with the above density value. Actually
one could state that no outcome emerges in the measurement. To
understand fully the meaning of this statement one could identify,
e.g., the final position of the test particle with a pointer
reading; the pointer would then not point to the middle position
(corresponding to equal densities in $A$ and $B$) but would be
split into ``two pointers of half density" pointing upward and
downward, respectively (compare with figure 7b).

If one takes an analogous attitude with reference to dynamical
reduction theories one does not meet the same difficulties because
they imply that linear superpositions of states corresponding to
far--apart macroscopic systems are dynamically suppressed in
extremely short times and measurements have outcomes. Therefore,
we can guess that, within the context of the dynamical reduction
program, the description of the world in terms of the mass density
function ${\mathcal M}({\bf r})$ is a good description precisely
because it becomes dynamically accessible at the macroscopic
level; moreover it is such as to allow one to base on it a
sensible psycho--physical correspondence.

Obviously, if one naively looks at the dynamical reduction models
by sticking to some sort of ``classical" ontology, one can be
tempted to claim that some fuzzy situations can occur also in this
context, when the mass density, as it may very well happen for a
microsystem, is not ``accessible", i.e. when (in the simplified
discretized version) criterion (\ref{acc}) is not satisfied.
However, as we are going to show, this does not give rise to any
difficulty whatsoever for the program we are furthering.

In order to show this we will examine, along the above lines, the
status of the mass density function ${\mathcal M}({\bf r})$ for
the various possible states which are not forbidden by the
reducing dynamics. We will discuss the cases of microsystems and
macro--systems, and, with reference to the latter, we will
identify two physically relevant classes of states which can
occur. As we have done previously we will deal with a discretized
space.

\subsubsection{Microscopic systems} \label{sec1231}

For the sake of simplicity, let us consider a single nucleon. As
it is well known, the reducing dynamics does not forbid the
persistence, for extremely long times, of linear superpositions of
far--away states of the particle, typically states like:
\begin{equation} \label{ecgz}
|\psi\rangle \; = \; \frac{1}{\sqrt{2}}\,\left[
|0,0,\ldots,1_{i},\ldots,0_{j},\ldots\rangle \, + \,
|0,0,\ldots,0_{i},\ldots,1_{j},\ldots\rangle \right]
\end{equation}
where $i$ and $j$ are two distinct and far apart cells. Such
microscopic states which are not eigenvectors of the operators
$M_{i}$ will be called ``microscopically non definite", the term
``non definite" making reference to the characteristic preferred
basis of the model. As is evident from (\ref{ecgz}) the mean
values of $M_{i}$ and $M_{j}$ are $(1/2)m_{0}$ and criterion
(\ref{acc}) is not satisfied at the space regions of both cells.
As it is (inconsistently) assumed within standard quantum
mechanics and as it is (rigorously) implied by CSL, a measurement
of the mass in one of these two cells would give the {\it definite
outcome} $0$ or $m_{0}$ with equal probability (corresponding to
the fact that wavepackets of microsystems diffuse, but however the
reaction of a detector devised to reveal them remains spotty) and
not $(1/2)m_{0}$, the value taken by the density function within
the considered cells. Accordingly, in the considered case the mass
in the cells is not accessible. This discrepancy, this {\it non
classical} character of ${\mathcal M}_{i}$ and ${\mathcal M}_{j}$,
cannot however be considered a difficulty for the theory with the
proposed interpretation, in particular, it does not forbid to take
seriously, i.e. to attach an objective status to the mass density
function; it simply amounts to the recognition  that we cannot
legitimately apply our {\it classical} pictures to the microworld.
On the contrary we must allow \cite{bellns} {\it microsystems to
enjoy the cloudiness of waves}. The crucial point is that within
the theory we are discussing, in spite of the mass density having
the value $(1/2)m_{0}$ in both regions, any attempt to detect its
value in one of them by an amplification process, implies, as a
rigorous consequence of the dynamical equation governing all
physical processes, that the outcome will be either $0$ or
$m_{0}$, in perfect agreement with our experience.

\subsubsection{Macroscopic systems} \label{sec1232}

The theory allows the persistence of two general classes of states
for macroscopic systems, i.e. those describing almost rigid bodies
with sharply defined (with respect to the characteristic  length
of the model) center of mass position, and those corresponding to
a macroscopic number of microsystems in microscopically non
definite states. Due to the fact that the center of mass of the
wavefunction has, in general, non compact support, the first class
obviously includes also states which, being brought in by the
reducing dynamics, have ``tails". The so called ``tail problem''
will be  discussed in the following section.

States of the first class have been extensively analyzed in
sections \ref{sec5}, \ref{sec6} and \ref{sec7};  we have seen
that superpositions of different macroscopic states are reduced
--- in a very short time --- to one of their terms;
correspondingly, the classical properties of
macroscopic systems are restored. Thus, for example, a state like
$|\psi^{\oplus}\rangle$ is spontaneously transformed, by the
reducing dynamics, either into the state $|\psi^{N}_{A}\rangle$ or
$|\psi^{N}_{B}\rangle$, i.e. into states which have an accessible
mass--density distribution.

Concerning states of the second class, it is of extreme relevance
to make  clear that they have a conceptual status which is very
different\footnote{This important difference has already been
appropriately stressed by A. Leggett \cite{legg80}, who, even
though in a different context, has introduced the mathematically
precise concept of disconnectivity to distinguish states of this
type from states like $|\psi^{\oplus}\rangle$.} from the one of
the superpositions of macroscopically distinguishable states like
$|\psi^{\oplus}\rangle$; moreover, they represent rather peculiar
situations which mainly have an ``academic'' character, since
states of this kind certainly do not appear often in practice.
However, it is worthwhile to discuss them in some detail.

Let us consider a system of $N$ nucleons and a discretization of
space in small cubes of linear dimensions $10^{-8}$ cm. We
consider again two macroscopic regions $A$ and $B$, and we label by the
indices $k_{A}$  and $k_{B}$  pairs of cubes within $A$ and $B$
respectively. For $k_{A} \neq \tilde k_{A}$  the two cubes are
disjoint and the union of all cubes $k_{A}$  ($k_{B}$) covers the
region $A$ ($B$). The index $k$ runs from $1$ to $N$, a very large
number; typically if $A$ and $B$ have volumes of the order of $1$
dm$^{3}$, $N$ will be of the order of $10^{27}$.

Let us denote by $|\psi_{k_{A}}\rangle$ and $|\psi_{k_{B}}\rangle$
the states of a particle whose coordinate representation are well
localized within $k_{A}$ and $k_{B}$, respectively. As an example
we could choose
\begin{equation}
\langle{\bf r}|\psi_{k_{A}}\rangle \quad = \quad \chi(k_{A})
\end{equation}
$\chi(k_{A})$  being the characteristic function of the cube $k_{A}$.
We now consider the following microscopically non definite state
for the $k$--th particle:
\begin{equation} \label{ge1}
|\psi^{k}\rangle \quad = \quad \frac{1}{\sqrt{2}}\, \left[
|\psi_{k_{A}}\rangle \; + \; |\psi_{k_{B}}\rangle \right]
\end{equation}
and the factorized state of the $N$ particles
\begin{equation} \label{ge2}
|\psi\rangle \quad  =\quad |\psi^{1}\rangle\otimes\ldots \otimes
|\psi^{k}\rangle \otimes\ldots\otimes |\psi^{N}\rangle.
\end{equation}
In spite of the fact that the state $|\psi\rangle$ is a direct
product of microscopically non definite states it is nevertheless
``almost" an eigenstate of the operators $M_{i}$ (remember that
the linear dimensions of the cell to which the index $i$ refers
are of the order of $10^{-5}$ cm so that one such cell contains
about $10^{9}$ cubes of the kind of $k_{A}$ ($k_{B}$)). In fact,
denoting by $n \simeq 10^{9}$  the number of $k_{A}$ ($k_{B}$)
small cubes contained in the $i$--th cell, one can easily see that
$|\psi\rangle$ gives rise to ``accessible" mass ${\mathcal M}_{i}$
in regions $A$ and $B$ respectively\footnote{In making the
computations we have identified the operators $M_{i}$ with the sum
of the projectors (multiplied by the nucleon mass $m_{0}$) of the
various particles in the $i$--th cell.}:
\begin{equation}
\langle M_{i(A,B)} \rangle \; \simeq \; \frac{1}{2}\, n \, m_{0},
\qquad\quad \langle M^{2}_{i(A,B)} \rangle \; \simeq \;
\frac{1}{4}\, (n^{2} + n)\, m^{2}_{0}
\end{equation}
hence
\begin{equation}
{\mathcal V}^{2}_{i(A,B)} \; \simeq \; \frac{1}{4}\, n \,
m^{2}_{0}, \qquad\quad\qquad {\mathcal R}_{i(A,B)} \; \simeq \;
\frac{1}{\sqrt{n}} \; \ll \; 1 \qquad
\end{equation}

To clarify the physical implications of the state $|\psi\rangle$,
from the point of view which interests us here, we can imagine
performing once more the gedanken experiment with a test particle
we have already considered in the previous subsection, assuming,
for simplicity, that the interactions between the test particle
and the considered $N$ particles do not change the
state\footnote{At any rate, possible changes in such a state would
be symmetrical with respect to the middle plane, so that the
subequent considerations would still hold true.} of the latter. By
substituting equation (\ref{ge1}) into equation (\ref{ge2}), we
see that $|\psi\rangle$ is a superposition of $2^{N}$ states in
which each particle is well localized. In such a superposition all
states have an equal amplitude $1/\sqrt{2^{N}}$  and almost all
states correspond to about $N/2$ particles being in regions $A$
and $B$ respectively. Therefore, in the language of dynamical
reduction models, the probability of occurrence of a realization
of the stochastic potential leading to the ``actualization" of an
almost completely undeflected trajectory for the test particle is
extremely close to one\footnote{It could be useful to remark that
if one would analyze the same experiment in terms of the  linear
quantum dynamics, the test particle would end up in the linear
superposition of an extremely large number of states. However,
since such states correspond to trajectories which are very near
and almost undeflected, the evaluation of the mass density
associated to  the final statevector would show that in the
``middle" region there would practically be  the total mass of the
test particle. Therefore, this represents a case in which even
without any reduction process the mass density referring to the
test particle would correspond to a precise outcome of the
measurement.}. This shows that the mass density function
${\mathcal M}({\bf r})$ corresponding to the state behaves in a
``classical way", so that no trouble arises in this case.

It has to be noted that, obviously, the mass ${\mathcal M}_{i}$
corresponding to the state (\ref{ge2}) coincides with the one
corresponding to the state $|\psi^{\otimes}\rangle$ of the
previous subsection, in spite of the fact that both states are
dynamically allowed and are quite different as physical states.
However, as we have shown, the masses ${\mathcal M}_{i}$ in the
two cases behave practically in the same way and are fully
unproblematic, contrary to what happens in the case of
$|\psi^{\oplus}\rangle$.

Concluding, we have made plausible that in the context of the
dynamical reduction program one can consistently describe the
world, at a given time, in terms of the mass density function
${\mathcal M}({\bf r})$ and that, due to the fact that such a function
becomes accessible at the macrolevel, such a description matches our
experience with the reality around us. Obviously, since with the elapsing
of time the state of the world changes, a complete description requires
the consideration of the motion picture of the density, i.e. of
${\mathcal M}({\bf r}, t)$ defined in equation (\ref{mdf}). We will
discuss in greater detail this crucial point in the next section.

\subsection{Defining an appropriate topology for the CSL model}
\label{sec124}

Let us consider a system $S$ of finite mass which will constitute
our ``universe" and its associated Hilbert space ${\mathcal
H}(S)$. We denote by $U(S)$ the unit sphere in ${\mathcal H}(S)$
and we consider the nonlinear map\footnote{To be rigorous, one
should consider the map ${\mathcal M}$ from the unit sphere of
${\mathcal H}(S)$ into the space $L^{2}$ of the square integrable
functions of ${\bf r}$. However, we can deal, without any loss of
generality, with the discretized version of the model.} ${\mathcal
M}$ associating to the element $|\varphi\rangle$ of $U(S)$ the
element ${\bf m} = \{ {\mathcal M}_{i}(|\varphi\rangle) \}$ of
$l_{2}$, $ {\mathcal M}_{i}(|\varphi\rangle)$ being the quantity
$\langle\varphi|M_{i}|\varphi\rangle$.

On $U(S)$ we define a topology by introducing a mapping $\Delta :
U(S) \otimes U(S) \rightarrow {\bf R}^{+}$ according to:
\begin{equation}
\Delta (|\varphi\rangle, |\psi\rangle) \; = \; d({\bf m}, {\bf n})
\; = \; \sqrt{\sum_{i}(m_{i} - n_{i})^{2}}
\end{equation}
where ${\bf m} = \{ {\mathcal M}_{i}(|\varphi\rangle) \}$, ${\bf
n} = \{ {\mathcal M}_{i}(|\psi\rangle) \}$. Such a mapping is not
a distance since, as it emerges clearly from the analysis of the
previous subsection, it may happen that $\Delta (|\varphi\rangle,
|\psi\rangle) = 0$ even though $|\varphi\rangle \neq
|\psi\rangle$. However $\Delta$ meets all other properties of a
distance:
\begin{equation}
\Delta (|\varphi\rangle, |\psi\rangle) \; = \; \Delta
(|\psi\rangle, |\varphi\rangle) \; \geq \; 0
\end{equation}
and
\begin{equation}
\Delta (|\varphi\rangle, |\psi\rangle) \; \leq \; \Delta
(|\varphi\rangle, |\chi\rangle) \; + \; \Delta (|\chi\rangle,
|\psi\rangle)
\end{equation}
as one  easily proves by taking into account the fact that $d$ is
a distance in $l_{2}$.

>From now on we will limit our considerations to the proper subset
$A(S)$ of $U(S)$ of those states which are allowed by the CSL
dynamics. In the previous subsection we have already identified,
even though in a rough way, the set $A(S)$. One could obviously be
very precise about such a set by adopting e.g. the following
criterion: let $|\varphi\rangle \in U(S)$, and let us consider the
ensemble $A(S)(|\varphi\rangle)$ of states which have a non
negligible (this obviously requires the definition of a threshold)
probability of being brought in by the reducing dynamics after a
time interval of the order of $10^{-2}$ sec, which is the
characteristic perception time of a human being\footnote{See the
discussion of section \ref{sec014}.}, for the given initial
condition $|\varphi\rangle$. The union of all subsets
$A(S)(|\varphi\rangle)$ for $|\varphi\rangle$ running over $U(S)$
is then $A(S)$. For our purposes, however, it is not necessary to
go through the cumbersome management of a very precise definition
of the set $A(S)$; the consideration of the cases we have
discussed in the previous subsection is sufficient to lead to the
interesting conclusions.

For any element $|\varphi\rangle$ of $A(S)$ we consider the set of
states of $A(S)$ for which $\Delta (|\varphi\rangle, |\psi\rangle)
\leq \epsilon$. Here the quantity $\epsilon$ has the dimensions of
a mass and is chosen of the order of $10^{18} m_{0}$, with $m_{0}$
the nucleon mass. From the properties of the map $\Delta$ it
follows that:
\begin{enumerate}
\item $ \{ \Delta (|\varphi\rangle, |\psi\rangle) \leq \epsilon$
and $\Delta (|\varphi\rangle, |\chi\rangle) \leq \epsilon \}$
implies $\Delta (|\chi\rangle, |\psi\rangle) \leq 2\epsilon.$
\item $ \{ \Delta (|\varphi\rangle, |\psi\rangle) \gg \epsilon$
and $\Delta (|\varphi\rangle, |\chi\rangle) \leq \epsilon \}$
implies $\Delta (|\chi\rangle, |\psi\rangle) \gg \epsilon.$
\end{enumerate}

We have introduced the parameter $\epsilon$ in such a way that it
turns out to be sensible to consider similar to each other states
whose ``distance" $\Delta$ is smaller than (or of the order of)
$\epsilon$. More specifically, when
\begin{equation} \label{ctu}
\Delta (|\varphi\rangle, |\psi\rangle) \quad \leq \quad \epsilon
\end{equation}
we will say that $|\varphi\rangle$ and $|\psi\rangle$ are
``physically equivalent". More about this choice in what follows.

To understand the meaning of this choice it is useful to compare
it with the natural topology of ${\mathcal H}(S)$. We begin by
pointing out the inappropriateness of the Hilbert space topology
to describe the concept of similarity or difference of two
macroscopic states. In fact suppose our system $S$ is an almost
rigid body and let us consider the following three states:
$|\varphi^{A}\rangle$, $|\varphi^{B}\rangle$ and
$|\tilde\varphi^{A}\rangle$. The state $|\varphi^{A}\rangle$
corresponds to a definite internal state of $S$ and to its center
of mass being well localized around $A$, the state
$|\varphi^{B}\rangle$ is simply the translated of
$|\varphi^{A}\rangle$ so that it is well localized in a distant
region $B$, the state $|\tilde\varphi^{A}\rangle$ differs from
$|\varphi^{A}\rangle$ simply by the fact that one or a microscopic
number of its ``constituents" are in states which are orthogonal
to the corresponding ones in $|\varphi^{A}\rangle$.

It is obvious that, on any reasonable assumption about similarity
or difference of the states of the universe,
$|\tilde\varphi^{A}\rangle$ must be considered very similar
(identical) to $|\varphi^{A}\rangle$ while $|\varphi^{B}\rangle$
must be considered very different from $|\varphi^{A}\rangle$. On
the other hand, according to the Hilbert space topology
\begin{equation}
\| |\varphi^{A}\rangle - |\tilde\varphi^{A}\rangle \| \; = \; \|
|\varphi^{A}\rangle - |\varphi^{B}\rangle \| \; = \; \sqrt{2}
\end{equation}
This shows with striking evidence that the Hilbert space topology
is totally inadequate for the description of the macroscopic
world. As a consequence such topology is also quite inadequate to
base on it any reasonable psycho-physical correspondence.

We now discuss the ``distorted" (with respect to the Hilbert space
one) topology associated to the ``distance" $\Delta$. First of all
we stress that the two states $|\varphi^{A}\rangle$ and
$|\tilde\varphi^{A}\rangle$ which are maximally distant in the
Hilbert space topology, turn out to be equivalent, i.e. to satisfy
condition (\ref{ctu}) in the new topology. This represents an
example showing how such a topology takes more appropriately into
account the fact that, under any sensible assumption, the
``universes" associated to the considered states are very similar.

Obviously, one problem arises. Criterion (\ref{ctu}) leads us to
consider as equivalent states which are quite different from a
physical point of view, even at the macroscopic level. To clarify
this statement we take into account two states $|\varphi\rangle$
and $|\psi\rangle$ corresponding to an almost rigid body located,
at $t = 0$, in the same position but with macroscopically
different momenta, let us say $P = 0$ and $P$, respectively. Even
though the two states are physically quite different, their
distance at $t = 0$ is equal to zero. However, if one waits up to
the time in which the state $|\psi\rangle$ has moved away from
$|\varphi\rangle$, the ``distance" $\Delta(|\varphi(t)\rangle,
|\psi(t)\rangle)$ becomes large and the two states are no longer
equivalent. We will discuss the now outlined problem in great
details in the next section.

Before concluding this part it is important to analyze the case of
two states $|\psi\rangle$ and $|\psi_{T}\rangle$ such that
$|\psi\rangle$ corresponds to an almost rigid body with a center
of mass wavefunction which is almost perfectly localized while
$|\psi_{T}\rangle$ corresponds to the same body with a ``tail" in a
distant region. As we have already discussed, the CSL dynamics allows the
existence of this latter type of states; however it tends to
depress more and more the tail in such a way as to make the mass
in the distant region extremely close to zero (much less than one
nucleon mass) in very short times. As a consequence, according to
the topology that we propose the two states $|\psi\rangle$ and
$|\psi_{T}\rangle$ turn out to be identical. This is quite
natural. In fact, in the same way in which taking away a single
particle from a macroscopic system would be accepted as being
totally irrelevant from a macroscopic point of view, when one
chooses, as we do, to describe reality in terms of mass density,
one must consider as equivalent situations in which their
difference derives entirely from the location of a small fraction
of the mass of a nucleon in the whole universe. We remark that
$|\psi\rangle$ and $|\psi_{T}\rangle$ are extremely close to each
other also in the standard Hilbert space topology.

\subsection{Deepening the proposed interpretation} \label{sec125}

We consider it appropriate to devote this subsection to discuss in
great generality the problem of giving an acceptable description
of the world within a given theory. Usually one tries to do so by
resorting to the notion of observable. As repeatedly remarked,
such an approach meets, within standard quantum mechanics, serious
difficulties since the formal structure of the theory allows only
probabilistic statements about measurement outcomes conditional
under the measurement being performed. In brief, the theory deals
with {\it what we find} not with {\it what is}. This is why J.S.
Bell has suggested \cite{bell84}  replacing  the notion of
observable with the one of ``beable", from the verb to be, to
exist. Obviously, the identification of the beables, of what is
real, requires the identification of appropriate formal
ingredients of the theory we are dealing with.

\subsubsection{The case of the Pilot--Wave theory}
\label{sec1251}

To clarify our point, it turns out to be useful to analyze the de
Broglie--Bohm Pilot--Wave theory. It describes the world in terms
of the wavefunction and of the actual positions of the particles
of our ``universe", each of which follows a definite trajectory.
Therefore, in such a theory it is quite natural to consider as the
beables the positions (which are the local elements accounting for
reality at a given instant) and the wavefunction (which is
nonlocal and determines uniquely the evolution of the positions).
It is important to remember that, within the theory under
discussion, all other ``observables" (in particular, e.g. the spin
variables) turn out, in general, to be contextual. This simply
means that {\it the truth value} of a statement about the outcome
of the measurement of one such observable (which in turn is simply
a statement about the future positions of some particles) may in
general depend (even nonlocally) on the {\it overall} context.
This obviously implies that the attribution of a value to the
considered observable cannot be thought as corresponding, in
general, to an ``objective property" of the system.

Before coming to discuss the problem of the beables within CSL we
would like to call attention to the fact that \cite{bellqg} within
the Pilot--Wave theory, one can construct, from the microscopic
variables ${\bf r}$, macroscopic variables ${\bf R}$ including
pointer positions, images on photographic plates etc. Obviously
this requires some fuzziness, but such a limitation is not
relevant for a consistent account of reality. Thus, in this theory
we are led to suppose that it is from the ${\bf r}$, rather than
from the wavefunction, that the observables we use to describe
reality are constructed. The positions are also the natural
candidates to be used in defining a psycho--physical parallelism,
if we want to go so far. An appropriate way to express the now
discussed features of the theory derives from denoting, as J.S.
Bell proposed, as ``exposed variables" the positions of particles
and as a ``hidden variable" the wavefunction $\psi$.

\subsubsection{The case of CSL} \label{sec1252}

Let us now perform a corresponding analysis for the dynamical
reduction models. Since, as should be clear from the discussion
given in the previous subsection, the most relevant feature of the
modified dynamics is that of suppressing linear superpositions
corresponding to different mass distributions, one is actually led
to identify as the local beables of the theory the mass density
function ${\mathcal M}({\bf r},t)$ at a given time. Obviously,
also within CSL just as for the Pilot--Wave the wavefunction plays
a fundamental role for the evolution so that it too acquires the
status of a nonlocal beable.

It has to be remarked that in the interpretation we are proposing,
even though the wavefunction is considered as one of the beables
of the theory, the ``exposed variables" are the values of the mass
density function at different points. It is then natural to relate
to them, as we have done in the previous section, the concept of
similarity or difference between universes.

In doing so one is led to consider equivalent, at a fixed time
$t$, two ``universes" which are almost identical in the exposed
beables (i.e. they satisfy the condition (\ref{ctu})). Obviously
the fact that the above condition holds at $t$ does by no means
imply that the two universes will remain equivalent as time
elapses. It has to be stressed that the just mentioned feature is
not specific of  the model and the interpretation we are
proposing, but is quite general and occurs whenever one tries to
make precise the idea of ``similarity" of physical situations. In
fact within all theories we know, and independently of the
variables we choose to use to define nearness, situations can
occur for which {\it nearby states} at a given time can evolve in
extremely short times in {\it distant states}.

To focus on this important fact we can consider even classical
mechanics with the assumption that both positions and momenta are
the beables of the theory\footnote{Obviously, within classical
mechanics any function of these variables can be considered as a
beable, but since all information about the system can be derived
from the positions and the momenta, consideration of such
variables is sufficient.}. As it is obvious, even if such an attitude
is taken there are at least two reasons for which nearby points in
phase space can rapidly evolve into distant ones. First of all one
must take into account that many systems exhibit dynamical
instability so that the distance between ``trajectories" grows
exponentially with time. Secondly, even for a ``dynamically
standard" situation one can consider cases in which just the
present conditions can give rise to completely different
evolutions depending on some extremely small difference in the
whole universe. Suppose in fact you consider two universes $A$,
$\tilde A$ differing only in the {\it direction} of propagation of
a single particle (such universes have to be considered as very
close in any sensible objective interpretation). If the trajectory
of the considered particle in $\tilde A$ is such that in a very
small time it triggers e.g. the discharge of a Geiger counter,
which in turn gives rise to some relevant macroscopic effect,
while in $A$ it does not, the evolved universes soon become quite
different. An analogous argument obviously holds for standard
quantum mechanics, the Pilot--Wave theory and, as previously
remarked, for CSL too.

It is appropriate to stress that, in a sense, the above
considerations favor taking a position about reality which can be
described in the following terms. One considers the sensible
``beables" for its theory at a fixed time and one distinguishes
similar or different universes on the basis of such a snapshot.
Obviously, one must then also pay attention to the way in which
the beables evolve, i.e. to compare snapshots at different
times\footnote{From this point of view, one could state that also
the classical world would be most appropriately described in terms
of positions at fixed time.}.

\subsubsection{The role of mass density} \label{sec1253}

The previous analysis has shown that the proposed interpretation
(mass density = exposed beable) can be consistently taken within
CSL. Obviously it gives an absolutely prominent role to the mass
in accordance with the fact that mass is the handle by which the
reduction mechanism induces macro--objectification.

Other features of natural phenomena, such as the effects related
to the charge are, in a sense, less fundamental since to become
objective they need mass as a support. To clarify this point we
remark that one could consider, e.g., a condenser with two plates
of about $1$ cm$^{2}$, at a distance of $1$ cm. The plates are
supposed to be perfectly rigid and in perfectly defined
positions\footnote{This assumption must be made because we are
just discussing the role of the charge with respect to the one of
the mass within the model. If one would allow deformations and/or
displacements of the plates, once more the ensuing reduction would
be due to the mass and not directly to the charge density
difference in states $|C_{0}\rangle$ and $|C_{c}\rangle$.}. Let us
also consider the following {\it gedanken} situation: the
condenser can be prepared in the superposition of two states,
$|C_{0}\rangle$ and $|C_{c}\rangle$, the first corresponding to
its plates being neutral, the second to its plates having being
charged by displacing $10^{12}$ {\it electrons} from one plate to
the other. We remark that for the two states the decoupling rate
(recall that electrons, which are very light, are quite
ineffective in suppressing superpositions) is about $10^{-8}$
sec$^{-1}$, i.e. that the superposition can persist for more than
ten years. The electric field within the plates is zero or about
$10^{8}$ V/m in the two states, respectively. Suppose now we
consider a small sphere of radius $10^{-5}$ cm and density
$10^{-2}$ gr/cm$^{3}$ carrying a charge corresponding to $10^{4}$
electrons. We send such a test particle through the plates of the
condenser. What
 happens? The final state is the entangled state
\begin{equation}
|\psi(t)\rangle \; = \; \frac{1}{\sqrt{2}}\,\left[|C_{0}\rangle\,
|\makebox{undeflected}\rangle \; + \; |C_{c}\rangle\,
|\makebox{deflected}\rangle \right]
\end{equation}
the location of the particle in the state
$|\makebox{undeflected}\rangle$ and $|\makebox{deflected}\rangle$
differing by macroscopic amounts. According to the CSL model of
section \ref{sec76}, one can easily evaluate the rate of
suppression of the superposition. As already remarked the
contribution of the electrons on the plates is totally negligible
so that the decoherence is governed mainly by the mass of the
particle. Then, with the above choices for the radius and the
density of the test particle, the superposition will persist for
more than one minute. In spite of the fact that macroscopically
relevant forces enter into play no reduction takes place for such
a time interval. On the contrary, if we put the same charge on a
particle of normal density and of radius $10^{-3}$ cm, we see that
the macroscopic force acting on it when the condenser is in the
state $|C_{c}\rangle$ leads to a displacement of the order of its
radius in about $10^{-5}$ sec and that within the same time the
reducing effect of the dynamics suppresses one of the two terms of
the superposition.

This example is quite enlightening since it shows that
superpositions of charge distributions generating different forces
which are relevant at the macroscopic level, are not suppressed
unless they induce displacements of masses. It goes without saying
that any attempt to relate reduction to charge is doomed to fail
since it will not suppress superpositions of macroscopically
different but electrically neutral mass distributions.

We hope to have made clear, with this perhaps tedious analysis,
the real significance of treating the mass function as the
``exposed beables" allowing one to describe reality.

\section{The ``tail problem'' in dynamical reduction models}
\label{sec013}

Dynamical reduction models have been repeatedly criticized because
the reducing dynamics does not lead, in the case of macroscopic
systems, to perfectly localized wavefunctions, i.e. to
wavefunctions having a compact support corresponding to a small
(with respect to the characteristic length $1/\sqrt{\alpha}$)
volume of space. On the contrary, wavefunctions describing
macroscopic systems always have (very small, as we shall see)
``tails'' spreading out to infinity.

Due to this feature, many authors \cite{shi90,alo1,lew,cli1,cli2}
have suggested that dynamical reduction models do not guarantee
the emergence of a objective (and classical) world at the
macroscopic level, a world in which macroscopic systems occupy a
precise position in space.

In the present section we will show why the ``tail problem'' is
not a problem within QMSL and CSL. After a brief introduction
(subsection \ref{sec131}), in subsection \ref{sec132} we list the
criticisms which have been put forward by the above quoted
authors. In subsection \ref{sec133} we give a quantitative
estimate of the order of magnitude of the tails, in the case of a
macroscopic system, showing that they represent an extremely small
portion of the wavefunction.

In subsections \ref{sec134} and \ref{sec135} we reply to the
criticisms, by means of the mass interpretation introduced in the
previous section: the tails of the wavefunction do not forbid the
(classical) description of macroscopic systems. However, this does
not mean that the tails do not have any  physical relevance, as we
will prove in subsection \ref{sec136}. We conclude the section
with a classical analog of the tail problem (subsection
\ref{sec137}).

\subsection{Historical remarks} \label{sec131}

J. Bell vividly described Schr\"odinger's trials to give a
consistent interpretation of the wavefunction \cite{bellam}:

\begin{quotation}
In the beginning Schr\"{o}dinger tried to interpret his
wavefunction as giving somehow the density of stuff of which the
world is made. He tried to think of an electron as represented by
a wavepacket ... a wavefunction appreciably different from zero
only over a small region in space. The extension of that region he
thought of as the actual size of the electron ... his electron was
a bit fuzzy. At first he thought that small wavepackets, evolving
according to the Schr\"{o}dinger equation, would remain small. But
that was wrong. Wavepackets diffuse, and with the passage of time
become indefinitely extended, according to the Schr\"{o}dinger
equation. But however far the wavefunction has extended, the
reaction of a detector to an electron remains spotty. So
Schr\"{o}dinger's ``realistic'' interpretation of his wavefunction
did not survive.

Then came the Born interpretation. The wavefunction gives not the
density of stuff, but gives rather (on squaring its modulus) the
density of probability. Probability of what, exactly? Not of the
electron being there, but of the electron being found there, if
its position is ``measured''.

Why this aversion to ``being'' and insistence on ``finding''? The
founding fathers were unable to form a clear picture of things on
the remote atomic scale. They became very aware of the intervening
apparatus, and of the need for a ``classical'' base from which to
intervene on the quantum system. And so the shifty split.
\end{quotation}

Some remarks having a direct connection with the central problem
of the present section emerge quite naturally. Schr\"{o}dinger
too, just as von Neumann, was certainly aware of the fact that the
electron wavefunction cannot have compact support, according to
his equation. In spite of that he was quite keen to interpret a
well localized wavefunction as describing a ``bit fuzzy''
electron, or better the ``stuff'' of which the electron is made
(its mass and charge density). The compelling reasons for
abandoning such a position did not come, as we all know, from the
fact that even extremely well localized wavefunctions have tails,
but, as appropriately stressed by Bell, from the fact that well
concentrated wavepackets become appreciably different from zero
over macroscopic regions in extremely short times. Thus,
Schr\"{o}dinger's realistic position had to be abandoned, to be
replaced by the probabilistic Born interpretation.

There is no doubt that Schr\"{o}dinger too was perfectly aware of
the fact that the integral over the whole space of the squared
modulus of the wavefunction does not change with time, as a very
consequence of Schr\"{o}dinger's equation. Consequently he
certainly had perfectly clear that by adopting the ``density of
stuff'' interpretation he had to accept also that, in spite of the
fact that his electron was there (in the small region in space in
which its wavefunction was concentrated), a negligible part of its
mass (or charge) would not be confined to that region. The
extremely relevant difference of the model theories we are
analyzing here with the case of Schr\"odinger, however, derives
from the fact that superpositions of functions of macro--systems,
appreciably different from zero over macroscopic distances, are
dynamically forbidden within QMSL, contrary to Schr\"odinger's
case.

\subsection{Criticisms about the ``tail problem''} \label{sec132}

As the reader has certainly grasped, the localization of the
wavefunction does not lead to an infinitely precise localization
of the pointer. Actually, after a localization the wavefunction
(like all conceivable wavefunctions, both within standard Quantum
Mechanics and CSL) unavoidably exhibits tails extending over the
whole space. In fact, since $|\psi\rangle$ has a noncompact
support in the position representation, multiplying it times a
Gaussian leaves it different from zero everywhere\footnote{We
stress that it would have been totally useless to make the
localization function of compact support, since the kinetic energy
part of the hamiltonian would immediately make the wavefunction
different from zero everywhere.} (recall the example at the end of
section \ref{sec51}). This fact is at the basis of the uneasiness
of various people who naively transfer the ontology of standard
quantum mechanics to the new theory. The first criticism of this
kind has been formulated by A. Shimony \cite{shi90} who has put
forward many desiderata for a modified quantum dynamics, one of
them being that:
\begin{quotation}
If a stochastic dynamical theory is used to account for the
outcome of a measurement, it should not permit excessive
indefiniteness of the outcome, where ``excessive'' is defined by
consideration of sensory discrimination. This desideratum
tolerates outcomes in which the apparatus variable does not have a
sharp value, but it does not tolerate ``tails'' which are so broad
that different parts of the range of the variable can be
discriminated by the senses, even if very low probability
amplitude is assigned to the tail.
\end{quotation}
It goes without saying that the perspective chosen by Shimony with
respect to the dynamical reduction models is entirely based on
{\it the standard probabilistic interpretation} of quantum
mechanics ({\it even if very low probability amplitude is assigned
to the tail}) about the possible `outcomes' of a {\it measurement}
process.

A quite similar criticism has been raised by Albert and Loewer
\cite{alo1}:
\begin{quotation}
Our worry is that GRW collapses almost never produce definite
outcomes even when outcomes are recorded in distinct positions of
macroscopically many particles. The reason is that a GRW jump does
not literally produce a collapse into an eigenstate of position. A
GRW collapse yields one of the states with tails in which almost
all the amplitude is concentrated in the region around one of the
two components but there is nonzero, though very small, amplitude
associated with other regions. ... This means that the post
collapse state is not an eigenstate of position and so does not
actually assign a definite position to the pointer.
\end{quotation}
Once more we stress the adherence of the authors to the standard
formalism: the only variables which {\it have values} are those of
which the statevector is an {\it eigenstate}. One should also note
that the request that the statevector be an eigenstate of position
is very peculiar: the localization mechanism should map a state of
the Hilbert space onto a non--normalizable state and, at any rate,
such a state would immediately spread everywhere, loosing what is
considered its fundamental feature of assigning {\it a definite
position to the pointer}. Subsequently, the same authors
\cite{alo1} have suggested that the GRW proposal could be saved
provided one would be keen to release the eigenvector--eigenvalue
link. More about this in what follows.

The tail problem leads in a straightforward way to the enumeration
anomaly which is the subject of the papers of Lewis \cite{lew} and
Clifton and Monton \cite{cli1,cli2} (who use the term `conjunction
introduction' instead of  `enumeration'). The idea is quite
simple. Lewis considers a macroscopic marble and a very large box,
he denotes the normalized eigenstates of the marble being inside
and outside the box as $|\makebox{in}\rangle$ and
$|\makebox{out}\rangle$, respectively, and he remarks that
starting from a state:
\begin{equation} \label{b}
\frac{1}{\sqrt{2}}\left( |\makebox{in}\rangle\; + \;
|\makebox{out}\rangle \right)
\end{equation}
the GRW dynamics will transform it, almost immediately, into a
state like
\begin{equation} \label{c}
a|\makebox{in}\rangle \; + \; b|\makebox{out}\rangle
\end{equation}
or into a state like
\begin{equation} \label{d}
b|\makebox{in}\rangle \; + \; a|\makebox{out}\rangle
\end{equation}
where $1> |a|^{2} \gg |b|^{2} > 0$ ($|a|^{2} + |b|^{2} = 1$).
Lewis recalls that the GRW theory requires us to interpret each
one of these states as one in which the marble is inside (in case
(\ref{c})) or outside (in case (\ref{d})) the box\footnote{We
point out that, with reference to the above states, it would have
been much more appropriate to assert that they represent a marble
which is located in the precise region where its wavefunction is
sharply peaked. Obviously that region is inside the box for state
(\ref{c}) and outside for state (\ref{d}).}. Then he considers a
system of $n$ non interacting marbles, each in a state like
(\ref{c}):
\begin{equation} \label{e}
|\Psi\rangle_{{\makebox{\tiny all}}} \, = \; \left(
a|\makebox{in}\rangle_{1} + b|\makebox{out}\rangle_{1}\right)
\otimes \left( a|\makebox{in}\rangle_{2} +
b|\makebox{out}\rangle_{2}\right) \otimes ...\otimes \left(
a|\makebox{in}\rangle_{n} + b|\makebox{out}\rangle_{n}\right).
\end{equation}
The counting anomaly is then easily derived: for a state like
(\ref{e}) the GRW theory allows us to claim that: `particle 1 is
in the box', `particle 2 is in the box', ..., `particle $n$ is in
the box' but for the same state the probability that all marbles
{\it be found} in the box is $|a|^{2n}$, which, for $n$
sufficiently large, can be made arbitrarily small.

Clifton and Monton \cite{cli1} agree, in principle, with Lewis,
and they also prove that the suggestion of releasing the
eigenvalue--eigenvector link put forward by Albert and Loewer
\cite{alo1} does not allow one to overcome the difficulty. In
fact, what Albert and Loewer propose is to weaken the
eigenvalue--eigenvector link for position according to a rule they
call PosR:
\begin{quotation}
`Particle {\it x} is in region R' if and only if the proportion of
the total squared amplitude of {\it x}'s wavefunction which is
associated with points in region R is greater or equal to $1-p,$
\end{quotation}
with an appropriately chosen (and small) $p$. Clifton and Monton
suggest, first of all, to generalize PosR for a multi--particle
system resorting to what they call the {\bf fuzzy link} criterion:
\begin{quotation}
`Particle $x$ lies in region $R_{x}$ and $y$ lies in $R_{y}$ and
$z$ lies in $R_{z}$ ...' if and only if the proportion of the
total squared amplitude of $\psi (t,r_{1},...,r_{N})$ that is
associated with points in $R_{x}\times R_{y} \times R_{z}$ ... is
greater than or equal to $1-p$.
\end{quotation}

It should be clear that, while according to the fuzzy link criterion for a
state like (\ref{e}) the propositions $A_{i}\equiv $ `particle $i$
is in the box' are true for any $i$, the conjuction $A_{1}\wedge
A_{2}\wedge ...\wedge A_{n}$ is false. In other words, the
proposal entails a failure of {\it conjunction introduction}.
Having remarked this, the above authors feel the necessity of
pointing out that the tail problem, even though present, can never
become manifest within the GRW theory. This is proved by {\it
operationalizing the procedure of counting marbles}, i.e. by
considering apparata aimed {\it to detect} whether particle 1 is
in the box, particle 2 is in the box and so on, and another
apparatus aimed {\it to detect} how many particles there are in
the box and comparing their outcomes. Then, by some assumptions
(which seem to us useless and inappropriate) one shows that a
situation resembling the one of the von Neumann chain emerges: in
spite of the different possible final outcomes there is always
consistency between the individual and global detections.

To be more precise we recall the reasoning of reference
\cite{cli1}. It goes as follows:
\begin{quotation}
An ideal measurement of whether marble 1 is in the box would
correlate orthogonal states of a macroscopic measuring apparatus
to the $|\makebox{in}\rangle$ and $|\makebox{out}\rangle$ states
of the marble.
\end{quotation}
Obviously one has to resort to $n$ such apparatuses. The evolution
leads to the state:
\begin{equation} \label{f}
(a|\makebox{in}\rangle_{1} |\makebox{`in'}\rangle_{M1} +
b|\makebox{out}\rangle_{1} |\makebox{`out'}\rangle_{M1}) \otimes
...\otimes (a|\makebox{in}\rangle_{n} |\makebox{`in'}\rangle_{Mn}
+ b|\makebox{out}\rangle_{n} |\makebox{`out'}\rangle_{Mn})
\end{equation}
where the states $|\makebox{`in'}\rangle_{Mk}$ (
$|\makebox{`out'}\rangle_{Mk})$ are eigenstates of the observable:
`the $k$--th apparatus has recorded that the marble is inside
(outside) the box'. At this stage the authors need a further
apparatus $M$ (again working ideally) to see how many marbles are
in the box. A new step in the chain is necessary and one ends up
with an entangled state of the kind:
\begin{equation} \label{g}
|\psi\rangle_{\makebox{\tiny count}} \; = \;
\sum_{k=0}^{n}a^{n-k}b^{k}\, |\phi
(n-k,\makebox{in};k,\makebox{out})\rangle \, |\makebox{`$O =
n-k$'}\rangle
\end{equation}
where the last factor refers to the eigenstate of $M$
corresponding to the eigenvalue `$n-k$ particles are inside the
box', and the states $|\phi
(n-k,\makebox{in};k,\makebox{out})\rangle$ are (in general) linear
superpositions of states in which there are $n-k$ factors of the
type $|\makebox{in}\rangle_{j}|\makebox{`in'}\rangle_{Mj}$ and $k$
factors $|\makebox{out}\rangle_{s}|\makebox{`out'}\rangle_{Ms}.$
And here comes the conclusion:
\begin{quotation}
The state $|\psi\rangle_{\makebox{\tiny count}}$ is highly
unstable given the GRW dynamics, ... since its various terms
differ as to the location of the pointer on $M$'s dial that
registers the value of O.
\end{quotation}
Moreover, even if the state collapses to a term $|\phi
(n-k,\makebox{in};k,\makebox{out})\rangle |\makebox{`$O =
n-k$'}\rangle$,  since the terms of the first factor differ as to
the location of at least one marble and since the marbles and the
apparata $M_{r}$ are macroscopic, then one would end up, e.g.,
with the state:
\begin{equation} \label{h}
(|\makebox{out}\rangle_{1} |\makebox{`out'}\rangle_{M1}
|\makebox{in}\rangle_{2} |\makebox{`in'}\rangle_{M2} \ldots
|\makebox{in}\rangle_{n} |\makebox{in}\rangle_{Mn}) |\makebox{`$O
= n - 1$'}\rangle
\end{equation}
in which the records of the various individual apparata $M_{s}$
agree with the one of $O$. Thus, the counting anomaly cannot
become manifest.

In what follows we will put forward precise motivations showing
that this operationalization process is useless. But we want to
call immediately the attention of the reader to an important fact.
Suppose we consider, for simplicity, only two terms of one of the
states $|\phi (n-k,\makebox{in};k,\makebox{out})\rangle$, e.g.:
\begin{equation} \label{i}
|\makebox{in}\rangle_{1} |\makebox{`in'}\rangle_{M1}
|\makebox{out}\rangle_{2} |\makebox{`out'}\rangle_{M2} \ldots +
|\makebox{out}\rangle_{1} |\makebox{`out'}\rangle_{M1}
|\makebox{in}\rangle_{2} |\makebox{`in'}\rangle_{M2}
\end{equation}
and recall that these authors denote as $|\makebox{in}\rangle_{1}$
the part of the GRW Gaussian--like wavefunction lying inside the
box and as $|\makebox{out}\rangle_{1}$ the tail outside the box.
Suppose particle 1 suffers a localization within the box. In
configuration space one sees immediately that multiplying the
wavefunction corresponding to (\ref{i}) times a Gaussian centered
within the box near the point at which the wavefunction is peaked
and normalizing the resulting statevector makes the second of the
two above terms of much smaller norm than the first, but in no way
suppresses it. Obviously the same argument applies for any
localization of the pointers of the individual apparata $M_{i}$
and $M$. So the state will never take precisely to form (\ref{h}).
The authors have implicitly {\it assumed} that at some level the
{\it standard quantum mechanical reduction} (and not the
spontaneous localization characterizing the GRW theory) takes
place\footnote{In \cite{cli2}, Clifton and Monton have tried to
de--emphasize their previous assertions by stating: {\it In our
exposition we were simply dividing the collapse process into
different stages for ease of exposition. Certainly we were aware
that the marbles themselves will be almost continually subject to
GRW collapses}. Our answer is quite simple: first, why not confine
the analysis to the marbles? What is the purpose of introducing
the apparata? Secondly, we invite the reader to introduce in the
game the appropriate description of the whole process by assuming
that, in turn, the registration of the outcome is given by the
location of a pointer whose wavefunction unavoidably has tails. He
will easily realize that there is no advantage in operationalising
the process of counting marbles: he will find himself back to
square one.}.

\subsection{A quantitative analysis} \label{sec133}

Let us consider a macroscopic object which is in an ``almost"
eigenstate of the mass operators $M_{i}$ (we consider once more
the discretized version of CSL) but which however have tails. Let
$|\psi\rangle$ be the normalized state
\begin{equation}
|\psi\rangle \quad  = \quad \alpha |\psi^{A}_{N}\rangle \; + \;
\beta |\psi^{B}_{N}\rangle
\end{equation}
where $|\psi^{A}_{N}\rangle$ and $|\psi^{B}_{N}\rangle$ are the
states appearing in (\ref{edcvf}) and $|\beta|^{2}$ is extremely
close to zero. In region $A$ we have
\begin{equation} \label{cplra}
{\mathcal M}_{i(A)} \; \simeq \; |\alpha|^{2}nm_{0}, \quad
{\mathcal V}_{i(A)} \; \simeq \; |\alpha|^{2}|\beta|^{2} n^{2}
m_{0}^{2}, \quad \makebox{and} \quad {\mathcal R}_{i(A)} \; \simeq
\; |\beta|^{2} \; \ll \; 1
\end{equation}
so that the masses ${\mathcal M}_{i(A)}$ are accessible and
practically equal to those corresponding to the state
$|\psi^{A}_{N}\rangle$. In region $B$ we have
\begin{equation} \label{cplrb}
{\mathcal M}_{i(B)} \; \simeq \; |\beta|^{2}nm_{0}, \quad
{\mathcal V}_{i(B)} \; \simeq \; |\alpha|^{2}|\beta|^{2} n^{2}
m_{0}^{2}, \quad \makebox{and} \quad {\mathcal R}_{i(A)} \; \simeq
\; |\beta|^{-2} \; \gg \; 1
\end{equation}
hence the masses ${\mathcal M}_{i(B)}$ are not accessible.

Our aim is to make a quantitative estimate of ${\mathcal
R}_{i(A)}$ and of the total mass in region $B$. To this purpose
(as it is evident from equations (\ref{cplra}) and (\ref{cplrb})) one
has to explicitly evaluate the order of magnitude of the parameter
$|\beta|^{2}$ implied by the  reducing dynamics. In order to do
this, to cover also the case of non homogeneous bodies, we
consider again two far apart regions $A$ and $B$, each containing
$K$ cells and a system of nucleons which at time $t=0$ is in a
(normalized) state of the type (the overall phase factor being
irrelevant)
\begin{eqnarray} \label{tps}
|\psi\rangle & = & \alpha(0)|n_{1(A)}, \ldots, n_{K(A)},
\ldots, 0, \ldots, 0\rangle \; + \nonumber \\
& & \beta(0) e^{i\gamma(0)} |0, \ldots, 0, \ldots, n_{1(B)},
\ldots, n_{K(B)} \rangle
\end{eqnarray}
where $\alpha(0)$ and $\beta(0)$ are comparable positive numbers
and $n_{i(A,B)}$  represents the occupation number in the $i$--th
cell in regions $A$ and $B$ respectively\footnote{We disregard the
cells which are not contained in regions $A$ and $B$ since they
are irrelevant for the following discussion.}. We then study the
ensemble of systems brought in by the reducing dynamics after a
time interval of the order of, e.g., $10^{-2}$ sec (the reason for
this choice will become clear in what follows).

According to the CSL model of section \ref{sec76}, after such a
time interval the normalized state corresponding to a definite
realization of the stochastic potential would be of the type
\begin{eqnarray} \label{fhs}
|\psi_{B}(t)\rangle & = & \alpha_{B}(t)|n_{1(A)}, \ldots,
n_{K(A)}, \ldots, 0, \ldots, 0\rangle \; + \nonumber \\
& & \beta_{B}(t) e^{i\gamma(0)} |0, \ldots, 0, \ldots, n_{1(B)},
\ldots, n_{K(B)} \rangle
\end{eqnarray}
with $\alpha_{B}(t)$ and $\beta_{B}(t)$ as positive numbers. The
ensemble of systems corresponding to all possible realizations of
the stochastic potential would be described by the statistical
operator
\begin{equation} \label{fhs2}
\rho(t) \quad = \quad \int dB_{1} \ldots dB_{2K}\,
P_{\makebox{\tiny cook}}[B(t)]\, |\psi_{B}(t)\rangle
\langle\psi_{B}(t)|
\end{equation}
satisfying\footnote{Even though we are using the CSL model
relating decoherence to the mass, the formulas of this subsection
coincide with the analogous ones of standard CSL. This is due to
the fact that we deal only with nucleons and that we have chosen
the coupling to the noise to be governed by the ratio
$\gamma/m_{0}^{2}$, taking the standard CSL value.}
\begin{eqnarray} \label{crcic}
\lefteqn{ \langle n_{1(A)}, \ldots, n_{K(A)}, \ldots, 0, \ldots,
0| \rho(t) |0, \ldots, 0, \ldots, n_{1(B)}, \ldots,
n_{K(B)}\rangle \; =}
\nonumber \\
& & e^{\displaystyle -\lambda\, t \sum_{i}^{K} n_{i}^{2}} \langle
n_{1(A)}, \ldots, n_{K(A)}, \ldots, 0, \ldots, 0| \rho(0) |0,
\ldots, 0, \ldots, n_{1(B)}, \ldots, n_{K(B)}\rangle \nonumber \\
& &
\end{eqnarray}
with $\lambda t \simeq 10^{-18}$. From (\ref{crcic}) we see that
the matrix elements of $\rho(t)$ between the considered states are
exponentially damped by a factor which is proportional to
$\sum_{i}^{K}n^{2}_{i}$.

In the following we consider only situations in which
$\sum_{i}^{K}n^{2}_{i}$ turns out to be much greater than
$10^{18}$, so that in the considered time interval of $10^{-2}$
sec the linear superposition (\ref{tps}) is actually suppressed,
i.e. either $\alpha_{B}(t)$ or $\beta_{B}(t)$ of equation
(\ref{fhs}) become very small. The states at time $t$ are then
typical states with ``tails", i.e. states whose existence is
considered as a drawback of the theory by the authors of
references \cite{shi90,alo1,lew,cli1,cli2}. Equation (\ref{crcic})
implies (taking into account equations (\ref{fhs}) and
(\ref{fhs2})) that
\begin{equation} \label{fhs3}
\int dB_{1} \ldots dB_{2K}\, P_{\makebox{\tiny cook}}[B(t)]\,
\alpha_{B}(t)\, \beta_{B}(t) \quad = \quad \alpha(0)\,\beta(0) \,
e^{\displaystyle -\lambda\, t \sum_{i}^{K} n^{2}_{i}}
\end{equation}
>From (\ref{fhs3}), since $\alpha_{B}(t)$ and $\beta_{B}(t)$ are
positive, one can easily deduce that the probability of occurrence
of realizations of the stochastic potential which would lead to a
value for the product $\alpha_{B}(t) \beta_{B}(t)$  much greater
than $e^{-\lambda\, t \sum_{i}^{K} n^{2}_{i}}$ must be extremely
small. Therefore, one can state that in practically all cases
\begin{equation}
\alpha_{B}(t) \beta_{B}(t) \quad \simeq \quad e^{\displaystyle
-\lambda\, t \sum_{i}^{K} n^{2}_{i}}
\end{equation}
If we assume that $\alpha_{B}(t) \simeq 1$, so that we consider an
individual case for which the reduction leads to the state
corresponding to the nucleons being in region $A$,
$|\beta_{B}(t)|^{2}$  must be of the order of $e^{-\lambda\, t
\sum_{i}^{K} n^{2}_{i}}$.  On the basis of this fact we can then
estimate the value of $|\beta|^{2}$, e.g., for a homogeneous
marble of normal density (so that $n_{i} = n \simeq 10^{9}$ is the
number of particles per cell) and of size 1 dm$^{3}$ (so that $K
\simeq 10^{18}$ is the number of cells in regions $A$ and $B$)
getting a figure of the order of $e^{-10^{18}}$. Correspondingly,
we have
\begin{equation} \label{efr}
{\mathcal R}_{i(A)} \quad \simeq \quad e^{\displaystyle -10^{18}}
\end{equation}
while for the total mass in region $B$ we get the value
\begin{equation} \label{efm}
{\mathcal M}_{B} \quad \simeq \quad e^{\displaystyle -10^{18}}
10^{27} m_{0}
\end{equation}
Equation (\ref{efr}) shows that the mass in region $A$ is
``accessible" to an extremely high degree of accuracy and equation
(\ref{efm}) shows that the total mass in region $B$ is incredibly
much smaller than the mass of a nucleon. If we consider a
situation in which $K$ or $n$ are greater than those of the
example we have discussed now, we find values for ${\mathcal
R}_{i(A)}$ and ${\mathcal M}_{B}$ which are even
smaller\footnote{Note that this holds also for objects like a
galaxy or a neutron star.} than those of equations (\ref{efr}) and
(\ref{efm}). This fact by itself (see also the analysis of the
following subsection) shows that the states with ``tails" allowed
by CSL cannot give rise to difficulties for the proposed
interpretation of the theory. For example, if we would perform the
usual gedanken experiment with the test particle it would be
deflected just as if in region A there would be the ``classical"
mass $Knm_{0}$.

\subsection{Mass density interpretation and marbles} \label{sec134}

Before replying to the criticisms previously mentioned
\cite{shi90,alo1,lew,cli1,cli2}, and in the light of the analysis
of the previous section, we call the reader's attention to the
following properties of the mass density interpretation within
dynamical reduction models which are relevant for the discussion
about the ``tail problem'':
\begin{enumerate}
\item In the case of a marble the mass density is accessible just
where the marble is located. Any test devised to `reveal' the mass
density distribution will agree with the statements which make
reference to the accessible mass density. One can easily evaluate
the contribution of the tails\footnote{In the previous subsection
it has been proved that the integral of the mass density extended
to all space exception made for the region in which the Gaussian
is centered amounts to an incredibly small fraction of the mass of
a nucleon.} to any possible gravitational test and conclude that
there would be no physically testable difference whatsoever
between the case of a marble whose wavefunction is a Gaussian of
width $10^{-11}$ cm, and one for which its wavefunction has
compact support.
\item This should make clear why we claim that there is no need to
operationalize the process of counting marbles. We already know
that the regime condition induced by the GRW dynamics corresponds
precisely to the statement that the particles are where they are,
i.e., in those regions in which there is an accessible mass
density, and that any test will confirm such statements.
\item It is important to stress a point which seems to have been
underestimated in \cite{cli1} and \cite{cli2}, i.e. the crucial
role played by the fact that the GRW theory leads to precise
regions in which the mass is accessible. For this reason it does
not seem a good choice to have schematized the `tail problem' by
resorting to the states $|\makebox{in}\rangle$ and
$|\makebox{out}\rangle$ (even though they are those which matter
for the counting anomaly). If the dynamics would allow a marble in
the box to be in the superposition of two states corresponding to
two Gaussians, or even (at a certain instant) to two wavefunctions
of comparable norm with disjoint and far apart (let us say 1
meter) compact supports, the associated mass density function
would not be accessible anywhere in ordinary space, and any talk
about the location of the marble would be devoid of any meaning.
We can speak of the position of a marble just because there is a
region coinciding (practically) with its location in which there
is an accessible mass density. If there is no such region, to be
allowed to say something about where the marble is one should
resort to `a measurement' and to some recipe leading to an
accessible mass density of the pointer of the measuring device.
But fortunately, within the GRW theory, when one reaches the
macroscopic level, there is no need to  invoke  observables or
measurement processes. This argument shows also how inappropriate
it is to make reference, as is repeatedly done in references
\cite{cli1} and \cite{cli2}, to microscopic systems replacing the
marbles: for such systems there is no mechanism forbidding the
spread of the wavefunction and consequently no talk about their
location has any meaning\footnote{Obviously, also the mass density
of an elementary particle can be accessible, if its wavefunction
is extremely well peaked in a region of, let us say, less tan
$10^{-6}$ cm. But as we all know, the hamiltonian evolution will
make it unacessible almost immediately due (not to the tails) but
to the extremely rapid increase of its spread without any
localization balancing  the hamiltonian spread. In our opinion this is
the appropriate way to read the statement of J.S. Bell: {\it
even for one particle I think one would have problems }with the mass
density and makes clear while, at any rate, one cannot  apply it to the
macroscopic case.}.
\end{enumerate}

As already mentioned, Clifton and Monton have pointed out that the
mass density interpretation requires a divorce of mass talk from
position talk.  We believe that what they want to stress is that
since the integral over the whole space of the operator $M(r)$
gives the the total mass operator of all particles in the
universe, and since the GRW theory does not contemplate creation
and/or annihilation processes, the mass associated to the region
in which it is accessible (i.e. in which, according to our views,
the marble is located) cannot coincide with the total mass of the
marble. This is true, but, as we have already remarked this
divorce amounts to an extremely small fraction of the mass of a
nucleon and fits perfectly the ontology of a theory which allows
microsystems \cite{bellns}:
\begin{quotation}
to enjoy the cloudiness of waves, while allowing tables and
chairs, and ourselves, and black marks on photographs, to be
rather definitely in one place rather than another, and to be
described in classical terms.
\end{quotation}

We do not want to be misunderstood. We are not playing once more
with the possibility of locating the split between micro and
macro, reversible and irreversible, quantum and classical, where
we consider it appropriate for our purposes. To clarify this point
we can consider a state of (the particles of) the marble in which,
for some physical interaction, a portion of $10^{-15}$ cm$^{3}$ of
the marble has been separated from the marble itself. We stress
that, within the GRW theory with the proposed interpretation this
situation is `objectively' different from that of an unsplit
marble and can be perfectly described in classical terms. In
particular, the superposition of a state of an `unbroken marble'
and the one of the marble which `has lost a piece' will not last
for the perception times, and, if reduction takes place to the
second state, the mass density will turn out to be accessible just
where the marble and where its fragment are located. But if,
instead of a fragment, the marble has been deprived of some
elementary particles, then, while the mass is still accessible
where the marble is, it is no more so for the region over which
the wavefunctions of the lost particles extends, and this
precisely for the reason that their wavefunction spread almost
instantaneously. We think that one has to keep clearly in mind
this situation  to understand why  the remarks of Clifton and
Monton about the location of the marble and the mass density are
not cogent (more about this later).

\subsection[Reply to criticisms]{Proof of the internal consistency of
the mass density interpretation: reply to criticisms}
\label{sec135}

>From the previous analysis it should be clear why we consider not
cogent the criticisms raised by Shimony, Albert and Loewer and
Lewis, and why we consider superfluous and inappropriate the
suggestion by Clifton and Monton to operazionalize the counting
process (see also \cite{bgsm1,bgsm2,bgsm3}). Let us be more
specific.
\begin{enumerate}
\item The tail problem is not a problem, it simply requires us to
abandon the idea that the presence of the tail implies that there
is a certain probability that a measurement gives the outcome:
`the marble has been found in the region of the tail'. In the GRW
theory there are no measurements. If we are interested in testing
the `properties' of a microsystem then we have to invent a device
correlating its different microstates to different regions in
which the mass density of the pointer becomes accessible. From the
knowledge of such an accessible mass density we can infer `the
outcome of the measurement'.  If the object we are interested in
is already macroscopic then there is nothing to measure about its
position: the object is where the associated mass density is
accessible. Measuring where the object is becomes then, in a
sense, tautological and has only a practical interest if, e.g., we
have no access to the object. In such a case the so called
`measurement procedure' consists simply in establishing a
correlation between the region where the mass density of the
measured object is accessible and the region in which the mass
density of the pointer is accessible, period.

\item According to the remarks under 1), in the state (\ref{e})
considered by Lewis the mass density is accessible only inside the
box, since one can easily check that the only points in space
where the mass is accessible lie inside it, while outside the box
there are no points where this occurs. This, in turn, implies that
in such a state any test aimed to ascertain where the marble is
will almost always give as a result `the marble is inside the
box'. We stress that we have added the specification `almost', not
because the regions in which the mass density is accessible are
(due to the tails) to some extent imprecise in a state like
(\ref{e}), but to take into consideration that any physical test
requires a certain time and that, when the number of marbles
increases beyond any limit, the peculiar rapid variations of the
centers of the Gaussians which define the accessibility regions
could lead us  to detect a particle outside the box, just because
it can make a sudden enormous jump during the test process. But
this is a story which has nothing to do with the enumeration
principle and with the problem of the tails: it originates
entirely from the dynamical structure of the theory. At any rate,
as we have already argued, even when the number of marbles which
one takes into account becomes enormous, at any given time one
will always be dealing with precisely $n$ Gaussians peaked around
$n$ precise positions. In brief: there is no probability of
finding marbles in different places than those where their mass
density is accessible and so there is no counting anomaly. The
statement that the probability of {\it finding} all the particles
within the box is $|a|^{2n}$ derives entirely from adopting the
standard {\bf probabilistic} interpretation and the standard
position about measurements.

\item For the above reasons, there is also no need to
operationalize the counting process: such a procedure, if
developed rigorously, will simply lead to more and more
macro--systems in states which are characterized by an accessible
mass density in the regions in which the theory allows us to say
they actually are, plus more and more tails spread out over the
whole universe. These tails require a divorce of position talk
from mass talk,  but contrary to what the remarks of reference
\cite{cli2} seem to suggest, this divorce is absolutely negligible
and experimentally undetectable. In fact even though one could
naively argue that when the marbles become more and more numerous
the mass density in the tails becomes more and more relevant (and
might amount to the mass of even 1000 marbles) one cannot forget
that increasing the number of marbles increases correspondingly
their gravitational effects (masking more and more the
gravitational contribution of the tails) and that the mass density
of the tails is never accessible. Stated in rather brute terms:
the tails cannot in any way conspire to produce an accessible mass
density which can be identified with `a marble'. Thus, asserting
that in a state like (\ref{e}) one could legitimately claim that
$n-$ 1000 marbles are in the box is completely
nonsensical\footnote{With reference to this point we would like to
point out that the same problem of the relations between position
and mass occurs even in Classical Mechanics and, in general, in
field theories. In fact the equivalence between mass and energy
implies that the mass of a classical object is not given only by
the masses of its constituents, but also by the energy of the
fields which keep the constituents together. And such fields, of
course, are spread out in space. Thus, if one wants to be very
pedantic, he could say that not all of the mass--energy of the
object is where we see the object to be, since a very small
portion of it is spread in outer space. From this point of view
the situation is quite similar to the one of the GRW theory: if
one considers an incredibly large number $n$ of macroscopic
classical objects inside a box and calculates the total mass
inside it, he could very well find as a result $m\,(n-1000)$. Do
we have to conclude that not only the GRW theory but also
Classical Mechanics and Field Theory violate the enumeration
principle? We think that this is not the case; we believe that it
is Lewis' and Clifton and Monton's points of view which are too
 strict and inadequate to the interpretation of such
theories. A field (quantum or classical), in general is never well
localized in space and thus concepts like `being located' are not
well suited to describe its properties. This is why when we want
to speak of a field as located in a certain region of space, we
have to accept a certain amount of fuzziness. Thus, for a field,
being confined to a certain region  means that almost all the
field is confined in such region, and if particles and matter have
to be described in terms of fields (and this is the trend in
modern physics), we have to accept some fuzziness, otherwise
nothing would be located anywhere in space.}. The analogy with a
classical situation we will present in the last subsection  will
allow us to deepen this point.

\item Clifton and Monton claim that the same arguments which
``serve to motivate the mass density criterion, also serve to
motivate speaking of a particle (or particles) as being located in
a region whenever its wavefunction assigns high probability to its
being detected in that region; that is they also motivate the
fuzzy link. Unfortunately, [Bassi and Ghirardi] never say why the
fuzzy link is `inappropriate' and not a `valid' way to understand
reduction theories''.

It is easy to find in this very sentence the reasons for the
inappropriateness of the Clifton and Monton analysis: the authors
use the terms {\it probability} and {\it being detected} which
shows that they are still bound to the orthodox interpretation.
But the whole sentence reveals that they have not grasped the real
meaning of the idea of accessibility: if a superposition of two
states of a marble with comparable weights and both with compact
support entirely within the box but separated by an appreciable
distance would be possible, then the mass would not be accessible.
It is just because of a lack of understanding of this point that
these authors feel comfortable in replacing a marble with a
particle, as  the above sentence shows. But this is totally
inappropriate and this is why the fuzzy link is inappropriate. In
brief: the main reasons to reject the fuzzy--link are that it puts
exactly on the same ground micro and macro--systems and it does
not take at all into account the most relevant feature of the GRW
theory, i.e., that what the dynamical evolution makes accessible
is precisely and solely the mass density at the appropriate macro level.
\end{enumerate}

\subsection{The stochastic nature of the evolution} \label{sec136}

So far we have discussed the description of the world allowed by
the CSL theory in terms of the values taken by the mass density
function ${\mathcal M}({\bf r})$ which have been recognized to
constitute the exposed beables of the theory. According to
equation (\ref{mdf}) it is the wavefunction associated to the
system which determines ${\mathcal M}({\bf r})$. It is useful to
analyze the evolution of the beables. Let us consider, for
convenience, the linear formulation of CSL: as we have discussed
in sections \ref{sec6} and \ref{sec7}, the dynamical evolution
equation for the wave function is fundamentally stochastic, being
governed by the stochastic processes $w({\bf r},t)$. The ``cooked"
probability of occurrence of such processes, based on the analog
of equation (\ref{edm10g}), depends on the wavefunction which
describes the system and this fact is of crucial importance for
getting the ``right" (i.e. the quantum) probabilities of
measurement outcomes. Therefore, in the CSL theory, the
wavefunction has both a descriptive (since it determines
${\mathcal M}({\bf r})$) and a probabilistic (since it enters in
the prescription for the cooking of the probability of occurrence
of the stochastic processes) role.

Also the ``tails" of the wavefunction have a precise role. In
fact, suppose our ``universe" is described at $t = 0$ by a
normalized state
\begin{equation}
|\psi(0)\rangle \quad = \quad \alpha(0)\,|a\rangle \; + \;
\beta(0)\, |b\rangle
\end{equation}
with $|\beta(0)|^{2}$ being extremely small. The ``reality" of the
universe at $t = 0$ is ``determined" by the state $|a\rangle$, as
we have explicitly shown in the previous subsection. However, one
cannot ignore the (extremely small) probability $|\beta(0)|^{2}$
that a realization of the stochastic potential occurs which, after
a sufficiently long time, leads to a normalized state
\begin{equation}
|\psi(t)\rangle \quad = \quad \alpha(t)\,|\tilde a\rangle \; + \;
\beta(t)\, |\tilde b\rangle
\end{equation}
with $|\alpha(t)|^{2}$  being extremely small and with $|\tilde
a\rangle$ and $|\tilde b\rangle$ two of the most probable states
at time $t$ for the initial conditions $|a\rangle$ and
$|b\rangle$, respectively. Then, the ``reality" at time $t$ is
that associated to the state $|\tilde b\rangle$ which has its
origin in the negligible component $|b\rangle$ at time $t = 0$.
Thus, some ``memory" of a situation which at time zero did not
correspond to the ``reality" of the world remains at time $t$.
Obviously, if such an extremely improbable case would occur one
would be tempted (wrongly) to retrodict that ``reality" at $t = 0$
was the one associated to $|b\rangle$ and not the one associated
to $|a\rangle$. However, we stress that such peculiar events,
which we could denote as the ``reversal of the status of the
universe", have absolutely negligible probabilities. As made
plausible by the estimate for the values of $\beta(t)$ given in
the previous section, the ``risk to be wrong" in retrodicting from
the present to the past ``status of the world" is comparable with
the probability of being wrong when, having observed now a table standing
on the floor, and knowing that it has been kept isolated, we claim
that it was standing there even one hour ago, in spite of the fact that
thermodynamically a very peculiar situation corresponding to its
``levitation" at that time could in principle have occurred.

\subsection{A classical analogue of the tail problem}
\label{sec137}

The criticisms which have been put forward concerning the tail
problem and which we have discussed in the previous subsections,
claim that the GRW theory, just due to the tail problem, is
fundamentally unsatisfactory from an ontological point of view. To
answer such criticisms we consider it appropriate to consider a
quite simple classical situation which shares many of the aspects
of the dynamical reduction models. Suppose we have a collection of
macroscopic objects of a precise shape and with precise mass
density: to be specific let us consider an assembly of identical
tables. Let us suppose that they are at rest and that the dynamics
allows processes in which each of the tables has an extremely
small probability (let us say of the order of $e^{-[10^{34}]})$ of
emitting just one nucleon or one electron in a certain time. The
specification {\it just one} is intended to be strict: a table can
emit one such particle but if the emission takes place, no more
particles  can be emitted. To go on, we {\it postulate} (this is
the analog of adopting the accessibility criterion within the GRW
theory) that a table which has lost just one nucleon or one
electron can still be called a table\footnote{Obviously, the
reader should have grasped why we allow the tables to loose just
one elementary constituent and we forbid the emission processes to
continue indefinitely. If one would have taken such an attitude
one would be compelled to define precisely up to which point one
is keen to consider the system to be still `a table', i.e. to
specify that a loss of, e.g., $10^{5}$ particles is acceptable but
that after that limit it is no longer legitimate  to call the
system `a table'. Since within GRW the accessibility derives from
the fact that the center--of--mass of a macro--system gets
localized with an extremely precise  accuracy (of about
$10^{-11}$ cm) and for ordinary objects like marbles or tables
this defines quite precisely the `mass in the tail' (which turns
out to be an extremely small fraction of a nucleon mass) the above
assumption is quite appropriate to develop the analogy we are
interested in.}. We now consider a universe made by an
unphysically large number $n$ (of the order of some
$10^{24}e^{[10^{34}]})$ of tables. After a while, a certain number
of nucleons and protons (proportional to $n$) will be emitted.
Thus the physical situation will be: we have $n-k$ tables in their
original status (with $k<<n)$, $k$ tables which have lost a
particle, and $k$ particles propagating in space. We think that
everybody would agree with the statement that there are still $n$
tables in our universe. And we also believe that nobody would
suggest that the emitted particles could, in principle, be used to
`build up a new table'\footnote{Taking the risk of being pedantic
we stress once more that both in the considered example as well as
within the GRW theory with the mass density interpretation, our
 `disregarding', in some sense, the emitted
particles (the tails) for what concerns the consideration of the
tables which are in our universe does not mean we wish  to deny to
them a real status and to ignore the physically relevant effects
they can trigger. One can think, e.g., of having in the universe
also many Geiger counters; there is no doubt that one of the
emitted particles could trigger a counter inducing relevant
dynamical changes. The same is true within the GRW theory: the
tails can trigger further appreciable effects, as we have already
stressed with particular emphasis.}. At any rate, if one would
take seriously such a possibility one would be lead to conclude
that he started with $n$ tables and he ended up with $n+1$ tables
without changing the total mass. How is this possible? Is there
something terribly wrong in the classical model we are envisaging?
Do we have to declare that the ontology of the model leads to
inconsistencies? Or does not this simply mean that the links
between objects, their masses and their locations is not so strict
as Shimony, Albert, Loewer, Clifton and Monton believe and that,
in particular, in the unrealistic and purely speculative case in
which the number of tables which enter into play is made
unphysically large one cannot avoid facing quite peculiar but
logically prefectly consistent situations?

\section{The psycho--physical parallelism within CSL}
\label{sec014}

The most characteristic and appealing feature of CSL and of its
interpretation consists in the fact that it allows one to give a
satisfactory account of reality, to take a realistic view about
the world, to talk about it as if it is really there even when it
is not observed. However, one cannot avoid raising the problem of
including also conscious observers in the picture, for
\cite{bellqg} {\it what is interesting if not experienced?} Thus
one is led to consider the problem of the psycho--physical
parallelism within dynamical reduction models: this is the subject
of the present section.

The section begins with a challenge for dynamical reduction
models, put forward by D. Albert and L. Vaidman \cite{avp,ap}: do
they always guarantee definite outcomes to measurement processes
(subsection \ref{sec141})? In order to answer this question, in
subsection \ref{sec142} we analyze what we expect a ``measurement
process'' and a ``measurement outcome'' to be. This leads us
(subsection \ref{sec143}) to consider how the process of
perception unfolds in time and how the reduction mechanism works
within the nervous system, thus proving that an observer always
has definite perceptions about measurement outcomes.

We conclude the section (subsections \ref{sec144} and
\ref{sec145}) with some further comments about the relation
between measurement outcomes and human perceptions.

\subsection{Introduction} \label{sec141}

Some years ago, two quite interesting papers, one by D. Albert and
L. Vaidman \cite{avp} and the other by D. Albert \cite{ap} (see
also \cite{al5}), challenging QMSL, have appeared. In them some
critical remarks have been put forward which, in the authors'
intention, should prove that dynamical reduction models suffers
from some serious limitations, in particular that they do not
satisfy fundamental requirements that conventional wisdom imposes
on a workable theory of collapse. The papers offer the opportunity
of a clarification about the model and its aims and deserve
various comments.

In reference \cite{avp} the authors start by listing the features
that any theory pretending to describe the collapse should
exhibit. Let us summarize them:
\begin{enumerate}
\item It ought to guarantee that {\it measurements always have
outcomes} after they are over, i.e., after a {\it recording} of
the outcomes exists in the measuring device.

\item It ought to imply that, for any given eigenvalue of the
measured observable, the probability that the statevector ends up
in  the associated eigenmanifold coincides with the probability
that standard quantum mechanics attributes to such an outcome.

\item It should not contradict any experimentally established
quantum mechanical predictions about physical systems, in
particular {\it the fact that isolated microscopic systems have
never yet been observed to undergo collapses}.
\end{enumerate}

As regards QMSL, Albert and Vaidman agree that, due to the
extremely low probability of occurrence of a localization and the
fact that the localization distance is large on the atomic scale,
the theory satisfies requirement 3). They also seem to agree on
the fact that, when consideration is given to a system containing
many particles which is in a superposition of two states
$|\psi_{1}\rangle$ and $|\psi_{2}\rangle$, such that in
$|\psi_{1}\rangle$ a large number $N$ of particles have spatial
positions which differ more than $1/\sqrt{\alpha}$ from those they
have in $|\psi_{2}\rangle$, then such a state is transformed
dynamically into either $|\psi_{1}\rangle$ or $|\psi_{2}\rangle$
in a time of the order of $10^{16} N^{-1}$ sec, which, for $N$ of
the order of Avogadro's number, means in less than a microsecond.
So, when consideration is given to cases in which the measurement
outcomes are indicated by some sort of a macroscopic pointer
taking macroscopically different spatial positions, QMSL does
satisfy also 1) and 2) besides 3).

Then, why do Albert and Vaidman assert that the theory runs into
difficulties with the first two requirements? The reason for this
is that they consider it  incorrect to assume that all measuring
instruments work in the above indicated way. They illustrate this
point with an example, which we briefly report here.

A spin--$1/2$ system is prepared in an eigenstate of $\sigma_{x}$
and then it is passed in a Stern--Gerlach arrangement with
non--uniform magnetic field in the $z$ direction. The two spin
states $|z\,+\rangle$ and $|z\,-\rangle$ are then correlated to
upward and downward ``trajectories," respectively. These
trajectories hit a fluorescent screen  at two different
points $A$ and $B$ which are separated by a macroscopic distance.
The screen is such that the particle impinging on it causes some
of the electrons of the atoms around the point of impact to jump
into excited orbitals. De--excitations of these electrons give
then rise to a luminous dot, which can be directly seen by an
experimenter. The situation, before any observer enters into play,
can then be described by the following statevector
\begin{eqnarray} \label{rsp}
|\psi(1,2)\rangle & = & \frac{1}{\sqrt{2}}\,\left[ |z+,
x=A\rangle_{MP} |1_{ex}, \ldots, n_{ex}; (n+1)_{gr}, \ldots,
2n_{gr}\rangle
\right.\nonumber \\
& + & \left. |z-, x=B\rangle_{MP} |1_{gr}, \ldots, n_{gr};
(n+1)_{ex}, \ldots, 2n_{ex}\rangle \right],
\end{eqnarray}
where $MP$ refers to the measured particle and the electrons from
$1$ to $n$ are near point $A$, the remaining ones are near $B$.
The indices $ex$ and $gr$ refer to excited and ground states for
the indicated electrons, which subsequently decay, emitting
photons from the indicated points, respectively.

With reference to this example, the authors argue then as follows:
since in the whole process few particles are involved and, in any
case, the displacements of the particles which are involved are
totally negligible with respect to $1/\sqrt{\alpha}$, the
reduction mechanism of QMSL cannot be effective in suppressing one
of the two states in equation (\ref{rsp}). The consideration of
the subsequent emission of photons does not change the situation
since, on one hand, the spontaneous localizations of QMSL do not
affect photons and, on the other hand, the photon wavefunctions
originating from $A$ and $B$ immediately overlap almost
completely. So, in spite of the fact that, corresponding to the
impinging of the particle on the screen having occurred at $A$ or
at $B$, there is a luminous record in different places which can
be perceived by a naked human eye, QMSL does not imply that the
suppression of one of the two states has occurred. Therefore QMSL
does not satisfy the basic requirement 1). To put it in different
words: while everybody would agree that the measurement has been
completed after the emission of the photons from the fluorescent
screen and before any observer actually looks at the light spot,
QMSL is not able to attribute a definite outcome to the
measurement.

The argument of the authors reduced to its essence consists in
stressing that not all conceivable processes which everybody would
agree to call measurement processes involve macroscopic
displacements of a macroscopic number of particles, and that,
under such conditions, dynamical reduction models cannot yield an
actual dynamical reduction of the wavepacket. With reference to
this pertinent remark and to the above example, we consider it
important to distinguish, for conceptual clarity, two mechanisms
which could make QMSL ineffective in inducing the reduction:
\begin{enumerate}
\item That the number of particles involved in the process (e.g.,
in the set--up considered, the electrons which have to be excited
in order that the emitted photons be perceivable by the unaided
eye of a human experimenter) be very small.

\item That, even if many particles are involved, the changes in
their states which occur as a consequence of their interactions
with the measured microsystem (in the example considered the
transitions from the ground to an excited state) involve position
changes which are much smaller than the localization distance
$1/\sqrt{\alpha}$ of the spontaneous process. The above two
alternatives will be discussed in the next subsection.
\end{enumerate}

\subsection{Measurements and outcomes} \label{sec142}

Let us begin by stating that we agree perfectly with the
conclusions drawn in \cite{avp} about the specific situation
discussed in it. Due to the extremely low rate of the spontaneous
localizations and the large value (on the atomic scale) of the
localization distance of QMSL, any superposition of states in
which only few particles are in appreciably different spatial
positions or in which many particles are only slightly displaced
is not dynamically suppressed by the theory.

We then agree that in the considered example the linear
superposition will persist for long times. Is this a feature which
shows the failure of the model in accounting for measurement
processes? We think that one should be careful in drawing such a
conclusion.

As a starting point it is useful to recall the lucid position of
J. Bell about such kind of problems. In \cite{bell84} he stated:
\begin{quotation}
The usual approach, centered on the notion of ``observable",
divides the world somehow into parts: ``system" and ``apparatus".
The ``apparatus" interacts from time to time with the ``system",
``measuring" ``observables"... There is nothing in the mathematics
to tell what is ``system" and what is ``apparatus", nothing to
tell, which natural processes have the special status of
``measurements".
\end{quotation}
Furthermore, in \cite{bellns}, he stressed:
\begin{quotation}
Surely the big and the small should merge smoothly with one
another. And surely in fundamental physical theory this merging
should be described not just by vague words but by precise
mathematics.
\end{quotation}

These quotations appropriately point out that, in what the author
would have considered a serious theory, the very mathematics of
the theory and not ``vague words" should define what is a
measurement. In this respect it seems to us that in \cite{avp}
what has to be considered a measurement, what plays the role of an
apparatus, is still defined by terms that are to some extent
vague. For instance, in footnote 1, the authors, trying to make
precise what is a recording by an instrument, require that it
consists in a change of the measuring device which is {\it
macroscopic, irreversible, and visible to the unaided eye of a
human experimenter}.

It is just to try to be very precise about the meaning of such
terms that we have chosen to make at the end of the previous
subsection a clear distinction between the two crucial points
which are at the basis of the arguments of Albert and Vaidman.

Let us consider point 1. If the number of particles which are
involved in the process under consideration is really very small
(as it is legitimate to assume due to the sensitivity of the human
eye to light quanta --- the threshold for visual perception being
of about 6 photons), then the changes of the measuring device are
surely not macroscopic, even though they can be irreversible and
visible to the human unaided eye. Is it correct to pretend that
also in such a case the process has to be considered as a genuine
measurement which should have an outcome before anything else
happens?

We can raise some doubts about the correctness of taking this
attitude by considering the following gedanken experiment. Suppose
we have an atom which can be prepared in an excited state
$|\makebox{Ex}\rangle$ of spin $1/2$ and with the following
characteristics: the excited state decays with a quite long
lifetime, of the order of several seconds or longer, in a state
$|\makebox{SLC}\rangle$ which is the ancestor of a short lived
sequential decay chain with various steps, all lifetimes in the
sequence being extremely short (of the order of nanoseconds). We
prepare the state $|\makebox{Ex}\rangle$ in a spin state that is
an eigenstate of $\sigma_{x}$, and we perform a Stern--Gerlach
experiment devised to measure the $z$ component of the spin. The
long lifetime of the initial state then allows us to correlate the
two $\sigma_{z}$ components to two different positions of the
atom, which we will denote by $A$ and $B$, respectively, just as
in the example considered by Albert and Vaidman. After a time of
the order of the lifetime for the first transition, the atom
decays; and, going through the cascade process, it emits several
photons, let us say a number larger than the perception threshold.
There is no doubt that the situation is very similar to the one
considered in \cite{avp}: the sequential decay is an irreversible
process in the same sense in which the excitation and decay of few
fluorescent electrons is. Moreover, the emission from two
macroscopically distant regions of a number of photons which is
sufficient to be perceived by the naked eye of a human
experimenter allows a direct detection of whether the atom has
been found with spin up or with spin down. Could anybody pretend
that a spin measurement has been performed on the atom before any
observer would look at it? We think that this would be a wrong
request and that actually nobody would feel embarrassed in
considering the system to be in the linear superposition of the
two final states. So, there is no reason to require that an
acceptable theory of collapse should guarantee that this specific
``measurement" has an outcome at this stage.

Let us now consider possibility 2. We think that it is perfectly
plausible to imagine interaction mechanisms between a microscopic
and a macroscopic system such that, on one hand, many constituents
of the macroscopic one can change their quantum state in a way
which depends on the state of the triggering system, and, on the
other hand, that such changes do not involve displacements of the
particles considered of amounts larger than $10^{-5}$ cm. The
simplest example which one can think of is just the one in which
many particles change their energy by amounts summing up to a
macroscopic energy change without appreciably changing their
position (on the relevant scale). However, the important question
is: can this change be really macroscopic and have no other
effect? In particular, would it not induce, at least indirectly,
macroscopic changes in the positions of a macroscopic number of
particles? It is not easy to exclude this, since one cannot
ignore, e.g., the interactions of the system with its
environment\footnote{It has to be stressed that here our resorting
to the environment is conceptually radically different from the
procedure followed in approaches like the one considered by Joos
and Zeh \cite{jz} to solve the measurement problems (see section
\ref{sec33}). There the environment plays the role of a system
whose correlations with the measured system and measuring
apparatus cannot be detected, so that, to make physical
predictions, one must take the partial trace on the environment
variables. As already discussed, the reduction does not then occur
at the individual level. Here the changes of the environment we
are interested in are those in which a macroscopic number of
particles are displaced. When they occur the universal dynamical
mechanism of QMSL actually induces reductions at the individual
level.}. In such a case, at least for reasons of thermal
equilibrium, changes of this sort must be recognized to occur.
Obviously one could object that the body could very well be kept
almost perfectly isolated. But how would then one check that it
has changed its state? The final answer would be: by direct
detection with an appropriate apparatus or by direct observation
(if this is possible, since, e.g., the body radiates) by a
conscious observer.

We can now raise our basic question: what would make it
unacceptable to consider that such a body remains in a
superposition of the two macroscopically different states under
consideration before it is detected and the result is recorded?

The above statement can obviously be meaningful only if one makes
precise the meaning of the expressions ``detected" and
``recorded". We completely agree on this, but we stress once more
that it is just the theory itself which must give a precise
meaning and make absolutely definite the significance of these
expressions. QMSL does this: the body is detected and the result
is recorded at the moment into which it evolves into, or, via its
interactions with other systems, it induces the occurrence of a
linear superposition of states containing a macroscopic number of
particles which are differently located in space by an amount of
the order of $10^{-5}$ cm or larger. At that moment it has to
choose one of its possible ways! And it is the dynamics of the
theory that guarantees that this will happen.

One may like or dislike this picture, but one has to recognize
that it is consistent and represents a step towards a possible
clarification of some of the puzzles of the quantum world.
Obviously the above position is tenable only provided one can
guarantee a fundamental fact: since our perceptions are definite,
at least in connection with any act of perception, a situation
must occur such that the reduction mechanism, whose taking place
is precisely defined by the theory, becomes effective. The next
subsection is devoted to making plausible that this is the case by
a discussion of visual perception.

\subsection{Reduction within the nervous system} \label{sec143}

As stressed in the previous subsection, the very possibility of
considering QMSL as yielding a unified description of all physical
phenomena rests on the fact that one can show that the physical
processes occurring in sentient beings, leading to definite
perceptions, involve a displacement of a sufficient number of
particles over appropriate distances to allow the reduction to
take place within the perception time.

We will describe now, in its essential aspects, the visual
perception process, to make plausible that the above situation
actually occurs. The reason to choose to discuss explicitly the
case of the visual perception should be obvious. First, vision is
directly involved in the example discussed in the previous
subsection. Secondly, the visual channel is presently the most
studied of all sensory channels and therefore the best understood.
Finally, other sensory modalities show similar characteristics,
both in the distal refined transduction mechanisms and in the
excitation pattern reaching the central nervous system.

We will divide our description into the three main cascades of
events that take place following the absorption of one photon in a
photoreceptor cell of the retina:
\begin{enumerate}
\item multiplicative chain in the photoreceptor cell,
\item transmission of the electrical signals along the fibers of
the optic nerve,
\item excitation of neurons in the cortical visual area.
\end{enumerate}
All these events are necessary for seeing. We will make rough
estimates of the number of particles moving in these processes.

An observation is relevant: In general, sensory cells have no
threshold in responding to external stimuli. In the case of
photoreceptors, the absorption of a photon by a pigment molecule
(retinene) determines a isomeric transition (cis--trans). But the
same transition can be also determined for example by thermal
excitation. The detection of external stimuli is therefore based
on statistical decision about the signal to noise ratio
\cite{bia}. For this reason, the psycho--physical threshold is set
at the average level of six absorbed photons.

We now analyze the multiplicative chain in the photoreceptor cell.
The excited state $R$ of the retinene, inside the rhodopsin (a
protein molecule with molecular weight $M_{w} = 39,000$) has a
lifetime sufficiently long on the disks of the rod to activate
about $100$ transducin molecules $T$, present in the interdisk
space (about of the same $M_{w}$). This is used to release the
inhibition of the enzyme PDE (phosphodiesterase, $M_{w} =
180,000$), able to hydrolyze very rapidly about $1,000$ cyclic
nucleotide c--GMP (guanosinmonophosphate). All together we obtain
a multiplication in molecule number of about $100,000$ for each
absorbed photon. The c--GMP molecules (cooperatively 3 of them)
normally keep open the channels of the plasma membrane, enclosing
the vertebrate rod outer segment. Their hydrolysis determines the
closure of the channels and the consequent hyperpolarization of
the cell and starts the electrical signal that will be transmitted
to a chain of neurons of the retina \cite{bpc}. The electric
current change through the membrane is of the order of a pA: one
can estimate that the ions (mostly Na) affected in their movement
are about $10^{8}$.

But we are interested in following the chain of events further on.
The hyperpolarization of the inner segment of the rod is followed
by the release of a chemical transmitter, kept in vesicles near
the membrane in such a way to diffuse rapidly (about 1 msec) in
the synaptic gap towards other neurons present in the retina: the
horizontal cells, the amacrine and bipolar cells. We will not try
to estimate how much these interactions will increase the number
of particles involved and their movement, keeping in mind that
those interactions develop not only in the forward direction but
also laterally and in the backward direction. To be very
conservative, we can indicate a factor 10 in the number of
molecules. The signals transmitted, both electrical or chemical,
in these stages are graded, modulated in linear and nonlinear
fashion. The last step is acted on the final output cells of the
retina, the ganglion cells. These cells, whose long axons (about
one million) form the optic nerve, send the electric impulses
(spikes) towards the CNS (central nervous system). The propagation
is regenerative and saltatory, going along a series of Ranvier
nodes, where the membrane of the nerve fiber is free of the
insulating sheets of the Schwann cells.

The above figures about the displaced particles are not sufficient
to guarantee by themselves that the reduction has already
occurred. However, we stress that all estimates we have made are
very conservative and correspond to having chosen for the numbers
of particles their minimal values. The high complexity of the
system and its connectivity would justify the introduction of
various amplifying factors which would remarkably raise the above
values. These considerations should have made plausible that the
number of particles and their displacements have reached the level
which is sufficient to make effective the reduction mechanism.
This makes correct the conviction of everybody working and
analyzing what is going on in the nervous system that, at this
stage, the quantum aspects of the phenomenology have already come
down to the level for which the classical description is adequate.

To be more specific and remove any possible doubt about the
definiteness of conscious perception, we show now, examining in
more details the signal propagation, that even if one limits his
considerations to this last step of the visual perception process,
the estimates of the number and of the displacements of the
particles which are involved would lead to the same conclusion.
For this purpose, we simply analyze what happens to a neuron
involved in the transmission of such a signal. The neuron has a
main cell body with a nucleus and a long tube, the axon, extending
from the cell body and having at its end a variety of hair--like
structures connecting it to other nerve cells. As already
remarked, in the case of the optic nerve, the axon is wrapped in a
series of small sheaths of myelin, an insulating substance,
separated at intervals of the order of one millimeter by nodes
referred to as Ranvier's nodes (see figure 3).
\begin{center}
\begin{picture}(350,100)(0,0)
\put(0,0){\line(1,0){350}} \put(0,100){\line(1,0){350}}
\put(0,0){\line(0,1){100}} \put(350,0){\line(0,1){100}}
\thicklines \put(60,50){\circle*{20}}
\qbezier(70,80)(70,52)(90,52) \qbezier(35,70)(58,55)(70,80)
\qbezier(25,45)(50,50)(35,70) \qbezier(40,20)(50,50)(25,45)
\qbezier(72,22)(55,45)(40,20) \qbezier(90,48)(75,48)(72,22)
\put(90,48){\line(1,0){200}} \put(90,52){\line(1,0){200}}
\put(290,48){\line(0,1){4}}
\put(140,52){\oval(30,8)[t]} \put(140,48){\oval(30,8)[b]}
\put(175,52){\oval(30,8)[t]} \put(175,48){\oval(30,8)[b]}
\put(210,52){\oval(30,8)[t]} \put(210,48){\oval(30,8)[b]}
\qbezier(290,52)(300,60)(320,72) \qbezier(290,51)(300,55)(320,62)
\qbezier(290,50)(300,50)(320,53) \qbezier(290,49)(300,47)(320,40)
\qbezier(290,48)(300,45)(320,28) \thinlines
\put(50,80){\vector(1,-3){7}} \put(22,82){\footnotesize Nucleus}
\put(90,65){\vector(-1,-1){16}} \put(91,65){\footnotesize Cell
body} \put(175,68){\vector(0,-1){10}} \put(150,71){\footnotesize
Myelin sheath} \put(192.5,31){\vector(0,1){15}}
\put(163,23){\footnotesize Node of Ranvier}
\put(260,65){\vector(0,-1){10}} \put(250,67){\footnotesize Axon}
\put(84,10){\vector(-1,1){10}} \put(85,10){\footnotesize Dendrite}
\end{picture}

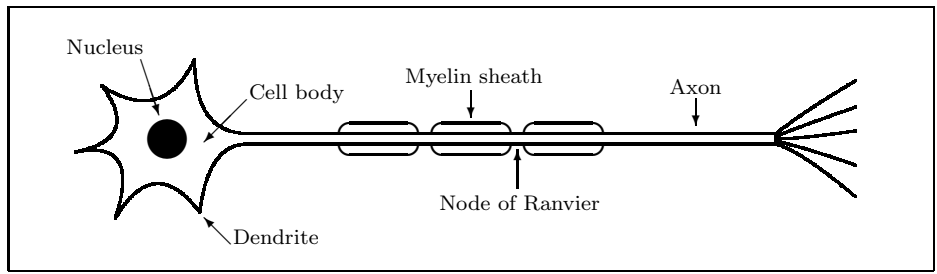
\vspace{0.2cm} \footnotesize \parbox{4.8in}{Figure 3: Drawing of a
nerve cell.} \normalsize
\end{center} \vspace{0.5cm}
To make the following discussion clearer, it is useful to mention
that axon's diameter is of the order of $10^{-4}$ cm, the myelin
sheath thickness is greater than $10^{-5}$ cm, and the membrane
thickness at Ranvier's node is of the order of $10^{-6}$ cm. The
transmission mechanism goes as follows: when an impulse is
generated, at Ranvier's nodes, ions channels open in the membrane
of the axon, through which Na$^{+}$ and K$^{+}$ ions flow. Thus,
in the course of the depolarization of the membrane, circular
currents, connecting two nearby Ranvier's nodes and closing
through the external conducting medium, arise (see figure 4).
\begin{center}
\begin{picture}(240,100)(0,0)
\put(0,0){\line(1,0){240}} \put(0,100){\line(1,0){240}}
\put(0,0){\line(0,1){100}} \put(240,0){\line(0,1){100}}
\thicklines \put(25,55){\line(1,0){185}}
\put(25,55){\oval(85,10)[tr]} \put(117.5,55){\oval(85,10)[t]}
\put(210,55){\oval(85,10)[tl]} \put(25,25){\line(1,0){185}}
\put(25,25){\oval(85,10)[br]} \put(117.5,25){\oval(85,10)[b]}
\put(210,25){\oval(85,10)[bl]}
\thinlines \qbezier(75,65)(117.5,85)(160,65)
\put(75,64.5){\vector(-2,-1){1}} \qbezier(75,48)(117.5,28)(160,48)
\put(160,48){\vector(2,1){1}}
\put(65,63){\tiny $R$} \put(165,63){\tiny $R$} \put(68.5,57){\tiny
$-$} \put(161,57){\tiny $+$} \put(213,56.5){\small $m$}
\put(68.5,48){\tiny $+$} \put(161,48){\tiny $-$}
\put(214,39){\small $a$} \put(68.5,28){\tiny $+$}
\put(161,28){\tiny $-$} \put(68.5,20){\tiny $-$}
\put(161,20){\tiny $+$} \put(213,21.5){\small $m$}
\end{picture}

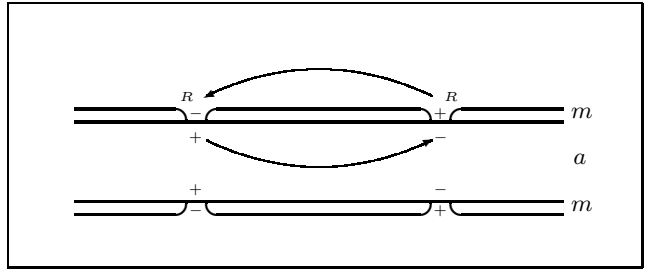
\vspace{0.2cm} \footnotesize \parbox{3.3in}{Figure 4: The scheme
of impulse transmission. $R$ = Ranvier node. $m$ = myelin. $a$  =
axoplasm.} \normalsize
\end{center} \vspace{0.5cm}
Important facts to be taken into account are the following: during
one impulse $\simeq 6\cdot 10^{6}$ sodium ions pass a Ranvier's
node; the time necessary to restore the resting potential in the
considered region of the axon is of the order of $10^{-3}$ sec;
finally the internal ion current flows near the axon membrane.
With these premises, we can try to evaluate the efficiency of the
reduction mechanism. To do this, in place of QMSL, we will make
use of CSL: in such a theory, as we have seen, the decoupling rate
for superpositions of states involving differently located
particles is approximately given by:
\begin{equation} \label{aqhil}
e^{\displaystyle -\lambda\,t\sum_{i}(n_{i} - {n'}_{i})^{2}},
\end{equation}
where $\lambda = 10^{-16}$ sec$^{-1}$, and $n_{i}$, ${n'}_{i}$ are
the numbers of particles present in the cell (of volume $10^{-15}$
cm$^{3}$) labeled by the index $i$, in the two states
$|\psi\rangle$ and $|\psi'\rangle$ whose superposition we are
considering, respectively. Obviously $|\psi\rangle$ is associated
with the occurrence of the impulse transmission, while
$|\psi'\rangle$ corresponds to no transmission.

In our case, we consider all cells surrounding the internal
membrane of the axon between two Ranvier's nodes. There are
$10^{5}$ such cells. Taking into account that the atomic number of
Na$^{+}$ is 11, so that one ion contains more than 30 particles
and that we have disregarded the K$^{+}$ ions, we have to fit
about $10^{9}$ particles in the $10^{5}$ cells. We then have that,
when the signal is transmitted, for about $10^{5}$ cells $|n_{i} -
{n'}_{i}|$ turns out to be $\sim 10^{4}$. These values, when
substituted in equation (\ref{aqhil}), taking into account that
the impulse lasts at least for $10^{-3}$ sec, give an exponent
$10^{-6}$.

To complete the description, we have to consider two further
steps, each involving a multiplicative factor. The optic fibers
arrive at lateral geniculate bodies (LGB), where they branch out
making contacts with many cells. The neurons in these bodies send
their axons again to the visual striate cortex, an essential step
for seeing. In both these stations a conservative estimate of the
multiplicative effect \cite{hub} can be a minimum of $10^{2}$,
therefore we can obtain at least a factor of $10^{4}$.

In our calculation we have completely disregarded further
displacements of particles induced by the macroscopic current
around the axon. In one of his books \cite{penenm}, R. Penrose has
considered an analogous problem, to reach the Planck mass level,
which in his approach would mark the setting up of the reduction.
To get the desired result, he needs to assume that such further
displacements imply a further amplification of a factor $10^{8}$,
a figure that he considers reasonable. We obviously need a much
smaller factor, e.g., one of the order of $10^{4}$ would be
largely sufficient. So, we have made perfectly plausible that the
number of displaced particles and the displacements which are
involved imply the dynamical reduction mechanism of QMSL (or
better of CSL) becomes fully effective in suppressing the
superposition of the two states (nervous signal)--(no signal)
before any act of conscious perception occurs. The fundamental
requirement which has to be imposed on the model to account for
our definite perceptions is therefore certainly satisfied.

\subsection{Does reduction require observers?} \label{sec144}

It has to be firmly stressed that the fact that we have felt the
necessity of performing an analysis of the visual perception
mechanism is not intended as the acceptance, on our part, of the
point of view that consciousness has a specific role in the
reduction process. QMSL states that reduction will take place
whenever the above indicated precise conditions for the reduction
mechanism to be effective occur. So, if in place of a human being
we put a spark chamber or a macroscopic pointer which is
displaced, or a device producing ink spots on a computer output,
reduction will take place. In the context of the previous
subsection, the human nervous system is simply something which has
the same function as one of these devices, if no such device
interacts with the system before the human observer does.

With respect to Albert and Vaidman's example, our attitude should
then be quite clear: the state is not reduced up to the moment in
which the dynamical evolution leads to the occurrence of a
superposition of states differing in the positions of a
macroscopic number of particles. Whether this happens because one
puts in front of the fluorescent screen a spark chamber or the
nervous system of a human observer is totally irrelevant.

With reference to the argument we have just developed, we think it
is appropriate to point out that some sentences of reference
\cite{ap} could turn out to be misleading. For instance, at the
beginning of section IV, the author, after having recalled the
situation of \cite{avp}, makes the following general statement:
``If we want to stick with QMSL, then we would have to insist ...
that no measurement is absolutely over, no measurement absolutely
requires an outcome, until there is a sentient observer who is
actually aware of that outcome''. As we have proven, this is not
true in general but only in very peculiar cases. Actually,
according to QMSL, not only practically all measuring experiments
of our laboratories but also all those measurement--like processes
which, as J. Bell \cite{bellam} has stressed, ``we are obliged to
admit ... are going on more or less all the time, more or less
everywhere", have definite ``outcomes" even in absence of any
sentient being.

\subsection[General comments on the impossibility proof of QMSL]{Some
general comments about an alleged general impossibility proof of
dynamical reduction models} \label{sec145}

We come now to examine further criticisms to QMSL, which have been
put forward in sections IV and V of \cite{ap} (see also
\cite{al5}). The whole argument can be briefly summarized as
follows: the author contemplates the possibility of the existence
of unusual sentient beings whose peculiar features consist in the
fact that their conscious beliefs are assumed to be $100\%$
correlated with the position of a single particle. To be more
precise, the author considers a science--fiction character, John,
who, due to a peculiar surgical implant in his brain, can perform
a measurement of a spin component, e.g. of an electron, by making
it cross a tunnel in his brain and interact with a microscopic
particle $P$ in the implanted device. John is such that his
``ready--to--perceive state" $|P_{0}\rangle$ corresponds to the
particle $P$ being in a certain position, while the perceptions
``spin up" and ``spin down" are uniquely correlated to $P$ being
up ($|P_{+}\rangle$) or down ($|P_{-}\rangle$) with respect to the
initial position.

It is stipulated:
\begin{enumerate}
\item that no other change in John's body occurs as a consequence
of the process (i.e., no other of John's atoms or electrons beside
$P$ is involved in this process);
\item that, nevertheless, John is {\it consciously aware as vividly
and as completely as he is ... or has ever been aware of anything
in his life} of what the value of the spin component is.
\end{enumerate}

Given these premises, provided one takes the attitude to believe
what John says, the argument is quite straightforward: if one
wants to account for the definite perceptions of John by a
dynamical reduction mechanism, one must consider a process which
suppresses the linear superpositions of position states of a
single particle, and this would surely imply an easily detectable
violation of the predictions of quantum mechanics for
microsystems. So, no physically acceptable reduction mechanism
whatsoever could do the desired job of making definite John's
perceptions.

The first obvious remark is that the very consideration of such a
being would upset any neurophysiologist. In fact, all we know
about consciousness points undoubtedly towards the necessity of
very complex systems like the human brain to support it. The
features of such systems which are commonly considered essential
for consciousness are, on the one hand, the fact that they contain
a gigantic number of transmitting units (like neurons) and, on the
other hand, that these units are wired up in an extremely
complicated network. For these reasons, one could disregard
objections which require as an essential ingredient the
consideration of sentient beings whose different beliefs are
correlated to states that are only microscopically different.

However, we can raise the question: even if one would accept that
``in principle" the possibility of the existence of ``microscopic"
sentient beings (whatever precise meaning one could attribute to
the word sentient) cannot be excluded, should one consider as
cogent the criticism put forward in the two last sections of
reference \cite{ap}? Answering the above question requires a
detailed analysis. First of all, it turns out to be appropriate to
recall the precise and relevant terms of the debate about the
foundations of quantum mechanics and the reasons which make
programs like the dynamical reduction one interesting. The
situation can be summarized as follows. The conceptual structure
of the quantum scheme does not allow, as it stands, the
elaboration of an articulate, systematic and coherent picture of
natural phenomena, i.e., what A. Shimony \cite{shipcqf} calls a
``philosophical world view". It is the desire to build such a
coherent picture which has led to the consideration of various
``alternative theories", which give for all practical purposes
(FAPP) the same predictions as quantum mechanics plus the
wavepacket reduction postulate. Each of them implies a certain
world view, and each of them is based on different specific
assumptions.

Let us start by considering two of the more appealing
``alternative theories" that have been proposed:
\begin{enumerate}
\item The dynamical reduction models;
\item The de Broglie--Bohm pilot wave theory, or, more generally,
the hidden--variable theories.
\end{enumerate}

It is useful to remark that both these theories aim at giving an
account of physical processes without having to make reference to
observers. The other alternative which deserves to be discussed,
and which Albert seems to prefer, is the many--minds theory. We
will consider it below.

Due to the fact that present--day technology does not allow us to
verify whether, at the macroscopic level and under appropriate
circumstances, wavepacket reduction occurs, there is no
experimental objective criterion to choose, e.g., between the two
above alternatives 1) and 2). Such a choice must then be based on
the identification of the assumptions which are essential for them
as well as on the consideration of possible difficulties they meet
or unpleasant features they exhibit. As concerns the assumptions,
we remark that, as appropriately stressed by J. Bell
\cite{bellam}, the characteristic element distinguishing program
1) from 2) consists in the fact that 1) assumes that the
wavefunction gives a complete description of natural phenomena,
while 2) accepts from the very beginning that additional variables
are necessary. Concerning what one could consider the limitations
of the corresponding programs, we mention that 1) requires the
acceptance of the fact that Schr\"odinger's equation is not
exactly true and the introduction of ``ad hoc" parameters; 2)
requires us to accept contextuality (i.e., that a complete
specification of a state assigns definite truth values to a
proposition only relative to a specific context).

Albert \cite{ap}, instead of following the above outlined
procedure to choose among the ``alternative schemes," seems to
adopt as a crucial criterion for accepting or rejecting one of
them the fact that it can accommodate beings like John. Various
comments are appropriate: first, we do not view it as legitimate
to adopt such a criterion to judge a theory, particularly in view
of the fact that, as already remarked, the very idea that such
beings exist seems to conflict with all we know to be needed to
support consciousness. Secondly, we want to stress that the
author's request amounts to plainly postulating that program 1)
has to be rejected. In fact, the very hypothesis that John exists
and is reliable leads to the denial of the fundamental assumption
lying at the basis of the dynamical reduction program, i.e., that
the wavefunction gives a complete description of natural
phenomena, since the same overall statevector
\begin{equation}
\frac{1}{\sqrt{2}}\,\left[ |z_{+}\rangle\, |P_{+}\rangle \; + \;
|z_{-}\rangle\, |P_{-}\rangle \right]
\end{equation}
would be associated in some cases with a situation in which John
believes that ``the spin is up" and in others to one in which John
believes that ``the spin is down\footnote{We do not want to be
misunderstood: we are not assuming that it is possible to give a
purely physical explanation of mental events, but we are simply
imposing the obvious condition that different beliefs must be
related to statevectors which somehow differ.}''. Finally, it is
worth noticing that even scheme 2) cannot accommodate, without
encountering serious conceptual difficulties concerning the
``status" that can be attributed to their perceptions, beings such
as John. The reason for this lies just in the fact that the
particle $P$ in John's device is a microscopic system. As a
consequence, the linear superpositions of states corresponding to
different positions of $P$ are not, even FAPP, equivalent to a
statistical mixture of states corresponding to $P$ having a
definite position. Then, the evolution leading the particle from
its ``ready state" to the final state is, not only in principle
but practically, reversible (so that one can ``undo" what has been
done). Moreover, observables which do not commute with the
position of $P$ must be admitted to be actually measurable,
contrary to what would happen if $P$ would be a macroscopic
system.

To illustrate the obvious fact that the process leading to John 's
definite perceptions about the spin value can be reversed, let us
assume, for simplicity, that the displacement of $P$ induced by
its interaction with the spin of the electron occurs as a
consequence of a $\sigma_{z} p_{z}$ coupling ($p_{z}$ being the
$z$ component of the linear momentum of $P$) lasting for the time
interval $\delta t$ taken by the electron to go through John's
implanted device. Then it follows that, if $P$ is in the up (down)
position, feeding into the device another electron with spin down
(up), will cause $P$ to return to its ready state position. We can
then consider the following experiment. We have two electrons in
the singlet state, which travel towards John's device with a
certain common velocity and which reach the device at times
differing by an amount $\Delta t$ larger than $\delta t$. Then
when the first electron crosses John's device, we have the
statevector evolution:
\begin{eqnarray} \label{sls}
\lefteqn{ \frac{1}{\sqrt{2}}\,\left[ |z_{2-}\rangle\,
|z_{1+}\rangle \; - \; |z_{2+}\rangle\, |z_{1-}\rangle \right]\,
|P_{0}\rangle \;
\longrightarrow \qquad\qquad} \nonumber \\
& \longrightarrow & \frac{1}{\sqrt{2}}\,\left[ |z_{2-}\rangle\,
|z_{1+}\rangle \, |P_{+}\rangle \; - \; |z_{2+}\rangle\,
|z_{1-}\rangle\, |P_{-}\rangle \right],
\end{eqnarray}
where $|P_{0}\rangle$, $|P_{+}\rangle$, $|P_{-}\rangle$ represent
the ready, up, and down states of $P$, respectively, and where we
have disregarded the spatial variables of the electrons. When the
state on the right hand side of equation (\ref{sls}) obtains, due
to the fact that within the de Broglie--Bohm theory all particles
at all times have definite positions, $P$ is either up or down
and, as a consequence, John can ``legimately" be said to have a
definite belief about the $z$ component of the spin of the first
electron. However, it has to be remarked that in the de
Broglie--Bohm theory the spin being a contextual variable is
not an observable quantity. As a consequence, in the case considered, the
``outcome" of John 's measurement cannot be related to any objective
property of the spin of the electron before the measurement (note that
this would hold also in the case in which $P$ would be a macroscopic
measuring device). Subsequently the second electron crosses John
's device, bringing $P$ back to its ready position $P_{0}$. As a
consequence, due to the assumption that no other change but the
displacement of $P$ has occurred in the first process, the state
after the passage of the second electron turns out to be again
exactly the initial state, i.e., the one on the left hand side of
equation (\ref{sls}). It is important to stress that in such a
state the spin variables are not correlated to the position of any
particle and that the process just described (i.e., the
interaction of the second electron with $P$) does not involve in
any way the first electron on which the ``measurement" of the spin
had been previously performed.

Let us now suppose that another observer performs a measurement of
the same spin component of the first electron. It is obvious that
also this second measurement turns out to be contextual, and
therefore its outcome is not related to the outcome of the
``measurement" performed by John. This puts into evidence that the
previous belief by John about the outcome of the spin
``measurement" he has performed cannot even be related to an
objective property of the electron after the ``measurement"; in
particular it cannot be used to foresee the outcomes of subsequent
measurements, i.e., in general it has no predictive value. Of course, the
situation would have been completely different if the particle $P$
in John 's device were a macroscopic ``pointer". In that case,
when the state on the right hand side of equation (\ref{sls})
obtains, also in the de Broglie--Bohm theory one can assert that
things go as if wavepacket reduction had occurred, the reason for
this being that it is practically impossible to completely reverse
the measuring process. In such a case, therefore, the belief of
the observer about the spin of the electron corresponds to an
``effective determinateness" of the spin of the electron after the
measurement, and all observers performing subsequent measurements
of the same component of the spin find the same result which has
been found by the first (macroscopic) observer\footnote{We want to
stress that this holds independently of the fact that at the
instants of the subsequent measurements the brain of the first
observer be still active and healthy or not.}. In our opinion,
this analysis puts into evidence that, even though in the de
Broglie--Bohm scheme John can be considered to have definite
perceptions as he claims, no validity, in the spirit of the
theory, can be attributed to them.

We come now to the consideration of the many--minds approach. As
already discussed in section \ref{sec3}, what a many--minds theory
takes physics to be ultimately about is what sentient observers
think. As a consequence, the adoption of a many--minds position
implies a dualistic attitude: there are a physical universe,
described by the statevector obeying an exact Schr\"odinger's
equation, and a mental universe made up of the impressions of
sentient observers (an expression which, in this context is
synonymous with observers with minds). In such a theory even the
beliefs of human observers are unavoidably not always ``true"
(i.e., reflecting objective properties of the physical universe)
but have a sort of ``effective validity" (in the sense that the
future evolution of the mental states of such beings will proceed,
in general, as if their beliefs were true). Obviously, within such
a theory, the possibility to attribute effective validity to a
belief requires a reliable memory recording the belief itself.

It seems to us that the fundamental criterion to judge whether the
many--minds theory can accommodate other kinds of sentient beings
has to be related to its being able to guarantee to their beliefs
the ``effective validity" which is required to characterize the
beliefs of human beings. Since, as already remarked, the
``effective validity" criterion requires some sort of memory, one
is compelled to choose between the following alternatives: either
he attributes to John a macroscopic memory, but in this case the
whole criticism to the dynamical reduction program breaks down;
or, alternatively, he assumes that not only the formation of
beliefs but also the memory storing process involves only
microscopic changes, e.g., the displacement of $P$, and nothing
else. In such a case, for obvious reasons (see the preceding
discussion and/or take into account that wavepackets unavoidably
spread), John's memory has not the level of reliability which is
necessary to allow the attribution of an effective validity to his
(ephemeral) beliefs.

Concluding, the above analysis has made plausible that even on the
basis of quantum mechanical considerations one is led to recognize
that consciousness and memory must be macroscopically supported in
order to exhibit those features of reliability and/or effective
validity that are necessary to make them meaningful and which are
characteristic of human consciousness and memory; thus, both QMSL
and CSL allow us to ``close the circle''.

\part{Relativistic Dynamical Reduction Models}

\section{White noise models: general framework and examples}
\label{sec9}

J. Bell \cite{bells}, in explicating QMSL, immediately identified
two aspects of it which required further investigations:
\begin{enumerate}
\item The model does not respect the symmetry requirements for systems
of identical particles.
\item The introduction of the localizations assigns a special role to
position and requires a smearing on space, which makes it quite
problematic to find a relativistic generalization of it.
\end{enumerate}
The first difficulty has been overcome by CSL, in which the sudden
localizations of QMSL have been replaced by a continuous
stochastic evolution of the statevector. Steps toward a solution
of the second problem have been made with the introduction of
relativistic CSL models \cite{rel,rel2,p62,rel3,rel4}, which are the
subject of this fourth part of the report.

In the present section we consider the general problem of
describing relativistic dynamical reductions in terms of
white--noise stochastic differential equations. In subsection
\ref{sec91} we analyze the issue of stochastic Galileian
invariance within non relativistic QMSL and CSL, which guarantees
that all the physical predictions of the above models are
invariant under the Galilei group of transformations. We will see
that stochastic invariance puts very severe limitations on the
type of stochastic processes which can be used to describe
spontaneous collapses.

In subsection \ref{sec92} we set up a relativistic (and
stochastically invariant under Lorentz transformation) model for
decoherence (i.e. for ensemble or von Neumann reductions) which is
useful as a first step towards the formulation of relativistic
dynamical reduction models: these will be considered in subsection
\ref{sec93}.

In subsections \ref{sec94} and \ref{sec95} we analyze the
invariance and reduction properties of such models, respectively.
In the final subsection, we discuss a specific relativistic model
of spontaneous collapse which has all the desired properties of
QMSL and CSL, except that it induces an infinite increase of
energy per unit time and unit volume on physical system. This
model, then, is not fully consistent; actually, this is a feature
of all models in which quantum fields are locally coupled to white
noises.

\subsection{Stochastic Galileian invariance of QMSL and CSL}
\label{sec91}

To bring out some concepts which will be useful in the following
subsections, it is appropriate to consider the transformation and
the invariance properties of non relativistic CSL.

Let us start by limiting our considerations to the evolution
equation for the statistical operator and let us consider two
observers $O$ and $O'$ related by a transformation of the Galilei
group. We take the so--called passive point of view according to
which the two observers look at the same physical situation. For
simplicity, let us suppose that the transformation connecting $O$
and $O'$ is a translation in space of an amount ${\bf a}$ and a
translation in time of an amount $\tau$, so that
\begin{equation} \label{qdelg}
{\bf r}' \; = \; {\bf r} - {\bf a}, \qquad\qquad t'\; = \; t -
\tau
\end{equation}
Let the observer $O$ describe the physical situation at his
subjective time $t$ by the statistical operator $\rho(t)$.
Observer $O'$, at the same objective time, i.e., at his subjective
time $t' = t - \tau$, will describe the physical situation by the
statistical operator
\begin{equation}
\rho'(t')\quad = \quad U({\bf a})\, \rho(t)\, U^{\dagger}({\bf a})
\end{equation}
where $U({\bf a}) = e^{i\,{\bf P}\cdot {\bf a}}$ is the usual
unitary operator inducing the space translation. The dynamical
equation for the statistical operator for observer $O'$ is then
\begin{equation} \label{teosf}
\frac{d\, \rho'(t')}{dt'} \quad = \quad U({\bf a})\, \frac{d\,
\rho(t)}{dt}\, U^{\dagger}({\bf a})
\end{equation}
Substituting equation (\ref{efso}), describing the evolution of
the statistical operator for the observer $O$, into the right hand
side of equation (\ref{teosf}), one gets\footnote{Throughout this
section, we set $\hbar = 1$.}
\begin{eqnarray} \label{aqds}
\frac{d\, \rho'(t')}{dt'} & = & - i\, U({\bf a})\, [H(t),
\rho(t)]\, U^{\dagger}({\bf a}) \; + \; \gamma \sum_{i} U({\bf
a})\, A_{i} \rho(t) A_{i}\, U^{\dagger}({\bf a}) \nonumber
\\
& & -\ \frac{\gamma}{2}\, U({\bf a})\, \sum_{i} \{ A_{i}^{2},
\rho(t)\} \, U^{\dagger}({\bf a}).
\end{eqnarray}
If $H$ is invariant under space and time translations
\begin{equation}
H'(t') \; \equiv \; U({\bf a})\, H(t)\, U^{\dagger}({\bf a})\; =
\; H(t')
\end{equation}
and if, moreover
\begin{equation}
\sum_{i} U({\bf a})\, A_{i}\, U^{\dagger}({\bf a})\, X\, U({\bf
a})\, A_{i}\, U^{\dagger}({\bf a}) \; = \; \sum_{i} A_{i}\, X\,
A_{i}
\end{equation}
for any bounded operator $X$, then equation (\ref{aqds}) implies
\begin{equation} \label{gtlr}
\frac{d\, \rho'(t')}{dt'} \quad = \quad -i\, [H, \rho'(t')] \; +
\; \gamma \sum_{i} A_{i} \rho'(t') A_{i} \; - \; \frac{\gamma}{2}
\sum_{i} \left\{ A_{i}^{2}, \rho'(t') \right\}.
\end{equation}
i.e., the theory is invariant for space and time translations. If
the same holds for all transformations of the restricted Galilei
group, we have invariance for the transformations of this group.
QMSL and CSL actually possess this invariance property.

Nonetheless, it is important to stress that there is a difference
between equations of the type we are considering and the usual
case in which one has a purely Hamiltonian evolution, with respect
to the connection between invariance and representations of the
symmetry group. This key difference arises from the fact that
while in the standard case one can always relate the statistical
operators used by $O$ and $O'$ to describe the physical situation
at the same subjective time t, in the present case this cannot be
done in general, when one considers negative values of $t$ in
equation (\ref{qdelg}). In fact, let us suppose that $O$, at his
own time $t = 0$, is dealing with a physical system described by a
pure state $\rho(0) = |\psi\rangle \langle\psi|$. Since the
dynamical evolution transforms pure states into statistical
mixtures, there is no way for $O$ to prepare a physical situation
at his own time $\tau < 0$ (corresponding to $t'=0$ for $O'$) such
that it evolves into the pure state $\rho(0)$ at $t = 0$.
Correspondingly, there is no way for $O'$ to prepare at his own
time $t' = 0$ a statistical operator such that its evolved state
at his time $- \tau > 0$ is $|\psi\rangle
\langle\psi|$\footnote{Concerning this point, we call the
attention of the reader to D. Albert's investigations \cite{alrd2}
on the impact of dynamical reduction models on statistical
mechanics and thermodynamics. He points out that, precisely due to
their fundamentally irreversible nature, such models allow, in his
opinion, a much more satisfactory derivation of the thermodynamical
tendency to equilibrium.}.

However, if the active point of view is taken and $O'$, at his
time $t' = 0$, prepares the same state $\rho(0)$, and the above
stated invariance requirements are satisfied, then $O$ and $O'$
will observe the same dynamical evolution at the ensemble level
for the same (subjective) initial situation.

Coming now to the group theoretic point of view, since for the
above reasons the map $\Sigma_{t}[\rho(0)]$, for a pure state
$\rho(0)$ is not defined for negative $t$, one has to consider the
proper Galilei semigroup $G_{+}$, with only forward time
translations \cite{grwctm}. Any transformation $g \in G_{+}$ can
be expressed as a transformation of the subgroup $G_{0}$ of
$G_{+}$ which does not contain time translations, times a forward
time translation
\begin{equation}
g \in G_{+}\,:\,\, g = g_{\tau}g_{0}
\end{equation}
The map on the Banach space of the trace class operators
\begin{equation}
g : \rho \longrightarrow \rho_{g}, \qquad\quad \rho_{g} \, = \,
\Sigma_{\tau}[U(g_{0})\, \rho\, U^{\dagger}(g_{0})]
\end{equation}
where $U(g_{0})$ is the usual unitary representation of $G_{0}$
and $\Sigma_{\tau}$ is such that, for $\tau > 0$,
$\Sigma_{\tau}[\rho(t)] = \rho(t+\tau)$ is the solution of
equation (\ref{efso}), is then easily checked to yield a
representation of $G_{+}$.

Up to now we have discussed the invariance properties of dynamical
reduction models from the point of view of the statistical
operator. However, since we are mainly interested in the evolution
equation for the statevector, it is appropriate to discuss the
problem of the invariance also at this level. For simplicity, we
will limit ourselves to the discussion of space translations.

Let us start by considering the simpler linear Stratonovich
equation (yielding only ensemble reduction, i.e. decoherence)
\begin{equation} \label{sicdup}
i\, \frac{d}{dt}\, |\psi_{V}(t)\rangle \quad = \quad V({\bf r},
t)\, |\psi_{V}(t)\rangle
\end{equation}
If we denote by $O'$ an observer whose reference frame is
translated by an amount ${\bf a}$ with respect to the frame of
$O$, he will experience the potential
\begin{equation}
V'({\bf r}', t) \quad = \quad V({\bf r}' + {\bf a}, t)
\end{equation}
so that, for a particular realization of $V$, there is no
invariance.

However, since we are dealing with a fundamentally stochastic
theory, the invariance requirement has to be formulated in the
appropriate way. We will say that the theory is {\bf
stochastically invariant} under space translations if, for all
observers $O'$, translated by any ${\bf a}$ with respect to $O$,
the stochastic ensemble of potentials is the same and
characterized by the same probabilities. This is equivalent to
requiring that, if $V({\bf r}, t)$ is a possible sample function
for $O$, then $V({\bf r} - {\bf a}, t)$, for any ${\bf a}$, is
also a possible sample function for him, having the same
probability of occurrence of $V({\bf r}, t)$, i.e.,
\begin{equation} \label{lwbc}
P_{\makebox{\tiny Raw}}[V({\bf r},t)] \quad = \quad
P_{\makebox{\tiny Raw}}[V({\bf r} - {\bf a},t)]
\end{equation}
Note that this is automatically guaranteed by the form
(\ref{sch1d2}) for the mean value and covariance function of the
gaussian noise.

In the case of the model based on equation (\ref{sch1d})
describing Heisenberg reduction processes, a separate discussion
is needed, since the stochastic invariance requirement has to be
referred to the cooked probabilities which depend on the initial
statevector. Let us therefore consider two observers $O$ and $O'$
and suppose they prepare the same (subjective) state
$|\psi(0)\rangle$ at time $t = 0$. The probability density of
occurrence of the same (subjective) potential $V({\bf r}, t)$ is,
for the two observers,
\begin{eqnarray}
P_{\makebox{\tiny Cook}}^{O'}[V({\bf r}, t)] & = &
P_{\makebox{\tiny Raw}}^{O'}[V({\bf r}, t)]\,
\| |\psi_{V}^{O'}(t)\rangle \|^{2} \nonumber \\
& & \\
P_{\makebox{\tiny Cook}}^{O}[V({\bf r}, t)] & = &
P_{\makebox{\tiny Raw}}^{O}[V({\bf r}, t)]\, \|
|\psi_{V}^{O}(t)\rangle \|^{2} \nonumber
\end{eqnarray}
Since $|\psi_{V}^{O'}(t)\rangle$ and $|\psi_{V}^{O}(t)\rangle$ are
the solutions of equation (\ref{sch1d}) with the same (subjective)
potential and satisfy the same initial conditions, they coincide.
Moreover, due to equation (\ref{lwbc}), $P_{\makebox{\tiny
Raw}}^{O'}[V({\bf r}, t)] = P_{\makebox{\tiny Raw}}^{O}[V({\bf r},
t)]$, implying
\begin{equation}
P_{\makebox{\tiny Cook}}^{O'}[V({\bf r}, t)] \quad = \quad
P_{\makebox{\tiny Cook}}^{O}[V({\bf r}, t)]
\end{equation}
This guarantees the invariance from the active point of view,
i.e., the observers cannot, by making physical experiments in
their own frames, discover that they are displaced. They agree on
the statistical distributions of future outcomes.

\subsection{Quantum Field Theory with a Hermitian stochastic
coupling: the case of decoherence} \label{sec92}

In trying to set up the framework for a relativistic
generalization of reduction models, we adopt the quantum field
theoretic point of view. We remark that the analogue of the idea
of considering, within a non--relativistic framework, a stochastic
potential $V({\bf r}, t)$ consists in assuming that the Lagrangian
density for fields contains a stochastic interaction term. In this
subsection we consider a model analogous to the non--relativistic
ones discussed in subsection \ref{sec22}, yielding only ensemble,
not individual reductions.

Let us consider, in the context of quantum field theory, the
Lagrangian density
\begin{equation}
L(x) \quad = \quad L_{0}(x) \; + \; L_{I}(x) V(x)
\end{equation}
where $L_{0}$ and $L_{I}$ are Lorentz scalar functions of the
fields (for the moment we do not need to specify the fields we
deal with). We assume that $L_{I}$ does not depend on the
derivatives of the fields, and that $V(x)$ is a $c$--number
stochastic process which is a scalar with respect to
transformations of the restricted Poincar\`e group, i.e., that
under the change of variables $x' = \Lambda x + b$, it transforms
according to
\begin{equation}
V'(x') \quad = \quad V[\Lambda^{-1}(x'-b)].
\end{equation}
We will also assume that $V(x)$ is a Gaussian noise with zero mean
and, to get a relativistic stochastically invariant theory, that
its covariance is an invariant function
\begin{equation} \label{qredm}
\llangle V(x)\, V(x') \rrangle \quad = \quad A(x-x')
\end{equation}
with $A(\Lambda^{-1} x) = A(x)$.

As discussed in the previous subsection, stochastic invariance
requires different observers to agree on the unfolding of physical
processes. This, in turn, is guaranteed by the condition that the
family of all sample functions $V(x)$ and the probability density
of occurrence of the same (subjective) sample function be the same
for all observers. This is achieved by requiring that, for a
single observer,
\begin{equation} \label{aqsen}
P_{\makebox{\tiny Raw}}[V(x)] \quad = \quad P_{\makebox{\tiny
Raw}}[V(\Lambda(x+b))]
\end{equation}
We stress that property (\ref{aqsen}) holds automatically if the
covariance is a relativistically invariant function. In fact, from
\begin{equation} \label{djbgb}
P_{\makebox{\tiny Raw}}[V(x)] \quad = \quad \frac{1}{N}\,
e^{\displaystyle -\frac{1}{2}\, \int dx\, dx'\, V(x) \tilde
A(x-x') V(x')}
\end{equation}
[where we have denoted by $\tilde A(x-x')$ the function satisfying
$\int dx'' A(x-x'')\tilde A(x''-x') = \delta(x-x')$], one gets
immediately, using the scalar nature of $A$ and consequently of
$\tilde A$, that
\begin{equation} \label{cderfv}
P_{\makebox{\tiny Raw}}[V(\Lambda(x+b))] \quad = \quad
P_{\makebox{\tiny Raw}}[V(x)].
\end{equation}
The most natural generalization of the case discussed in the
previous subsection is obtained by assuming that $V(x)$ is a white
noise in all variables, i.e.,
\begin{equation} \label{ulilb}
\llangle V(x)\, V(x') \rrangle \quad = \quad A(x-x') \quad = \quad
\lambda \delta(x-x').
\end{equation}

We study, first of all, the physical consequences of the
stochastic coupling $L_{I}(x) V(x)$. In Schr\"odinger's picture we
have, for a given $V(x)$, the evolution equation:
\begin{equation} \label{msdli}
i\frac{d}{dt}\, |\psi_{V}(t)\rangle \quad = \quad \left[ H_{0} \;
- \; \int d^{3}x\, L_{I}({\bf x}, 0) V({\bf x}, t) \right]
|\psi_{V}(t)\rangle
\end{equation} where $H_{0}$ is the Hamiltonian corresponding to $L_{0}$.
Equation (\ref{msdli}) implies
\begin{equation} \label{cito}
|\psi_{V}(t)\rangle \quad = \quad T\, e^{\displaystyle -i H_{0}t
\; + \; i\int_{0}^{t}d\tau \int d^{3}x\, L_{i}({\bf x}, 0) V({\bf
x}, \tau)} |\psi(0)\rangle
\end{equation}
This equation shows how, for a given initial state
$|\psi(0)\rangle$, one gets an ensemble of states
$|\psi_{V}(t)\rangle$ at time $t$, according to the particular
realization of the stochastic process. The statistical ensemble
can then be described by the statistical operator obtained by
averaging over the sample functions. In the case under
consideration one gets a closed evolution equation for the
statistical operator. In fact, we observe that, due to the fact
that $L_{I}(x)$ does not depend on the derivatives of the fields
\begin{equation}
[L_{I}({\bf x}, 0), L_{I}({\bf x}', 0)] \quad = \quad 0.
\qquad\quad \forall\; {\bf x},\, {\bf x}'
\end{equation}
We then
have:
\begin{eqnarray} \label{mcl}
\rho(t+\epsilon) & = & \LLangle\left[ 1 - iH_{0}\epsilon + i
\int_{t}^{t+\epsilon} d\tau \int d^{3} x\, L_{I}({\bf x}, 0)
V({\bf x}, \tau) \right.\right.\right. \nonumber \\
& & \left. -\frac{1}{2}\int_{t}^{t+\epsilon} d\tau \, d\tau' \int
d^{3} x \, d^{3} x'\, L_{I}({\bf x}, 0) L_{I}({\bf x}', 0) V({\bf
x}, \tau) V({\bf x}', \tau') \right]
\nonumber \\
& & |\psi_{V}(t)\rangle\langle\psi_{V}(t)| \left[ 1 +
iH_{0}\epsilon - i \int_{t}^{t+\epsilon} d\tau \int d^{3} x\,
L_{I}({\bf x}, 0) V({\bf x}, \tau)  \right. \nonumber \\
& & \left.\left.\left. -\frac{1}{2}\int_{t}^{t+\epsilon} d\tau\,
d\tau' \int d^{3} x \, d^{3} x'\, L_{I}({\bf x}, 0) L_{I}({\bf
x}', 0) V({\bf x}, \tau) V({\bf x}', \tau') \right] \RRangle
\quad\qquad
\end{eqnarray}
We recall now the properties associated with a zero mean gaussian
probability distribution
\begin{eqnarray}
\llangle V({\bf x}_{1}, t_{1})\ldots V({\bf x}_{n}, t_{n})
\rrangle & = & 0 \quad\qquad\qquad\qquad\qquad\qquad\quad
\makebox{for $n$ odd} \nonumber \\
& & \\
\llangle V({\bf x}_{1}, t_{1})\ldots V({\bf x}_{n}, t_{n})
\rrangle & = & \sum_{\makebox{\tiny all pairs}} \llangle V({\bf
x}_{i}, t_{i}) V({\bf x}_{j}, t_{j})
\rrangle\cdot \nonumber \\
& & \qquad\;\; \cdot \llangle V({\bf x}_{k}, t_{k}) V({\bf x}_{l},
t_{l}) \rrangle \quad \makebox{for $n$ even.} \nonumber
\end{eqnarray}
>From (\ref{mcl}) we then have
\begin{equation} \label{erop}
\frac{d}{dt}\,\rho(t) \; = \; -i [H_{0}, \rho(t)] + \lambda \int
d^{3}x\, L_{I}({\bf x}, 0) \rho(t) L_{I}({\bf x}, 0) -
\frac{\lambda}{2} \int d^{3}x\, \left\{ L_{I}^{2}({\bf x}, 0),
\rho(t) \right\}.
\end{equation}
Note that the obtained equation is of the Lindblad type.

The non--Hamiltonian terms in equation (\ref{erop}) imply a
suppression of the off--diagonal elements of the statistical
operator in the basis of the common eigenstates of the commuting
operators $L_{I}({\bf x}, 0)$. Putting
\begin{equation}
L_{I}({\bf x}, 0)\, |\ldots \nu \ldots \rangle \quad = \quad
\nu({\bf x})\, |\ldots \nu \ldots \rangle
\end{equation}
one gets, when the Hamiltonian term in (\ref{erop}) is
disregarded,
\begin{equation}
\langle\ldots\nu\ldots|\rho(t)|\ldots\nu'\ldots\rangle\; = \;
e^{\displaystyle -\frac{\lambda}{2}\,t\int d^{3}x\, [\nu({\bf x})
- \nu'({\bf x})]^{2}}
\langle\ldots\nu\ldots|\rho(0)|\ldots\nu'\ldots\rangle
\end{equation}

As in the non--relativistic case, however, for a single
realization of the stochastic potential $V({\bf x}, t)$, the
statevector is not driven into one of the eigenmanifolds
characterized by a given $\nu({\bf x})$, since
$|\langle\ldots\nu\ldots|\psi_{V}(t) \rangle|^{2}$ does not change
with time. These considerations point out that, in order to have
Heisenberg reductions, one has to resort to a skew--hermitian
coupling with the noise.

Equation (\ref{erop}) for the statistical operator is not
manifestly covariant, even though, from the procedure which has
been followed to derive it, we know that the theory is
stochastically invariant. To obtain a manifestly covariant
description of the statistical operator evolution, we note that
the model presented above is obviously equivalent to the following
scheme:
\begin{enumerate}
\item Assume that the fields are solutions of the Heisenberg
equations obtained in the standard way from the Lagrangian density
$L_{0}(x)$ (note that we do not require $L_{0}(x)$ to describe
free fields).
\item Assume that the evolution of the statevector is governed by
the Tomonaga--Schwinger equation
\begin{equation} \label{etsr}
i\,\frac{\delta |\psi_{V}(\sigma)\rangle}{\delta\sigma(x)} \quad =
\quad - L_{I}(x)V(x) |\psi_{V}(\sigma)\rangle,
\end{equation}
$L_{I}(x)$ being a function of the fields considered in 1) which
does not involve their derivatives. As a consequence of the
assumptions about $L_{I}(x)$, for any two points $x, x' \in
\sigma$, $\sigma$ being a spacelike surface, $[L_{I}(x),
L_{I}(x')] = 0$, and consequently equation (\ref{etsr}) is
integrable.
\end{enumerate}

Let us consider the formal solution of equation (\ref{etsr}):
\begin{equation}
|\psi_{V}(\sigma)\rangle \quad = \quad T\, e^{\displaystyle
i\int_{\sigma_{0}}^{\sigma} d^{4}x\, L_{I}(x) V(x)}
|\psi(\sigma_{0})\rangle.
\end{equation}
Defining
\begin{equation} \label{tsos}
\rho(\sigma) \quad = \quad \llangle|\psi_{V}(\sigma)\rangle
\langle\psi_{V}(\sigma)|\rrangle
\end{equation}
using (\ref{tsos}), and following the procedure outlined in
equations (\ref{mcl})--(\ref{erop}), we get the
Tomonaga--Schwinger equation for the statistical operator
\begin{equation} \label{cuat}
\frac{\delta\, \rho(\sigma)}{\delta\sigma(x)} \quad = \quad
\lambda\, L_{I}(x) \rho(\sigma) L_{I}(x) \; - \;
\frac{\lambda}{2}\, \left\{ L_{I}^{2}(x), \rho(\sigma) \right\}
\end{equation}
which is manifestly covariant.

\subsection{White noise relativistic dynamical reduction models}
\label{sec93}

In this subsection we present a stochastically invariant theory
yielding Heisenberg reductions. To this purpose we keep assumption
1) of the previous subsection and we replace 2) by the requirement
that $|\psi_{V}(\sigma)\rangle$, instead of being governed by
equation (\ref{etsr}), obeys the following equation of the
Tomonaga--Schwinger type:
\begin{equation} \label{qeqg}
\frac{\delta |\psi_{V}(\sigma)\rangle}{\delta\sigma(x)} \quad =
\quad [L_{I}(x)V(x)\; - \; \lambda\, L_{I}^{2}(x)]
|\psi_{V}(\sigma)\rangle.
\end{equation}
The main difference between the two equations (\ref{etsr}) and
(\ref{qeqg}) derives from the skew--hermitian character of the
coupling to the stochastic $c$--number field. On the right hand
side of (\ref{qeqg}) a term guaranteeing the conservation of the
average value of the square norm of the state appears. It is
important to remark that equation (\ref{qeqg}), for a given sample
potential, does not conserve the norm of the statevector.

Let $|\psi_{V}(\sigma)\rangle$ be the solution of equation
(\ref{qeqg}) for a given realization of the stochastic potential
\begin{equation} \label{olhc}
|\psi_{V}(\sigma)\rangle \quad = \quad T\, e^{\displaystyle\;
\int_{\sigma_{0}}^{\sigma} d^{4}x\, [L_{I}(x) V(x) - \lambda\,
L_{I}^{2}(x)]} |\psi_{V}(\sigma_{0})\rangle
\end{equation}
and let us define the stochastic average
\begin{equation} \label{kjh}
\rho(\sigma) \quad = \quad \llangle|\psi_{V}(\sigma)\rangle
\langle\psi_{V}(\sigma)|\rrangle.
\end{equation}
Following the same procedure of the previous subsection one sees
that $\rho(\sigma)$ still satisfies equation (\ref{cuat}) derived
in the Hermitian case.

As in the non--relativistic case we have then two conceptually
different dynamical evolutions for the statevector, i.e.,
(\ref{etsr}) and (\ref{qeqg}), which give rise to the same
dynamics for the statistical operator and therefore to the same
physical predictions at the ensemble level. The very definition
(\ref{kjh}) of the statistical operator, when confronted with the
fact that the equation for the statevector does not preserve the
norm, implies the adoption of the point of view that a cooking
procedure, analogous to the one discussed in section \ref{sec61},
is necessary. This means that one has to consider normalized
vectors $|\psi_{V}(\sigma)\rangle/\| |\psi_{V}(\sigma)\rangle\|$
and has to attribute to the considered realization $V(x)$ of the
stochastic potential, having support in the spacetime region lying
between the two spacelike hypersurfaces $\sigma_{0}$ and $\sigma$,
not the probability density $P[V(x)]$ given by (\ref{djbgb}), but
a cooked probability density $P_{\makebox{\tiny Cook}}[V(x)]$
given by
\begin{equation}
P_{\makebox{\tiny Cook}}[V(x)] \quad = \quad P_{\makebox{\tiny
Raw}}[V(x)]\, \|  |\psi_{V}(\sigma)\rangle\|^{2}
\end{equation}
In the above equation $|\psi_{V}(\sigma)\rangle$ is the solution
of equation  (\ref{qeqg}) satisfying
\begin{equation}
|\psi_{V}(\sigma_{0})\rangle \quad = \quad |\psi_{0}\rangle.
\end{equation}
Before discussing the cooking procedure, the role of the
counterterm, and the relativistic invariance of the theory, an
important remark is necessary. As  has been discussed in
\cite{rel}, at the level of the statistical operator the map
$\Sigma_{t}$ does not exist when $t < 0$. For this reason, even at
the statevector level, we will only consider equation (\ref{qeqg})
as yielding the evolution from the statevector associated to a
given spacelike surface $\sigma_{0}$ to spacelike surfaces lying
entirely in the future of $\sigma_{0}$.

For what concerns the properties of the cooking procedure one can
immediately see that equation (\ref{cuat}) preserves the trace of
$\rho$,  which amounts to the statement that equation (\ref{qeqg})
preserves the average of the square norm of the statevector. In
particular, this implies
\begin{equation}
\int {\mathcal D}[V]\, P_{\makebox{\tiny Cook}}[V(x)] \quad =
\quad \int {\mathcal D}[V]\, P[V(x)] \, \|
|\psi_{V}(\sigma)\rangle \|^{2} \quad = \quad 1
\end{equation}
which shows that the requirement that the cooked probability
density sums to 1, is satisfied.

\subsection{Transformation properties and invariance of the theory}
\label{sec94}

We discuss now the transformation properties of the theory for a
given realization of the stochastic potential, in going from a
given reference frame $O$ to another one $O'$ related to it by a
transformation of the restricted Poincar\'e group
\begin{equation} \label{flcdc}
(\Lambda,b):\, x\; \rightarrow \; x'\, = \, \Lambda x\, + \, b
\end{equation}

We remind the reader that in the Tomonaga--Schwinger formalism of
conventional quantum field theory each reference frame $O$ is able
to assign a statevector to each spacelike hypersurface. Our first
concern is to demonstrate that the consistency of the composition
law for Lorentz transformations remains intact when one resorts to
the Tomonaga--Schwinger formalism.

Suppose that the transformation (\ref{flcdc}) involves a boost and
consider a given spacelike surface $\sigma$ for $O$. The surface
which is subjectively the same for $O'$ involves points which lie
in the past of the surface $\sigma$ for $O$. Our previous
discussion has pointed out that we will only use the
Tomonaga--Schwinger equation to go from a given spacelike surface
$\sigma$ to surfaces lying entirely in the future of $\sigma$.
Therefore, contrary to the standard case, we are not allowed to
raise here the following question: which state vector
$|\psi'(\sigma)\rangle$ would $O'$ associate to his subjective
surface $\sigma$ to describe the same physical situation described
by $O$ who assigns the statevector $|\psi(\sigma)\rangle$ to his
subjective surface $\sigma$?

We can, however, legitimately consider subjective surfaces
$\sigma{}^{\sim\prime}$ for $O'$, such that they lie in the future
of the surface $\sigma$ for $O$. Suppose the observer $O$
associates the statevector $|\psi_{O}(\sigma)\rangle$ to his
subjective surface $\sigma$ to describe the physical situation.
Let us denote by $\sigma{}^{\sim}$ the surface of $O$ which is
objectively the same as the above--mentioned surface
$\sigma{}^{\sim\prime}$ for $O'$. Then $O$ associates to
$\sigma{}^{\sim}$ the state $|\psi_{O}(\sigma{}^{\sim})\rangle$
obtained by solving equation (\ref{qeqg}) with the initial
condition that it reduces to $|\psi_{O}(\sigma)\rangle$ on
$\sigma$. We have
\begin{equation}
|\psi_{O}(\sigma{}^{\sim})\rangle \quad = \quad \frac{S_{V}
(\sigma{}^{\sim}, \sigma) |\psi_{O}(\sigma)\rangle}{ \|
S_{V}(\sigma{}^{\sim}, \sigma) |\psi_{O}(\sigma)\rangle\|}
\end{equation}
with
\begin{equation} \label{fqfl}
S_{V}(\sigma{}^{\sim}, \sigma) \quad = \quad T\, e^{\displaystyle
\; \int_{\sigma}^{\sigma{}^{\sim}} d^{4}x\, [L_{I}(x) V(x) -
\lambda L_{I}^{2}(x)]}
\end{equation}

Then the observer $O'$ will associate to his surface
$\sigma{}^{\sim\prime}$ the statevector
\begin{equation} \label{qtc}
|\psi_{O'}(\sigma{}^{\sim\prime})\rangle \; = \; U(\Lambda, b)
|\psi_{O}(\sigma{}^{\sim})\rangle \; = \; \frac{ U(\Lambda, b)
S_{V}(\sigma{}^{\sim}, \sigma) |\psi_{O}(\sigma)\rangle}{\|
S_{V}(\sigma{}^{\sim}, \sigma) |\psi_{O}(\sigma)\rangle\|}
\end{equation}
In equation (\ref{qtc}), $U(\Lambda, b)$ is the unitary operator
whose infinitesimal generators $P^{\mu}$ and $J^{\mu\nu}$ are
obtained in the standard way from the Lagrangian density
$L_{0}(x)$. Let now $\sigma$, $\sigma{}^{\sim}$,
$\sigma{}^{\sim}{}^{\sim}$ be three spacelike surfaces for $O$
each lying entirely in the future of the previous ones. Let us
consider two other observers $O'$ and $O''$ related by two
successive Lorentz transformations (the generalization to
Poincar\'e transformations is straightforward): $O' = \Lambda_{1}
O$, $O'' = \Lambda_{2} O'$, and let us denote by $\sigma'$,
$\sigma{}^{\sim\prime}$, $\sigma{}^{\sim}{}^{\sim\prime}$ and
$\sigma''$, $\sigma{}^{\sim\prime\prime}$,
$\sigma{}^{\sim}{}^{\sim\prime\prime}$ the above surfaces as seen
by $O'$ and $O''$, respectively.

The map (\ref{fqfl}), for a given realization of the stochastic
potential, has the following property. Suppose $O$ assigns the
state $|\psi_{O}(\sigma)\rangle$ to the surface $\sigma$. Then
$O'$ assigns the state (\ref{qtc}) to the surface
$\sigma{}^{\sim\prime}$. For $O'$ this state evolves according to
the Tomonaga--Schwinger equation (\ref{qeqg}) with $V'(x') =
V(\Lambda_{1}^{-1}x')$ from $\sigma{}^{\sim\prime}$ to
$\sigma{}^{\sim}{}^{\sim\prime}$:
\begin{equation}
|\psi_{O'}(\sigma{}^{\sim}{}^{\sim\prime})\rangle \quad = \quad
\frac{ {S'}_{V'}(\sigma{}^{\sim}{}^{\sim\prime},
\sigma{}^{\sim\prime} |\psi_{O'}(\sigma{}^{\sim\prime})\rangle}{\|
{S'}_{V'}(\sigma{}^{\sim}{}^{\sim\prime}, \sigma{}^{\sim\prime})
|\psi_{O'}(\sigma{}^{\sim\prime})\rangle\|}.
\end{equation}
The observer $O''$ will describe the final situation by assigning
the statevector
\begin{equation}
|\psi_{O''}(\sigma{}^{\sim}{}^{\sim\prime\prime})\rangle \quad =
\quad U(\Lambda_{2})
|\psi_{O'}(\sigma{}^{\sim}{}^{\sim\prime})\rangle
\end{equation}
to the surface $\sigma{}^{\sim}{}^{\sim\prime\prime}$. On the
other hand, one can consider the evolution from $\sigma$ to
$\sigma{}^{\sim}{}^{\sim}$ as seen from $O$,
\begin{equation}
|\psi_{O}(\sigma{}^{\sim}{}^{\sim})\rangle \quad = \quad \frac{
S_{V}(\sigma{}^{\sim}{}^{\sim}, \sigma)
|\psi_{O}(\sigma)\rangle}{\| S_{V}(\sigma{}^{\sim}{}^{\sim},
\sigma) |\psi_{O}(\sigma)\rangle\|}.
\end{equation}
and then look at it from $O'' = \Lambda_{2}\Lambda_{1} O$, getting
the state
\begin{equation}
|\psi^{*}_{O''}(\sigma{}^{\sim}{}^{\sim\prime})\rangle \quad =
\quad U(\Lambda_{2}\Lambda_{1})
|\psi_{O}(\sigma{}^{\sim}{}^{\sim})\rangle.
\end{equation}
For consistency,
$|\psi^{*}_{O''}(\sigma{}^{\sim}{}^{\sim\prime})\rangle$ must
coincide with
$|\psi_{O''}(\sigma{}^{\sim}{}^{\sim\prime})\rangle$. This can be
easily proved to hold.

Although we have just seen that the theory implies an assignment
of a statevector to a hypersurface by any observer that fulfills
the Lorentz (also Poincar\'e) group requirements, this does not
mean that the description is Lorentz invariant. In fact, because a
particular realization of the stochastic potential $V$ looks
different from two different reference frames, the map
$S_{V}(\sigma{}^{\sim}, \sigma)$ obviously depends upon the
reference frame $O$. This shows that, at the individual level, the
theory does not posses the property of standard (i.e.,
non--stochastic) Lorentz invariance. However, for stochastic
Lorentz invariance, one must consider the ensemble of possible
sample potentials. When one takes into account the Lorentz
invariance of the requirement (\ref{qredm}) for the correlation
function $\llangle V(x)\, V(x') \rrangle$, and the invariance of
the cooking procedure that must be performed to get the physics of
the problem, one can easily prove, along the same lines as in the
non relativistic case, that there is stochastic invariance in the
statevector language, i.e., the stochastic ensemble of evolution
operators $S_{V}(\sigma{}^{\sim}, \sigma)$ is the same in each
reference frame.

In the language of the statistical operator, invariance is evident
from the manifestly covariant Tomonaga--Schwinger form
(\ref{cuat}) of the evolution equation.

\subsection{Reduction properties} \label{sec95}

Once we have guaranteed the invariance of the formalism by using
its Tomonaga--Schwinger formulation, in order to discuss specific
features of the process, we can consider $t$ = const hyperplanes
in the Schr\"odinger picture. In so doing, the equation
corresponding to (\ref{qeqg}) is
\begin{equation} \label{slsre}
\frac{d\,|\psi_{V}(t)\rangle}{dt}\; = \; \left[ -i H_{0}\, + \,
\int d^{3}x\, \left( L_{I}({\bf x}, 0) V({\bf x}, t) \, - \,
\lambda L_{I}^{2}({\bf x}, 0) \right)\right] |\psi_{V}(t)\rangle
\end{equation}
This is a Stratonovich equation for the statevector. By standard
procedures one can consider the corresponding It\^o stochastic
dynamical equation
\begin{equation} \label{slire}
d\,|\psi_{V}(t)\rangle\; = \; \left[\left( -i H_{0}\, - \,
\frac{\lambda}{2}\int d^{3}x\, L_{I}^{2}({\bf x}, 0) \right) dt \,
+ \, \int d^{3}x\, L_{I}({\bf x}, 0) dV({\bf x}) \right]
|\psi_{V}(t)\rangle,
\end{equation}
where $dV({\bf x})$ is a real Wiener process satisfying
\begin{equation}
\llangle dV({\bf x}) \rrangle = 0 \qquad\quad \llangle dV({\bf x})
dV({\bf y}) \rrangle = \lambda \delta({\bf x} - {\bf y}) dt.
\end{equation}
Note that both equations (\ref{slsre}) and (\ref{slire}) do not
conserve the norm of the statevector but they conserve the average
of its squared norm.

As discussed in section \ref{sec6} one can take two equivalent
attitudes to describe the physics of the process. One can solve
equation (\ref{slsre}) or (\ref{slire}) for a given initial
condition, and then one can consider the normalized vectors
$|\psi_{V}(t)\rangle/||\psi_{V}(t)\rangle\|$ at time $t$ and
assume that the probability of their occurrence is obtained by
cooking the probability density of occurrence of $V(x)$, i.e., by
multiplying it by $\||\psi_{V}(t)\rangle\|^{2}$. Alternatively,
one can consider the nonlinear stochastic dynamical equation
\begin{eqnarray} \label{snlire}
d\,|\psi_{V}(t)\rangle & = & \left[\left( -i H_{0}\, - \,
\frac{\lambda}{2}\int d^{3}x\, [L_{I}^{2}({\bf x}, 0) \, - \,
\langle L_{I}({\bf x}, 0) \rangle]^{2} \right) dt \right.
\, + \nonumber \\
& & \left. + \int d^{3}x \left( L_{I}({\bf x}, 0) \, - \, \langle
L_{I}({\bf x}, 0) \rangle \right) dV({\bf x}) \right]
|\psi_{V}(t)\rangle,
\end{eqnarray}
(where $\langle L_{I}({\bf x}, 0) \rangle = \langle\psi_{V}(t)|
L_{I}({\bf x}, 0) |\psi_{V}(t)\rangle$), without cooking, i.e.,
using just the probability weighting of $V(x)$.

As we know from the discussion of the previous sections, when one
disregards the Hamiltonian term in (\ref{snlire}), the evolution
leads the state vector to enter one of the common eigenmanifolds
of the commuting operators $L_{I}({\bf x}, 0)$. The theory induces
therefore Heisenberg reductions, as required.

\subsection{The models so far proposed} \label{sec96}

In this subsection we will consider some specific choices for the
Lagrangian densities $L_{0}$ and $L_{I}$ which, when used in
connection with the formalism presented in the previous
subsections, yield stochastically invariant relativistic reduction
models. The goal is to build up a framework leading to
localization in position of the basic constituents of all matter.

The simplest and most immediate idea would be to introduce a
fermion field for particles of mass $M$ and to choose for the
Lagrangian density the expressions
\begin{equation}
L_{0}(x) \; = \; \overline{\psi}(x)\left(
i\gamma^{\mu}\partial_{\mu} - M\right)\psi(x), \qquad\quad
L_{I}(x) \; = \; \overline{\psi}(x)\,\psi(x)
\end{equation}
However, in the non relativistic limit, the dynamics induced by
the above choice would lead to infinitely sharp position
localizations for the fermions, and this, as is well known
\cite{gr62}, is unacceptable.

We have then to enrich the formalism. This can be done by
following the proposal put forward in reference \cite{p62}. One
considers a fermion field coupled to a real scalar meson field and
chooses
\begin{eqnarray} \label{mct}
L_{0}(x) & = & \frac{1}{2}\left[ \partial_{\mu}\phi(x)
\partial^{\mu}\phi(x) - m^{2}\phi^{2}(x)\right] \; + \;
\overline{\psi}(x)\left( i\gamma^{\mu}\partial_{\mu} -
M\right)\psi(x) \; +
\nonumber \\
& &
\eta \overline{\psi}(x)\,\psi(x)\,\phi(x) \nonumber \\
L_{I}(x) & = & \phi(x).
\end{eqnarray}
The introduction of the meson field coupled to the fermion field
allows one to overcome the difficulty of the infinitely sharp
localization for fermions met by the previous model. In the
Schr\"odinger representation the evolution equation for the state
vector corresponding to the choice (\ref{mct}) is
\begin{equation} \label{mtre}
\frac{d\,|\psi_{V}(t)\rangle}{dt}\; = \; \left[ -i H_{0}\, + \,
\int d^{3}z\, \left( \phi({\bf z}, 0) V({\bf x}, t) \, - \,
\lambda \phi^{2}({\bf z}, 0) \right)\right] |\psi_{V}(t)\rangle
\end{equation}

Let us consider now the non--relativistic infinite mass limit for
fermions and let us confine our discussion to the sector
containing one fermion (note that in the limit the fermion number
is a conserved quantity). The state of a fermion at position ${\bf
q}$ is the ``dressed'' state
\begin{equation}
|1{\bf q}\rangle \quad = \quad a^{\dagger}({\bf q})\, A({\bf q})\,
|0\rangle
\end{equation}
where $a^{\dagger}({\bf q})$ is the creation operator for a
fermion at ${\bf q}$ and $A({\bf q})\, |0\rangle = |m_{\bf
q}\rangle$ is a coherent state which can be characterized as
either the common eigenstate of the annihilation operators of
physical mesons with eigenvalue zero or as the common eigenstate
of the annihilation operators $b({\bf k})$ of bare mesons with
momentum ${\bf k}$, with eigenvalues $(\eta/\sqrt{2})e^{-i{\bf
k}\cdot {\bf q}}/(2\pi k_{0})^{3/2}$.

To be rigorous, in the three--dimensional case, one should
introduce an ultraviolet cut--off on the momentum of mesons in the
interaction term to avoid ultraviolet singularities. In the limit
in which the cut--off is removed the meson states $|m_{\bf
q}\rangle$, $|m_{{\bf q}'}\rangle$ tend to become orthogonal for
${\bf q} \neq {\bf q}'$. In this way, due to the coupling of the
fermion field to the meson field, the ``position'' of one fermion
turns out to be strictly correlated to states of the meson field
which are approximately orthogonal.

We note that the mean value of $\phi({\bf z}, 0)$ in the state
$|m_{\bf q}\rangle$ turns out to be
\begin{equation}
\langle m_{\bf q}| \phi({\bf z}, 0) |m_{\bf q}\rangle \quad =
\quad f({\bf z} - {\bf q}) \quad = \quad \frac{\eta e^{-m|{\bf z}
- {\bf q}|}}{4\pi |{\bf z} - {\bf q}|}.
\end{equation}
In what follows, in order to discuss the localization properties
of the model for physical fermions, we make a gross simplification
(which coincides with the non relativistic approximation), i.e.,
we treat the states $|m_{\bf q}\rangle$ as eigenstates of
$\phi({\bf z}, 0)$ pertaining to the eigenvalue $f({\bf z} - {\bf
q})$. Let us then consider the physical state for one fermion
\begin{equation} \label{enn}
|\psi(t)\rangle \quad = \quad \int d^{3}q\, \psi({\bf q}, t)
|1{\bf q}\rangle.
\end{equation}
By substituting (\ref{enn}) into equation (\ref{mtre}) and
disregarding the standard Hamiltonian $H_{0}$, we get the equation
for $\psi({\bf q}, t)$:
\begin{equation} \label{pmse}
\frac{\partial \psi_{V}({\bf q}, t)}{\partial t} \;  = \; \int
d^{3}z\, f({\bf z} - {\bf q}) V({\bf z}, t)\, \psi_{V}({\bf q},
t)\, - \, \frac{\lambda \eta^{2}}{8\pi m}\, \psi_{V}({\bf q}, t),
\end{equation}
i.e.,
\begin{equation} \label{oss}
\frac{\partial \psi_{V}({\bf q}, t)}{\partial t} \;  = \;
V^{\sim}({\bf q}, t)\, \psi_{V}({\bf q}, t)\, - \, \frac{\lambda
\eta^{2}}{8\pi m}\, \psi_{V}({\bf q}, t),
\end{equation}
with $V^{\sim}({\bf q}, t)$ a Gaussian noise with zero mean and
covariance
\begin{equation}
\llangle V^{\sim}({\bf q}, t)\, V^{\sim}({\bf q}', t') \rrangle
\quad = \quad \frac{\lambda\eta}{8\pi m}\, e^{-m|{\bf q} - {\bf
q}'|} \delta(t-t').
\end{equation}
Equation (\ref{oss}) is essentially the same as equation
(\ref{sch1d}) of CSL for the case of a single particle. If one
considers the sector with $N$ fermions, in the above
approximations, one gets an equation of the CSL type [see
equations (\ref{imef}) and (\ref{ando})] with the operator
\begin{equation} \label{ndor}
D({\bf x}) \quad = \quad \frac{m^{2}}{4\pi}\,\int d^{3}z\,
\frac{e^{-m|{\bf x} - {\bf z}|}}{|{\bf x} - {\bf z}|}\,
a^{\dagger}({\bf z}) a({\bf z})
\end{equation}
taking the place of $N({\bf x})$, and $(\lambda\eta^{2})/m^{4}$
taking the place of $\gamma$.

Thus it appears reasonable that the model (\ref{mct}) possesses
the desired localizing features. However, it also presents a
serious difficulty. The evolution equation (\ref{erop}) for the
statistical operator, specialized to the Lagrangian (\ref{mct}),
is:
\begin{equation} \label{eropm}
\frac{d}{dt}\,\rho(t) \; = \; -i [H_{0}, \rho(t)] + \lambda \int
d^{3}z\, \phi({\bf z}, 0) \rho(t) \phi({\bf z}, 0) -
\frac{\lambda}{2} \int d^{3}z\, \left\{ \phi^{2}({\bf z}, 0),
\rho(t) \right\}.
\end{equation}\
Let us consider the Hamiltonian $H$ for the free meson field; by
using (\ref{eropm}) one can evaluate the increase per unit time of
the mean value of $H$, getting
\begin{equation}
\frac{d\, \langle H \rangle}{dt} \quad = \quad - \frac{\lambda}{2}
\int d^{3}z\, \langle [ \phi({\bf z}, 0), [\phi({\bf z}, 0), H]]
\rangle
\end{equation}
i.e.,
\begin{equation}
\frac{d\, \langle H \rangle}{dt} \quad = \quad \frac{\lambda}{2}
\int d^{3}z\, \delta(0)
\end{equation}
Therefore, the increase per unit time and per unit volume of the
mean value of the energy of the meson field turns out to be
infinite. So, in addition to the desired reduction behaviour, the
model displays an undesidered additional behaviour: because the
white noise source is locally coupled to the meson field, it
copiously produces mesons out of the vacuum\footnote{A similar
conclusion has been reached also by Adler and Brun in recent
investigations on relativistic statevector collapse models
\cite{ad3}.}. We note that the now outlined difficulty does not
show up in the non--relativistic approximation of the model
discussed above [equations (\ref{pmse})--(\ref{ndor})] due to the
gross simplification of treating the states $|m_{\bf q}\rangle$ as
eigenstates of $\phi({\bf z}, 0)$.

Since the divergences originate from the local coupling between the
quantum fields and the white noise, the natural way to cure the
infinite vacuum fluctuations is to replace the white noise with a
more general Gaussian stochastic process; this possibility has
been explored by P. Pearle \cite{ppo1, ppo2}. He considers an
evolution equation for the statevector which is the
straightforward generalization of the white noise equation
(\ref{olhc}). He then proves that --- in the lowest order in
perturbation theory --- only the time--like components of the
spectrum of the noise are responsible for the vacuum excitations;
accordingly, he chooses as the spectrum of the noise that of a
tachion  of mass $\mu = \hbar/\alpha c \sim 1$ eV, where $\alpha$
is the QMSL localization parameter. Anyway, at higher orders in
perturbation theory there are still vacuum excitations; to avoid
such excitations, Pearle proposes to remove the time--ordering
product from Feyman diagrams.

Recently, Nicrosini and Rimini \cite{bgrel} have proposed a
different solution to the  problem arising from the appearance of
divergences. They couple the noise not to the quantum fields, but to
``macroscopic operators'' which are defined as the integral of the usual
quantum fields over appropriately chosen (Lorenz invariant) spacetime
surfaces. In other words, the coupling between the quantum fields and the
white noise is not local anymore.

Both the attempts by Pearle and by Nicrosini and Rimini are
promising; anyway, they still have to be studied in detail. In
particular, it is not clear yet whether the evolution can be
expressed in terms of a integrable Tomonaga--Schwinger  equations:
this is a necessary requirement in order to put on solid grounds
any relativistic theory of dynamical reductions.

\section{Local and nonlocal features of relativistic CSL}
\label{sec10}

As is well known, the quantum theory of measurement, in addition
to the difficulties discussed in section \ref{sec1} which
constitute the main motivation for the consideration of dynamical
reduction models, presents some further difficulties arising
specifically from the assumed instantaneous nature of the collapse
of the wavefunction. In particular, at the individual level of
description, nonlocal features as well as odd aspects (from the
relativistic point of view) emerge. Such problems have already
been extensively discussed in the literature
\cite{de,bl1,hkd,aa1}, in the case of standard quantum mechanics:
we will review them in subsection \ref{sec101}. It is interesting
to look at them from the perspective of the relativistic dynamical
reduction models, analyzed in the previous section: this will be
the subject of subsection \ref{sec102}.

In the final section we discuss the problem of parameter
dependence in dynamical reduction models. We will show that in the
non linear model there is a set of realizations of the stochastic
process for which parameter independence is violated for parallel
spin components in a EPR--Bohm--like setup. Such a set has an
appreciable probability of occurrence ($\simeq 1/2$). On the other
hand, we will prove that the linear model exhibits only extremely
small parameter dependence effects.

\subsection{Quantum theory with the reduction postulate}
\label{sec101}

\subsubsection{Objective properties of individual systems}
\label{sec1011}

Suppose one accepts it as meaningful, within standard quantum
theory, to consider an individual level of description with the
possibility of attributing objective properties to a quantum
system. As discussed in section \ref{sec14}, a natural attitude
corresponding to the one first introduced in the celebrated EPR
paper \cite{epr} is to assume the following. If an individual
physical system $S$ is associated to a definite statevector
$|\psi\rangle$ which is an eigenstate of an observable $A$
pertaining to the eigenvalue $a$, then one can state that ``$S$
has the property $A=a$" or that ``there exists an element of
physical reality" referring  to the considered observable. We
remark that if we denote by $P_{a}$ the projection operator on the
closed linear manifold of the eigenstates of $A$ belonging to the
eigenvalue $a$, then
\begin{equation} \label{cno}
\langle\psi | P_{a} |\psi\rangle \quad = \quad 1
\end{equation}
We want to stress, however, that even within non--relativistic
standard quantum mechanics, one is compelled to take the attitude
of attributing objective properties to a system even when
condition (\ref{cno}) is valid only to an extremely high degree of
accuracy. To clarify this statement, we can think, for example, of
the spin measurement of a spin--$1/2$ particle by a Stern--Gerlach
apparatus. In such a case, the two spin values are strictly
correlated to two states $\psi_{1}$ and $\psi_{2}$ describing the
spatial degrees of freedom. Even though these wavefunctions are
appreciably different from zero in two extremely narrow and
distant regions, their supports cannot have a void intersection.
As a consequence even an arbitrarily precise measurement of the
position cannot reduce the statevector exactly to an eigenstate of
the spin component. The final state unavoidably exhibits an
(extremely slight) entanglement of position with spin variables
and as such cannot be an eigenstate of a spin component operator.

Incidentally we remark that the above considerations are even more
appropriate in the case of dynamical reduction models. In fact, on
the one hand, such models, with the requirement that they induce
Heisenberg reductions, are introduced just with the purpose of
implying, at the individual level, the emergence of objective
properties for macroscopic objects (in particular the property of
being in one place rather than in another). Correspondingly, they
induce indirectly the appearance of objective properties also for
microscopic systems, at least when they interact with macroscopic
measuring--like devices. On the other hand, as is well known and
as has been repeatedly stressed in references
\cite{rel,gprfqm,p62}, within dynamical reduction models, the
unavoidable persistence of the tails, the tiny but nonzero terms
corresponding to the parts of a linear superposition which have
been suppressed by the spontaneous localization process, prevents
us from asserting with absolute certainty that the ``macroscopic
pointers" are in a definite space region, if one adopts the
standard probability interpretation.

The conclusion is the same we have drawn in Section 11 concerning the
tails problem, i.e., that within the dynamical reduction program is
perfectly consistent to accept that it makes
sense to attribute appropriate objective properties to individual
systems  even
when the mean value of the projection operator on the eigenmanifold
associated to the eigenvalue corresponding to the attributed property is
not exactly equal to $1$, but is extremely close to it.

\subsubsection{Non locality} \label{sec1012}

Nonlocal features\footnote{For an exhaustive discussion, the
reader is referred to the excellent book by M. Redhead
\cite{redb}.} of quantum mechanics arise from the fact that, due
to the instantaneous nature of the collapse of the wavefunction,
possible actions performed in a certain space region can, under
specific circumstances, induce immediate changes in distant
regions. In this connection two important questions arise: first,
do these changes correspond to some modifications of the physical
situation in the distant region? Secondly, are these modifications
detectable, so that one can take advantage of them to send faster
than light signals?

To make precise and unambiguous these questions, it is necessary
to specify the level of description of physical processes one is
considering. In particular, it is important to make a clear
distinction between the ensemble and the individual levels of
description.

To understand the above situation, one can make reference either
to the well--known EPR--Bohm type setup for an ``entangled" state
of a composite system $S = S_{1}+ S_{2}$, the components being far
apart and non--interacting, or to the position measurement of a
particle whose state is the linear superposition of two distant
packets. In the first case, as is well known, at the level of the
individual members of the ensemble, the far away system (let us
say $S_{2}$) is \cite{sche} ``steered or piloted into one or the
other type of state" according to the measurement which is
performed on $S_{1}$ and the specific result which is obtained. In
the second case, let us write $\psi (x, t) = \psi_{1}(x, t) +
\psi_{2}(x, t)$, where the states $\psi_{1}(x, t)$ and
$\psi_{2}(x, t)$ have equal norms and are appreciably different
from zero only in two far apart regions $\alpha_{1}$ and
$\alpha_{2}$, respectively. Then, a measurement aimed to test
whether the particle is in $\alpha_{1}$ and yielding, for example,
the answer ``no" (``yes"), instantaneously collapses $\psi (x, t)$
to $\psi_{2}(x, t)$ ($\psi_{1}(x, t)$). Correspondingly the
quantity $\int_{\alpha_{2}} dx\, |\psi (x, t)|^{2}$ (i.e., ``the
mean value"\footnote{We are using the common phrase ``mean value''
to represent diagonal matrix elements like (\ref{cno}), even
though the statistical connotation of this phrase has no meaning
in our discussion.} of the projection operator on region
$\alpha_{2}$), changes from $1/2$ to either $1$ or $0$ according
to the outcome of the position measurement at $\alpha_{1}$. This
puts into evidence how, if interpreted as a theory describing
individual systems, quantum mechanics exhibits nonlocal features.

The situation is quite different when looked at from the ensemble
point of view. In fact, as is well known
\cite{heb,grwfl,gwfl,shi1}, no measurement procedure in a given
region can change the statistical distribution of prospective
measurement results in a distant region. Of course, this does not
mean that, from the ensemble point of view, Quantum Mechanics
displays a local character. The theory is still highly nonlocal
(for example, Bell's inequalities hold), but this non locality
cannot be used in any way to send faster that light signals to
distant observers.

These remarks, although made in the context of ordinary quantum
theory with the reduction postulate, are not essentially modified
(i.e., the word ``instantaneously" must be changed to ``in a split
second" \cite{bells}) in the case of the CSL theory with its
reduction dynamics.

\subsubsection{Relativistic oddities with observations}
\label{sec1013}

In the above analysis we have discussed a measurement process in a
given reference frame $O$. The consideration of the instantaneous
change of the statevector induced by a measurement raises
interesting questions when looked at by different observers. Since
the distance between the two space regions $\alpha_{1}$ and
$\alpha_{2}$ mentioned above can be arbitrarily large, even the
passage to a reference frame which is moving with respect to $O$
with an arbitrarily small velocity can change the time order of
simultaneous (for $O$) events occurring in the two regions.

To illustrate briefly the main points of the problem, we consider
the observer $O$ looking at a system of one particle in the state
$\psi (x, t) = \psi_{1}(x, t) + \psi_{2}(x, t)$ which is a
superposition of two well--localized wavepackets propagating in
opposite directions with respect to the origin $x = 0$.
Disregarding the extension and the spreading of the wavepackets,
we can represent the situation by the spacetime diagram of figure
5,
\begin{center}
\begin{picture}(240,150)(0,0)
\put(0,0){\line(1,0){240}} \put(0,150){\line(1,0){240}}
\put(0,0){\line(0,1){150}} \put(240,0){\line(0,1){150}}
\put(30,20){\vector(1,0){180}} \put(120,20){\vector(0,1){110}}
\put(215,18){$x$} \put(120,135){$t$}
\thicklines \put(120,20){\line(2,3){60}}
\put(120,20){\line(-2,3){60}} \thinlines \put(180,115){1}
\put(55,115){2}
\put(90,65){\circle*{3}} \put(160,80){\circle*{3}}
\put(40,65){$B(x_{2}, t_{2})$} \put(170,80){$C(x_{1}, t_{1})$}
\end{picture}

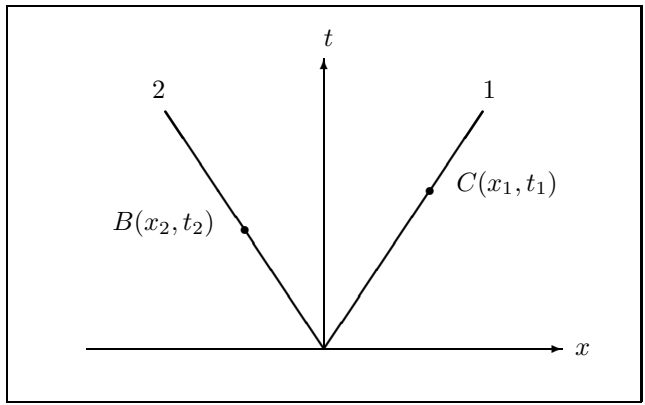
\vspace{0.2cm} \footnotesize \parbox{3.3in}{Figure 5: World lines
of two well--localized wavepackets 1 and 2, belonging to a single
particle which is detected at event $C$.} \normalsize
\end{center} \vspace{0.5cm}
in which the two world lines $1$ and $2$ are associated with
$\psi_{1}$ and $\psi_{1}$, respectively. Suppose that, at the
spacetime point $C = (x_{1}, t_{1})$ there is a device designed to
test whether the particle is there, and let us suppose that, in
the specific individual case we are considering, the result of the
test is ``yes". This is a covariant statement on which all
observers must agree. If one adopts the wavepacket reduction
postulate of standard quantum theory and one assumes that the
collapse occurs for each reference frame along the hyperplane $t'
=$ const, where $t'$ is the subjective time of the event $C$ for
such a frame, one meets a puzzling situation. Let us in fact
consider an objective point $B$ on world line $2$, which is
spacelike separated from $C$ and which is labeled by $(x_{2},
t_{2})$ (see figure 5). For $O$, $t_{2} < t_{1}$ and, by the above
assumption, no reduction has occurred at time $t_{2}$ and the
statevector is $\psi (x, t_{2})$. If one considers the projection
operator $P_{2}$ on the space region around $x_{2}$, one has
$\langle\psi|P_{2}|\psi\rangle \simeq 1/2$. Accordingly, we could
say that the situation is such that, at time $t_{2}$, $O$ cannot
attribute to the particle the ``property" of being or not being in
the region around $x_{2}$.

However, there exists an observer $O'$ such that $t_{2}' >
t_{1}'$, where $t_{2}'$ and $t_{1}'$ are the time labels
attributed by $O'$ to the events $B$ and $C$, respectively. For
$O'$ the particle has triggered the detector in $C$ at $t_{1}'$.
Therefore at $t_{2}'$ the state of the system is $\psi_{1}$. Then,
for $O'$, the mean value of the projection operator $P_{2}$ at
$t_{2}'$ is zero\footnote{Obviously, to be rigorous, both the
statement that the state is $\psi_{1}$ or $\psi_{2}$, as well the
consideration of the projection operators $P_{1}$ and $P_{2}$, are
not correct, because one should consider a relativistic
description of the system and of the observables. However, since
$O'$ is moving with a very small velocity $v \ll c$ with respect
to $O$, the above approximations are appropriate.}. Observer $O'$
can then state that the particle has the property of ``not being
around $B$". Thus, $O$ and $O'$ do not agree on a statement
referring to a local property at an objective spacetime point.

It is useful to note that this ambiguity occurs only for the
points of the world line $2$ which are spacelike with respect to
$C$; for a point $B$ in the past of $C$ all observers agree in
stating that the particle has no definite location while for a
point $B$ in the future of $C$ all observers agree in saying that
the particle ``is not around $B$".

The above discussion follows essentially the one given in
reference \cite{bl1}. The consideration of these kinds of
difficulties have led various authors to take different attitudes.
Bloch \cite{bl1} and Aharonov and Albert \cite{aa1} derive from
this the conclusion that one cannot attach an objective meaning to
wavefunctions for individual systems. Hellwig and Kraus \cite{hkd}
have tried to solve the ambiguity about the wavefunction at a
given objective spacetime point by requiring that the collapse of
the state vector due to the measurement at $C$ takes place along the past
light cone originating from $C$. Thus, at points outside the past light
cone the statevector is reduced, while at points inside the past light
cone the statevector is unreduced. This is a covariant statement and leads
the authors to the identification of a unique statevector to be
associated to any given spacetime point. However, such a
prescription implies that there are spacelike surfaces (those
crossing the past cone of $C$) to which it is not possible to
associate a definite state vector. This, as nicely illustrated by
Aharonov and Albert \cite{aa1}, forbids the consideration of
nonlocal observables on these hypersurfaces; for example it does
not allow one to speak consistently of the total charge of the
system. Moreover, the assumption that the reduction occurs on the
hypersurface delimiting the past light cone raises conceptual
difficulties with the cause--effect relation.

\subsection{Relativistic reduction models} \label{sec102}

We discuss here the local and nonlocal features of reduction
models in the relativistic case. In order to investigate whether
the dynamics presented in section \ref{sec94} has nonlocal
effects, we make reference to the procedure outlined in reference
\cite{hrf}, i.e., we consider whether a modification of the
Lagrangian density in a spacetime region $C$ can have effects in a
region $B$ which is spacelike separated from it (this will be
discussed in subsections \ref{sec1021} and \ref{sec1022}). In
particular, since we want to study the possibility of nonlocal
effects due to the reducing character of the dynamics, we will
take into account modifications of the Lagrangian density $L_{I}$
coupled to the noise.

The problems which we want to discuss require the consideration of
``local observables." By this expression we mean the integral of a
function of the fields and their derivatives in the interaction
picture:
\begin{equation}
A_{I}(\sigma) \quad = \quad \int_{\sigma} dx'\, f_{\alpha}(x')\,
F[\phi_{I}(x'), \partial_{\mu}\phi_{I}(x')]
\end{equation}
with $f_{\alpha}(x')$ a function of class $C^{\infty}$ with
compact support $\alpha$ on the spacelike surface $\sigma$. The
physically interesting quantities, for our analysis, are the mean
values of such local observables. As usual it is necessary to make
precise the level at which the nonlocality problem is discussed.
We will consider it, as before, both at the ensemble and at the
individual level.

At this last level, we will discuss also questions analogous to
those considered in subsection \ref{sec1013} which originate from
looking at the wavepacket reduction postulate, taking the point of
view of relativity theory. In the present context, they emerge
naturally from the relativistic dynamics described by the
Tomonaga--Schwinger equation. In particular, it turns out that,
for all Tomonaga--Schwinger surfaces coinciding on $\alpha$, the
mean value of the local observable depends upon the specific
Tomonaga--Schwinger surface on which it is evaluated (see
subsection \ref{sec1023}). This is not the case with the
Tomonaga--Schwinger description of an ordinary relativistic
quantum field theory, and such a difference gives rise to
interesting questions about the possibility of attributing objective
properties to the systems  which we will discuss in subsection
\ref{sec1024}.

\subsubsection{Ensemble level} \label{sec1021}

As already emphasized, at the ensemble level, the statistical
operator and therefore the physics of the two models considered in
sections \ref{sec92} and \ref{sec93} coincide. Thus, to
investigate properties referring to the statistical ensemble, one
can make reference to the stochastic dynamics with hermitian
coupling, which can be easily handled by familiar methods.

With reference to the model of section \ref{sec93}, we consider
the mean value of a local observable $A_{I}(\sigma)$:
\begin{equation}
\langle A_{I}(\sigma) \rangle \quad = \quad \makebox{Tr}\, [
A_{I}(\sigma) \rho_{I}(\sigma) ]
\end{equation}
Let us denote by $U_{V}(\sigma, \sigma_{0})$ the evolution
operator
\begin{equation}
U_{V}(\sigma, \sigma_{0}) \quad = \quad T\,e^{\displaystyle
i\int_{\sigma_{0}}^{\sigma} dx\, L_{I}(x) V(x)}
\end{equation}
and by $A_{HV}(\sigma) = U_{V}^{\dagger}(\sigma, \sigma_{0})
A_{I}(\sigma) U_{V}(\sigma, \sigma_{0})$ the observable in the
Heisenberg picture which corresponds to $A_{I}(\sigma)$ when the
realization $V$ of the stochastic potential occurs. Let
$A_{H}(\sigma)$ be the stochastic average over V of
$A_{HV}(\sigma)$
\begin{equation}
A_{H}(\sigma) \quad = \quad \int{\mathcal D}[V]\,
P_{\makebox{\tiny Raw}}[V]\, A_{HV}(\sigma).
\end{equation}
We then have
\begin{equation} \label{dcvlp}
\langle A_{I}(\sigma) \rangle \quad = \quad \makebox{Tr}\, [
A_{H}(\sigma)\, \rho(\sigma_{0}) ].
\end{equation}
The support of $A_{I}(\sigma)$ defines a partition of spacetime
into three regions: the future, the past, and the set of points
which are spacelike separated from all points belonging to this
support (see figure 6).
\begin{center}
\begin{picture}(240,150)(0,0)
\put(0,0){\line(1,0){240}} \put(0,150){\line(1,0){240}}
\put(0,0){\line(0,1){150}} \put(240,0){\line(0,1){150}}
\put(20,20){\vector(1,0){190}} \put(20,20){\vector(0,1){110}}
\put(225,18){$x$} \put(20,135){$t$}
\thicklines \put(95,80){\line(-1,1){50}}
\put(95,80){\line(-1,-1){50}} \put(135,80){\line(1,1){50}}
\put(135,80){\line(1,-1){50}} \qbezier(95,80)(105,90)(115,80)
\qbezier(115,80)(125,70)(135,80) \thinlines
\put(105,100){future} \put(110,50){past} \put(50,78){3}
\put(180,78){3}
\end{picture}

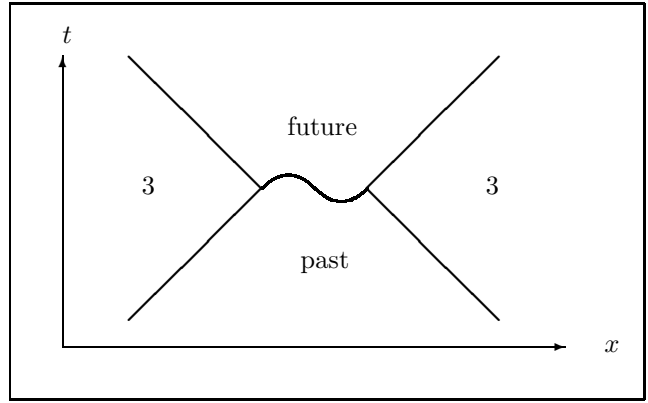
\vspace{0.2cm} \footnotesize \parbox{3.3in}{Figure 6: The support
of the local observable $A_{I}(\sigma)$, and the set (3) of points
bearing a spacelike relation to this support.} \normalsize
\end{center} \vspace{0.5cm}
We choose now a spacetime region $C$ entirely contained in region
3 and we consider a modification of the Lagrangian density
$L_{I}(x)$ coupled to the noise. We replace $L_{I}(x)$ with a new
density $L_{I}^{\sim}(x) = L_{I}(x) + \Delta L_{I}(x)$, with
$\Delta L_{I}(x)$ different from zero only for $x \in C$. If
$A_{HV}^{\sim}(\sigma)$ denotes the local observable in the
Heisenberg picture, when we replace $L_{I}(x)$ with
$L_{I}^{\sim}(x)$, we have
\begin{equation}
A_{HV}^{\sim}(\sigma) = \left[ T\,e^{ i\int_{\sigma_{0}}^{\sigma}
dx\, \Delta L_{I}(\phi_{HV}(x)) V(x)} \right]^{\dagger}
A_{HV}(\sigma) \left[ T\,e^{ i\int_{\sigma_{0}}^{\sigma} dx\,
\Delta L_{I}(\phi_{HV}(x)) V(x)} \right].
\end{equation}
The fields $\phi_{HV}(x)$ which appear in $\Delta L_{I}(x)$ are
the fields in Heisenberg picture for the original Lagrangian
density $L_{0}(x) + L_{I}(x) V(x)$. The appearance of $\Delta
L_{I}(x)$ actually restricts the integration in the exponential to
the spacelike region $C$, which is spacelike separated with
respect to the support of $A_{HV}(\sigma)$. It follows that the
exponential commutes with $A_{HV}(\sigma)$, and therefore
\begin{equation} \label{rccgm}
A_{HV}^{\sim}(\sigma) \quad = \quad A_{HV}(\sigma)
\end{equation}
for any given realization of the stochastic potential. One then
has
\begin{equation}
A_{H}^{\sim}(\sigma) \quad = \quad A_{H}(\sigma)
\end{equation}
i.e., due to equation (\ref{dcvlp}), at the level of the
statistical ensemble any modification of $L_{I}(x)$ in a spacetime
region $C$ cannot cause physical changes in regions which are
spacelike separated from it. We stress that this conclusion is
true for the case of non hermitian coupling as well as for the
case of hermitian coupling, even though the argument was carried
out in terms of the hermitian coupling alone, as it depends solely
upon the consideration of the statistical operators which are
identical for both cases.

\subsubsection{Individual Level} \label{sec1022}

>From the result (\ref{rccgm}) it is also evident that, in the case
of hermitian coupling [i.e. for (\ref{etsr})] a variation of the
Lagrangian density $L_{I}(x)$ in a region $C$ has no effect on the
mean value of any local observable with support which is spacelike
separated from $C$, even at the level of an individual system
(i.e., for any realization of the stochastic potential). This
property is related to the fact that, in this case, no Heisenberg
reduction takes place.

The situation is quite different in the case of a non hermitian
coupling. In fact, let us consider equation (\ref{qeqg}) and the
operator $S_{V}(\sigma, \sigma_{0})$ given by (\ref{fqfl}). The
mean value of a local observable $A_{I}(\sigma)$ is then
\begin{eqnarray}
\langle A_{I}(\sigma) \rangle & = & \frac{\langle
\psi_{V}(\sigma)| A_{I}(\sigma) | \psi_{V}(\sigma) \rangle}{
\| |\psi_{V}(\sigma)\rangle \|^{2}} \nonumber \\
& = & \frac{\langle \psi(\sigma_{0})| S_{V}^{\dagger}(\sigma,
\sigma_{0}) A_{I}(\sigma) S_{V}(\sigma, \sigma_{0}) |
\psi(\sigma_{0}) \rangle}{ \| S_{V}(\sigma, \sigma_{0})
|\psi(\sigma_{0})\rangle \|^{2}}
\end{eqnarray}
We now replace in (\ref{qeqg}) $L_{I}(x)$ by $L_{I}(x) + \Delta
L_{I}(x)$, $\Delta L_{I}(x)$ being different from zero only for $x
\in C$, and we denote by $S_{V}^{\Delta}(\sigma, \sigma_{0})$ the
corresponding evolution operator. The mean value $\langle
A_{I}^{\Delta}(\sigma) \rangle$ of the same local observable, for
the same initial condition, is now
\begin{equation}
\langle A_{I}^{\Delta}(\sigma) \rangle = \frac{\langle
\psi(\sigma_{0})| S_{V}^{\Delta\dagger}(\sigma, \sigma_{0})
A_{I}(\sigma) S_{V}^{\Delta}(\sigma, \sigma_{0}) |
\psi(\sigma_{0}) \rangle}{ \| S_{V}^{\Delta}(\sigma, \sigma_{0})
|\psi(\sigma_{0})\rangle \|^{2}}.
\end{equation}
Note that in general
\begin{equation}
\langle A_{I}^{\Delta}(\sigma) \rangle \quad \neq \quad \langle
A_{I}(\sigma) \rangle
\end{equation}
in spite of the fact that $[\Delta L_{I}(x), A_{I}(\sigma)] = 0$,
$\forall\, x$.

\subsubsection{Mean values of local observables and oddities in
relativistic reduction models} \label{sec1023}

Let us consider a physical system satisfying the initial condition
$|\psi(\sigma_{0})\rangle = |\psi_{0}\rangle$ on the spacelike
surface $\sigma_{0}$, the local observable $A$, and two arbitrary
spacelike surfaces $\sigma_{1}$ and $\sigma_{2}$ coinciding on the
support $\alpha$ of $A$ (see figure 7).
\begin{center}
\begin{picture}(240,150)(0,0)
\put(0,0){\line(1,0){240}} \put(0,150){\line(1,0){240}}
\put(0,0){\line(0,1){150}} \put(240,0){\line(0,1){150}}
\put(20,20){\line(1,0){190}} \put(220,18){$\sigma_{0}$}
\thicklines \qbezier(20,60)(40,70)(60,60)
\qbezier(60,60)(70,55)(90,60) \qbezier(90,60)(130,70)(160,90)
\qbezier(160,90)(180,103.3)(210,110)
\qbezier(90,60)(120,67.5)(140,70) \qbezier(140,70)(180,75)(210,90)
\thinlines \put(220,88){$\sigma_{1}$} \put(220,108){$\sigma_{2}$}
\put(30,55){\line(0,1){15}} \put(80,50){\line(0,1){15}}
\put(50,45){$\alpha$}
\end{picture}

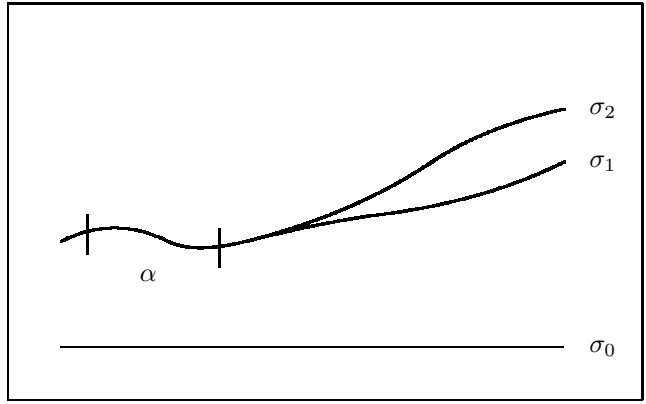
\vspace{0.2cm} \footnotesize \parbox{3.3in}{Figure 7: The
spacelike surfaces $\sigma_{1}$ and $\sigma_{2}$ coinciding on the
support $\alpha$ of local observable $A$.} \normalsize
\end{center} \vspace{0.5cm}
When the dynamics (\ref{etsr}) ruled by a hermitian interaction is
considered, for any given realization of the stochastic potential,
as is well known, the mean value of $A$ in the state
$|\psi(\sigma_{1})\rangle$ coincides with the one in the state
$|\psi(\sigma_{2})\rangle$. It  follows, at the individual level
and for the case of hermitian coupling and, as a consequence, at
the ensemble level for both cases of hermitian and skew--hermitian
coupling, that the mean value of a local observable does not
depend on the particular spacelike surface which one chooses among
all those coinciding on its support (and therefore on the specific
statevector $|\psi(\sigma_{1})\rangle$ or
$|\psi(\sigma_{2})\rangle$ which describes the physical situation
concerning the two considered surfaces). Incidentally, this
represents a different proof that also in the case of dynamical
reduction models, at the ensemble level, one can consistently
define, as in standard quantum field theory, local observables.

Again, the situation at the individual level is quite different in
the skew--hermitian case. In fact, for a given realization of the
stochastic potential, one has
\begin{equation}
\frac{\langle \psi_{V}(\sigma_{2})| A | \psi_{V}(\sigma_{2})
\rangle}{\| |\psi_{V}(\sigma_{2})\rangle \|^{2}} \; = \;
\frac{\langle \psi_{V}(\sigma_{1})| S_{V}^{\dagger}(\sigma_{2},
\sigma_{1}) A S_{V}(\sigma_{2}, \sigma_{1}) | \psi_{V}(\sigma_{1})
\rangle}{\| S_{V}(\sigma_{2}, \sigma_{1})
|\psi_{V}(\sigma_{1})\rangle \|^{2}}
\end{equation}
which, in general, is different from $\langle
\psi_{V}(\sigma_{1})| A | \psi_{V}(\sigma_{1}) \rangle/\|
|\psi_{V}(\sigma_{1})\rangle \|^{2}$ even though the spacetime
region spanned in tilting $\sigma_{1}$ into $\sigma_{2}$ is
spacelike separated from the support $\alpha$ of $A$, and,
consequently
\begin{equation}
[A, S_{V}(\sigma_{2}, \sigma_{1})] \quad = \quad 0.
\end{equation}
This dependence, at the individual level, of the mean value of a
local observable upon the spacelike surface (among those
coinciding on the support) over which it is evaluated is not {\it
per se} a difficulty of the theory. It becomes, however, a
difficulty if one wishes to claim that such a mean value
corresponds to an objective property of an individual system.

Before facing this problem (see next subsection), a deeper
analysis of the implications of relativistic reduction models for
microscopic [case (a) below] and macroscopic [case (b)] systems is
necessary.
\\ \\
\noindent {\bf Case (a).} Let us start by reconsidering the case
(subsection \ref{sec1013}) of a microscopic system coupled to a
macroscopic one which acts as a ``measuring apparatus" in the
sense of dynamical reduction models. Let $A_{1}$ and $A_{2}$ be
two local observables of the microsystem whose supports
$\alpha_{1}$ and $\alpha_{2}$ are spacelike separated, and suppose
the macroscopic system is devised to measure $A_{1}$. For our
purposes we can ignore the hamiltonian evolution for the operators
and we consider the Tomonaga--Schwinger evolution equation of the
statevector, for a specific realization of the stochastic
potential
\begin{equation} \label{qeqgbis}
\frac{\delta |\psi_{V}(\sigma)\rangle}{\delta\sigma(x)} \quad =
\quad [i L_{1-S}(x) \; + \; L_{I}(x)V(x)\; - \; \lambda\,
L_{I}^{2}(x)] |\psi_{V}(\sigma)\rangle.
\end{equation}
Here $L_{1-S}(x)$ (describing the local system--apparatus
interaction) and $L_{I}(x)$ may be taken as different from zero
only in a spacetime region $C$ which is spacelike with respect to
$\alpha_{2}$ (see figure 8).
\begin{center}
\begin{picture}(240,150)(0,0)
\put(0,0){\line(1,0){240}} \put(0,150){\line(1,0){240}}
\put(0,0){\line(0,1){150}} \put(240,0){\line(0,1){150}}
\put(20,20){\line(1,0){190}} \put(220,18){$\sigma_{0}$}
\put(20,60){\line(1,0){190}} \put(220,58){$\sigma_{1}$}
\put(20,100){\line(1,0){190}} \put(220,98){$\sigma_{2}$}
\put(60,55){\line(0,1){10}} \put(90,55){\line(0,1){10}}
\put(160,55){\line(0,1){10}} \put(190,55){\line(0,1){10}}
\put(70,50){$\alpha_{2}$} \put(170,50){$\alpha_{1}$}
\put(95,60){\line(2,1){115}} \put(55,60){\line(-2,-1){35}}
\put(220,123){$\tilde\sigma_{2}$}
\put(180,78){\oval(19,15)} \put(177,75){$C$}
\end{picture}

\vspace{0.2cm} \footnotesize \parbox{3.3in}{Figure 8: A
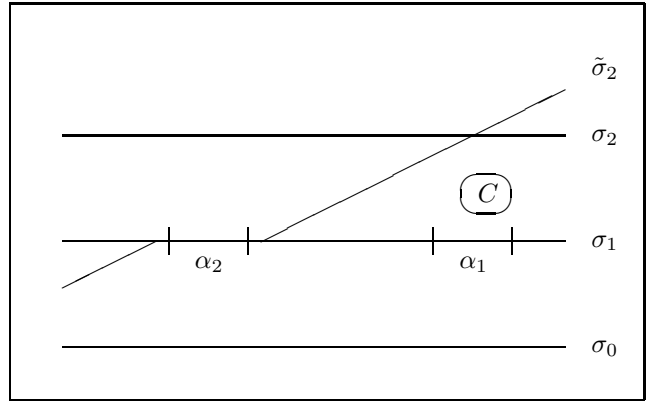
macroscopic apparatus measures local observable $A_{1}$ in
spacetime region $C$. $A_{1}$'s support $\alpha_{1}$ is spacelike
separated with respect to $\alpha_{2}$, the support of another
local observable $A_{2}$.} \normalsize
\end{center} \vspace{0.5cm}
Let us assume that the local observables $A_{1}$ and $A_{2}$ have
a purely point spectrum with eigenvalues 0 and 1, and let us
consider the initial state
\begin{equation}
|\psi(\sigma_{0})\rangle \quad = \quad \frac{1}{\sqrt{2}}\, [
|\psi_{1}\rangle \; + \; |\psi_{2}\rangle ]\, |\chi_{1}\rangle
\end{equation}
with
\begin{equation}
A_{i}\, |\psi_{j}\rangle \quad = \quad \delta_{ij}\,
|\psi_{j}\rangle, \qquad\quad i, j = 1, 2
\end{equation}
$|\chi_{1}\rangle$ being the untriggered apparatus state. Let us
furthermore assume that the particular realization of the
stochastic potential $V(x)$ is one of those ``yielding the result
1 for the measurement of $A_{1}$". The situation is then the
following:
\begin{enumerate}
\item The state associated to $\sigma_{0}$ and $\sigma_{1}$ is
$|\psi(\sigma_{0})\rangle$.
\item The state associated to $\sigma_{2}$ is ($N$ being a
normalization factor)
\begin{equation} \qquad\quad
|\psi_{V}(\sigma_{2})\rangle \; = \; \frac{1}{N}\,
e^{\displaystyle \int_{\sigma_{1}}^{\sigma_{2}} dx\, [i L_{1-S}(x)
\; + \; L_{I}(x)V(x)\; - \; \lambda\, L_{I}^{2}(x)]}
|\psi(\sigma_{0})\rangle
\end{equation}
which, under the assumptions which have been made, is
approximately an eigenstate of $A_{2}$ pertaining to the
eigenvalue zero.
\item The state associated to $\sigma_{2}^{\sim}$ is also
$|\psi_{V}(\sigma_{2})\rangle$.
\end{enumerate}
Indeed, the relativistic CSL dynamics considered in section
\ref{sec96} is such that, when a spacelike hypersurface crosses
the region $C$ toward the future, no matter what is the behaviour
in regions far apart from $C$, the statevector associated to this
hypersurface collapses to the eigenstate of $A_{1}$ corresponding
to the eigenvalue which has been found.

Looking at the problem from the point of view of the evolution
from $\sigma_{1}$ to $\sigma_{2}$, one could be tempted to say
that, since the mean value of $A_{2}$ has become practically zero
as a consequence of the ``measurement" in the spacetime region
$C$, an element of physical reality associated with $A_{2}$ has
emerged. This is a nonlocal effect of the type of those occurring
in an EPR setup.

However, one must realize that the same change of the mean value
of $A_{2}$ occurs when one considers the Tomonaga--Schwinger
evolution from $\sigma_{1}$ to $\sigma_{2}^{\sim}$, in accordance
with point 3. This gives rise to an ambiguity in the mean value of
$A_{2}$, i.e., in a quantity that, when the support $\alpha_{2}$
shrinks to zero, refers to a unique objective spacetime point.
This is not surprising; it corresponds simply to the emergence,
within the relativistic reducing dynamics, of the aspects
discussed in subsection \ref{sec1013} for the standard quantum
theory with the reduction postulate. In fact, one can remark that
$\sigma_{1}^{\sim}$ can be approximately identified with a $t'$ =
const hyperplane for a boosted observer for which the interaction
with the macro--object has already taken place\footnote{The
bending of the surface at the left of $\alpha_{2}$ shown in figure
5 is allowed since, under the assumptions we have made, $L_{I}(x)
= 0$ in that region.}.
\\ \\
\noindent {\bf Case (b).} Let us discuss now the same problem for
macroscopic systems. We consider a situation analogous to the
previous one but in which there are two macroscopic systems
performing measurements of the observables $A_{1}$ and $A_{2}$.
The initial condition is given by assigning to the surface
$\sigma_{0}$ the state:
\begin{equation}
|\psi(\sigma_{0})\rangle \quad = \quad \frac{1}{\sqrt{2}}\, [
|\psi_{1}\rangle \; + \; |\psi_{2}\rangle ]\, |\chi_{1}\rangle\,
|\chi_{2}\rangle
\end{equation}
where $|\chi_{1}\rangle$ and $|\chi_{2}\rangle$ refer to the
untriggered apparatuses. The evolution equation, with the usual
approximation, is now
\begin{eqnarray} \label{qeqgter}
\frac{\delta |\psi_{V}(\sigma)\rangle}{\delta\sigma(x)} & = & [i
L_{1-S}(x) \; + \; i L_{2-S}(x) \; + \; L_{I1}(x)V(x)
\; + \; L_{I2}(x)V(x) \nonumber \\
& & \; - \; \lambda\, L_{I1}^{2}(x) \; - \; \lambda\,
L_{I2}^{2}(x) ] |\psi_{V}(\sigma)\rangle.
\end{eqnarray}
where the meaning of the symbols is obvious. To clearly define the
situation from the physical point of view, we assume that the time
which is necessary in order that the microsystem triggers the
apparatus is sensibly shorter than the typical reduction time for
the apparatus. This means that in the above equation we can
consider $L_{1-S}(x)$ and $L_{2-S}(x)$ to be different from zero
only in the regions $C_{1}$ and $B_{1}$, respectively, and
$L_{I1}(x)$ and $L_{I2}(x)$ in the regions $C_{2}$ and $B_{2}$,
respectively, as shown in figure 9.
\begin{center}
\begin{picture}(240,150)(0,0)
\put(0,0){\line(1,0){240}} \put(0,150){\line(1,0){240}}
\put(0,0){\line(0,1){150}} \put(240,0){\line(0,1){150}}
\put(15,25){\line(1,0){190}} \put(210,23){$\sigma(t_{0})$}
\put(15,47){\line(1,0){190}} \put(210,45){$\sigma(t_{1})$}
\put(15,69){\line(1,0){190}} \put(210,67){$\sigma(t_{2})$}
\put(205,91){\line(-2,-1){132}}
\put(210,89){$\tilde\sigma(t_{0})$}
\put(205,113){\line(-2,-1){132}}
\put(210,111){$\tilde\sigma(t_{1})$}
\put(205,135){\line(-2,-1){132}}
\put(210,133){$\tilde\sigma(t_{2})$}
\put(25,20){\line(0,1){10}} \put(55,20){\line(0,1){10}}
\put(36,15){$\alpha_{2}$} \put(165,20){\line(0,1){10}}
\put(195,20){\line(0,1){10}} \put(176,15){$\alpha_{1}$}
\put(40,36){\oval(19,15)} \put(35,33){$B_{1}$}
\put(40,58){\oval(19,15)} \put(35,55){$B_{2}$}
\put(180,36){\oval(19,15)} \put(175,33){$C_{1}$}
\put(180,58){\oval(19,15)} \put(175,55){$C_{2}$}
\end{picture}

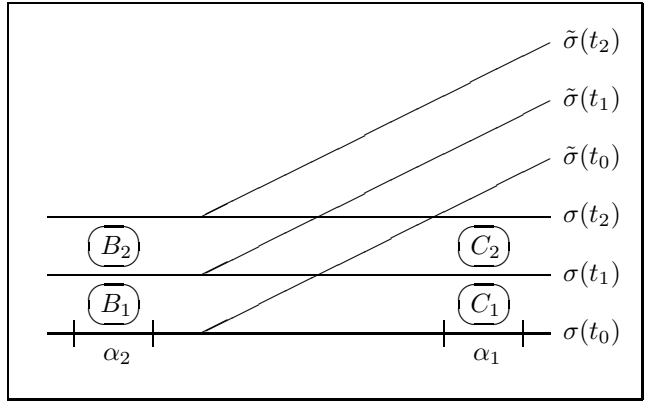
\vspace{0.2cm} \footnotesize \parbox{3.3in}{Figure 9: Measurements
take place in $C_{1}$ and $B_{1}$, followed by reduction dynamics
in $C_{2}$ and $B_{2}$, of local observables $A_{1}$ and $A_{2}$,
respectively.} \normalsize
\end{center} \vspace{0.5cm}
Let us also assume that the specific realization of the stochastic
potential is one leading to the value 1 for $A_{1}$. We are
interested in discussing the states of the macro--system used to
measure $A_{2}$ and the mean values of its observables on various
hypersurfaces. In particular, let $A_{2}^{\sim}$ be the observable
of the apparatus corresponding to the yes--no experiment asking
whether the result 0 has been found in a measurement of $A_{2}$.
We consider $t$ = const hypersurfaces $\sigma(t)$ and also the
bent hypersurfaces $\sigma^{\sim}(t)$ containing the spatial
support of $A_{2}^{\sim}$ at time $t$ (see figure 9). The
situation can now be summarized as follows:
\begin{enumerate}
\item For $t < t_{0}$ the state associated to any surface $\sigma(t)$
or $\sigma^{\sim}(t)$ has always the form of a factorized state;
one of the factors refers to the apparatus 2 and it is
$|\chi_{2}\rangle$. Note that what changes in going from
$\sigma(t)$ to $\sigma^{\sim}(t)$ is the state of the system +
apparatus 1.

\item For $t = t_{1}$ the state associated to $\sigma(t_{1})$ is
\begin{equation} \label{edli}
\qquad\quad |\psi_{V}(\sigma(t_{1}))\rangle \; = \;
\frac{1}{\sqrt{2}}\, [|\psi_{1}\rangle\,
|\chi_{1}^{1}\rangle\,|\chi_{2}^{0}\rangle \; + \;
|\psi_{2}\rangle\, |\chi_{1}^{0}\rangle\, |\chi_{2}^{1}\rangle],
\end{equation}
where, obviously, the superscripts identify the states of the
macroscopic apparatuses which have been triggered by the
interaction with the microsystem, these states being labeled by
the eigenvalues which have been found.

>From (\ref{edli}) one sees that the state
$|\psi_{V}(\sigma(t_{1}))\rangle$ is not a factorized state and as
a consequence it cannot be an eigenstate of the relevant
observable of apparatus 2. In particular, the mean value of
$A_{2}^{\sim}$ in the state (\ref{edli}) is 1/2.

However, it is important to remark that the state to be associated
to the surface $\sigma^{\sim}(t_{1})$ of figure 6 is, for the
particular realization of the stochastic potential,
\begin{equation}
\qquad\quad |\psi_{V}(\sigma^{\sim}(t_{1}))\rangle \quad \simeq
|\psi_{1}\rangle\, |\chi_{1}^{1}\rangle\,|\chi_{2}^{0}\rangle.
\end{equation}
This state is factorized and it is an eigenstate of
$A_{2}^{\sim}$.
\item The state to be associated with any surface $\sigma(t)$ and
$\sigma^{\sim}(t)$ when $t > t_{2}$, is once more a factorized
state with the factor $|\chi_{2}^{0}\rangle$ for the apparatus 2.
\end{enumerate}
The conclusion is that, even though the dependence of the mean
value of a local observable upon the spacelike surface on which it
is evaluated is present also in the case of macro--objects, this
dependence occurs only for a time interval of the order of the one
which is necessary for the reduction to take place\footnote{Here the
argument has been presented with reference to the evolution from one
space-like surface lying below the regions $C_{1}$ and/or $C_{2}$ to one
which has ``crossed" these regions. Obviously analogous considerations
hold with reference to the regions $B_{1}$ and $B_{2}$ and to the
apparatus which is present there.}.

\subsubsection{Objective properties of micro and macroscopic systems}
\label{sec1024}

We started this subsection \ref{sec102} by relating the
possibility of attributing objective properties to individual
systems to requirement (\ref{cno}) being satisfied to an extremely
high degree of accuracy. In the relativistic case, however, as
shown with great detail in the previous subsection, the mean value
of a projection operator associated to a local observable is
affected by an ambiguity depending on the spacelike surface used
to evaluate it, and, under specific circumstances, by changing the
surface its value can vary from, for example, $1/2$ to almost
exactly 1. This shows that the above definition of objective
properties for individual systems is inadequate, and must be made
more precise.

In accordance with the discussion of Section \ref{sec012} we think that
the appropriate attitude is the following: when considering a local
observable $A$ with its associated support we say that an individual
system has the accessible property $a$ ($a$ being an eigenvalue of $A$),
only when the mean value of $P_{a}$ is extremely close to one, when
evaluated on all spacelike hypersurfaces containing the support of $A$.

Thus, according to this prescription, one cannot attribute an
objective property to an individual system when there is an
appreciable dependence of the mean value of the local observable
upon the surface used to evaluate it.

Let us analyze the implications of this attitude in the cases of
microscopic and macroscopic systems. For a microsystem, with
reference to case (a) of the previous subsection, we observe that
no objective property corresponding to a local observable can
emerge as a consequence of a ``measurement process" performed in a
region which is spacelike separated from the support of the
considered observable. This does not mean that microsystems cannot
acquire objective local properties as a consequence of a
measurement performed in another spacetime region; in fact, with
reference to the discussion in (a) and to an EPR--Bohm--like setup
one can remark that if one considers the spin component of
particle 2, when the particle is in the future of the region in
which the spin of particle 1 has been measured, then one can
attribute to particle 2 the objective local property of having its
spin ``up" or ``down" and that such a property has emerged just
due to the measurement which has been performed.

We wish to emphasize again that in the case of macrosystems the
discussion under (b) has shown that the impossibility of
associating local properties to them lasts only for a time
interval of the order of the one which is necessary for the
``spontaneous dynamical reduction" to take place. In fact, before
macroapparatus 2 interacts with the microsystem the state of the
apparatus is obviously well defined and corresponds to the
untriggered state, independently of the considered surface. After
the reduction ensuing from the interaction of the microsystem with
it, apparatus 2 is again in a well--defined state, corresponding
to the result which it has registered. Moreover, this result is
``correctly" correlated to the result registered by apparatus
1\footnote{Perhaps it is worth noticing that it would be possible
to give another covariant prescription for the attribution of
objective local properties to physical systems. More precisely one
could, for any local observable $A$, consider the mean value of
the projection operator $P_{a}$ on one of $A$'s eigenmanifolds
evaluated for the state vector associated to the surface which
delimits the future light cone of the support of $A$. Then, if
this mean value is extremely close to 1, one asserts that the
system has the objective property $a$. This is quite different
from the previously considered criterion (i.e., that the mean
value be extremely close to one on all hypersurfaces containing
the support of $A$) and would, in case (a) of the previous
subsection, lead to the assignment of the objective property
corresponding to the value zero for the observable $A_{2}$ to the
microsystem, contrary to what would occur by the adoption of the
previous criterion.

This attitude would correspond to the following particular
interpretation, at the relativistic level, of the EPR criterion
for elements of physical reality: ``if there exists {\it at least
one} observer who can {\it predict}, almost (in the above
specified sense) with certainty and without disturbing a system in
any way, the value of a physical quantity, then there exists an
element of physical reality corresponding to that quantity".

We do not want to enter here into a detailed discussion of the
conceptual implications involved in adopting the above
prescription. We believe that they lead to some conceptual
difficulties in connection with the cause--effect relation. This
is not surprising since the considered prescription is analogous,
in the present context, to the Hellwig--Kraus \cite{hkd} postulate
about wavepacket reduction. For these reasons we drop the
criterion considered in this footnote.}.

In conclusion, relativistic dynamical reduction models, together
with the prescription for the attribution of objective properties
to physical systems proposed in this section, allows one to
overcome the difficulties discussed in subsection \ref{sec1013}.
The theory assigns a statevector to any spacelike hypersurface,
and the dependence, at the individual level, of the mean value of
a local observable upon the specific spacelike surface used to
evaluate it, does not constitute a difficulty. It simply requires
a precise and appropriate criterion for relating the objective
properties of a physical system to the mean values of local
observables: in particular, this criterion permits practically
always the attribution of objective local properties to
macro--objects, at the individual level. In a sense, the above
analysis has proven once more that dynamical reduction models meet
the requirement put forward by J. S. Bell \cite{bellns} for an
exact and serious formulation of quantum mechanics, i.e., that it
should ``allow electrons to enjoy the cloudiness of waves, while
allowing tables and chairs, and ourselves, and black marks on
photographs, to be rather definitely in one place rather than
another, and to be described in classical terms".

\subsection{Parameter dependence in dynamical reduction models}
\label{sec011}

As is well known, the locality assumption needed to prove Bell's
theorem \cite{bell71} is equivalent to the conjunction of two
other assumptions, viz., in Shimony's terminology, parameter
independence and outcome independence \cite{sz,vf82,ja84,shi1}; in
view of the experimental violation of the Bell inequality, one has
to give up either or both of these assumptions. We now analyze
these issues within the framework of dynamical reduction models.

To start with, let us fix our notation. We will denote by
$\lambda$ all parameters (which may include the quantum mechanical
statevector or even reduce to it alone) that completely specify
the state of an individual physical system. For simplicity we will
refer to a standard EPR--Bohm setup and we will denote by
\begin{equation}
p^{LR}_{\lambda}(x,y;{\bf n}, {\bf m})
\end{equation}
the joint probability of getting the outcome $x$ ($x = \pm 1$) in
a measurement of the spin component along ${\bf n}$ at the left
($L$) and $y$ ($y = \pm 1$) in a measurement of the spin component
along ${\bf m}$ at the right ($R$) wing of the apparatus. We
assume that the experimenter at $L$ can make a free--will choice
of the direction ${\bf n}$; and similarly for the experimenter at
$R$ and the direction ${\bf m}$. Both experimenters can also
choose not to perform the measurement. Finally, we assume that the
micro--macro interactions taking place at $L$ and $R$ that trigger
the reduction are governed by appropriate coupling constants
$g_{L}$ and $g_{R}$; in particular, the situations in which one of
the coupling constants is made equal to zero corresponds to no
measurement being performed.

Bell's locality assumption can be expressed as
\begin{equation} \label{bla}
p^{LR}_{\lambda}(x,y;{\bf n}, {\bf m}) \quad = \quad
p^{L}_{\lambda}(x;{\bf n}, *)\, p^{R}_{\lambda}(y; *, {\bf m})
\end{equation}
where the symbol $*$ appearing on the right hand side denotes that
the corresponding measurement is not performed. Condition
(\ref{bla}) has been shown \cite{ja84,shi1} to be equivalent to
the conjunction of two logically independent conditions:
\begin{eqnarray} \label{pi}
p^{L}_{\lambda}(x;{\bf n}, {\bf m}) & = & p^{L}_{\lambda}(x;{\bf
n}, *)
\nonumber \\ & &  \\
p^{R}_{\lambda}(y;{\bf n}, {\bf m}) & = & p^{R}_{\lambda}(y; *,
{\bf m}) \nonumber
\end{eqnarray}
and
\begin{equation} \label{oi}
p^{LR}_{\lambda}(x,y;{\bf n}, {\bf m}) \; = \;
p^{L}_{\lambda}(x;{\bf n}, {\bf m})\, p^{R}_{\lambda}(y;{\bf n},
{\bf m})
\end{equation}
where we have denoted, e.g., by the symbol $p^{L}_{\lambda}(x;{\bf
n}, {\bf m})$ the probability of getting, for the given settings
${\bf n}$, ${\bf m}$, the outcome $x$ at $L$.

Conditions (\ref{pi}) express {\bf parameter independence}, i.e.,
the requirement that the probability of getting an outcome at $L$
($R$) is independent of the setting chosen at $R$ ($L$), while
equation (\ref{oi}) ({\bf outcome independence}) expresses the
requirement that the probability of an outcome at one wing does
not depend on the outcome obtained at the other wing.

\subsubsection{The case of the nonlinear CSL Model}
\label{sec1131}

To simplify the discussion, we assume that the initial state
$|\psi(0)\rangle$ is the singlet state and we confine our
attention to the case in which both spin measurements are in the
same direction, i.e., ${\bf n} = {\bf m}$. We assume that the
measurement at $R$, if it takes place (i.e., if $g_{R} \neq 0$),
occurs at an earlier time than the one at $L$.

Consider now the realizations $\tilde w_{L}({\bf x}, t)$ of
$w_{L}({\bf x}, t)$ that give rise to the outcome $+ 1$ for the
left apparatus when it is triggered by $|\psi(0)\rangle$. The
probability of occurrence of such processes is $1/2$. We will
denote by $p^{L}_{|\psi(0)\rangle}(-1; g_{R} = 0|w_{L})$ and
$p^{L}_{|\psi(0)\rangle}(-1; g_{R} \neq 0|w_{L})$ the conditional
probability, given $w_{L}$, of the outcome $-1$ at left when the
initial state is $|\psi(0)\rangle$ and the $R$ apparatus is
switched off or on, respectively. We then have
\begin{equation}
p^{L}_{|\psi(0)\rangle}(-1; g_{R} = 0|\tilde w_{L}) \quad = \quad
0.
\end{equation}
We now evaluate the probability $p^{L}_{|\psi(0)\rangle}(-1; g_{R}
\neq 0|\tilde w_{L})$. Since $g_{R} \neq 0$ and the measurement at
$R$ occurs before the one at $L$, we have to take into account the
possible realizations of the stochastic process at $R$. Let us
consider the realizations $\tilde w_{R}({\bf x}, t)$ of
$w_{R}({\bf x}, t)$ that, when triggered by the singlet state,
yield the outcome $+ 1$ at $R$. When one of these processes
$\tilde w_{R}$ occurs, the outcome at $L$ turns out to be $-1$
irrespective of the particular realization of the stochastic
process $w_{L}$ and therefore also for all processes $\tilde
w_{L}$ considered above. To understand this, recall that within
the nonlinear model, the same stochastic process at $L$ can give
rise to different outcomes, depending on the statevector which
triggers the apparatus at $L$. As a consequence one has
\begin{equation}
p^{L}_{|\psi(0)\rangle}(-1; g_{R} \neq 0|\tilde w_{L} \& \tilde
w_{R}) \quad = \quad 1.
\end{equation}
Since the probability of occurrence of a process $\tilde w_{R}$ is
equal to $1/2$ and is independent of the particular realization
$\tilde w_{L}$, and since if $w_{R}$ is not one of the $\tilde
w_{R}$, then outcome $-1$ on the left cannot occur (barring
improbable exceptions), one has
\begin{equation}
p^{L}_{|\psi(0)\rangle}(-1; g_{R} \neq 0|\tilde w_{L}) \quad =
\quad 1/2.
\end{equation}
We stress that the difference of the probabilities is appreciable,
\begin{equation}
0\; = \; p^{L}_{|\psi(0)\rangle}(-1; g_{R} = 0|\tilde w_{L}) \;
\neq \;  p^{L}_{|\psi(0)\rangle}(-1; g_{R} \neq 0|\tilde w_{L})\;
= \; 1/2
\end{equation}
and that the probability of occurrence of these realizations
$\tilde w_{L}$ is also appreciable ($= 1/2$). Thus the nonlinear
CSL model exhibits parameter dependence.

\subsubsection{The case of the linear CSL model} \label{sec1132}

For the linear model, we can easily solve the evolution equation,
and thereby show parameter independence in the $t \rightarrow
\infty$ limit, once we simplify the description by considering
only the spin Hilbert space.

Thus one has, in the case in which both apparatuses are switched
on ($g_{R} \neq 0$ and $g_{L} \neq 0$), a linear dynamical
equation analogous to (\ref{smefsa}):
\begin{equation} \label{wacy}
\frac{d|\psi_{w_{L},w_{R}}(t)\rangle}{dt} \; = \; \left\{ [ ({\bf
\sigma}^{L}\cdot {\bf n})w_{L}(t) - \gamma] + [ ({\bf
\sigma}^{R}\cdot {\bf m})w_{R}(t) - \gamma] \right\}
|\psi_{w_{L},w_{R}}(t)\rangle
\end{equation}
with
\begin{equation}
\llangle w_{L,R}(t)\rrangle \; = \; 0 \qquad\quad \llangle
w_{L}(t)\, w_{R}(t')\rrangle \; = \; \gamma\, \delta_{L,R}\,
\delta(t - t').
\end{equation}
The probability distribution of the stochastic processes is
obtained through the cooking procedure. To compare this case with
the one in which $g_{R} = 0$, one has to consider another
stochastic equation, i.e.,
\begin{equation} \label{ycaw}
\frac{d|\psi_{w_{L}}(t)\rangle}{dt} \; = \; \left\{ ({\bf
\sigma}^{L}\cdot {\bf n})w_{L}(t) - \gamma  \right\}
|\psi_{w_{L}}(t)\rangle
\end{equation}
The solutions of equations (\ref{wacy}) and (\ref{ycaw}) at time
$t$ for the same initial conditions are
\begin{equation} \label{etpo}
|\psi_{B_{L}, B_{R}}(t)\rangle \quad = \quad e^{\displaystyle
F_{L\, B_{L}}(t)} e^{\displaystyle F_{R\, B_{R}}(t)}
|\psi(0)\rangle
\end{equation}
and
\begin{equation} \label{etpo2}
|\psi_{B_{L}}(t)\rangle \quad = \quad e^{\displaystyle F_{L\,
B_{L}}(t)} |\psi(0)\rangle
\end{equation}
respectively\footnote{In equation (\ref{etpo}) and following, we
change notation for the same reason as we did in equation
(\ref{ccv}).}. In equations (\ref{etpo}) and (\ref{etpo2}) we have
put
\begin{equation}
F_{L\, B_{L}}(t) \; = \; {\bf \sigma}^{L}\cdot{\bf n}\, B_{L}(t) -
\gamma t, \qquad F_{R\, B_{R}}(t) \; = \; {\bf
\sigma}^{R}\cdot{\bf n}\, B_{R}(t) - \gamma t,
\end{equation}
where
\begin{equation}
B_{L}(t) \; = \; \int_{0}^{t}d\tau\, w_{L}(\tau), \qquad\quad
B_{R}(t) \; = \; \int_{0}^{t}d\tau\, w_{R}(\tau).
\end{equation}
We come back now to equation (\ref{wacy}) and we evaluate the
cooked probability density of occurrence of the Brownian processes
$B_{L}(t)$ and $B_{R}(t)$ by multiplying the raw probability
density by the square of the norm of the statevector (\ref{etpo}).
As usual we have
\begin{equation} \label{etpdd}
P_{\makebox{\tiny Cook}}[B_{L}(t)\,\&\, B_{R}(t)] \; = \;
P_{\makebox{\tiny Raw}}[B_{L}(t)\,\&\, B_{R}(t)]\, \|
|\psi_{B_{L}, B_{R}}(t)\rangle \|^{2}
\end{equation}
and
\begin{equation}
P_{\makebox{\tiny Raw}}[B_{L}(t)\,\&\, B_{R}(t)] \quad = \quad
P_{\makebox{\tiny Raw}}[B_{L}(t)]\, P_{\makebox{\tiny
Raw}}[B_{R}(t)].
\end{equation}
Taking into account equation (\ref{etpo}), one then gets from
(\ref{etpdd})
\begin{eqnarray}
P_{\makebox{\tiny Cook}}[B_{L}(t)\,\&\, B_{R}(t)] & = &
P_{\makebox{\tiny Raw}}[B_{L}(t)]\, P_{\makebox{\tiny
Raw}}[B_{R}(t)] \| |\psi_{B_{L}, B_{R}}(t)\rangle \|^{2} \; =
\nonumber \\ & = & P_{\makebox{\tiny Raw}}[B_{L}(t)]\, \|
e^{F_{L\, B_{L}}(t)}|\psi(0)\rangle \|^{2} \cdot \nonumber \\ & &
P_{\makebox{\tiny Raw}}[B_{R}(t)] \left\| \frac{ e^{F_{R\,
B_{R}}(t)} e^{F_{L\, B_{L}}(t)} |\psi(0)\rangle}{\| e^{F_{L\,
B_{L}}(t)}|\psi(0)\rangle\|} \right\|^{2}
\end{eqnarray}
Let us consider the marginal cooked probability density of
$B_{L}(t)$
\begin{eqnarray} \label{fqelg}
P^{\#}_{\makebox{\tiny Cook}}[B_{L}(t)] & = & \int {\mathcal
D}[B_{R}(t)]
P_{\makebox{\tiny Cook}}[B_{L}(t)\,\&\, B_{R}(t)] \nonumber \\
& = & P_{\makebox{\tiny Raw}}[B_{L}(t)]\, \| e^{F_{L\,
B_{L}}(t)}|\psi(0)\rangle \|^{2} \times \nonumber \\ & & \int
{\mathcal D}[B_{R}(t)] P_{\makebox{\tiny Raw}}[B_{R}(t)] \left\|
\frac{ e^{F_{R\, B_{R}}(t)} e^{F_{L\, B_{L}}(t)}
|\psi(0)\rangle}{\| e^{F_{L\, B_{L}}(t)}|\psi(0)\rangle\|}
\right\|^{2}. \qquad
\end{eqnarray}
Since the equation
\begin{equation}
\frac{d|\psi_{w_{R}}(t)\rangle}{dt} \; = \; \left\{ ({\bf
\sigma}^{R}\cdot {\bf n})w_{R}(t) - \gamma  \right\}
|\psi_{w_{R}}(t)\rangle
\end{equation}
preserves the stochastic average of the square of the norm of the
statevector, the last integral in equation (\ref{fqelg}) takes the
value 1. This means that $P^{\#}_{\makebox{\tiny Cook}}[B_{L}(t)]$
turns out to equal $P_{\makebox{\tiny Cook}}[B_{L}(t); *]$, i.e.,
the cooked probability density of occurrence of the Brownian
process $B_{L}(t)$ for the same initial condition if the process
were described by equation (\ref{ycaw}), i.e., if the apparatus at
$R$ were switched off.

But now recall from section \ref{sec64} that within linear CSL
there is a one--to--one correspondence between the outcome at left
(right) at $t = \infty$ and the specific value taken by the
Brownian process $B_{L}(t)$ [$B_{R}(t)$] for $t \rightarrow
\infty$. So the above proof that $P^{\#}_{\makebox{\tiny
Cook}}[B_{L}(t)]$ equals $P_{\makebox{\tiny Cook}}[B_{L}(t);*]$
amounts to a proof that linear CSL exhibits parameter independence
at the $t = \infty$ limit.

When one considers a {\it finite} time $t$ of the order of or
greater than the characteristic reduction time $\Delta t$, the
situation is more complicated: the one--to--one correspondence
between the outcomes and the values taken by the Brownian process
is only approximate (though valid to an extremely high degree of
accuracy). As a consequence, linear CSL does not enjoy strict
parameter independence at finite times. To clarify this point,
consider the values $B_{L}(t) = 2\gamma t$ and $B_{R}(t) = 4\gamma
t$ for the Brownian processes at time $t$. The cooked probabi1ity
density of occurrence of such values at the {\it finite} time $t$,
though extremely small, is not exactly zero. One can show
\cite{ngbf} that these values lead, through equation (\ref{etpo}),
to a statevector at $t$ which corresponds to the outcomes $+1$ at
right and $-1$ at left, respectively. On the other hand, for the
case in which $g_{R} = 0$, the substitution of $B_{L}(t) = 2
\gamma t$ in equation (\ref{etpo2}) leads, at time $t$, to a
statevector corresponding to the outcome $+1$ at left. Thus, there
are values of the Brownian process $B_{L}(t)$ for which the
outcome at left depends on whether $g_{R}$ is equal to zero or
not. Accordingly, there is parameter dependence at the level of
individual $B(t)$'s. However, given $B_{L}(t)$, this happens only
for values $B_{R}(t)$ of the Brownian process at right such that
the cooked conditional probability $P_{\makebox{\tiny
Cook}}[B_{R}(t)\,|\,B_{L}(t)]$ is extremely small. This in turn
implies that the model exhibits only negligibly small parameter
dependence effects.

To conclude, although the linear CSL model exhibits parameter
dependence at finite times, these effects are at any rate
extremely small with respect to those of the nonlinear CSL
model\footnote{Actually, explicit evaluations of such effects show
that they are characterized by probabilities which are smaller,
e.g., than those of classical thermodynamical processes which
violate the second law of thermodynamics.}.

\part{Dynamical Reduction Models and Experiments}

\section{Decoherence, quantum telegraph, proton decay, and
superconducting devices} \label{sec15}

Dynamical reduction models require a precise change of quantum
dynamics, so that they constitute a theory genuinely different
from standard quantum mechanics. It becomes then interesting to
analyze the conceptual and practical possibility of testing them
versus quantum mechanics.

In subsection \ref{sec151}, we analyze the role of decoherence in
experiments, and how it can mask the physical consequences of the
localization mechanism of dynamical reduction models. In
subsections \ref{sec152}, \ref{sec153} and  \ref{sec154} we
discuss three specific experiments, the quantum telegraph, the
nucleon decay and dissipation in superconducting devices
respectively, and the role played by dynamical reductions.

\subsection[Decoherence and dynamical reductions]{Decoherence and
the possibility of testing dynamical reductions} \label{sec151}

We have discussed in sections \ref{sec22} and \ref{sec4} that
decoherence --- i.e., the interaction between a given physical
system and the surrounding environment --- by itself does not
constitute a solution of the macro--objectification problem of
Quantum Mechanics, since it yields only an apparent collapse of
the wavefunction, not a real one. Nonetheless, decoherence effects
on quantum measurements are very important and often pose serious
limitations to the possibility of measuring specific properties of
physical systems, in particular to put into evidence the
superposition of different states of mesoscopic and macroscopic
systems, i.e. systems whose interaction with the environment is
more difficult to control.

As regards the possibility of testing dynamical reduction models
versus standard quantum mechanics, the role played by decoherence
is very tricky. In fact, in order to observe dynamical reductions,
experiments must be performed on quantum systems containing a
sufficiently large number of particles --- this is the case of
mesoscopic or macroscopic systems --- otherwise the reduction
mechanism would be ineffective for too a long time. On the other
hand, mesoscopic and macroscopic systems are very rapidly affected
by decoherence in such a way that, given a superposition of
different states, what would appear to be a spontaneous reduction
into one of such states might be attributed only to the
interactions with the surrounding environment. It is then
important to compare the ``reduction rates" and the physical
consequences of specific examples of decoherence mechanisms  with
those of QMSL and CSL, in order to understand whether there are
situations in which a possibly observed reduction process is
{\it real} --- thus confirming the predictions of dynamical
reduction models --- or only {\it apparent}, i.e. it is a result of
decoherence. Such a comparison has been presented in an
interesting  paper by Tegmark \cite{teg}, which we are going to
discuss.

In the just quoted paper the environment is felt by the physical system
of interest as a background noise due to the (instantaneous) scattering of
photons, neutrinos or air molecules off a system, the effect on
the compound initial state $\rho_{S+E}(t_{i})$ being determined by
a transition matrix $T$
\begin{equation}
\rho_{S+E}(t_{i}) \quad \longrightarrow \quad \rho_{S+E}(t_{f}) \;
= \; T\,\rho_{S+E}(t_{i})\, T^{\dagger}.
\end{equation}
Let ${\bf p}$, ${\bf k}$ be the momenta of the system and of a
background incident particle, respectively, and $a_{\bf p k}({\bf
q})$ the probability amplitude that the momentum transferred to
the system is ${\bf q}$. The author makes the following reasonable
assumptions  and approximations about the nature of the scattering
processes:
\begin{enumerate}
\item Conservation of energy and momentum:
\[
\langle {\bf p}', {\bf k}'|T|{\bf p}, {\bf k} \rangle \quad =
\quad \delta({\bf p}' + {\bf k}' - {\bf p} - {\bf k})\, a_{\bf p
k}({\bf p}' - {\bf p}).
\]
\item Independence of $a_{\bf p k}({\bf q})$ from the motion of
the system due to the high velocity of the incident particle:
\[
a_{\bf p k}({\bf q}) \quad = \quad a_{\bf k}({\bf q}).
\]
\item The system is supposed to be exposed to a constant particle
flux $\Phi$ per unit area and unit time scattered off with a total
scattering cross section $\sigma$ and a temporal distribution
modeled by a Poisson process with intensity $\Lambda = \sigma
\Phi$. Furthermore, the background incident particles are supposed
to be in momentum eigenstates or incoherent mixtures of them with
probability momentum distribution $\mu({\bf k})$. \item The
momentum of the incident particles is isotropically distributed
with
\begin{equation}
\mu({\bf k}) \quad = \quad \frac{1}{4\pi{\bf k}^{2}}\,
\lambda_{0}\nu(\lambda_{0}|{\bf k}|),
\end{equation}
where $\lambda_{0}$ is a typical wavelength of the hitting
particle and $\nu(x)$ is a probability distribution on the
positive real axis.
\end{enumerate}

If $\rho_{S}(t_{i})$ and $\rho_{S}(t_{f})$are the initial (before)
and final (after a scattering process) states of the system
obtained by tracing out the environmental  degrees of freedom, and
\begin{eqnarray} \label{pgiu}
P_{\bf k}({\bf q}) & = & |a_{\bf k}({\bf q})|^{2} \\
\hat{P}_{\bf k}({\bf x}) & = & \frac{1}{(2\pi)^{3}} \int_{{\bf
R}^{3}} d^{3}q\, e^{\displaystyle -i {\bf q x}} P_{\bf k}({\bf q})
\end{eqnarray}
are the probability distributions of momentum transfer and its Fourier
transform,
respectively,it follows that, in coordinate representation,
\begin{eqnarray} \label{rfsw}
\langle {\bf x}| \rho_{S}(t_{f}) | {\bf y} \rangle & = &
\hat{P}({\bf x} - {\bf y})\,
\langle {\bf x}| \rho_{S}(t_{i}) | {\bf y} \rangle \\
\hat{P}({\bf x} - {\bf y}) & = & \int_{{\bf R}^{3}} d^{3}k\,
\mu({\bf k})\, \hat{P}_{\bf k}({\bf x} - {\bf y}).
\end{eqnarray}

Taking into account  assumption 3 above and denoting by
$T_{\makebox{\tiny Sca}}[\rho]$ the $\rho_{S}(t_{f})$ in
(\ref{rfsw}), in the time interval $[t, t + dt]$ a scattering
induced process occurs with probability $\Lambda dt$. Consequently
the unitary Schr\"odinger evolution becomes a master equation
very much similar to (\ref{meqmsl}):
\begin{equation} \label{ftjn}
\rho(t + \delta t) \quad = \quad -\frac{i}{\hbar}\, [H, \rho(t)]
\delta t + (1 - \Lambda \delta t) \rho(t) + \Lambda \delta t
T_{\makebox{\tiny Sca}}[\rho(t)].
\end{equation}

A comparison of $T_{\makebox{\tiny Sca}}[\rho]$ and
$T_{\makebox{\tiny GRW}}[\rho]$ (i.e. the reduction operator given
by equation (\ref{meqmsl})) is possible by looking at the
expansions of the Gaussian damping factor that appears in
(\ref{maup}) and the factor $\hat{P}({\bf x})$, which plays an
analogous role in (\ref{rfsw}) ($|\hat{P}({\bf x})| \leq 1$), in
powers of ${\bf x} = (x_{1}, x_{2}, x_{3})$:
\begin{eqnarray}
\langle {\bf x}|T_{\makebox{\tiny GRW}}[\rho]|{\bf y} \rangle &
\simeq & \left[1 - \frac{\alpha}{4}\, |{\bf x} - {\bf y}|^{2} +
\ldots
\right]\, \langle {\bf x}|\rho|{\bf y} \rangle, \\
\langle {\bf x}|T_{\makebox{\tiny Sca}}[\rho]|{\bf y} \rangle &
\simeq & \left[1 - i\sum_{j=1}^{3}(x_{j} - y_{j}) \int_{{\bf
R}^{3}}
d^{3}q\, q_{j}\, P({\bf q}) \; - \right. \label{mmqe} \\
& & \left. -\frac{1}{2}\sum_{j,k=1}^{3}(x_{j} - y_{j})(x_{k} -
y_{k}) \int_{{\bf R}^{3}} d^{3}q\, q_{j}q_{k}\, P({\bf q}) +
\ldots \right] \langle {\bf x}|\rho|{\bf y} \rangle \nonumber
\end{eqnarray}

According to assumption 4, the linear term in (\ref{mmqe}) vanishes
and the quadratic one (covariance matrix) is completely determined
by
\begin{equation}
l^{-2}_{\makebox{\tiny eff}} \; = \; \int_{{\bf R}^{3}} d^{3}q\,
q_{i}^{2}\, P({\bf q}) \; = \; \left. \frac{\partial}{\partial
x_{i}^{2}} \hat{P}({\bf x}) \right|_{{\bf x} = {\bf 0}} \qquad i =
1,2,3,
\end{equation}
which has dimension cm${}^{-2}$ and defines a characteristic
length $l_{\makebox{\tiny eff}}$ of the reduction processes which has  to
be compared with $(\alpha/2)^{-1/2}$ of QMSL. Moreover, a natural
time scale is given by $\tau = \Lambda^{-1}$. Solving (\ref{ftjn})
for short times $\Lambda \delta t \gg 1$, yields:
\begin{equation}
\langle {\bf x}| \rho(t + \delta t) |{\bf y} \rangle \quad \simeq
\quad e^{\displaystyle -\Lambda \delta t (1 - \hat{P}({\bf x} -
{\bf y}))}\langle {\bf x}| \rho(t) |{\bf y} \rangle.
\end{equation}
Off the diagonal ($|{\bf x} - {\bf y}|l^{-1}_{\makebox{\tiny eff}}
\gg 1$) the damping is dominated by $e^{-\Lambda \delta t}$,
whereas near the diagonal ($|{\bf x} - {\bf
y}|l^{-1}_{\makebox{\tiny eff}} \ll 1$) the damping goes as
$e^{-\Lambda \delta t |{\bf x} - {\bf y}|^{2} / 2
l^{2}_{\makebox{\tiny eff}}}$. We can then introduce:
\begin{eqnarray} \label{ctqmsl}
\makebox{\bf Decoherence time:} & & \tau  \; = \; \Lambda^{-1} \\
\makebox{\bf Decoherence rate:} & & \Delta \; = \;
\frac{\Lambda}{l^{2}_{\makebox{\tiny eff}}}.
\end{eqnarray}

The decoherence time $\tau$ is fixed by the total scattering cross
section $\sigma$ and the flux of incident particles per unit area
and unit time, while the decoherence rate $\Delta$ by the
differential cross section that enters the expression of $P_{\bf
k}({\bf q})$ in (\ref{pgiu}). Tegmark also calculates the values
of $l_{\makebox{\tiny eff}}$ and $\tau$ for a microsystem
(electron) in different physical backgrounds which we report in Table 1.
\\ \\ \begin{center}
\begin{tabular}{||l|lll||} \hline
Cause of collapse & $l_{\makebox{\tiny eff}}$ [cm] & $\Phi$
[cm${}^{2}$ sec${}^{-1}$] & $\tau_{\makebox{\tiny electron}}$
[sec]
\\ \hline
300K air at 1 atm  & $10^{-9}$ & $10^{24}$ & $10^{-13}$ \\
300K air in lab vacuum  & $10^{-9}$ & $10^{11}$ & $1$ \\
Sunlight on earth  & $9\times 10^{-5}$ & $10^{17}$ & $10^{7}$ \\
300K photons  & $2\times 10^{-3}$ & $10^{19}$ & $10^{5}$ \\
Background radioactivity  & $10^{-12}$ & $10^{-4}$ & $10^{18}$ \\
Quantum gravity  & $10^{5}$ -- $10^{12}$ & $10^{109}$ & $30$ \\
\hline
GRW effect  & $10^{-5}$ &   & $10^{16}$ \\
\hline Cosmic microwave background  & $2\times 10^{-1}$ &
$10^{13}$ &
$10^{11}$ \\
Solar neutrinos & $10^{-9}$ & $10^{11}$ & $10^{33}$ \\
Comic background neutrinos  & $3\times 10^{-1}$ & $10^{13}$ &
$10^{51}$ \\ \hline
\end{tabular}
\\ \vspace{.3cm}
{\small {\bf Table} 1: $l_{\makebox{\tiny eff}}$ for various
scattering processes and $\tau_{\makebox{\tiny electron}}$.}
\end{center}

The absence of a second figure in the row describing the GRW situation
emphasizes the deep conceptual difference between collapse and
decoherence: there is no particle flux scattered off  the system inducing
the localization mechanism on the length scale $l_{\makebox{\tiny
eff}} = (\alpha/2)^{-1/2}$, the effect being due to a completely
new dynamics and not to the environment. Analogously, Tegmark
gives estimates of $\Delta$ for various objects in different
backgrounds.
\\ \\ \begin{center}
\begin{tabular}{||l|lll||} \hline
Cause of apparent & Free & Dust & Bowling \\
wavefunction collapse & electron & particle & ball \\ \hline
300K air at 1 atm  & $10^{31}$ & $10^{37}$ & $10^{45}$ \\
300K air in lab vacuum  & $10^{18}$ & $10^{23}$ & $10^{31}$ \\
Sunlight on earth  & $10$ & $10^{20}$ & $10^{28}$ \\
300K photons  & $1$ & $10^{19}$ & $10^{27}$ \\
Background radioactivity  & $10^{-4}$ & $10^{15}$ & $10^{23}$ \\
Quantum gravity  & $10^{-25}$ & $10^{10}$ & $10^{22}$ \\
\hline
GRW effect  & $10^{-6}$ & $10^{9}$ & $10^{21}$ \\
\hline Cosmic microwave background  & $10^{-10}$ & $10^{6}$ &
$10^{17}$ \\
Solar neutrinos & $10^{-15}$ & $10$ & $10^{13}$ \\
\hline
\end{tabular}
\\ \vspace{.3cm}
{\small {\bf Table} 2: $\Delta$ in cm${}^{-2}$sec${}^{-1}$ for
various scattering processes.}
\end{center}

As is evident from table 2, the GRW effect, e.g. for a free
electron, is weaker than those of air molecules in lab vacuum or
photons on earth by a factor in between $10^{26}$ and $10^{6}$,
respectively, hence masked by them.
There follows that, to put into evidence effects due to spontaneous
localization mechanisms one should isolate the physical system of
interest from the environment to a presently hardly attainable
degree of accuracy.
However, the figures in table 2 might lead to erroneous
conclusions, if not correctly understood. In fact, e.g. for bound
electrons, QMSL can, as we shall discuss later, induce, as a
result of the localization mechanism, transitions (which are not
considered in the preceding analysis of environment induced
decoherence) leading to excitations or dissociations of the
composed systems to which the electrons belong.

As an example, in a recent paper \cite{prc}, it is argued that the
Lyman--$\alpha$ ultraviolet radiation emitted by hydrogen atoms
(about 1 -- 10 photons per second per mole) as a consequence of
the spontaneous localizations suffered by electrons could be
detected by an appropriate experimental setup. On the other hand,
on the basis of energy balance considerations, it can be shown
that an analogous effect due to collisions of an atom with 300
Kelvin air molecules at 1 atmosphere is, in
comparison, much smaller .

\subsection{Dynamical reduction and the quantum telegraph}
\label{sec152}

In this subsection we do not examine a suggested experimental test
for the dynamical reduction program, but, rather, some recent
claims that there exists some already available empirical evidence
that spontaneous collapses, but of a nature
different from either QMSL or CS, are necessary. We quote directly from
A. Shimony \cite{shi90}:
\begin{quotation}
A great weakness of the investigations carried out so far in
search of modifications of quantum dynamics is the absence of
empirical heuristic. To be sure there is a grand body of empirical
fact which motivates all the advocates of nonlinear modifications:
that is, the occurrence of definite events, and, in particular,
the achievement of definite outcomes of measurement. But this body
of fact is singularly unsuggestive of the details of a reasonable
modification of Quantum Mechanics. What is needed are phenomena
which are suggesting and revelatory ...

No more promising phenomena for this purpose have been found than
the intermittency of resonant fluorescence of a three--level atom.
\end{quotation}

\subsubsection{The phenomenology of the quantum telegraph}
\label{sec1521}

The physical system consists of two laser beams of intensities
$I_{1}$, $I_{2}$ scattered off a single trapped atomic system
which can be treated as a three--level system with a ground state
$|0\rangle$ and two excited  states: a higher level $|1\rangle$
and a metastable lower level $|2\rangle$, with mean lives
\begin{equation}
\beta^{-1}_{1} \; \simeq \; 10^{-8} \, \makebox{sec} \;\; \ll \;\;
\beta^{-1}_{2} \; \simeq \; 1 \, \makebox{sec}.
\end{equation}
\begin{center}
\begin{picture}(240,150)(0,0)
\put(0,0){\line(1,0){240}} \put(0,150){\line(1,0){240}}
\put(0,0){\line(0,1){150}} \put(240,0){\line(0,1){150}}
\thicklines \put(25,110){\line(1,0){70}}
\put(145,70){\line(1,0){70}} \put(85,30){\line(1,0){70}}
\thinlines
\put(50,115){\small $|1\rangle,\, \beta_{1}$} \put(170,75){\small
$|2\rangle,\, \beta_{2}$} \put(115,18){\small $|0\rangle$}
\put(105,40){\vector(-2,3){40}} \put(105,40){\vector(2,-3){1}}
\put(135,40){\vector(3,2){35}} \put(135,40){\vector(-3,-2){1}}
\end{picture}

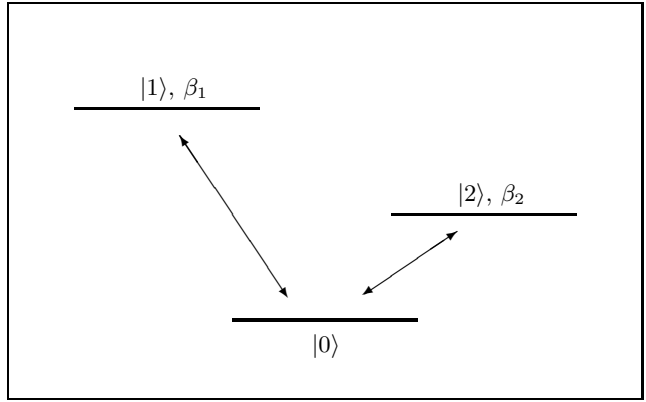
\vspace{0.2cm} \footnotesize \parbox{3.3in}{Figure 10: Atomic
system in the quantum telegraph phenomenon.} \normalsize
\end{center} \vspace{0.5cm}

The laser beams are tuned so that the one of intensity $I_{1}$
excites the atom from the ground state $|0\rangle$ to $|1\rangle$,
that of intensity $I_{2}$ provokes the transition $|0\rangle
\rightarrow |2\rangle$, followed by emissions of blue,
respectively, red photons and return to $|0\rangle$.

The emission pattern of an experiment conducted with $I_{1} \gg
I_{2}$, nearly $10^{8}$ vs. $10$ photons per second, reveals an
intermittent blue fluorescence randomly interrupted by periods of
darkness.
\begin{center}
\begin{picture}(240,120)(0,0)
\put(0,0){\line(1,0){240}} \put(0,120){\line(1,0){240}}
\put(0,0){\line(0,1){120}} \put(240,0){\line(0,1){120}}
\put(20,15){\vector(1,0){190}} \put(20,15){\vector(0,1){80}}
\put(215,13){\tiny $t$} \put(20,100){\tiny photo--detection
signal}
\put(20,35){\oval(5,40)[br]} \put(25,35){\oval(5,40)[t]}
\put(30,35){\oval(5,40)[b]} \put(35,35){\oval(5,40)[t]}
\put(40,35){\oval(5,40)[b]} \put(45,35){\oval(5,40)[t]}
\put(50,35){\oval(5,40)[b]} \put(55,35){\oval(5,40)[t]}
\put(60,35){\oval(5,40)[b]} \put(65,35){\oval(5,40)[t]}
\put(70,35){\oval(5,40)[bl]}
\put(100,35){\oval(5,40)[br]} \put(105,35){\oval(5,40)[t]}
\put(110,35){\oval(5,40)[b]} \put(115,35){\oval(5,40)[t]}
\put(120,35){\oval(5,40)[b]} \put(125,35){\oval(5,40)[t]}
\put(130,35){\oval(5,40)[b]} \put(135,35){\oval(5,40)[t]}
\put(140,35){\oval(5,40)[bl]}
\put(170,35){\oval(5,40)[br]} \put(175,35){\oval(5,40)[t]}
\put(180,35){\oval(5,40)[b]} \put(185,35){\oval(5,40)[t]}
\put(190,35){\oval(5,40)[bl]}
\put(90,75){\tiny periods of darkness}
\qbezier(95,70)(85,70)(85,50) \put(85,50){\vector(0,-1){1}}
\qbezier(145,70)(155,70)(155,50) \put(155,50){\vector(0,-1){1}}
\end{picture}

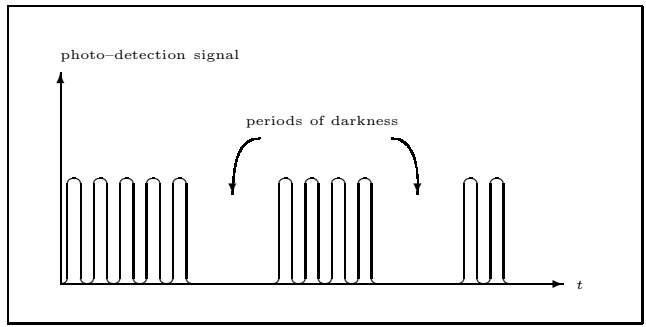
\vspace{0.2cm} \footnotesize \parbox{3.3in}{Figure 11: Emission
pattern in the quantum telegraph phenomenon.} \normalsize
\end{center} \vspace{0.5cm}

\subsubsection{Quantum mechanical interpretations} \label{sec1522}

At  first glance, the experimental evidence seems to be
explainable by a naive argument based on the concepts of photons
and of transitions among energy levels.

{\bf A.} Because of the higher intensity of the beam $I_{1}$, the
atom is most of the time excited to the short--lived level
$|1\rangle$ from which it jumps down to the ground state in
approximately $10^{-8}$ seconds with the emission a blue photon.
But, every now and then, a red photon from the beam $I_{2}$ sneaks
in and the atom is excited to the metastable state $|2\rangle$,
where it gets shelved for approximately 1 second before emitting a
red photon and starting again a period of blue fluorescence.

However, these conclusions are far too classical (\`a la Bohr) and
underestimate a relevant quantum effect, namely the interference
between blue and red photons in the laser beams which is
propagated to the atom by the linearity of the quantum evolution
and results in the emergence of linear superpositions of the
atomic levels.

{\bf B.} After interacting with the laser beams, the atom,
initially in its ground state $|0\rangle$, evolves in $t$ seconds
into a new state $|\phi(t)\rangle$ which is the coherent
superposition of the three levels:
\begin{equation}
|0\rangle \; \longrightarrow \; |\phi(t)\rangle \; =
c_{0}(t)|0\rangle + c_{1}(t)|1\rangle + c_{2}(t)|2\rangle,
\end{equation}
the probability $P_{i}$ of a spontaneous emission corresponding to
the jump $|i\rangle \rightarrow |0\rangle$ being $P_{i} =
\beta_{i} |c_{i}(t)|^{2}$.

Since the amplitude $|c_{2}(t)|^{2} \ll |c_{1}(t)|^{2}$ for almost
all $t$, one expects an emission pattern consisting in continuous
fluorescence and, sometimes, the emission of a red photon, periods
of darkness resulting extremely unlikely.
\begin{center}
\begin{picture}(240,120)(0,0)
\put(0,0){\line(1,0){240}} \put(0,120){\line(1,0){240}}
\put(0,0){\line(0,1){120}} \put(240,0){\line(0,1){120}}
\put(20,15){\vector(1,0){195}} \put(20,15){\vector(0,1){80}}
\put(220,13){\tiny $t$} \put(20,100){\tiny photoemission}
\put(24,15){\line(0,1){40}} \put(27,15){\line(0,1){40}}
\put(30,15){\line(0,1){40}} \put(33,15){\line(0,1){40}}
\put(36,15){\line(0,1){40}} \put(39,15){\line(0,1){40}}
\put(42,15){\line(0,1){40}} \put(45,15){\line(0,1){40}}
\put(48,15){\line(0,1){40}} \put(51,15){\line(0,1){40}}
\thicklines \put(53.8,15){\line(0,1){40}}
\put(54,15){\line(0,1){40}} \put(54.2,15){\line(0,1){40}}
\thinlines \put(54,15){\line(0,1){40}} \put(57,15){\line(0,1){40}}
\put(60,15){\line(0,1){40}} \put(63,15){\line(0,1){40}}
\put(66,15){\line(0,1){40}} \put(69,15){\line(0,1){40}}
\put(72,15){\line(0,1){40}} \put(75,15){\line(0,1){40}}
\put(78,15){\line(0,1){40}} \put(81,15){\line(0,1){40}}
\put(84,15){\line(0,1){40}} \put(87,15){\line(0,1){40}}
\put(90,15){\line(0,1){40}} \put(93,15){\line(0,1){40}}
\put(96,15){\line(0,1){40}} \put(99,15){\line(0,1){40}}
\put(102,15){\line(0,1){40}} \put(105,15){\line(0,1){40}}
\put(108,15){\line(0,1){40}} \thicklines
\put(110.8,15){\line(0,1){40}} \put(111,15){\line(0,1){40}}
\put(111.2,15){\line(0,1){40}} \thinlines
\put(114,15){\line(0,1){40}} \put(117,15){\line(0,1){40}}
\put(120,15){\line(0,1){40}} \put(123,15){\line(0,1){40}}
\put(126,15){\line(0,1){40}} \put(129,15){\line(0,1){40}}
\put(132,15){\line(0,1){40}} \thicklines
\put(134.8,15){\line(0,1){40}} \put(135,15){\line(0,1){40}}
\put(135.2,15){\line(0,1){40}} \thinlines
\put(138,15){\line(0,1){40}} \put(141,15){\line(0,1){40}}
\put(144,15){\line(0,1){40}} \put(147,15){\line(0,1){40}}
\put(150,15){\line(0,1){40}} \put(153,15){\line(0,1){40}}
\put(156,15){\line(0,1){40}} \put(159,15){\line(0,1){40}}
\put(162,15){\line(0,1){40}} \put(165,15){\line(0,1){40}}
\put(168,15){\line(0,1){40}} \put(171,15){\line(0,1){40}}
\put(174,15){\line(0,1){40}} \put(177,15){\line(0,1){40}}
\put(180,15){\line(0,1){40}} \put(183,15){\line(0,1){40}}
\put(186,15){\line(0,1){40}} \put(189,15){\line(0,1){40}}
\put(192,15){\line(0,1){40}} \put(195,15){\line(0,1){40}}
\thicklines \put(197.8,15){\line(0,1){40}}
\put(198,15){\line(0,1){40}} \put(198.2,15){\line(0,1){40}}
\thinlines
\end{picture}

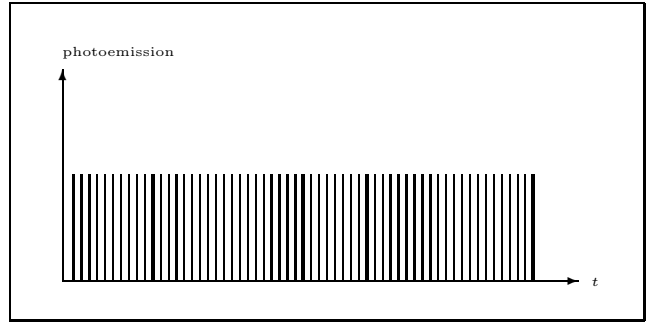
\vspace{0.2cm} \footnotesize \parbox{3.3in}{Figure 12: Emission
pattern according to argument B. Thick lines correspond to {\it
red} photons.} \normalsize
\end{center} \vspace{0.5cm}

The above attempt to embody linearity does not explain the
occurrence of intermittency in the emission pattern and in
reference \cite{mp} (compare also corresponding references in
\cite{shi90}) it is suggested that an explanation is only possible
if a reduction mechanism corresponding to null measurements
(seeing no photons) is introduced into the game.

In view of these facts A. Shimony concludes \cite{shi90}:
\begin{quotation}
Two propositions seem to me to suggest themselves quite strongly.
The first is that a stochastic modification of quantum dynamics is
a natural way to accommodate the jumps from a period of darkness
to a period of fluorescence. The second is that the natural locus
of the jumps is the interaction of a physical system with the
electromagnetic vacuum. Whether stochasticity is exhibited when
the system in question is simple and microscopic like a single
atom, or only when it is macroscopic and complex like the phosphor
of a photo--detector, is not suggested preferentially by the
quantum telegraph, for the simple reason that the single trapped
atom and the photo--detector are both essential ingredients in the
phenomenon ...

But, whichever choice is made points to a stochastic modification
of quantum dynamics which has little to do with spontaneous
localization.
\end{quotation}

\subsubsection{The correct quantum argument} \label{sec1523}

Concerning the alleged impossibility of explaining the
intermittent fluorescence of the quantum telegraph by resorting to
a dynamical reduction model with localization, it must be stressed
that

{\bf C.} The presence of periods of darkness in the emission
pattern can be deduced within a purely unitary quantum mechanical
scheme \cite{ctd}, by taking into account the whole system
\[
\makebox{ATOM} \; + \; \makebox{RADIATION FIELD}
\]
without any need of invoking reduction processes induced by
detecting no photons.

To be correct, the analysis  must consider states
$|\psi(t)\rangle$ of the form
\begin{equation}
|\psi(t)\rangle \quad = \quad \sum_{i=0}^{2} \sum_{\{ n \}} c_{i,
\{n\}}(t)\, |i\rangle \otimes |\{n\}\rangle,
\end{equation}
where $|\{n\}\rangle$ is a state with $n$ scattered photons. Then,
with
\begin{equation}
P \quad = \quad \left[ \sum_{i=0}^{2} |i\rangle\langle i|
\right]\otimes |\{0\}\rangle\langle\{0\}|
\end{equation}
the orthogonal projection onto the Fock sector with no scattered
photons, the probability $P(t)$ of periods of darkness extending
in the interval of time $[0, t]$ when, initially, $|\psi(0)\rangle
= |0\rangle\otimes |\{0\}\rangle$, is:
\begin{equation}
P(t) \quad = \quad \| P |\psi(t)\rangle \|^{2} \quad = \quad
\sum_{i=0}^{2} |c_{i, \{0\}}(t)|^{2}.
\end{equation}
The study of $P(t)$ leads to the correct prediction of periods of
darkness in a purely quantum dynamical context. Moreover, during a
period of darkness the state of the system ATOM + RADIATION FIELD
is
\begin{equation}
\left[ c_{0}(t) |0\rangle + c_{1}(t) |1\rangle + c_{2}(t)
|2\rangle \right] \otimes |\{0\}\rangle,
\end{equation}
so that periods of darkness can end with the emission of both red
and blue photons with an emission pattern like the one of the following
picture:
\begin{center}
\begin{picture}(240,120)(0,0)
\put(0,0){\line(1,0){240}} \put(0,120){\line(1,0){240}}
\put(0,0){\line(0,1){120}} \put(240,0){\line(0,1){120}}
\put(20,15){\vector(1,0){195}} \put(20,15){\vector(0,1){80}}
\put(220,13){\tiny $t$} \put(20,100){\tiny photoemission}
\put(24,15){\line(0,1){40}} \put(27,15){\line(0,1){40}}
\put(30,15){\line(0,1){40}} \put(33,15){\line(0,1){40}}
\put(36,15){\line(0,1){40}} \put(39,15){\line(0,1){40}}
\put(42,15){\line(0,1){40}} \put(45,15){\line(0,1){40}}
\put(48,15){\line(0,1){40}} \put(51,15){\line(0,1){40}}
\thicklines \put(74.8,15){\line(0,1){40}}
\put(75,15){\line(0,1){40}} \put(75.2,15){\line(0,1){40}}
\thinlines \put(78,15){\line(0,1){40}} \put(81,15){\line(0,1){40}}
\put(84,15){\line(0,1){40}} \put(87,15){\line(0,1){40}}
\put(90,15){\line(0,1){40}} \put(93,15){\line(0,1){40}}
\put(96,15){\line(0,1){40}} \put(99,15){\line(0,1){40}}
\put(102,15){\line(0,1){40}} \put(105,15){\line(0,1){40}}
\put(108,15){\line(0,1){40}} \put(111,15){\line(0,1){40}}
\put(114,15){\line(0,1){40}}
\thicklines \put(143.8,15){\line(0,1){40}}
\put(144,15){\line(0,1){40}} \put(144.2,15){\line(0,1){40}}
\thinlines \put(147,15){\line(0,1){40}}
\put(150,15){\line(0,1){40}} \put(153,15){\line(0,1){40}}
\put(156,15){\line(0,1){40}} \put(159,15){\line(0,1){40}}
\put(162,15){\line(0,1){40}}
\put(186,15){\line(0,1){40}} \put(189,15){\line(0,1){40}}
\put(192,15){\line(0,1){40}} \put(195,15){\line(0,1){40}}
\put(198,15){\line(0,1){40}} \put(201,15){\line(0,1){40}}
\put(204,15){\line(0,1){40}}
\end{picture}

\vspace{0.2cm} \footnotesize \parbox{3.3in}{Figure 13: Emission
pattern according to argument C.} \normalsize
\end{center} \vspace{0.5cm}

As a further remark, it must be noted that a complete account of
the quantum telegraph experiment ought to include the macroscopic
detectors that are involved in the measurement of the emission
pattern. Consequently, the physical system to be dealt with is
\[
\makebox{ATOM} \; + \; \makebox{RADIATION FIELD} \; + \;
\makebox{MACROSCOPIC DETECTORS}
\]

Within the dynamical reduction program the actualization of the
different macro\-sta\-tes of the detecting apparatuses is
accounted for by the new dynamics and the corresponding
objectification of macroproperties is thus obtained. One could
raise the question whether this can have any appreciable influence
on the quantum telegraph phenomenon.

It is sufficient to observe that, according to the analysis of the
previous subsection, the effects of the reduction mechanism are
comparable with those of environment--induced reductions that
occur at the detectors level.
Indeed, on the basis of the agreement of the correct quantum
mechanical computations of the probability of occurrence of
periods of darkness and of their duration with the experimental
results, one can safely conclude that:
\begin{enumerate}
\item QMSL and CSL dynamics play, for the process under
consideration, exactly the same role as for any macroscopic
detection process, namely they objectify macro--properties.
\item There is no need to require that new nonlinear and stochastic
modifications of standard quantum mechanics become effective at
the microscopic level to account for the quantum telegraph
phenomenology.
\item In particular, nothing, in the considered experiments, suggests
that reductions take place with respect to an ``energy" rather
than to a ``position" preferred basis.
\end{enumerate}

\subsection{Dynamical reduction and the nucleon decay} \label{sec153}

The presence in nature of a mechanism that localizes particles
would be accompanied by a corresponding spreading in their
momenta. It is thus interesting to study its effect on the
stability of atoms and nuclei. It is possible to get a rough idea
of the consequences of QMSL by modeling atoms and nuclei as one
dimensional systems moving in a harmonic potential so that their
ground states can be approximated by Gaussian wavefunctions
$\psi_{G}(q)$ of appropriate width $\gamma^{-1}$:
\begin{equation}
\psi_{G}(q) \; = \; \left[\frac{2\gamma^{2}}{\pi}\right]^{1/4}
e^{\displaystyle -\gamma^{2}q^{2}}, \qquad
\begin{array}{ll}
\gamma \, \simeq \, 10^{8} \makebox{cm}^{-1} & \makebox{for an
atom}, \\
\gamma \, \simeq \, 10^{12} \makebox{cm}^{-1} & \makebox{for a
nucleus}.
\end{array}
\end{equation}
If the particle undergoes a localization around $x$ as in as in
(\ref{rc}), $\psi_{G}(q)$ changes into:
\begin{equation}
\frac{\phi_{x}(q)}{\|\phi_{x}(q)\|}, \qquad \phi_{x}(q) \; = \;
\left[\frac{\alpha}{\pi}\right]^{1/4} e^{\displaystyle
-\frac{\alpha}{2}(x - q)^{2}} \psi_{G}(q).
\end{equation}
>From (\ref{pd}), the probability density that a localization takes
place at $x$ is given by $\|\phi_{x}\|^2$. Accordingly, the
probability that, if a hitting process occurs, the state of the
system is still $\psi_{G}(q)$ is given by:
\begin{equation}
P_{ND} \; = \; \int d^{3}x\, |\langle \psi_{G}|\phi_{x}
\rangle|^{2} \; = \; \frac{1}{\sqrt{1+ \alpha/4\gamma^{2}}} \;
\simeq \; \left\{
\begin{array}{ll}
1 - 10^{-7} & \makebox{for an atom}, \\
1 - 10^{-15} & \makebox{for a nucleus}.
\end{array} \right.
\end{equation}
Since microsystems are supposed to undergo one localization every
$10^{16}$ seconds, the transition rate $Q_{E+D}$ to an excited or
dissociated state is:
\begin{equation} \label{pdkt}
Q_{E+D} \; = \; \lambda (1 - P_{ND}) \; \simeq \; \frac{\lambda
\alpha}{8\gamma^{2}} \; = \; \left\{
\begin{array}{ll}
10^{-23} & \makebox{for an atom}, \\
10^{-31} & \makebox{for a nucleus}.
\end{array} \right.
\end{equation}
In \cite{prc} Pearle has considered the case of the hydrogen atom and
has compared the flux of Lyman--$\alpha$ ultraviolet photons emitted
by intergalactic hydrogen with the one expected if a GRW mechanism
were at work, the latter turning out to be much weaker than the
one observed.

Applying the same argument to the quark model of a proton one would
get a decay time of the same order of magnitude as the one of a
nucleus ($10^{31}$ sec), whereas the proton lifetime is estimated
longer than $10^{31}$ years, that is $10^{38}$ sec.
This fact seems to indicate that the reduction rate $\Delta$
should be decreased by a factor $10^{7}$.

However, the consequences of $\Delta \simeq 10^{-13}$ cm$^{-2}$
sec$^{-1}$ would be unacceptable. In fact, since $\alpha^{-1/2} =
10^{-5}$ cm is a reasonable value for the localization length
($l_{\makebox{\tiny eff}}$ in Table 1), it would yield the value
$\tau_{N} \simeq 10^{23}/N$ sec for the macroscopic decoherence
time in (\ref{ctqmsl}). Thus, linear superpositions of spatially
separated states of any reasonable macroscopic ``pointer" ($N
\simeq 10^{23}$) would typically take times of the order of the
second to be suppressed.

\subsubsection{Reconsidering the argument within the CSL approach}
\label{sec1531}

In \cite{pes} the authors have considered an initial bound state
$\rho_{B} = |\psi_{B}\rangle\langle\psi_{B}|$ which evolves into
$\rho(t)$ according to the dynamics (\ref{cslso}) of CSL and have
studied the transition probability $P(t)$ to an excited orthogonal
state $|\Psi_{E}\rangle$:
\begin{equation}
\left.\frac{d}{dt}\, P(t)\right|_{t = 0} \quad = \quad \left.
\frac{\partial}{\partial_{t}} \langle \psi_{E} | \rho(t)
|\psi_{E}\rangle \right|_{t = 0}.
\end{equation}

The only contribution is that from the reducing term in
(\ref{cslso}). By developing up to the first order in $\alpha$ one gets:
\begin{equation} \label{davs}
\left. \frac{d}{dt}\, P(t) \right|_{t = 0} \; \simeq \; \sum_{j}
\frac{\alpha\lambda N_{j}^{2}}{2}\, |\langle \psi_{E} | {\bf
Q_{j}} | \psi_{B} \rangle |^{2}, \qquad {\bf Q}_{j} \;  = \;
\frac{1}{N_{j}} \sum_{i = 1}^{N_{j}} {\bf q}_{ij},
\end{equation}
$j$ numbering the species of identical particles making up the
system, ${\bf q}_{ij}$ being the position operators of the  particles of
type $j$ and ${\bf Q}_{j}$ their center of mass.

For just one nucleon the result agrees with that of QMSL, while,
for macro--systems, due to the more efficient decoupling rate, it
appears that a correction of $\Delta = \alpha\lambda/2$ which
would lead to no conflict with the proton lifetime is possible.
However, such a change of the value of $\Delta$ would lead  to
the limit of acceptability for small objects: the dynamics will
not reduce within the perception time an object containing
$10^{15}$ particles like a carbon particle of radius $10^{-3}$ cm.
Similarly, the argument of \cite{abp} discussed in section
\ref{sec014} about the perception mechanism, would no longer be
correct.

\subsubsection{The Pearle and Squires argument} \label{sec1532}

Considering a macroscopic body of total mass $M$ made up of
different types of identical particles of mass $m_{j}$, one may
think of relating the reduction process to the mass density
operator
\begin{equation}
M({\bf x}) \quad = \quad \sum_{k} m_{k}
\left[\frac{\alpha}{2\pi}\right]^{3/2} \int_{{\bf R}^{3}} d^{3}y\,
e^{\displaystyle -\frac{\alpha}{2} ({\bf y} - {\bf x})^{2}}
a^{\dagger}_{k}({\bf y})\, a_{k}({\bf y}),
\end{equation}
(see the discussion of section \ref{sec76}) rather than
considering independent stochastic processes for the various
kinds of particles.

The first order analysis that has led to (\ref{davs}) can be
similarly carried out in this case;  one merely  has to consider
the total center  of mass ${\bf Q}$ of the system and not only the
centers of mass ${\bf Q}_{j}$ of the various species
\begin{equation}
{\bf Q} \;  = \; \frac{1}{M} \sum_{j} m_{j} \sum_{i = 1}^{N_{j}}
{\bf q}_{ij}.
\end{equation}
\noindent as a consequence the rate of internal excitation and/or
dissociation appears as a second order effect. Indeed, the total center
of mass ${\bf Q}$ cannot excite any internal degree of freedom:
\begin{equation}
\left. \frac{d}{dt}\, P(t) \right|_{t = 0} \; \simeq \; \sum_{j}
\frac{M^{2}\alpha\lambda}{2m_{0}^{2}}\, |\langle \psi_{E} | {\bf
Q_{j}} | \psi_{B} \rangle |^{2} \; = \; 0
\end{equation}

If one takes the QMSL value of $\lambda$ for nucleons, that is if the
reference mass $m_{0}$ is identified with the nucleon mass, one
has a remarkable decrease of the QMSL rate $Q_{D+E}$ in
(\ref{pdkt}) for atoms: from one atom per mole being either
excited or dissociated every second to  every $10^{12}$ seconds ($10^{5}$
years). Nevertheless, all the important features of QMSL are preserved:
\begin{itemize}
\item The decoupling of macroscopically distinguishable states is
taken care of by the nucleons of the macroscopic bodies.
\item The energy increase is almost identical to  acceptable
one which is implied by standard QMSL.
\item The collapse induced decay probability of a proton is
depressed by a factor $10^{-16}$ making the CSL predicted lifetime
well compatible with the experimental data.
\end{itemize}

Some concluding comments are in order at this point: the above
analysis has appropriately pointed out the nice features deriving
from relating reduction to the mass of the elementary particles.
However one cannot avoid mentioning that:
\begin{enumerate}
\item The dynamical reduction program has made clear that one can
try to follow a new line to solve the conceptual problems that
Quantum Mechanics meets with macro--objects and measurement
processes, namely, modifying the dynamics so that the physics of
microsystems remains unaltered, while macro--systems exhibit an
acceptable behaviour. However one cannot forget that for the time
being the program still requires many improvements, in particular
the crucial problem to be faced is to work out reasonable
relativistic generalizations of it. Being the quark dynamics
fundamentally relativistic, and due to the great difficulties
haunting the so called ``non--relativistic quark models",
applying directly the specific models of QMSL and CSL to nucleons
seems a little bit too hazardous. \item Other collapse theories
are still under consideration. The model presented in section
\ref{sec8} tries to relate the decoherence mechanism to gravity
and to reduce the number of new fundamental constants
characterizing CSL. In particular, the reduction mechanism is
linked to the mass of the systems involved, already meeting the
basic request of \cite{pes}. \item The difficulties connected with
nucleon decay might also be avoided by slight modifications of the
standard CSL, for instance, by using a higher power $N^{4/3}({\bf
x})$ of the smeared number operator appearing in (\ref{ando}).
\end{enumerate}

\subsection[Superconducting devices]{Spontaneous localizations
in superconducting devices} \label{sec154}

We conclude the analysis of the experimental implications of
dynamical reduction models mentioning the work of Rae \cite{rae},
of Rimini \cite{rimg} and of Buffa, Nicrosini and Rimini
\cite{bnr} on the effects of spontaneous localizations on
superconducting devices. The argument of Rae \cite{rae} goes as
follows: consider the BCS wavefunction \cite{bcs} of a
superconducting state
\begin{equation}
\psi \; = \; \psi_{\bf k_{1}}\, \psi_{\bf k_{2}}\, \ldots
\psi_{\bf k_{n}}\, e^{iS({\bf x})},
\end{equation}
where $\psi_{\bf k_{i}}$ represents the wavefunction of a Cooper
pair of electrons with wavevectors $+ {\bf k_{i}}$ and $- {\bf
k_{i}}$, and $S({\bf x})$ is the macroscopic phase associated with
the supercurrent. The most relevant effect of spontaneous
reductions of an electron in a superconducting device is to break
one of the Cooper pairs, which would result in the supercurrent
being reduced by about one part in $10^{20}$. Assuming that a
reduction happens every $10^{-5}$ sec, the resulting decay would
remain well below the experimental detection limits which are of
the order of one part in $10^{13}$ per second.

A more realistic model in which the possibility of re--creation of
Cooper particles is taken into account (a phenomenon which lowers
the effects of spontaneous localizations) shows that dynamical
reduction models are even more compatible with the existence of
superconductivity, something which is not at all trivial.

A much more detailed and mathematical precise analysis of the
effect of spontaneous localizations on superconducting devices has
been performed in refs. \cite{rimg,bnr}, within the framework of
CSL: the conclusion is that, by taking into account the
indistinguishability of electrons, the effects are even smaller
than those predicted by Rae. We refer the reader to the above
cited papers for the complete analysis of the problem.

\section{Conclusions}

In this paper we have analyzed in detail a quite radical proposal
which, at the non relativistic level, allows one to circumvent the
conceptual difficulties that standard quantum mechanics meets with
the macro--objectification problem. Obviously, even though the
theory is not a reinterpretation (as many of the attempts we have
discussed in part I) of the standard theory, but qualifies itself
as a rival theory of it, up to the moment in which technological
improvements will make experimental tests actually feasible, to
accept it is, to a large extent, a matter of taste and of the
attitude one has with respect to the foundational problems of
quantum theory. The theory we have reviewed in this report in its
present formulation still has  a phenomenological character and
necessitates further hard work before it can be taken as a
fundamental theory of natural processes. As we have already
remarked,  the real (and, we believe, relevant) merits which
characterize it  derive from the fact that it represents a
conceptually new proposal for overcoming the embarrassing
questions raised by the macro--objectification problem. However,
finding a consistent relativistic generalization of the dynamical
reduction theories, remains, as Bell has stressed, {\it the big
problem} to be faced.

Having made clear the perspective we consider appropriate for  the
dynamical reduction program and its limitations,  it seems useful to
conclude this report by  recalling the nice features which
characterize it.

First of all, according to the theory all natural processes,  the
microscopic and the macroscopic ones, as well as those involving
interactions between micro and macro systems, are governed by the
universal evolution equation of the theory. Such an equation has
never to be disregarded, contrary to what happens for
Schr\"odinger's evolution  of the standard scheme, on the basis of
supplementary, imprecise, verbal prescriptions. All the
embarrassing ambiguities of the standard theory concerning macro
processes are only momentary in the new scheme. Again, in Bell's
words \cite{bells},  within the GRW theory {\it the cat is not
both dead and alive for more that a split second}.

Another feature of the theory which deserves to be stressed is its
structural difference from the (in our opinion) unique other
consistent and fully worked out proposal to solve the measurement
problem, i.e. Bohmian mechanics. The GRW theory is a genuine
Hilbert space theory and does not add any kind of variables to
standard quantum mechanics. However, by introducing mathematically
precise modifications to it, it allows one  to answer in an
unambiguous way to all the fundamental questions which characterize
the debate on quantum mechanics since its birth: which systems and
processes must be treated as classical and which ones as quantum,
which are essentially reversible and for which ones
irreversibility plays a fundamental role, and so on. Moreover the
definite mathematical description of reductions makes also precise
the action at a distance of ordinary quantum mechanics, throwing a
new light on EPR--like situations and on quantum non locality. The
nice features of the proposal we have reviewed have been
summarized by Bell in  a very concise sentence \cite{bells} {\it
for myself, I see the GRW model as a very nice illustration of how
quantum mechanics, to become rational, requires only a change
which is very small (on some measures!)}.

Coming to the relativistic aspect we recall that the theory, even
though no consistent relativistic generalization of it has  been
fully worked out, presents some nice aspects which, once more, can
be taken as interesting hints for the elaboration of a
relativistic theory inducing reductions, an old problem which, as
we have discussed in this report, has drawn a lot of attention. In
this respect, it is useful to stress the different conceptual
status of the dynamical reduction theories with respect, e.g.
hidden variable theories. We are making reference here to the fact
that the locality requirement can be split in the two conditions
of parameter and outcome independence and that  the linear version of
dynamical reduction theories  exhibits only outcome dependence, a fact
that conflicts less than parameter dependence with a relativistic point
of view.  Actually, what J.S.  Bell has has proved in ref.\cite{bells} is
equivalent to checking that the GRW theory does not present parameter
dependence. This analysis led him to state: {\it the model is as Lorentz
invariant as it could be in the non relativistic version. It takes away
the grounds of my fear that any exact formulation of quantum mechanics
must conflict with fundamental Lorentz invariance}. Finally, we would
like to conclude by stressing that the natural interpretation of the
theories we have reviewed implies that they do not deal, as does  the
standard theory, with the probabilities of something occurring provided
some specific action (a measurement) is performed by a conscious
observer, i.e, with {\it what we would find},  but they speak directly of
{\it what  is}  (i.e. an objective mass distribution), at the appropriate
macroscopic level.

Concerning the philosophical implications of these approaches, if
one is interested also in these aspects of scientific knowledge,
it has to be remarked that they allow one to {\it close the circle}
in the precise sense of Shimony \cite{shipcqf}, i.e., to build up
a coherent worldview which can accommodate  at the same time what
we know about the peculiar behaviour of microscopic systems and
the definiteness of the macroscopic world and of our perceptions
about it. In particular, the theory makes fully legitimate a
macro--realistic position about nature and has no need whatsoever
to attribute any peculiar role to conscious observers, an
unavoidable fact within the standard formalism and most of the
proposed interpretations aiming to solve its conceptual
difficulties.

\section*{Acknowledgements}

As the reader would have certainly grasped, the cooperation of
many researchers has been necessary for the elaboration and the
full exploitation of the ideas which are the subject of this
review. In particular one of us (G.C.G.) wants to express his deep
gratitute to his collegue  A. Rimini, the collaboration with whom
has been so fundamental for the working out of the dynamical
reduction program. G.C.G. wants to express similar feelings to his
colleagues T. Weber, P. Pearle and to his student R. Grassi who
has been an essential collaborator in furthering the program, in
particular for the elaboration of its interpretation and for the
attempts at a relativistic generalization. Finally he is deeply
indebted to J.S. Bell, whose interest in the dynamical reduction
program, whose stimulating remarks and whose friendship has meant
much more for him than words can say. Finally, both the authors
are grateful to Prof. S. Goldstein who has made an incredible job
acting as a referee for this report and has contributed in a
fundamental way in improving it with his many deep and stimulating
remarks.

\end{document}